\documentclass[11pt,a4,twoside]{article}
\pdfoutput=1
\usepackage[dvips,a4paper,left=2.49cm,right=2.49cm,top=2.50cm,bottom=2.50cm,headsep=1em]{geometry}
\usepackage{fancyhdr}
\usepackage{titlesec}
\usepackage[latin1]{inputenc}
\usepackage{amsmath,amsfonts,amssymb,amstext,amsthm}
\usepackage{enumitem}
\usepackage{graphicx}
\usepackage{float}
\usepackage{subcaption}
\usepackage{rotating}
\usepackage[font={small}]{caption}
\usepackage{natbib}
\usepackage{lmodern,slantsc}
\usepackage{relsize}
\usepackage[breaklinks, colorlinks=true, linkcolor=blue,
citecolor=black, urlcolor=blue, pdfborder={0 0 0},
pdfpagelabels]{hyperref}
\usepackage{tikz}
\DeclareMathAlphabet{\mathsfbf}{OT1}{cmss}{bx}{n}
\DeclareMathAlphabet{\mathmibf}{OT1}{cmr}{bx}{it}
\DeclareMathAlphabet\mathbfcal{OMS}{cmsy}{b}{n}

\def\vc{\mathbf c}

\def\vk{\mathbf k}

\def\vM{\mathbf M}

\def\v0{\boldsymbol{0}}

\newcommand{\hyf}{\!\!\phantom{.}_1\hspace{-0.25mm}F_1}
\newlength{\FigureHeight}
\newlength{\FigureHeightHalf}
\newcommand{\FigureXYLabel}[5]
{\settoheight{\FigureHeight}{#1}
\setlength{\FigureHeightHalf}{0.5\FigureHeight}
\begin{center}
\raisebox{\FigureHeightHalf}{\makebox{#4\makebox[#5]{}}}
#1\\
\vspace{#3}
#2\\
\end{center}}
\numberwithin{equation}{section}
\titleformat{\section}
{\large\bfseries}{\thetitle.}{0.5em}{}
\titleformat{\subsection}
{\normalfont\itshape\filcenter}{\normalfont\thetitle.}{0.5em}{}
\titleformat{\subsubsection}
{\normalfont\itshape}{\normalfont\thetitle.}{0.5em}{}
\begin{document}

\title{On physically redundant and irrelevant features\\ when applying Lie-group symmetry analysis\\ to hydrodynamic stability analysis}
\author{Michael Frewer$\,$\thanks{Email address for correspondence:
frewer.science@gmail.com}\\
Tr\"ubnerstr.$\,$42, 69121 Heidelberg, Germany\\}
\date{{\small\today}}
\clearpage \maketitle \thispagestyle{empty}

\vspace{0em}
\begin{abstract}
\noindent Every linear system of partial differential equations
(PDEs) admits a scaling symmetry in~its dependent variables. In
conjunction with other admitted symmetries that only exhibit a linear dependence of
the dependent variables in their infinitesimals, the associated differential condition to
generate invariant solutions poses a linear eigenvalue
problem. If the combined operator of these appendant symmetries is further structured such
that the spectral theorem\linebreak[4] applies, then the general solution of the
considered linear PDE system is obtained by summing or integrating
the invariant eigenfunctions (modes) over all eigenvalues,
depending on whether the spectrum of the operator is discrete or
continuous respectively. By first studying the one-dimensional
(1-D) diffusion equation as a demonstrating example, this method is then applied
to a relevant 2-D problem from hydrodynamic stability
analysis. The aim of this study is to draw attention to the
following two independent facts that need to be addressed in future studies when constructing
particular or general solutions
for linear dynamical PDEs with the method of Lie-symmetries: {\it (i)}~Although each new
symmetry leads~to a mathematically different spectral decomposition, they may all be physically
redundant\linebreak to a common set of symmetries that either induces a purely temporal or a purely spatial decomposition,
and thus, besides these common ones, do not reveal a new physical mechanism behind the overall considered dynamical process,
as incorrectly asserted, for example, in the recent studies by the group of Oberlack {\it et al.}
Hence, with regard to linear stability analysis, no physically ``new" or ``more general" modes are generated by this method
than the ones already established. {\it (ii)}~Next to the eigenvalue parameters, which act as
complementary coordinates to the ones in physical space,
each single mode can also acquire non-system parameters,
so-called regularization parameters, which can be picked up during
the construction process depending on the choice of its underlying symmetry.
As will be demonstrated,
these symmetry-induced parameters are all irrelevant in physical space, since their effect on
a single mode will cancel when considering all modes collectively. In particular, the collective
action of all single modes is identical for all symmetry-based decompositions
and thus indistinguishable when considering the full physical fields.

\vspace{0.5em}\noindent{\footnotesize{\bf Keywords:} {\it Partial
Differential Equations, Lie-Group Theory, Lie Algebras, Invariant
Solutions, Equivalent Systems,
Linear Eigenvalue Problems, Diffusion Equation, Acoustics, Stability Analysis}}
\end{abstract}

\newpage
\thispagestyle{empty}
\tableofcontents
\newpage

\pagenumbering{arabic}\setcounter{page}{1}

\section{Preliminary information\label{S1}}

To grasp the essence of this study, it is necessary to distinguish between two types of invertible transformations:
\vspace{-0.25em}
\begin{itemize}
\itemsep0em
\item[--] {\it Change of reference frame}, by transforming the coordinates (independent variables) of the considered system,
\vspace{-0.15em}
\item[--] {\it Change of representation}, by transforming the field solutions (dependent variables) of the considered system.
\end{itemize}
\vspace{-0.25em}
The former one, the change of reference frame, is a transformation only in the coordinates, dictating then the transformation for the
fields according to the rules of tensor analysis. In other words, the field components of the system in this case do not transform independently to its coordinates; they either
transform covariantly, contravariantly or invariantly, depending on whether the field is vectorially decomposed by a contravariant or covariant basis, or whether the field is
a scalar, respectively (see e.g. \cite{Schroedinger50}). A change of reference frame changes our perception of the evolving dynamics in a physical system, since all perceived motion is relative.
Typical and well-known examples are the Galilei transformation, connecting inertial reference frames, or the transformation that switches to a rotating frame, which, as an accelerated frame, gives
rise to additional (fictitious) forces which do not exist in the inertial frame,~etc.
When solving the systems' equations to yield the dynamics, the aim is to find the optimal frame, either for reasons that can be of mathematical nature, namely to find
a frame where the initial and boundary conditions along with the resulting solution reduces to a minimum in complexity, or for pure physical reasons, for example, to find a frame where external forces
can be transformed away. Such an optimal frame is standardly realized by co-moving with the systems' center of mass or with the mean bulk motion perceived in the inertial frame, or by moving along the field lines of the external forces.
Typical examples are the co-moving rotating frame when simulating rotating flow, where the implementation of the boundary conditions and the resulting solutions are technically as well as
functionally more simple than those in the inertial frame, or the free-falling frame moving along the gravitational field lines, a frame where the gravitational force (at least locally) has been
transformed away to zero. This free-falling system (approximately induced by parabolic flights or on the international space station ISS) is optimal to study
the internal dynamics of physical, chemical and biological systems, without the distracting influence of the external or superimposed force of gravitation.

The second type of transformation mentioned above, is the change of representation. Here only the field variables are transformed, with the effect to give a different
picture of solution space. In contrast to the aforementioned coordinate transformation, a change in representation may give complementary information or a complementary understanding
of the systems' dynamics, in particular if the transformation is of non-local (integral) nature. Typical examples of such transforms, are the local log-transform which converts the harder operation
of multiplication and division into that of addition and subtraction, forming thus the basis of all logarithm tables used in computations prior to the advent of computers and calculators, or the
non-local Fourier or Laplace integral transform which converts differential operators into algebraic expressions, leading thus to a complementary space of solutions allowing for an alternative
insight into the systems' dynamics.

On the basis of the above it is clear that the first type of transformations cannot provide new or alternative information about the systems' dynamics, as the transformation of the second type can do.
When changing the frame of reference, only our perception of the dynamics changes, however, for accelerating frames,
sometimes in a highly non-intuitive way due to the appearance of fictitious forces acting on the system. Particularly for unbounded dynamics, if the solution is known in one coordinate frame, it is known in all coordinate frames, simply because
the solution in each frame is connected to the solution in every other frame by a bijective (one-to-\linebreak[4] one)\hfill mapping\hfill of\hfill the\hfill independent\hfill coordinates.\hfill Hence,\hfill no\hfill information\hfill is\hfill lost,\hfill but\hfill also\hfill no\hfill new

\newgeometry{left=2.49cm,right=2.49cm,top=2.50cm,bottom=1.975cm,headsep=1em}

\noindent information is gained when changing the frame of reference. The solution is mostly sought in the optimal frame, which aims to lay open only the relevant internal dynamics of the system, being ideally free of any external superimposed and thus
non-relevant motion.

Although for the second type of transformation no information is lost too, it nevertheless changes the solution space in itself, in that the dependent variables of physical space get transformed
into a new or complementary space, thus allowing to probe and to extract new or complementary information about the systems' dynamics. Yet it should be clear that a change in representation does not induce new physics.
When finding or constructing different representations for a specific physical system, one can only expect in each case a different perspective on the dynamical mechanisms and principles at play, but not new physics.
A prominent example is the modal and non-modal approach to linear stability analysis \citep{Schmid01,Schmid07}: Both representations, the modal (when also including the analysis of pseudo-spectra \citep{Trefethen05}) as well as the non-modal one, describe the same physics and give the same conclusions whether a system is asymptotically stable or not, or whether it can experience transient growth or not, however, in each case, represented in a mathematically different way that can be regarded as complementary. Hence, it is desirable to discover as many {\it non}-redundant representations as possible to gain full understanding and insight into the physical system being considered, as once remarked by Richard \cite{Feynman65}:~``...~every theoretical physicist who is any good knows six or seven different theoretical representations of exactly the same physics"~[p.~168].

In this study both types of transformation will be considered and applied concurrently, the change of reference frame as well as the change of representation, with the aim to reformulate a proposed change of representation in its optimal frame of reference.
This will demonstrate that the new or more general invariant solutions proposed in the recent publication by \cite{Hau17},
and in its precursor by \cite{Nold13}, do not induce
physically ``new" or physically more ``general" modes for unbounded linear shear flow as claimed, but only physically redundant ones, not going beyond the two different and already established representations
of the modal and non-modal (Kelvin mode)\label{171202:1451}\footnote[2]{In \cite{Hau17}, and for the first time in \cite{Chagelishvili94,Chagelishvili97},
the Kelvin modes are also coined as ``non-modal" to differentiate them from the modes of the usually applied temporal modal approach. Within a temporal framework to stability analysis (see e.g. \cite{Schmid01}), the usual (single) modal approach is based on a (complex valued) temporal mode that is separated from the spatial structure of the wavelike perturbation, which itself stays unchanged as time progresses. In the Kelvin approach no such temporal mode exists, hence the name ``non-modal". But this naming can be misleading, and is not to be confused with the proper non-modal approach as originally defined e.g. in \cite{Schmid01} and \cite{Schmid07}. Because, also the Kelvin approach is based on modes, namely on purely spatial modes, the so-called Kelvin modes, but which in the temporal framework are standardly forced to take arbitrarily but fixed (purely imaginary) values such that, as in the modal approach, the spatial structure of the wavelike perturbation does not grow or decay as time progresses. This disables the (generalized) eigenvalue problem for the spatial Kelvin modes and thus turns it into a (non-modal) initial value problem. However, in this specific Kelvin approach the spatial structure is only really time independent in the optimal frame of reference, that properly co-moves with the base flow, and {\it not} in the inertial frame.\label{Kelvin}} solution approaches.

Said differently, the aim in \cite{Hau17} and \cite{Nold13} is to discover and explore some alternative ways of describing the linearized dynamics of perturbed unbounded
shear flow that should go beyond the classical modal and Kelvin mode approach. But the new representations obtained therein, by employing different combinations of symmetries of the underlying dynamical equations, are,
not mathematically, but physically equivalent to the classical ones. The decisive reason is that in the respective optimal reference frame, the associated modes of these new representations
all reduce to the classical Laplace or Fourier modes of the modal and Kelvin mode approach, namely either to the temporal-spatial or to the purely spatial modes, respectively. The new representations are just mathematically more intricate than those two already established representations. Any new physical insight into the dynamical processes is not revealed, making them thus, in this sense, not mathematically but physically redundant.

Hence, reformulating a proposed invariant solution into its optimal frame of reference, will give a method at hand, to identify whether the
induced representation will lead to new or more general stability modes, or whether they turn out to be physically redundant. For unbounded linear shear flow, the result to perform an analysis on stability is final:
Next to the modes of the modal and Kelvin approach no other physical modes exist.

To close this section, it is important to note two things: (i) Although both terms carry the word ``optimal", the notion ``optimal system", as used in \cite{Hau17}, is not to be confused with
the notion ``optimal frame of reference" in this study. The former is a pure mathematical notion from group theory, in particular from Lie-group symmetry theory to ensure that
within the process of constructing group-invariant solutions only inequivalent functions are considered which cannot be connected by any of the group's symmetry transformations, while
the latter is a notion from the physical concept of relativity, that all perceived motion is relative, which makes it necessary to find that particular frame among all frames
where the complexity of motion is reduced to a minimum, to effectively clear the internal dynamics of the considered system from any superimposed and thus irrelevant motion. (ii) The concept of
finding and using an optimal frame of reference as defined herein, is only applicable if the dynamical rules governed by a set of equations are already explicitly known and need not to be modelled.
For modelling dynamical equations, the aim is different and not focussed on finding the optimal frame. Quite the contrary: Using a specific frame is even counterproductive for modelling, since one of the
key aims in the theory of modelling is to formulate the model equations form-invariantly and in some cases even frame-indifferently (see e.g. \cite{Frewer09.1,Frewer09.2,Frewer09.3,Frewer16}
for more details on these notions). In other words,
one of the key properties of modelled equations is not to change their form when changing the frame of reference\footnote[2]{If the change of frame is
time-dependent, then general form-invariance can always be naturally achieved within a 4-D covariant formalism; see e.g. \cite{Frewer09.1,Frewer09.2,Frewer09.3,Frewer16} and the references therein.}, i.e., to be valid in all frames and not only in one particular frame (which is one of the key ideas in the theory of general relativity). But, when solving these modelled equations, the goal is different from that when modeling these,
because now a specific frame needs to be chosen in order to explicitly formulate the solution. Hence, the natural question is: What is the optimal frame to solve these equations? For unbounded linear shear flow in the context of a linear stability analysis, this question will be answered in Section~\ref{S3}. To this point, the next section serves as an introduction to demonstrate the basic concepts that are needed to solve a linear stability problem with the method of Lie-symmetries and to elucidate the interpretational issues that go along with it, in particular the two issues of physical redundancy when specific invariant solutions are chosen and that of structural irrelevance when considering their collective interaction. To feature all key properties of this method while keeping the mathematical formalism at a minimum, a 1-D linear PDE will be solved.

\section{Introduction: The 1-D diffusion equation as a demonstrating example\label{S2}}

The general solution for the spatially unbounded linear 1-D
scalar diffusion equation, when stated as the following Cauchy problem where the diffusivity has been normalized to unity,\footnote[3]{Note that normalizing the diffusivity $\nu$ to unity can also be interpreted as defining either a new time variable $\hat{t}:=\nu\cdot t$ that has units of square meters, or, alternatively, as a new space variable $\hat{x}:=x/\sqrt{\nu}$ that, when squared, has units of time. By inverting these transformation rules, one can always go back to the original time or space variable, respectively. This dimensional re-definition in \eqref{151113:1225}-\eqref{151113:1713} has been done to simplify the expressions in this section, where for notational convenience the transformation symbol has been suppressed.}
\begin{equation}
\partial _t u(t,x) = \partial_x^2 u(t,x),\quad\;\;
u(t,x)\big|_{t=0}=u_0(x),\quad\;\;
(t,x)\in\mathbb{R}^+_0\times\mathbb{R},\label{151113:1225}
\end{equation}
is, for any\footnote[9]{The general solution \eqref{151113:1713}
also allows for initial conditions $u_0$ which are {\it not}
decaying at space infinity. The only restriction is that $u_0$ may
not increase faster than $e^{x^2}$ for $|x|\rightarrow\infty$. In
general, to study non-decaying initial conditions for a Cauchy
problem can be an interesting option, since in certain practical
situations there exist solutions which are unbounded at space
infinity, e.g. as in unbounded homogeneous shear flows.} initial
condition $u_0$, given as (see e.g. \cite{Polyanin02})
\begin{equation}
u(t,x)=\frac{1}{\sqrt{4\pi t}}\int_{-\infty}^\infty dx^\prime
u_0(x^\prime)e^{-\frac{(x-x^\prime)^2}{4t}}.\label{151113:1713}
\end{equation}

\restoregeometry

\noindent The aim now is to generate or to reproduce this general solution \eqref{151113:1713} with the method of Lie-symmetry groups. For that a symmetry analysis of the governing equation has to be performed.
The complete set of independent Lie-point symmetries admitted by
the unrestricted\footnote[2]{Unrestricted PDE means excluding the initial condition in \eqref{151113:1225}.} PDE \eqref{151113:1225} has the resulting form (see e.g.
\cite{Stephani89,Olver93,Bluman96})
\begin{equation}
\left.
\begin{aligned}
\mathsf{S}_\mathsf{1}\!:&\;\;\; \tilde{t}=t+\varepsilon,\;\;\;
\tilde{x}=x,\;\;\;\tilde{u}=u,\\[0.25em]
\mathsf{S}_\mathsf{2}\!:&\;\;\; \tilde{t}=t,\;\;\;
\tilde{x}=x+\varepsilon,\;\;\;\tilde{u}=u,\\[0.25em]
\mathsf{S}_\mathsf{3}\!:&\;\;\; \tilde{t}=t,\;\;\;
\tilde{x}=x,\;\;\;\tilde{u}=e^\varepsilon u,\\[0.25em]
\mathsf{S}_\mathsf{4}\!:&\;\;\; \tilde{t}=e^{2\varepsilon}t,\;\;\;
\tilde{x}=e^{\varepsilon}x,\;\;\;\tilde{u}=u,\\[0.25em]
\mathsf{S}_\mathsf{5}\!:&\;\;\; \tilde{t}=t,\;\;\; \tilde{x}=x+
2t\varepsilon,\;\;\;\tilde{u}=e^{-x\varepsilon
-t\varepsilon^2}\, u,\\[0.25em]
\mathsf{S}_\mathsf{6}\!:&\;\;\; \tilde{t}={\textstyle \frac{1}{1-
4t\varepsilon}}\, t,\;\;\; \tilde{x}={\textstyle \frac{1}{1-
4t\varepsilon}}\, x,\;\;\;\tilde{u}={\textstyle \sqrt{1-
4t\varepsilon}}\, e^{-\frac{\varepsilon x^2}{1- 4t\varepsilon}}\, u,\\[0.25em]
\mathsf{S}_\mathsf{7}\!:&\;\;\; \tilde{t}=t,\;\;\; \tilde{x}=
x,\;\;\;\tilde{u}=u+ f(t,x),\;\;\; \partial_t f-\partial_x^2
f=0,\label{171026:1545}
\end{aligned}
~~~~\right\}
\end{equation}
where each Lie-group action in $\varepsilon$ is
generated by its corresponding local tangent field
\begin{equation}
\left.
\begin{aligned}
X_1&=\partial_t,\\[0.25em]
X_2&=\partial_x,\\[0.25em]
X_3&=u\partial_u,\\[0.25em]
X_4&=2t\partial_t+x\partial_x,\\[0.25em]
X_5&=2t\partial_x-x u\partial_u,\\[0.25em]
X_6&=4t^2\partial_t+4tx\partial_x-\left(2tu+x^2u\right)\partial_u,\\[0.25em]
X_7&=f(t,x)\partial_u,\;\;\;\partial_t f-\partial_x^2 f=0.\label{171026:1546}
\end{aligned}
~~~~~~~~~~~~~~~~~~~~~~~~~~~~\right\}
\end{equation}
The finite dimensional sub-group
$\mathsf{S}_\mathsf{1}$-$\mathsf{S}_\mathsf{6}$ forms a closed 6-dimensional
Lie-algebra relative to the commutator $[X_i,X_j]=X_iX_j-X_jX_i$,
with its relations given as \citep{Olver93}
\begin{table}[h]
\centering
\renewcommand{\arraystretch}{2.0}
\begin{equation}
\begin{tabular}{c|c c c c c c}
$[\,\cdot\, ,\cdot\,]$ & $X_1$ & $X_2$ & $X_3$ & $X_4$ & $X_5$ &
$X_6$
\\\hline
$X_1$ & $0$ & $0$ & $0$ & $2X_1$ & $2X_2$ & $-2X_3+4X_4$
\\
$X_2$ & $0$ & $0$ & $0$ & $X_2$ & $-X_3$ & $2X_5$
\\
$X_3$ & $0$ & $0$ & $0$ & $0$ & $0$ & $0$
\\
$X_4$ & $-2X_1$ & $-X_2$ & $0$ & $0$ & $X_5$ & $2X_6$
\\
$X_5$ & $-2X_2$ & $X_3$ & $0$ & $-X_5$ & $0$ & $0$
\\
$X_6$ & $2X_3-4X_4$ & $-2X_5$ & $0$ & $-2X_6$ & $0$ & $0$
\end{tabular}
\label{151124:1023}
\end{equation}
{\small Table 1: Commutator table of the generators $X_1$-$X_6$
\eqref{171026:1546}.}
\end{table}

\noindent When later proposing different invariant functions as
solutions to \eqref{151113:1225}, it is interesting to examine in
how far these functions are equivalent\footnote[3]{In group theory
two elements are equivalent if they are conjugate two each other.}
to each other. For that we also need the adjoint representation of
the above Lie-algebra \eqref{151124:1023} (see Table 2
\eqref{151124:1041}-\eqref{151124:1042}), where the adjoint
mapping is defined in terms of the Lie series as \citep{Olver93}
\begin{equation}
\text{Ad}(e^{\varepsilon
X_i})X_j=X_j-\varepsilon[X_i,X_j]+{\textstyle\frac{1}{2}}
\varepsilon^2[X_i,[X_i,X_j]]-\cdots
\end{equation}

\vspace{-1em}
\begin{table}[h]
\centering
\renewcommand{\arraystretch}{2.0}
\begin{equation}
\hskip-2.375cm\begin{tabular}{c|c c c c c c}
$\text{Ad}$ & $X_1$ & $X_2$ & $X_3$ & $\;\;\cdots$
\\\hline
$X_1$ & $X_1$ & $X_2$ & $X_3$
\\
$X_2$ & $X_1$ & $X_2$ & $X_3$
\\
$X_3$ & $X_1$ & $X_2$ & $X_3$
\\
$X_4$ & $e^{2\varepsilon}X_1$ & $e^{\varepsilon}X_2$ & $X_3$
\\
$X_5$ & $X_1+2\varepsilon X_2-\varepsilon^2 X_3$ &
$X_2-\varepsilon X_3$ & $X_3$
\\
$X_6$ & $X_1-2\varepsilon X_3+4\varepsilon X_4+4\varepsilon^2X_6$
& $X_2+2\varepsilon X_5$ & $X_3$
\end{tabular}
\label{151124:1041}
\end{equation}
{\small Table 2a: First part of the adjoint representation for the
Lie-algebra \eqref{151124:1023}.\hspace{2.05cm}}
\end{table}

\begin{table}[h]
\centering
\renewcommand{\arraystretch}{2.0}
\begin{equation}
\hskip-0.825cm\begin{tabular}{c|c c c c c c}
$\text{Ad}$ & \quad$\cdots$ & $X_4$ & $X_5$ & $X_6$
\\\hline
$X_1$ & & $X_4-2\varepsilon X_1$ & $X_5-2\varepsilon X_2$ &
$X_6-4\varepsilon X_4+2\varepsilon X_3+4\varepsilon^2 X_1$
\\
$X_2$ & & $X_4-\varepsilon X_2$ & $X_5+\varepsilon X_3$ &
$X_6-2\varepsilon X_5-\varepsilon^2 X_3$
\\
$X_3$ & & $X_4$ & $X_5$ & $X_6$
\\
$X_4$ & & $X_4$ & $e^{-\varepsilon}X_5$ & $e^{-2\varepsilon}X_6$
\\
$X_5$ & & $X_4+\varepsilon X_5$ & $X_5$ & $X_6$
\\
$X_6$ & & $X_4+2\varepsilon X_6$ & $X_5$ & $X_6$
\end{tabular}
\label{151124:1042}
\end{equation}
{\small Table 2b: Second, continued part of the adjoint
representation for the Lie-algebra \eqref{151124:1023}.}
\end{table}

\vspace{1.75em}\noindent The adjoint mapping can also be naturally represented in
matrix form
\begin{equation}
\text{Ad}(e^{\varepsilon X_i})X_j = \sum_{k=1}^6
M^{(i)}_{jk}X_k,\;\;\;\; 1\leq i,j\leq 6,
\end{equation}
where the matrix elements then depend on the applied group
parameter $M^{(i)}_{jk}=M^{(i)}_{jk}(\varepsilon)$. To study
equivalence classes of the considered Lie-algebra, it is necessary
to construct the adjoint mapping for the general group element
$g=\prod_{i=1}^6 e^{\varepsilon_i X_i}$, which consequently has
the structure\footnote[2]{Note that when the aim is only to study
equivalence classes within the considered Lie-algebra, the
ordering as how to combine the adjoint transformation matrices
$\vM^{(i)}$ in \eqref{151129:1653} is unimportant and does not
matter, since for each different ordering we would get a
different resultant $\vM$, due to the non-commutativity of the
matrix product. But all these differently combined resultants are
in the end equivalent to each other, which means that it is
sufficient to consider only one particular ordering.}
\begin{align}
\text{Ad}(g)X_j & \,=\,\text{Ad}\!\left(\prod_{i=1}^6
e^{\varepsilon_i X_i}\right)X_j= \text{Ad}(e^{\varepsilon_1
X_1})\circ
 \text{Ad}(e^{\varepsilon_2 X_2})\circ\cdots\circ \text{Ad}(e^{\varepsilon_6
 X_6})X_j\nonumber\\
 &\, =\, \sum_{k=1}^6 M_{jk} X_k,\;\; \text{where}\;\;
 M_{jk}=(\vM)_{jk}=(\vM^{(1)}\cdot \vM^{(2)}\cdots
 \vM^{(6)})_{jk}.\label{151129:1653}
\end{align}
The matrix elements then depend on all group parameters
$M_{jk}=M_{jk}(\varepsilon_1,\varepsilon_2,\dotsc ,
\varepsilon_6)$, and, in this particular case for the Lie-algebra
of the 1-D diffusion equation, they are explicitly given~as
\begin{equation}
\left.
\begin{aligned}
&\hspace{-0.30cm}
M_{11}=e^{2\varepsilon_4},\;
M_{12}=2\varepsilon_5e^{2\varepsilon_4},\;
M_{13}=-e^{2\varepsilon_4}(\varepsilon_5^2+2\varepsilon_6),\;
M_{14}=4e^{2\varepsilon_4}\varepsilon_6,\\
&\hspace{-0.30cm}
M_{15}=4e^{2\varepsilon_4}\varepsilon_5\varepsilon_6,\;
M_{16}=4e^{2\varepsilon_4}\varepsilon_6^2,\;
M_{21}=0,\;
M_{22}=e^{\varepsilon_4},\;
M_{23}=-e^{\varepsilon_4}\varepsilon_5,\;
M_{24}=0,\\
&\hspace{-0.30cm}
M_{25}=2e^{\varepsilon_4}\varepsilon_6,\;
M_{26}=0,\;
M_{31}=0,\;
M_{32}=0,\;
M_{33}=1,\;
M_{34}=0,\;
M_{35}=0,\;
M_{36}=0,\\
&\hspace{-0.30cm}
M_{41}=-2e^{2\varepsilon_4}\varepsilon_1,\;
M_{42}=-e^{\varepsilon_4}(\varepsilon_2+4e^{\varepsilon_4}\varepsilon_1\varepsilon_5),\;
M_{43}=e^{\varepsilon_4}(\varepsilon_2\varepsilon_5+2e^{\varepsilon_4}
       \varepsilon_1(\varepsilon_5^2+2\varepsilon_6)),\\
&\hspace{-0.30cm}
M_{44}=1-8e^{2\varepsilon_4}\varepsilon_1\varepsilon_6,\;
M_{45}=\varepsilon_5-2e^{\varepsilon_4}\varepsilon_2\varepsilon_6
       -8e^{2\varepsilon_4}\varepsilon_1\varepsilon_5\varepsilon_6,\;
M_{46}=2\varepsilon_6(1-4e^{2\varepsilon_4}\varepsilon_1\varepsilon_6),\\
&\hspace{-0.30cm}
M_{51}=0,\;
M_{52}=-2e^{\varepsilon_4}\varepsilon_1,\;
M_{53}=\varepsilon_2+2e^{\varepsilon_4}\varepsilon_1\varepsilon_5,\;
M_{54}=0,\;
M_{55}=e^{-\varepsilon_4}-4e^{\varepsilon_4}\varepsilon_1\varepsilon_6,\\
&\hspace{-0.30cm}
M_{56}=0,\;
M_{61}=4e^{2\varepsilon_4}\varepsilon_1^2,\;
M_{62}=4e^{\varepsilon_4}\varepsilon_1(\varepsilon_2+2e^{\varepsilon_4}
       \varepsilon_1\varepsilon_5),\\
&\hspace{-0.30cm}
M_{63}=-\varepsilon_2^2+\varepsilon_1(2-4e^{\varepsilon_4}\varepsilon_2\varepsilon_5)
       -4e^{2\varepsilon_4}\varepsilon_1^2(\varepsilon_5^2+2\varepsilon_6),\\
&\hspace{-0.30cm}
M_{64}=4\varepsilon_1(-1+4e^{2\varepsilon_4}\varepsilon_1\varepsilon_6),\;
M_{65}=2e^{-\varepsilon_4}(\varepsilon_2+2e^{\varepsilon_4}\varepsilon_1\varepsilon_5)
       (-1+4e^{2\varepsilon_4}\varepsilon_1\varepsilon_6),\\
&\hspace{-0.30cm}
M_{66}=e^{-2\varepsilon_4}(1-4e^{2\varepsilon_4}\varepsilon_1\varepsilon_6)^2.
\end{aligned}
~\right\}\label{151129:1900}
\end{equation}
Two $r$-dimensional Lie-subalgebras $\mathcal{L}=\{Y_1,Y_2,\dotsc,
Y_r\}$ and $\mathcal{L}^\prime=\{Y^\prime_1,Y^\prime_2,\dotsc,
Y^\prime_r\}$, where in the present case $1\leq r\leq 6$, are
defined to be equivalent if one can find a mapping
$g=\prod_{i=1}^6 e^{\varepsilon_i X_i}$ and some constants
$c_{ij}$, for $1\leq i,j\leq 6$, such that the following system of
equations\footnote[3]{Depending on the dimension $r$ of the
Lie-subalgebra, system \eqref{151129:1835} can either represent an
over- or underde-\linebreak termined set of equations: $6r+1$
determining equations (including the regularity constraint) for
$r^2+6$ unknowns, which are the constants $c_{ij}$, and, in this
case, the six group parameters $\varepsilon_i$. The reason for
having $6r$ determining equations (excluding the regularity
constraint) is that the elements $Y_{i}=\sum_{k=1}^6\alpha_{ik}
X_k$ and $Y^\prime_{j}=\sum_{k=1}^6\alpha^\prime_{jk} X_k$ need to
be matched against each other, where $\alpha_{ij}$ and
$\alpha^\prime_{ij}$ are arbitrary but fixed expansion
coefficients representing these elements.} is satisfied
\citep{Ovsiannikov82,Olver93}
\begin{equation}
\begin{pmatrix}
\, Y^\prime_1\,\\Y^\prime_2\\\vdots\\ Y^\prime_r
\end{pmatrix}
=
\begin{pmatrix}
c_{11} & c_{12} & \cdots & c_{1r}\\
c_{21} & c_{22} & \cdots & c_{2r}\\
\vdots & \vdots & \ddots & \vdots\\
c_{r1} & c_{r2} & \cdots & c_{rr}
\end{pmatrix}
\begin{pmatrix}
\,\text{Ad}(g)Y_1\,\\
\text{Ad}(g)Y_2\\
\vdots\\
\text{Ad}(g)Y_r
\end{pmatrix},\label{151129:1835}
\end{equation}
but, also such that the coefficient matrix $(\vc)_{ij}=c_{ij}$ is
regular, i.e. $\text{det}\,\vc\neq 0$, simply to ensure the
necessary invertibility of the equivalence relation. The purpose
of this definition is obvious, because if two Lie-subalgebras are
equivalent then so are their corresponding two invariant functions
which can be generated from them, that is, these two functions are
then just connected by a group (symmetry) transformation.

A natural question which now arises is whether the general
solution \eqref{151113:1713} can be generated from particular
invariant solutions to those symmetries under which the underlying
linear PDE stays invariant. The answer is yes, and the crucial
part to achieve this, is to recognize that every {\it linear} PDE admits
the particular scaling as given here by $X_3$ \eqref{171026:1546}
for the scalar diffusion equation, which, when included into the
construction process for generating invariant functions

\newgeometry{left=2.49cm,right=2.49cm,top=2.50cm,bottom=2.34cm,headsep=1em}

\noindent from any combination of the remaining symmetries $X_i$, $1\leq i\leq 6,i\neq3$ \eqref{171026:1546}, turns
each associated invariance condition into an ordinary linear eigenvalue
problem.\footnote[2]{The reason that always an ordinary linear eigenvalue problem is posed for the case considered here, is that all defining symmetries $X_i,1\leq i\leq 6$ \eqref{171026:1546} exhibit here at most only a linear dependence in the dependent variable~$u$.} To illustrate this, we
consider four different examples, by choosing the following four 1-D subalgebras
\begin{equation}
\left.
\begin{aligned}
\mathfrak{f}_1=\{X_1+\alpha X_3\},\quad\; \mathfrak{f}_2&=\{X_2+\alpha X_3\},\hspace{1.17cm}\\[0.25em]
\mathfrak{f}_3=\{\lambda X_2+X_5+\alpha X_3\},\quad\; \mathfrak{f}_4&=\{2\tau X_1+X_4+\alpha X_3\}.\label{171027:1229}
\end{aligned}
~~~\right\}
\end{equation}
For $\tau\neq 0$ and $\lambda\neq 0$, all subalgebras considered above are inequivalent to each other, except for~$\mathfrak{f}_3$ which is equivalent to $\mathfrak{f}_2$, as can be readily seen when applying
the procedure as outlined above in \eqref{151129:1653}-\eqref{151129:1835} for $r=1$, with $Y_1^\prime=X_2+\alpha^\prime X_3$ and $Y_1=\lambda X_2+X_5+\alpha X_3$:
\begin{align}
X_2+\alpha^\prime X_3 & \, =\, c_{11}\text{Ad}(g)\Big(\lambda X_2
+
\alpha X_3 +X_5\Big)\label{171026:1907}\\[0.25em]
& \,=\,
c_{11}e^{\varepsilon_4}\big(\lambda-2\varepsilon_1\big)X_2+c_{11}\big(\alpha+\varepsilon_2+
2e^{\varepsilon_4}\varepsilon_1\varepsilon_5-e^{\varepsilon_4}\lambda\varepsilon_5\big)X_3\hspace{4cm}
\nonumber\\[0.25em]
&\hspace{6.65cm}
+c_{11}e^{\varepsilon_4}\big(2\lambda\varepsilon_6
-4\varepsilon_1\varepsilon_6+e^{-2\varepsilon_4}\big)X_5,
\nonumber
\end{align}
which has several (infinitely many) different solutions
such that the right-hand of \eqref{171026:1907} matches with the
left-hand side. The most simplest solution is given by\footnote[3]{Note that although $c_{11}=1$, it nevertheless carries a physical dimension, namely that of inverse time $1/t$. In~order to constantly track the physical dimensions during the analysis in this section, a dimensional factor $\gamma$ with the unit of time has been introduced, which in the end of any calculation then can be put to $\gamma=1$.}
\begin{equation}
\begin{aligned}
c_{11}&={\textstyle\frac{1}{\gamma}}=1, &
\alpha^\prime&={\textstyle\frac{\alpha}{\gamma}}, &
\varepsilon_1&={\textstyle\frac{1}{2}}(\lambda-\gamma), &
\varepsilon_2&=\varepsilon_3=\varepsilon_4=\varepsilon_5=0, &
\varepsilon_6&=-{\textstyle\frac{1}{2\gamma}}.\label{171026:1908}
\end{aligned}
\end{equation}
Hence, the invariant function that will be generated from $\mathfrak{f}_3$ (see \eqref{171027:2244} in Sec.~\ref{S23})
can also be
obtained by just transforming the invariant function generated from $\mathfrak{f}_2$ (see \eqref{171027:1900} in Sec.~\ref{S22})
via the combined symmetry transformation
$\mathsf{S}_\mathsf{1}\circ\mathsf{S}_\mathsf{6}$ \eqref{171026:1545} for the parameters as given in \eqref{171026:1908}\linebreak[4] --- this equivalence is explicitly shown in \eqref{171031:1101} in the appendix.
Although the general solution ansatz generated by $\mathfrak{f}_3$ is equivalent to the one proposed by
$\mathfrak{f}_2$, we still want to study this case, to see in how far this equivalence expresses itself
when constructing the general Cauchy solution~\eqref{151113:1713}
from~it. In particular, $\mathfrak{f}_3$ should serve as an example to explicitly illustrate the following remarkable difference to $\mathfrak{f}_4$ when relating both to the more simple subalgebras $\mathfrak{f}_1$ and~$\mathfrak{f}_2$: While the former algebra $\mathfrak{f}_3$ is mathematically equivalent and thus physically\footnote[9]{We assume here that the considered system \eqref{151113:1225} has next to its pure mathematical (platonic) formulation also a true physical interpretation, for example, as that it can be applied to model the conductive heat transfer in a sufficiently long and thin solid rod. This physical accessibility allows us then to view or to examine all occurring dynamical processes from different physical reference frames, to then choose the optimal one in which the dynamics evolves with a minimum of mathematical complexity.} redundant to $\mathfrak{f}_2$, the latter algebra $\mathfrak{f}_4$ is physically redundant to $\mathfrak{f}_1$, although being mathematically inequivalent~to~it.

\subsection{Subalgebra $\mathfrak{f}_1$ --- Temporal approach\label{S21}}

To derive an invariant solution from $\mathfrak{f}_1$ \eqref{171027:1229} as a basis for the general solution \eqref{151113:1713}, it is necessary to realize that this 1-D subalgebra is spanned by the single element basis
\begin{equation}
Y_1=X_1+\alpha X_3,
\end{equation}
which, due to the occurrence of the linearly induced symmetry $X_3=u\partial_u$ \eqref{171026:1546}, turns its corresponding invariant-solution condition (see e.g. \cite{Bluman96})
\begin{equation}
Y_1 F\Big|_{F=0}=0,\;\;\text{defined on the auxiliary function
$F=u-u(t,x)$},\label{171027:1234}
\end{equation}
into the following ordinary linear eigenvalue problem
\begin{equation}
X_1 u=\alpha\, u,\label{171027:1314}
\end{equation}

\newgeometry{left=2.49cm,right=2.49cm,top=2.50cm,bottom=1.90cm,headsep=1em}

\noindent where the group parameter $\alpha$ is then identified as the eigenvalue to the continuous operator $X_1=\partial_t$ \eqref{171026:1546}. The most general solution of system \eqref{171027:1314} is a continuous spectrum in the complex plane ($\alpha\in\mathbb{C}$), associated to the complete set of eigenvectors\footnote[2]{Note that the single-mode representation \eqref{171026:1323}, as it stands, is physically inconsistent. The left-hand side of \eqref{171026:1323} is always real-valued, while the right-hand side generally is complex-valued. However, since the governing equation \eqref{151113:1225} is linear and real, this inconsistency can easily be removed by adding next to this single mode its corresponding complex conjugated solution. This will yield a double-mode, specifically a complex pair representation consistent then to the physical, real-valued field solution. This procedure is possible and permissible because (i) linear equations allow for the superposition principle, and (ii) complex eigenvalues of a real system always occur in complex conjugate pairs.\label{171202:2041}}
\begin{equation}
u(t,x)=u_1(\alpha, x)\, e^{\alpha t},\label{171026:1323}
\end{equation}
where $u_1$ is an arbitrary spatial integration function depending in general on the eigenvalue~$\alpha$. For some specified value of $\alpha$,
the invariant solution \eqref{171026:1323} can be regarded as a particular solution ansatz for the PDE system \eqref{151113:1225},
which in turn will reduce it to an ODE system in the spatial variable $x$. By making use of the linear superposition principle, a more general solution
ansatz can be obtained by summing over all eigenvalues (or modes) in the complex plane
\begin{equation}
u(t,x)=\int_{\mathbb{C}}d\alpha\, u_1(\alpha,x)\, e^{\alpha t},\label{171205:1827}
\end{equation}
which then, ultimately, also determines the physical dimension of the new field $u_1$. Now, when choosing the contour of the $\alpha$-integration
along a straight line at a position $\sigma\in\mathbb{R}\hspace{0.5mm}$ in the direction of the imaginary axis, such that
it includes all singularities of the integrand when closing the contour around infinity (e.g. by a half circle), one obtains the effect that \eqref{171205:1827} turns into an
{\it invertible} integral relation. Hence, under these conditions, the invariant solution \eqref{171026:1323} to the 1-D subalgebra $\mathfrak{f}_1$ induces the following non-local change of
representation in the dependent solution variable\footnote[3]{The redefined function $\hat{u}_1$ in \eqref{171205:1837}-\eqref{171205:1834} is related to the
original one in \eqref{171205:1827} by: $\hat{u}_1(\alpha,x)\equiv 2\pi i\cdot u_1(\alpha,x)$.}
\begin{equation}
u(t,x)= \int_{\sigma-i\infty}^{\sigma+i\infty}\frac{d\alpha}{2\pi i}\, \hat{u}_1(\alpha,x)\, e^{\alpha t},\;\;
\sigma\in\mathbb{R},\: \alpha\in\mathbb{C},\: t\geq 0,\label{171205:1837}
\end{equation}
which constitutes the Laplace transform of the field $u$ in physical $t$-space into its complementary field $\hat{u}_1$
in complex $\alpha$-space, with the inverse transformation then given~as
\begin{equation}
\hat{u}_1(\alpha,x)=\int_0^\infty\! dt \, u(t,x)\, e^{-\alpha t}.\label{171205:1834}
\end{equation}
Note that in order to obtain an invertible relation from \eqref{171205:1827}, the contour of the $\alpha$-integration has to be constrained to the path as explained above. The reason is
that since the complementary variable to $\alpha$, namely the physical time coordinate $t$, is only defined in the half-domain $t\geq 0$, an invertible relation to \eqref{171205:1827} has to respect and satisfy
this constraint, as can be seen then in the result \eqref{171205:1834}.

In summary, the algebra $\mathfrak{f}_1$ thus induces the local (single-mode) representation \eqref{171026:1323} as well as the non-local (general) representation \eqref{171205:1837} of the dependent solution variable $u$ relative to its time coordinate $t\geq 0$. Hence, $\mathfrak{f}_1$ is a temporal\footnote[9]{The $\mathfrak{f}_1$-induced eigenfunction \eqref{171026:1323} consists of the defining temporal mode $e^{\alpha t}$ having spatial shape~$u_1(\alpha,x)$.} invariant-solution approach to system \eqref{151113:1225}.

Now, to generate the general solution of the Cauchy problem
\eqref{151113:1225} we have to solve the invariantly reduced ODE, which can be directly obtained by
either inserting the local \eqref{171026:1323} or the non-local representation \eqref{171205:1837} into
the PDE \eqref{151113:1225}. However, this approach is not advantageous to take, since it
does not incorporate the initial condition $u_0$ when solving it.
A better option is to Laplace transform the PDE directly via its defining relation \eqref{171205:1834}
\begin{align}
0=\partial_t u(t,x)-\partial_x^2 u(t,x)\;\;\Leftrightarrow\;\; 0 &
= \int_0^\infty dt\, \partial_t u(t,x)\, e^{-\alpha t} -\int_0^\infty
dt\, \partial_x^2 u(t,x)\, e^{-\alpha t}\nonumber\\[0.5em]
& = \left[u(t,x)\, e^{-\alpha t}\right]_{t=0}^{t=\infty}
+\alpha\,\hat{u}_1(\alpha,x)-\partial_x^2\hat{u}_1(\alpha,x)\nonumber\\[0.75em]
& = -u_0(x) +\alpha\,\hat{u}_1(\alpha,x)-\partial_x^2\hat{u}_1(\alpha,x),
\label{151130:2331}
\end{align}

\restoregeometry

\noindent which now explicitly involves the initial condition
$u_0$ of the Cauchy problem \eqref{151113:1225}, where we assumed that the solution $u$ temporally decays faster asymptotically than $e^{\text{Re}(\alpha)t}$. The general
solution of this ODE \eqref{151130:2331} is given as
\begin{align}
\hat{u}_1(\alpha,x)&\,=\, C_1(\alpha)\,e^{x\sqrt{\alpha}}+C_2(\alpha)\, e^{-x\sqrt{\alpha}}
\nonumber\\[0.25em]
&\hspace{0.6cm}
-\frac{1}{2\sqrt{\alpha}}\left(e^{x\sqrt{\alpha}}\int_{-\infty}^x
e^{-x^\prime\sqrt{\alpha}}\,u_0(x^\prime)\,dx^\prime-e^{-x\sqrt{\alpha}}\int_{-\infty}^x
e^{x^\prime\sqrt{\alpha}}\,u_0(x^\prime)\,dx^\prime \right),\;\;
\label{151115:1817}
\end{align}
where $C_1(\alpha)$ and $C_2(\alpha)$ are two arbitrary
integration functions which can only be determined by putting
spatial boundary conditions. But since the considered Cauchy
problem \eqref{151113:1225} is free of any spatial boundary
conditions, except for the very weak restriction that the initial
condition $u_0$ may not increase faster than $e^{x^2}$ at space
infinity (see footnote to solution \eqref{151113:1713}), we can
specify $C_1(\alpha)$ and $C_2(\alpha)$ freely. A reasonable specification
is of course to choose these functions such that no singularity in
the solution \eqref{151115:1817} arises when
$|x|\rightarrow\infty$. However, to study the asymptotics of
\eqref{151115:1817} it is necessary to first rewrite it into a
symmetrized form such that no integration limit is privileged. The
equivalent but symmetrized form of \eqref{151115:1817} reads
\begin{align}
\hat{u}_1(\alpha,x) & \,=\, C_1(\alpha)\,e^{x\sqrt{\alpha}}+C_2(\alpha)\, e^{-x\sqrt{\alpha}}
\nonumber\\[0.5em]
&\hspace{0.6cm}
-\frac{1}{4\sqrt{\alpha}}\left(e^{x\sqrt{\alpha}}\int_{-\infty}^x
e^{-x^\prime\sqrt{\alpha}}\,u_0(x^\prime)\,dx^\prime-e^{-x\sqrt{\alpha}}\int_{-\infty}^x
e^{x^\prime\sqrt{\alpha}}\,u_0(x^\prime)\,dx^\prime
\right)\nonumber\\[0.5em]
&\hspace{0.6cm}
-\frac{1}{4\sqrt{\alpha}}\left(e^{-x\sqrt{\alpha}}\int_{x}^\infty
e^{x^\prime\sqrt{\alpha}}\,u_0(x^\prime)\,dx^\prime-e^{x\sqrt{\alpha}}\int_{x}^\infty
e^{-x^\prime\sqrt{\alpha}}\,u_0(x^\prime)\,dx^\prime \right),
\label{151201:0104}
\end{align}
which, of course, is still a solution to equation
\eqref{151130:2331}. The symmetrized form \eqref{151201:0104} now
reveals that in order to ensure a regular solution at
$|x|\rightarrow\infty$, the integration functions has to be
chosen~as
\begin{equation}
C_1(\alpha)=\frac{1}{4\sqrt{\alpha}}\int_{-\infty}^\infty
e^{-x^\prime\sqrt{\alpha}}\,u_0(x^\prime)\,dx^\prime,\quad\;\;
C_2(\alpha)=\frac{1}{4\sqrt{\alpha}}\int_{-\infty}^\infty
e^{x^\prime\sqrt{\alpha}}\,u_0(x^\prime)\,dx^\prime.
\end{equation}
Inserting the above result then back into the inverted
Laplace transform \eqref{171205:1837}, we finally yield the
general solution \eqref{151113:1713}
\begin{align}
u(t,x)&\,=\,\frac{1}{2\pi i}\int_{\sigma-i\infty}^{\sigma+i\infty}
d\alpha\, \hat{u}_1(\alpha,x)\,e^{\alpha t}\nonumber\\[0.5em]
&\,=\,\frac{1}{2\pi i}\int_{-\infty}^\infty dx^\prime\,
u_0(x^\prime)\int_{\sigma-i\infty}^{\sigma+i\infty}d\alpha
\left(\frac{e^{(x-x^\prime)\sqrt{\alpha}}}{4\sqrt{\alpha}}
+\frac{e^{-(x-x^\prime)\sqrt{\alpha}}}{4\sqrt{\alpha}}\right)e^{\alpha t}\nonumber\\[0.5em]
&\hspace{0.6cm}-\frac{1}{2\pi i}\int_{-\infty}^x dx^\prime\,
u_0(x^\prime)\int_{\sigma-i\infty}^{\sigma+i\infty}d\alpha
\left(\frac{e^{(x-x^\prime)\sqrt{\alpha}}}{4\sqrt{\alpha}}
-\frac{e^{-(x-x^\prime)\sqrt{\alpha}}}{4\sqrt{\alpha}}\right)e^{\alpha t}\nonumber\\[0.5em]
&\hspace{0.6cm}-\frac{1}{2\pi i}\int_{x}^\infty dx^\prime\,
u_0(x^\prime)\int_{\sigma-i\infty}^{\sigma+i\infty}d\alpha
\left(\frac{e^{-(x-x^\prime)\sqrt{\alpha}}}{4\sqrt{\alpha}}
-\frac{e^{(x-x^\prime)\sqrt{\alpha}}}{4\sqrt{\alpha}}\right)e^{\alpha t}\nonumber\\[0.5em]
&\,=\, \frac{1}{\sqrt{4\pi t}}\int_{-\infty}^\infty dx^\prime
u_0(x^\prime)e^{-\frac{(x-x^\prime)^2}{4t}},
\end{align}
which means that dependent on the initial condition we
ultimately cover {\it all} solutions of the general Cauchy
solution with the ansatz \eqref{151115:1817}, emerging again from the invariant-solution approach of the 1-D subalgebra $\mathfrak{f}_1$ \eqref{171027:1229}. Hence, the general solution
\eqref{151113:1713} can be generated by the invariant functions of the symmetries $X_1$ and $X_3$ as incorporated by $\mathfrak{f}_1$.

\newgeometry{left=2.49cm,right=2.49cm,top=2.50cm,bottom=2.275cm,headsep=1em}

\subsection{Subalgebra $\mathfrak{f}_2$ --- Spatial approach\label{S22}}

The aim in this subsection is to independently repeat the previous approach, however, now with~the (to $\mathfrak{f}_1$ inequivalent) 1-D subalgebra $\mathfrak{f}_2$ \eqref{171027:1229}.
This algebra is spanned by the single element basis
\begin{equation}
Y_2=X_2+\alpha X_3,\label{171103:1045}
\end{equation}
which turns the corresponding invariant surface condition $(Y_2 F)|_{F=0}=0$, for $F=u-u(t,x)$, into the following ordinary linear eigenvalue problem
\begin{equation}
X_2 u =\alpha\, u,\label{171027:1857}
\end{equation}
where the group parameter $\alpha$ is then again identified as the eigenvalue, however, now to the continuous operator $X_2=\partial_x$ \eqref{171026:1546}. As before for the eigenvalue problem \eqref{171027:1314} associated to $\mathfrak{f}_1$, the general solution of \eqref{171027:1857} is again a continuous spectrum of eigenvalues in the complex plane ($\alpha\in\mathbb{C}$), however, now associated to the complete
set of eigenvectors complementary to the unbounded spatial variable $x$
\begin{equation}
u(t,x)=u_2(t,\alpha)\, e^{\alpha x},\label{171027:1900}
\end{equation}
where $u_2$ is an arbitrary temporal integration function depending in general on the eigenvalue~$\alpha$.
For some specified value of $\alpha$,
the invariant solution \eqref{171027:1900} can again be regarded as a particular solution ansatz for the PDE system \eqref{151113:1225},
which now will reduce it to an ODE system in the temporal variable $t$. By making again use of the linear superposition principle, a more general solution
ansatz can be obtained by summing again over all eigenvalues (or modes) in the complex plane, which then, ultimately, also determines the physical dimension of the new field $u_2$,
\begin{equation}
u(t,x)=\int_{\mathbb{C}}d\alpha\, u_2(t,\alpha)\, e^{\alpha x}.\label{171027:1929}
\end{equation}
When choosing the contour of the $\alpha$-integration exactly along the imaginary axis, i.e., $\alpha=i k$ with $k\in \mathbb{R}$, one obtains the following invertible integral relation\footnote[2]{The redefined function $\hat{u}_2$ in \eqref{171027:1940}-\eqref{171027:1943} is related to the original function $u_2$ \eqref{171027:1929} by: $\hat{u}_2(t,k)\equiv i\cdot u_2(t,ik)$.}
\begin{equation}
u(t,x)= \int_{-\infty}^{\infty}dk\, \hat{u}_2(t,k)\, e^{ik x},\;\,
k\in\mathbb{R};\: x\in\mathbb{R},\label{171027:1940}
\end{equation}
which constitutes the Fourier transform of the field $u$ in physical $x$-space into its complementary field $\hat{u}_2$
in $k$-space, with the inverse transformation then given~as
\begin{equation}
\hat{u}_2(t,k)=\int_{-\infty}^\infty \frac{dx}{2\pi} \, u(t,x)\, e^{-ik x}.\label{171027:1943}
\end{equation}
Thus, next to the already obtained representations \eqref{171026:1323} and \eqref{171205:1837}, the 1-D subalgebra $\mathfrak{f}_2$ induces a further physically non-redundant local (single-mode) representation \eqref{171027:1900} as well as a non-local (general) representation \eqref{171027:1940} of the dependent solution variable $u$, however, now relative to its unbounded spatial coordinate $-\infty\leq x\leq\infty $. Hence, $\mathfrak{f}_2$ is a spatial\footnote[3]{The $\mathfrak{f}_2$-induced eigenfunction \eqref{171027:1900} consists of the defining spatial mode $e^{\alpha x}$ having temporal shape~$u_2(t,\alpha)$.} invariant-solution approach to system \eqref{151113:1225}.

Now, to generate the general solution of the Cauchy problem
\eqref{151113:1225} based on $\mathfrak{f}_2$, we have to solve the following invariantly reduced ODE, which can be directly obtained by
either inserting the local \eqref{171027:1900} or the non-local representation \eqref{171027:1940} into the governing PDE and by using the inverse relation \eqref{171027:1943} to accordingly adjust to the initial condition. This reduced system reads
\begin{equation}
\partial_t \hat{u}_2(t,k)=-k^2 \hat{u}_2(t,k), \quad\;\;
\hat{u}_2(t,k)\big|_{t=0}=\int_{-\infty}^\infty \frac{dx}{2\pi} \, u_0(x)\, e^{-ik x},
\end{equation}
with its general solution given as
\begin{equation}
\hat{u}_2(t,k)=C(k)e^{-k^2t}, \;\;\text{where}\;\; C(k)=\int_{-\infty}^\infty \frac{dx}{2\pi} \, u_0(x)\, e^{-ik x}.
\end{equation}

\restoregeometry

\noindent Then, by inserting the above result
back into in the inverted Fourier transform \eqref{171027:1940}, we finally obtain the general
solution \eqref{151113:1713} of the Cauchy problem
\eqref{151113:1225}
\begin{align}
u(t,x) & = \int_{-\infty}^\infty dk\,C(k)\, e^{-k^2 t}\,
e^{ik x} = \frac{1}{2\pi} \int_{-\infty}^\infty dx^\prime dk\,
u_0(x^\prime)\, e^{ik(x-x^\prime)-k^2t}\nonumber\\[0.5em]
& = \frac{1}{\sqrt{4\pi t}}\int_{-\infty}^\infty dx^\prime
u_0(x^\prime)e^{-\frac{(x-x^\prime)^2}{4t}}.
\end{align}
Hence, the general solution
\eqref{151113:1713} can also be generated by the invariant functions of the symmetries $X_2$ and $X_3$ as incorporated by the 1-D subalgebra $\mathfrak{f}_2$.

\subsection{Subalgebra $\mathfrak{f}_3$ --- Equivalent and redundant approach\label{S23}}

As in the two previous subsections, the aim here again is to derive the general solution \eqref{151113:1713} of the governing system \eqref{151113:1225}, however, now with the invariant-solution approach of the more complex 1-D subalgebra $\mathfrak{f}_3$ \eqref{171027:1229}. This algebra is spanned by the single element basis
\begin{equation}
Y_3=\lambda X_2+X_5+\alpha X_3,\label{171028:1021}
\end{equation}
which turns the corresponding invariant surface condition $(Y_3 F)|_{F=0}=0$, for $F=u-u(t,x)$, into the following linear eigenvalue problem
\begin{equation}
\big(x+(2t+\lambda)X_2\big) u =\alpha\, u,\label{171027:2240}
\end{equation}
where the group parameter $\alpha$ is then again identified as the eigenvalue, however, now to the $\lambda$-dependent continuous operator $x+(2t+\lambda)X_2$, where $X_2=\partial_x$ \eqref{171026:1546}. The general solution of~\eqref{171027:2240} is given by\footnote[2]{To note is the conceptual difference between the two parameters $\alpha$ and $\lambda$. While the former is identified as an eigenvalue of the problem \eqref{171027:2240}, the latter serves as a regulating parameter in the single mode solution \eqref{171027:2244} to ensure a regular result at the initial point $t=0$. In particular, since the eigenvalue defining operator of the problem \eqref{171027:2240} is $\lambda$-dependent, the eigenvalue itself can in general be a function of the regulator $\lambda$, i.e., $\alpha=\alpha(\lambda)$.}
\begin{equation}
u(t,x)=u_3(t,\alpha;\lambda)\,e^{-\frac{x(x-2\alpha)}{2(\lambda+2t)}},\label{171027:2244}
\end{equation}
but where at this stage it is not clear yet, whether these eigenfunctions form a complete set when assuming a true continuous spectrum of $\alpha$ in the complex plane ($\alpha\in\mathbb{C}$). However, since the invariant solution \eqref{171027:2244} is mathematically, or more precise, group-theoretically equivalent\footnote[3]{See Appendix \ref{SA}, where it is explicitly shown how the currently considered invariant function \eqref{171027:2244} of $\mathfrak{f}_3$ is connected to the previously derived invariant function \eqref{171027:1900} of $\mathfrak{f}_2$ via the symmetry group chosen in \eqref{171026:1908}.} to the invariant solution \eqref{171027:1900}, as shown in \eqref{171026:1907}-\eqref{171026:1908}, the eigenfunctions form indeed a complete continuous set over $\alpha\in\mathbb{C}$. Hence, using the linear superposition principle, the single mode solution \eqref{171027:2244} can be summed over all (continuous) eigenvalues $\alpha\in\mathbb{C}$ to yield the general solution, which then, ultimately, also determines the physical dimension of the new field~$u_3$,\footnote[9]{Note that $\lambda$ in the global representation \eqref{171028:0947} constitutes an irrelevant parameter. That means, the integral evaluation on the right-hand of this relation does not depend on $\lambda$, simply because the general solution $u$ on its left-hand side cannot depend on it, as can be clearly seen in the result \eqref{151113:1713}, which ultimately is based on the fact that the originally formulated Cauchy problem \eqref{151113:1225} is independent to $\lambda$. In other words, the parameter $\lambda$ is neither mathematically (as an auxiliary parameter) nor physically (as a modelling parameter) part of the original problem \eqref{151113:1225}.}
\begin{equation}
u(t,x)=\int_{\mathbb{C}}d\alpha\, u_3(t,\alpha;\lambda)\,e^{-\frac{x(x-2\alpha)}{2(\lambda+2t)}}.\label{171028:0947}
\end{equation}
The question which now arises is whether this integral relation can be inverted when aiming at a further non-local (general) representation of the dependent solution variable $u$. In other words, how should the integration path in \eqref{171028:0947} be chosen such that this integral relation is invertible in order to constitute a true representation of the solution? For the particular subalgebra $\mathfrak{f}_3$ \eqref{171028:1021} considered here, this question can be solved in two ways: Either by trial-and-error, or, more systematically, by changing the frame of reference to yield a more optimal symmetry algebra than the non-optimal one considered here by $\mathfrak{f}_3$ \eqref{171028:1021}:

{\it (1) Trial and error:} This method aims to rename and to redefine the variables in the integral relation \eqref{171028:0947} such that it reduces to a form for which its inverse relation is known. This method is random
and in general one has no guarantee that it is successful. For the particular relation~\eqref{171028:0947}, however, it is successful and easy to achieve, because we know that its defining invariant solution~\eqref{171027:2244} is mathematically {\it equivalent} or at least physically {\it redundant} to the more simple invariant solution~\eqref{171027:1900} induced by the symmetry algebra $\mathfrak{f}_2$, as explicitly shown in~\eqref{171031:1101},
or more general in \eqref{171101:0928}. This equivalence readily allows us to reduce the integral relation \eqref{171028:0947} to the more accessible one~\eqref{171027:1929} in a very straightforward way by just redefining the variables, yielding thus a relation for which its inverse is known by \eqref{171027:1943} when choosing the contour of the $\alpha$-integration along the imaginary axis, i.e., $\alpha=i k$, for $k\in \mathbb{R}$:
\begin{align}
u(t,x)& =\int_{\mathbb{C}}d\alpha\, u_3(t,\alpha;\lambda)\,e^{-\frac{x(x-2\alpha)}{2(\lambda+2t)}}\nonumber\\[0.25em]
&=e^{-\frac{x^2}{2(\lambda+2t)}}\int_{\mathbb{C}}d\alpha\, u_3(t,\alpha;\lambda)\,e^{\frac{\alpha x}{\lambda+2t}}\nonumber\\[0.25em]
&=e^{-\frac{1}{2}(\lambda+2t)\hat{x}^2}\int_{\mathbb{C}}d\alpha\, u_3(t,\alpha;\lambda)\,e^{\alpha\hat{x}},\;\;\text{where $\,\hat{x}={\textstyle\frac{x}{\lambda+2t}}$, $\,[\hat{x}]=[1/x]$},\nonumber\\[0.25em]
\Leftrightarrow\;\;\: \hat{u}(t,\hat{x};\lambda) &=\int_{\mathbb{C}}d\alpha\, u_3(t,\alpha;\lambda)\,e^{\alpha\hat{x}},\;\;\text{where $\,\hat{u}(t,\hat{x};\lambda)=e^{\frac{1}{2}(\lambda+2t)\hat{x}^2} u\big(t, (\lambda+2t)\hat{x}\big)$}.\;\;
\label{171102:0825}
\end{align}
Since the re-defined spatial variable $\hat{x}$ of inverse length is also, as $x$, in both directions unbounded, the above relation~\eqref{171102:0825} can be identified as the regular 1-D Fourier transform in this new variable, with its inverse then given~as
\begin{align}
u_3(t,\alpha;\lambda)&=\int_{-\infty}^\infty \frac{d\hat{x}}{2\pi}\hat{u}(t,\hat{x};\lambda)e^{-\alpha\hat{x}}\nonumber\\[0.25em]
&=\int_{-\infty}^\infty \frac{d\hat{x}}{2\pi}e^{\frac{1}{2}(\lambda+2t)\hat{x}^2} u\big(t, (\lambda+2t)\hat{x}\big)e^{-\alpha\hat{x}}.
\label{171101:2136}
\end{align}
Hence, if the contour path in \eqref{171028:0947} is chosen along the imaginary axis then this relation can be inverted and is finally given by \eqref{171101:2136} as
\begin{equation}
u_3(t,\alpha;\lambda)=\int_{-\infty}^\infty \frac{dx}{2\pi(\lambda+2t)}e^{\frac{x^2}{2(\lambda+2t)}} u(t,x) e^{-\alpha\frac{x}{\lambda+2t}},\;\;\;\; \alpha=ik,\; k\in\mathbb{R},\label{171103:1148}
\end{equation}
where the 1-D integration variable in \eqref{171101:2136} has been transformed back to the original variable~$x$. Note that when just renaming the integration (dummy) variable in \eqref{171101:2136} to $x$ will also give an inverse relation but one that before its evaluation will not be dimensionally consistent in its physical units to \eqref{171028:0947}, simply because $\hat{x}\neq x$.

{\it (2) Change of reference frame:} This method is systematic and has a physical interpretation of why the above trial-and-error method for the considered case was successful. The aim of this method is to change the physical point of view by looking for a new reference frame where a non-optimal symmetry algebra reduces to a mathematically more simpler structure for which the inverse of its induced invariant modal solution is known when considering all modes collectively. For the particular in the inertial frame formulated symmetry algebra $\mathfrak{f}_3$ considered here, the already performed examination in Appendix \ref{SA} shows that there exists an infinite family of different reference frames for which $\mathfrak{f}_3$ reduces to the significantly simpler and already well-established symmetry algebra $\mathfrak{f}_2$ explored in Section \ref{S22}. According to the corresponding result \eqref{171101:1447}, the frame change with the lowest degree of complexity is given by the particular choice
\begin{equation}
\mathsf{T}\!:\;\;\; \tilde{t}=t,\;\;\; \tilde{x}={\textstyle \frac{\xi x}{\lambda+2t}},\;\;\; \tilde{u}=e^{\frac{x^2}{2(\lambda+2t)}}\cdot u,\label{171103:1007}
\end{equation}
which, by its intended construction \eqref{171101:1426}, is based on a pure coordinate transformation going along with a subsequent local re-gauging of the scalar field $u$, where $\xi$ is a constant factor to ensure
dimensional consistency. Indeed, according to $\mathsf{T}$ \eqref{171103:1007}, the symmetry algebra $\mathfrak{f}_3$ \eqref{171028:1021}, formulated in the inertial (non-tilded) frame, reduces in the new~(tilded) frame to $\mathfrak{f}_2$ \eqref{171103:1045}
\begin{equation}
\mathfrak{f}_3=\{\lambda X_2+X_5+\alpha X_3\}\overset{\mathsf{T}}{\underset{\eqref{171103:1025}}=}\{\tilde{X}_2+\tilde{\alpha}\tilde{X}_3\}=\tilde{\mathfrak{f}}_2,\;\;\text{with $\,\tilde{\alpha}=\alpha/\xi$},\label{171120:1439}
\end{equation}
where it is to note that $\mathsf{T}$ \eqref{171103:1007} is {\it not} a symmetry transformation of the underlying equation~\eqref{151113:1225}. Nevertheless, with this change of frame we have the result that in the new frame the $\mathfrak{f}_3$-induced representation \eqref{171028:0947} reduces to the $\mathfrak{f}_2$-induced representation \eqref{171027:1929}\footnote[2]{Note that within the new frame the general solution $\tilde{u}$ explicitly depends on the regulator $\lambda$ and hence also its complementary field $\tilde{u}_2$, simply because the defining transformation $\mathsf{T}$ \eqref{171103:1007} connecting the inertial frame to this new frame is dependent on $\lambda$.}
\begin{equation}
\tilde{u}(\tilde{t},\tilde{x};\lambda)=\int_{\mathbb{C}}d\tilde{\alpha}\, \tilde{u}_2(\tilde{t},\tilde{\alpha};\lambda)\, e^{\tilde{\alpha} \tilde{x}},\label{171103:1127}
\end{equation}
which, if the contour path of the $\tilde{\alpha}$-integration is chosen along the imaginary axis, i.e., $\tilde{\alpha}=i \tilde{k}$, for $\tilde{k}\in \mathbb{R}$, has the inverse relation
\begin{equation}
\tilde{u}_2(\tilde{t},\tilde{\alpha};\lambda)=\int_{-\infty}^\infty \frac{d\tilde{x}}{2\pi}\tilde{u}(\tilde{t},\tilde{x};\lambda)e^{-\tilde{\alpha}\tilde{x}}.\label{171106:1524}
\end{equation}
Now, when transforming this result back to the original inertial (untilded) frame according to the defining transformation $\mathsf{T}$ \eqref{171103:1007}, and using the single-mode result of \eqref{171101:0928}, namely that
$\tilde{u}_2(\tilde{t},\tilde{\alpha};\lambda)= \xi u_3(t,\alpha;\lambda)$, due to that $f_1(t)=t$, $f_2(t)=0$, $f_3(t)=1$ and $f_4(t)=0$ for the considered transformation $\mathsf{T}$~\eqref{171103:1007}, we obtain the wanted inverse relation of \eqref{171028:0947}
\begin{equation}
u_3(t,\alpha;\lambda)=\int_{-\infty}^\infty \frac{dx}{2\pi(\lambda+2t)}e^{\frac{x^2}{2(\lambda+2t)}} u(t,x) e^{-\alpha\frac{x}{\lambda+2t}},\;\;\;\; \alpha=ik,\; k\in\mathbb{R},\label{171103:1217}
\end{equation}
which, of course, coincides with the result \eqref{171103:1148} obtained before by the preceding trail-and-error method. However, to note here is that there are actually two dimensionally consistent ways of how \eqref{171106:1524} can be transformed: Either by transforming {\it or} by renaming the integration element, where the latter approach is admissible since here the integration (dummy) variable $\tilde{x}$ in \eqref{171106:1524} carries the same physical dimension as the original spatial variable $x$. The particular structure \eqref{171103:1217} obtained is the result of transforming the integration element in \eqref{171106:1524}, while when renaming it, we obtain the following alternative and also dimensionally consistent representation for the inverse relation of \eqref{171028:0947}
\begin{equation}
u_3(t,\alpha;\lambda)=\int_{-\infty}^\infty \frac{dx}{2\pi\xi}e^{\frac{(\lambda+2t)x^2}{2\xi^2}}u\!\left(t,{\textstyle\frac{(\lambda+2t)x}{\xi}}\right)e^{-\alpha\frac{x}{\xi}},\;\;\;\; \alpha=ik,\; k\in\mathbb{R},
\label{171106:1637}
\end{equation}
which obviously, when transforming the integration variable in \eqref{171106:1637} accordingly, is mathematically identical to the alternative representation \eqref{171103:1217}. The integral relation \eqref{171106:1637} explicitly shows that the parameter $\xi$ is inessential, since the evaluation of the integral does not depend on it, as also the $\xi$-independent left-hand side in \eqref{171106:1637} is clearly showing.

Now that we are in the possession of the inverse relation to the global representation \eqref{171028:0947} induced by the symmetry algebra $\mathfrak{f}_3$, we can finally start to generate the $\mathfrak{f}_3$-based solution procedure of the general solution for the underlying Cauchy problem \eqref{151113:1225}. This will be presented here in three different ways: Once directly in the originally formulated inertial frame, once in a changed frame defined by a symmetry transformation, and once in a changed frame defined by a non-symmetry transformation.

\subsubsection{Direct solution procedure within the originally formulated inertial frame\label{S231}}

To solve is the unbounded 1-D diffusion problem, which in an inertial\footnote[2]{Note that the inertial-frame-connecting Galilei transformations for the 1-D diffusion equation are given by the~modified Galilei transformations $\mathsf{S}_\mathsf{5}$ \eqref{171026:1545}, consisting of a Galilean boost and a subsequent local re-gauging in~the scalar field $u$.} frame of reference is mathematically formulated as the following Cauchy problem \eqref{151113:1225}
\begin{equation}
\partial _t u(t,x) = \partial_x^2 u(t,x),\quad\;\;
u(t,x)\big|_{t=0}=u_0(x),\quad\;\;
(t,x)\in\mathbb{R}^+_0\times\mathbb{R}.\label{171103:1332}
\end{equation}
The aim here is to derive its general solution~\eqref{151113:1713} on the basis of the equation's symmetry algebra $\mathfrak{f}_3$-induced non-local representation \eqref{171028:0947}\footnote[3]{To constitute a true representation, the integral relation \eqref{171028:0947} needs to be invertible, which, as has been shown before, can only be achieved if the integration path is chosen along the imaginary axis, i.e., if $\alpha=i k$, $k\in \mathbb{R}$.\linebreak[4] The redefined function $\hat{u}_3$ in \eqref{171103:1503} is then related to the original function $u_3$ \eqref{171028:0947} by: $\hat{u}_3(t,k;\lambda)\equiv i\cdot u_3(t,ik;\lambda)$.}
\begin{equation}
u(t,x)=\int_{-\infty}^\infty dk\, \hat{u}_3(t,k;\lambda)\,e^{-\frac{x(x-2ik)}{2(\lambda+2t)}},\label{171103:1503}
\end{equation}
with its inverse \eqref{171103:1217}
\begin{equation}
\hat{u}_3(t,k;\lambda)=\int_{-\infty}^\infty \frac{dx}{2\pi(\lambda+2t)}e^{\frac{x^2}{2(\lambda+2t)}} u(t,x) e^{-\frac{ikx}{\lambda+2t}},\label{171103:1521}
\end{equation}
without leaving the inertial frame of reference, that is, in other words, to solve the system \eqref{171103:1332} by the symmetry algebra $\mathfrak{f}_3$ without changing the frame of reference.

Since \eqref{171103:1503}-\eqref{171103:1521} is based on a true {\it spatial} representation, the procedure evolves exactly as already outlined~in Section \ref{S22} for the symmetry algebra $\mathfrak{f}_2$. The first step is to insert the symmetry-induced representation \eqref{171103:1503} into the PDE \eqref{171103:1332} to obtain, by construction, the invariantly reduced ODE for the complementary field $\hat{u}_3$, which here takes the form
\begin{equation}
\partial_t \hat{u}_3(t,k;\lambda)+\frac{\lambda+2t+k^2}{(\lambda+2t)^2}\hat{u}_3(t,k;\lambda)=0,\label{171103:1628}
\end{equation}
while its associated initial condition is given by \eqref{171103:1521} as
\begin{equation}
\hat{u}_3(t,k;\lambda)\big|_{t=0}=\int_{-\infty}^\infty \frac{dx}{2\pi\lambda}e^{\frac{x^2}{2\lambda}} u_0(x) e^{-\frac{ikx}{\lambda}}.\label{171103:1629}
\end{equation}
The second step is then to explicitly solve the reduced initial-value problem \eqref{171103:1628}-\eqref{171103:1629}, which has the general solution
\begin{equation}
\hat{u}_3(t,k;\lambda)=C(k;\lambda)\frac{e^{\frac{k^2}{2(\lambda+2t)}}}{\sqrt{\lambda+2t}},
\;\;\text{where}\;\; C(k;\lambda)=\frac{e^{-\frac{k^2}{2\lambda}}}{\sqrt{\lambda}}\int_{-\infty}^\infty \frac{dx}{2\pi}e^{\frac{x^2}{2\lambda}} u_0(x) e^{-\frac{ikx}{\lambda}}.\label{171108:0840}
\end{equation}
The third and last step is then to insert this result back into the representation \eqref{171103:1503} to finally obtain the general
solution of the Cauchy problem \eqref{171103:1332}, as it is given by \eqref{151113:1713}
\begin{align}
u(t,x)&=\int_{-\infty}^\infty \!dk\: C(k;\lambda)\frac{e^{\frac{k^2}{2(\lambda+2t)}}}{\sqrt{\lambda+2t}} \,e^{-\frac{x(x-2ik)}{2(\lambda+2t)}}\nonumber\\[0.25em]
&= \frac{1}{2\pi\sqrt{\lambda(\lambda+2t)}}\int_{-\infty}^\infty \! dx^\prime u_0(x^\prime) e^{\frac{x^{\prime 2}}{2\lambda}-\frac{x^{2}}{2(\lambda+2t)}}
\!\int_{-\infty}^\infty \! dk\, e^{-\frac{k^2t}{\lambda(\lambda+2t)}}
e^{-ik\big(\frac{x^\prime}{\lambda}-\frac{x}{\lambda+2t}\big)}\nonumber\\[0.0em]
&\hspace{-0.4cm}\underset{t\geq 0,\lambda>0}{=}\;
\frac{1}{2\pi\sqrt{\lambda(\lambda+2t)}}\int_{-\infty}^\infty \! dx^\prime u_0(x^\prime) e^{\frac{x^{\prime 2}}{2\lambda}-\frac{x^{2}}{2(\lambda+2t)}}
\cdot 2\pi{\textstyle \sqrt{\lambda(\lambda+2t)}}\,\frac{e^{-\frac{(2tx^\prime+(x^\prime-x)\lambda)^2}{4t\lambda(\lambda+2t)}}}{\sqrt{4\pi t}}\hspace{0.75cm}\nonumber\\[0.75em]
&=\frac{1}{\sqrt{4\pi t}}\int_{-\infty}^\infty dx^\prime
u_0(x^\prime)e^{-\frac{(x-x^\prime)^2}{4t}}.\label{171105:1604}
\end{align}
\vspace{-0.5em}\phantom{x}

\subsubsection{Solution procedure within a new reference frame defined by a symmetry transformation\label{S232}}

In this approach we exploit the fact that the symmetry algebra $\mathfrak{f}_3$ is mathematically equivalent to~$\mathfrak{f}_2$. Hence, to construct the solution of the Cauchy problem \eqref{171103:1332} based on  $\mathfrak{f}_3$, no new equations need to be derived and to be solved as those already presented and discussed for the symmetry algebra $\mathfrak{f}_2$ in Section \ref{S22}. Based on the result \eqref{171026:1907}-\eqref{171026:1908}, let us consider a change of reference frame defined by the following coordinate transformation that also goes along with the following local re-gauging of the scalar field
\begin{equation}
\mathsf{T}\!:\;\;\; \tilde{t}={\textstyle \frac{\gamma}{2}-\frac{\gamma^2}{2(\lambda+2t)}},\;\;\; \tilde{x}={\textstyle \frac{\gamma x}{\lambda+2t}},\;\;\;
\tilde{u}={\textstyle \sqrt{\frac{\lambda+2t}{\gamma}}}\, e^{\frac{x^2}{2(\lambda+2t)}} u,\;\;\text{with}\;\;\gamma=1,\;\; [\gamma]=[t],\label{171104:1222}
\end{equation}
which represents the particular symmetry transformation $\mathsf{S}_\mathsf{1}\circ\mathsf{S}_\mathsf{6}$ \eqref{171026:1908} derived from the equivalence condition \eqref{171026:1907}. The explicit construction of this transformation is given in Appendix~\ref{SA}, where in the above relation a dimensional factor $\gamma$ has been included in order to make the dimensional consistency explicitly visible, thus demonstrating that the transformed variables $(\tilde{t},\tilde{x},\tilde{u})$ carry the same dimension as the original untransformed ones $(t,x,u)$, respectively.

Now, since the change of frame \eqref{171104:1222} is chosen such that as a symmetry transformation it not only leaves the defining dynamical equation of the Cauchy problem \eqref{171103:1332} invariant
\begin{equation}
\partial _{\tilde{t}} \tilde{u}(\tilde{t},\tilde{x};\lambda) = \partial_{\tilde{x}}^2 \tilde{u}(\tilde{t},\tilde{x};\lambda),\quad\;\;
\tilde{u}(\tilde{t},\tilde{x};\lambda)\big|_{\tilde{t}=\frac{\gamma}{2}-\frac{\gamma^2}{2\lambda}}={\textstyle \sqrt{\frac{\lambda}{\gamma}}}\, e^{\frac{\lambda \tilde{x}^2}{2\gamma^2}}u_0\big({\textstyle \frac{\lambda}{\gamma}\tilde{x}}\big),\quad\;\;
(\tilde{t},\tilde{x})\in\mathbb{R}_\lambda\times\mathbb{R},\label{171104:1615}
\end{equation}
but that it also reduces the equation's symmetry algebra from $\mathfrak{f}_3$ to $\mathfrak{f}_2$
\begin{equation}
\mathfrak{f}_3=\{\lambda X_2+X_5+\alpha X_3\}\overset{\mathsf{T}}{\underset{\eqref{171104:1640}}=}\{\tilde{X}_2+\tilde{\alpha}\tilde{X}_3\}=\tilde{\mathfrak{f}}_2,
\;\;\text{with $\,\tilde{\alpha}=\alpha/\gamma$},\label{171104:1644}
\end{equation}
we thus yield in this new frame for $\mathfrak{f}_3$ the same mathematical equations (up to the initial condition) as for $\mathfrak{f}_2$ in the originally formulated inertial frame. Hence, according to the already performed derivation in Section \ref{S22}, the general solution of the Cauchy problem within the new frame \eqref{171104:1615} based on the reduced symmetry algebra \eqref{171104:1644} is thus given by
\begin{align}
\tilde{u}(\tilde{t},\tilde{x};\lambda)&=\int_{-\infty}^\infty d\tilde{k}\, C(\tilde{k};\lambda)\, e^{-\tilde{k}^2\tilde{t}} e^{i\tilde{k}\tilde{x}},\;\:\text{where}\;\:
C(\tilde{k};\lambda)=e^{\tilde{k}^2\big(\frac{\gamma}{2}-\frac{\gamma^2}{2\lambda}\big)}\!\int_{-\infty}^\infty \frac{d\tilde{x}}{2\pi}{\textstyle\sqrt{\frac{\lambda}{\gamma}}}
\,e^{\frac{\lambda\tilde{x}^2}{2\gamma^2}}u_0\big({\textstyle \frac{\lambda}{\gamma}\tilde{x}}\big)e^{-i\tilde{k}\tilde{x}}\nonumber\\[0.25em]
&={\textstyle\sqrt{\frac{\lambda}{\gamma}}}\int_{-\infty}^\infty \frac{d\tilde{x}^\prime}{2\pi}e^{\frac{\lambda\tilde{x}^{\prime 2}}{2\gamma^2}}u_0\big({\textstyle \frac{\lambda}{\gamma}\tilde{x}^\prime}\big)
\!\int_{-\infty}^\infty d\tilde{k}\, e^{-\tilde{k}^2\big(\tilde{t}-\frac{\gamma}{2}+\frac{\gamma^2}{2\lambda}\big)}e^{i\tilde{k}(\tilde{x}-\tilde{x}^\prime)}\nonumber\\[0.25em]
&\hspace{-0.10cm}\underset{\gamma=1}{=} {\textstyle\sqrt{\lambda}}\int_{-\infty}^\infty \frac{d\tilde{x}^\prime}{2\pi}e^{\frac{\lambda\tilde{x}^{\prime 2}}{2}}u_0\big({\textstyle \lambda\tilde{x}^\prime}\big)
\cdot {\textstyle\frac{\sqrt{2\pi\lambda}}{\sqrt{1+\lambda(2\tilde{t}-1)}}}e^{-\frac{\lambda (\tilde{x}-\tilde{x}^\prime)^2}{2-2\lambda+4\lambda\vphantom{A^{a^.}}\tilde{t}}}\nonumber\\[0.25em]
&= {\textstyle\frac{1}{\sqrt{2\pi+2\pi\lambda(2\tilde{t}-1)}}}\int_{-\infty}^\infty d\tilde{x}^{\prime\prime} u_0(\tilde{x}^{\prime\prime}) e^{\frac{\tilde{x}^{\prime\prime 2}}{2\lambda}}
e^{-\frac{(\lambda\tilde{x}-\tilde{x}^{\prime\prime})^2}{2\lambda-2\lambda^2+4\lambda^2\tilde{t}}},
\end{align}
where in the last line the integration variable $\tilde{x}^\prime$ has been renamed to $\tilde{x}^{\prime\prime}=\lambda\tilde{x}^\prime$ having the same integration range as $\tilde{x}^\prime$. If this result is now transformed back to the inertial frame according to the defining transformation rule \eqref{171104:1222}, along with renaming the integration (dummy) variable also back to the inertial frame notation $x^\prime$,
\begin{equation}
{\textstyle\sqrt{\lambda+2t}}\, e^{\frac{x^2}{2(\lambda+2t)}}u(t,x)=\frac{\sqrt{\lambda+2t}}{\sqrt{4\pi t}}\int_{-\infty}^\infty dx^\prime u_0(x^\prime)\, e^{\frac{x^{\prime 2}}{2\lambda}}\,
e^{-\frac{(\lambda x-(\lambda+2t)x^\prime)^2}{(\lambda+2t)4\lambda t}},
\end{equation}
we obtain, after simplifying this relation, the expected inertial frame result of the Cauchy problem \eqref{171103:1332}, as it is given by \eqref{151113:1713}
\begin{equation}
u(t,x)=\frac{1}{\sqrt{4\pi t}}\int_{-\infty}^\infty dx^\prime
u_0(x^\prime)e^{-\frac{(x-x^\prime)^2}{4t}}.\label{171105:1605}
\end{equation}

\subsubsection{Solution procedure within a new reference frame defined by a non-symmetry transformation\label{S233}}

In this approach we exploit the fact that the symmetry algebra $\mathfrak{f}_3$ is physically redundant to~$\mathfrak{f}_2$. Based on the already discussed result \eqref{171101:1447} in Appendix \ref{SA}, there exists an infinite family of different reference frames to which $\mathfrak{f}_3$ is redundant to $\mathfrak{f}_2$. As considered before in \eqref{171103:1007}, let us again choose that change to a new frame which has the lowest degree of mathematical complexity
\begin{equation}
\mathsf{T}\!:\;\;\; \tilde{t}=t,\;\;\; \tilde{x}={\textstyle \frac{\xi x}{\lambda+2t}},\;\;\; \tilde{u}=e^{\frac{x^2}{2(\lambda+2t)}}\cdot u,\label{171107:0823}
\end{equation}
where again the constant dimensional factor $\xi$ has been included to ensure that the transformed variables $(\tilde{t},\tilde{x},\tilde{u})$ carry the same physical dimension as the original untransformed ones $(t,x,u)$, respectively. Since \eqref{171107:0823} is not a symmetry transformation, the Cauchy problem \eqref{151113:1225} will thus transform into the new frame as the following mathematical formulation
\begin{equation}
\partial _{\tilde{t}} \tilde{u}(\tilde{t},\tilde{x};\lambda)+\frac{\tilde{u}(\tilde{t},\tilde{x};\lambda)}{\lambda+2\tilde{t}} = \frac{\xi^2\partial_{\tilde{x}}^2 \tilde{u}(\tilde{t},\tilde{x};\lambda)}{(\lambda+2\tilde{t})^2},\;\;\;
\tilde{u}(\tilde{t},\tilde{x};\lambda)\big|_{\tilde{t}=0}=e^{\frac{\lambda \tilde{x}^2}{2\xi^2}}u_0\big({\textstyle \frac{\lambda}{\xi}\tilde{x}}\big),\;\;\;
(\tilde{t},\tilde{x})\in\mathbb{R}^+_0\times\mathbb{R}.\;\label{171107:0951}
\end{equation}
Although mathematically different it still describes the same physical mechanism of a 1-D diffusion process, however, just in a new physical frame of reference. By construction, the considered symmetry algebra $\mathfrak{f}_3$ now reduces in this new frame to $\mathfrak{f}_2$ where the solution process has already been established in Section \ref{S22}. Indeed, since
\begin{equation}
\mathfrak{f}_3=\{\lambda X_2+X_5+\alpha X_3\}\overset{\mathsf{T}}{\underset{\eqref{171103:1025}}=}\{\tilde{X}_2+\tilde{\alpha}\tilde{X}_3\}=\tilde{\mathfrak{f}}_2,\;\;\text{where $\,\tilde{\alpha}=\alpha/\xi$},\label{171120:1433}
\end{equation}
the general solution of \eqref{171107:0951} can be represented as the spatial 1-D Fourier transform \eqref{171027:1940}
\begin{equation}
\tilde{u}(\tilde{t},\tilde{x};\lambda)= \int_{-\infty}^{\infty}d\tilde{k}\, \hat{\tilde{u}}_2(\tilde{t},\tilde{k};\lambda)\, e^{i\tilde{k}\tilde{x}},\label{171107:1044}
\end{equation}
with its inverse
\begin{equation}
\hat{\tilde{u}}_2(\tilde{t},\tilde{k};\lambda)=\int_{-\infty}^\infty \frac{d\tilde{x}}{2\pi} \, \tilde{u}(\tilde{t},\tilde{x};\lambda)\, e^{-i\tilde{k}\tilde{x}}.\label{171107:1045}
\end{equation}
Hence, since the symmetry algebra $\mathfrak{f}_3$ formulated in the inertial frame induces the same modal approach as $\mathfrak{f}_2$ in the new frame defined by $\mathsf{T}$ \eqref{171107:0823}, the symmetry algebra $\mathfrak{f}_3$ is physically redundant to~$\mathfrak{f}_2$. In other words, no new physical modes are generated by the symmetry algebra~$\mathfrak{f}_3$ than those already established by $\mathfrak{f}_2$. In particular, when inserting the symmetry-induced representation \eqref{171107:1044} into the governing system \eqref{171107:0951}, we yield for the shape $\hat{\tilde{u}}_2$ of the $\mathfrak{f}_2$-induced Fourier-mode $e^{i\tilde{k}\tilde{x}}$ the following invariantly reduced ODE-system
\begin{equation}
\partial_{\tilde{t}} \hat{\tilde{u}}_2(\tilde{t},\tilde{k};\lambda)+\frac{\lambda+2\tilde{t}+\xi^2\tilde{k}^2}{(\lambda+2\tilde{t})^2}\hat{\tilde{u}}_2(\tilde{t},\tilde{k};\lambda)=0,
\;\;\;
\hat{\tilde{u}}_2(\tilde{t},\tilde{k};\lambda)\big|_{\tilde{t}=0}=\int_{-\infty}^\infty \frac{d\tilde{x}}{2\pi} e^{\frac{\lambda \tilde{x}^2}{2\xi^2}}u_0\big({\textstyle \frac{\lambda}{\xi}\tilde{x}}\big)e^{-i\tilde{k}\tilde{x}},\;\;
\label{171107:1852}
\end{equation}
which, including the initial condition, is mathematically exactly the same invariantly reduced ODE-system (up~to the dimensional factor $\xi$) as for the shape $\hat{u}_3$ \eqref{171103:1628}-\eqref{171103:1629} of the $\mathfrak{f}_3$-induced mode in the inertial frame, which is clear since both shapes only differ by this normalization constant: $u_3(t,\alpha;\lambda)= \tilde{u}_2(\tilde{t},\tilde{\alpha};\lambda)/\xi$ (see \eqref{171101:0928} for the specific choice \eqref{171107:0823}), or, equally, by $\hat{u}_3(t,k;\lambda)=\hat{\tilde{u}}_2(\tilde{t},\tilde{k};\lambda)/\xi$, since $\alpha=ik$. Nevertheless, it is a remarkable and surprising result when recognizing here the fact that the connection $\mathsf{T}$ \eqref{171107:0823} is {\it not} a symmetry transformation. Hence, corresponding to the inertial frame solution \eqref{171108:0840}, the general solution in the new frame to~\eqref{171107:1852} is thus readily given by
\begin{equation}
\hat{\tilde{u}}_2(\tilde{t},\tilde{k};\lambda)=C(\tilde{k};\lambda)\frac{e^{\frac{\xi^2\tilde{k}^2}{2(\lambda+2\tilde{t})}}}{\sqrt{\lambda+2\tilde{t}\,}},
\;\;\text{where}\;\; C(\tilde{k};\lambda)=\sqrt{\lambda}\, e^{-\frac{\xi^2\tilde{k}^2}{2\lambda}}\int_{-\infty}^\infty \frac{d\tilde{x}}{2\pi} e^{\frac{\lambda \tilde{x}^2}{2\xi^2}}u_0\big({\textstyle \frac{\lambda}{\xi}\tilde{x}}\big)e^{-i\tilde{k}\tilde{x}},
\end{equation}
and, according to \eqref{171107:1044}, the general solution of the Cauchy problem \eqref{171107:0951} in the new frame thus by
\begin{align}
\tilde{u}(\tilde{t},\tilde{x};\lambda) &= \int_{-\infty}^\infty d\tilde{k}\,
C(\tilde{k};\lambda)\frac{e^{\frac{\xi^2\tilde{k}^2}{2(\lambda+2\tilde{t})}}}{\sqrt{\lambda+2\tilde{t}\,}}\, e^{i\tilde{k}\tilde{x}}\nonumber\\[0.25em]
&= \frac{\sqrt{\lambda}}{2\pi\sqrt{\lambda+2\tilde{t}\,}}
\int_{-\infty}^\infty d\tilde{x}^\prime\, e^{\frac{\lambda \tilde{x}^{\prime 2}}{2\xi^2}}u_0\big({\textstyle \frac{\lambda}{\xi}\tilde{x}^\prime}\big)
\int_{-\infty}^\infty d\tilde{k}\, e^{-\frac{\xi^2\tilde{k}^2}{2\lambda}+\frac{\xi^2\tilde{k}^2}{2(\lambda+2\tilde{t})}}\, e^{i\tilde{k}(\tilde{x}-\tilde{x}^\prime)}\nonumber\\[0.25em]
&\hspace{-0.7cm}\underset{\tilde{t}\geq 0,\lambda>0 ,\xi>0}{=}\;
\frac{\sqrt{\lambda}}{2\pi\sqrt{\lambda+2\tilde{t}\,}}
\int_{-\infty}^\infty d\tilde{x}^\prime\, e^{\frac{\lambda \tilde{x}^{\prime 2}}{2\xi^2}}u_0\big({\textstyle \frac{\lambda}{\xi}\tilde{x}^\prime}\big)\cdot
2\pi{\textstyle\sqrt{\smash[b]{\lambda(\lambda+2\tilde{t})}}}\,\frac{e^{-\frac{\lambda(\lambda+2\tilde{t})(\tilde{x}-\tilde{x}^\prime)^2}{4\xi^2\tilde{t}}}}{\xi\sqrt{4\pi\tilde{t}\,}}\;\;\;\;\nonumber\\[0.25em]
&= \frac{1}{\sqrt{4\pi\tilde{t}\,}}\int_{-\infty}^\infty d\tilde{x}^{\prime\prime}\, u_0(\tilde{x}^{\prime\prime})\,
e^{\frac{\tilde{x}^{\prime\prime 2}}{2\lambda}}\, e^{-\frac{(\lambda+2\tilde{t})(\lambda\tilde{x}/\xi-\tilde{x}^{\prime\prime})^2}{4\lambda\tilde{t}}},
\end{align}
where in the last line the integration variable $\tilde{x}^\prime$ has been renamed to $\tilde{x}^{\prime\prime}=\lambda\tilde{x}^\prime/\xi$ having the same dimension and integration range as $\tilde{x}^\prime$. If this result is now transformed back to the inertial frame according to the defining transformation rule \eqref{171107:0823}, along with renaming the integration (dummy) variable also back to the inertial frame notation $x^\prime$,
\begin{equation}
e^{\frac{x^2}{2(\lambda+2t)}}\, u(t,x) = \frac{1}{\sqrt{4\pi t}}\int_{-\infty}^\infty dx^{\prime}\, u_0(x^{\prime})\,
e^{\frac{x^{\prime 2}}{2\lambda}}\, e^{-\frac{(\lambda x-(\lambda+2t)x^\prime)^2}{(\lambda+2t)4\lambda t}},
\end{equation}
we obtain again, after simplification, the same expected inertial frame result \eqref{151113:1713} of the Cauchy problem \eqref{151113:1225} as in the two preceding $\mathfrak{f}_3$-induced solution approaches done before:
\begin{equation}
u(t,x)=\frac{1}{\sqrt{4\pi t}}\int_{-\infty}^\infty dx^\prime
u_0(x^\prime)e^{-\frac{(x-x^\prime)^2}{4t}}.\label{171109:1355}
\end{equation}

\subsubsection{Short r\'esum\'e of the $\mathfrak{f}_3$-induced solution approach  \label{S234}}

This Section \ref{S23} has shown in detail that the invariant solution \eqref{171027:2244} induced by the symmetry algebra $\mathfrak{f}_3$ did not reveal any new insight into the Cauchy problem \eqref{151113:1225}, neither mathematically nor physically: Section \ref{S232} demonstrated that $\mathfrak{f}_3$ is mathematically equivalent and Section~\ref{S233} that it is physically redundant to the structurally simpler symmetry algebra $\mathfrak{f}_2$. Hence, no new physical modes are generated by the symmetry algebra $\mathfrak{f}_3$ through its invariant solution~\eqref{171027:2244} than those already established by $\mathfrak{f}_2$, which, for $\alpha=ik$ ($k\in\mathbb{R}$), are just the spatial Fourier modes~\eqref{171027:1900}. In particular, Section \ref{S231} showed that if within $\mathfrak{f}_3$ no connection is made to $\mathfrak{f}_2$, then no mathematical guideline or physical interpretation is provided when trying to {\it fully} solve the Cauchy problem \eqref{151113:1225}. The reason is that the necessary inverse relation \eqref{171103:1521} of the general solution
within $\mathfrak{f}_3$ was only establishable with the help of the Fourier modes \eqref{171103:1127}-\eqref{171106:1524} of $\mathfrak{f}_2$ --- even the trial-and-error method \eqref{171102:0825}-\eqref{171103:1148} had to make use of the Fourier modes of $\mathfrak{f}_2$. Moreover, in contrast to $\mathfrak{f}_2$, the symmetry algebra $\mathfrak{f}_3$ introduces an additional parameter $\lambda$ that in the end even turns out to be irrelevant when considering all of its induced modes collectively. In other words, although $\lambda$ is part of the single-mode solution \eqref{171027:2244} with the role as a regulating parameter to avoid at the initial point $t=0$ a singularity, the general solution \eqref{171028:0947} describing the full dynamics of the originally formulated physical 1-D diffusion problem \eqref{151113:1225}, however, does not depend on it. A trying, therefore, to interpret $\lambda$ as a parameter of physical relevance would be delusive, since this parameter is nothing else than just a mathematical regulator on which the physics may not depend.

Hence, up to now, only the two common symmetry algebras $\mathfrak{f}_1$ and $\mathfrak{f}_2$ provide a {\it non}-redundant approach to solve the Cauchy problem \eqref{151113:1225}. The former reveals a temporal insight (via Laplace modes) and the latter a spatial insight (via Fourier modes) into the dynamical process of the 1-D diffusion problem. At this stage, however, it already is clear that $\mathfrak{f}_1$ and $\mathfrak{f}_2$ are the only two symmetry algebras that are non-redundant, simply because the physical problem \eqref{151113:1225} stated in this section is described by only two independent variables: a time and a 1-D space coordinate.

\subsection{Subalgebra $\mathfrak{f}_4$ --- Redundant approach\label{S24}}

As before, the aim here is again to derive the general solution \eqref{151113:1713} of the governing system \eqref{151113:1225}, however, now with the invariant-solution approach of the 1-D subalgebra $\mathfrak{f}_4$ \eqref{171027:1229}, which is group-theoretically inequivalent to all previously considered symmetry algebras. It is spanned by the single element basis
\begin{equation}
Y_4=2\tau X_1+X_4+\alpha X_3,\label{171110:1215}
\end{equation}
which turns the corresponding invariant surface condition $(Y_4 F)|_{F=0}=0$, for $F=u-u(t,x)$, into the following linear eigenvalue problem
\begin{equation}
\big(2\tau X_1+X_4\big) u =\alpha\, u,\label{171110:1216}
\end{equation}
where the group parameter $\alpha$ is then again identified as the eigenvalue, however, now to the $\tau$-dependent continuous operator $2\tau X_1+X_4=2(t+\tau)\partial_t+x\partial_x$ \eqref{171026:1546}. The general solution of~\eqref{171110:1216} is given by\footnote[2]{To note is again the conceptual difference between the two parameters $\alpha$ and $\tau$. While the former is identified as the eigenvalue of the problem \eqref{171110:1216}, the latter only serves as a regulating parameter in the single-mode solution~\eqref{171110:1217} to ensure a regular result at the initial point $t=0$.}
\begin{equation}
u(t,x)=u_4(\rho,\alpha;\tau)\,(1+t/\tau)^{\alpha/2},\;\;\;\text{where}\;\; \rho=\frac{x}{\sqrt{1+t/\tau}},\label{171110:1217}
\end{equation}
but where at this stage it is not clear yet, whether the spectrum $\alpha$ is continuous or discrete, or even whether it is real ($\alpha\in\mathbb{R}$) or not. For an answer, one option is to change the frame of reference and use the fact that the symmetry algebra $\mathfrak{f}_4$ is physically redundant to $\mathfrak{f}_1$, as explicitly shown in Appendix \ref{SB}. From an infinite set of coordinate transformations \eqref{171110:1322} to choose from, the one with the lowest degree of complexity is given~by
\begin{equation}
\mathsf{T}\!:\;\;\;\; \tilde{t}={\textstyle\frac{1}{2}}\tau\ln(1+t/\tau),\quad\;\;
\tilde{x}=\frac{x}{\sqrt{1+t/\tau}},\quad\;\;
\tilde{u}=u,\label{171110:1537}
\end{equation}
having the inverse
\begin{equation}
\mathsf{T}^{\mathsf{-1}}\!:\;\;\;\; t=\tau e^{2\tilde{t}/\tau}-\tau,\quad\;\;
x=e^{\tilde{t}/\tau}\tilde{x},\quad\;\;
u=\tilde{u}.\label{171110:1538}
\end{equation}
Hence, within this new frame the invariant solution \eqref{171110:1217} reduces to the far more simpler and already established invariant solution \eqref{171026:1323} of $\mathfrak{f}_1$:
\begin{align}
u(t,x)=u_4(\rho,\alpha;\tau)\,(1+t/\tau)^{\alpha/2}\;\; &\Leftrightarrow\;\;
\tilde{u}(\tilde{t},\tilde{x};\tau)=u_4(\tilde{x},\alpha;\tau)\, e^{\frac{\alpha\tilde{t}}{\tau}}\nonumber\\[0.25em]
&\hspace{2.25cm}\equiv {\textstyle\frac{1}{\tau}}\,\tilde{u}_1(\tilde{x},\tilde{\alpha};\tau)\, e^{\tilde{\alpha}\tilde{t}},
\,\;\text{where $\,\tilde{\alpha}=\alpha/\tau$,}\label{171110:1433}
\end{align}
for which its corresponding non-local representation is given by the Laplace transformation, as it was already derived and discussed in Section \ref{S21},
\begin{equation}
\tilde{u}(\tilde{t},\tilde{x};\tau)= \int_{\sigma-i\infty}^{\sigma+i\infty}\frac{d\tilde{\alpha}}{2\pi i}\, \hat{\tilde{u}}_1(\tilde{x},\tilde{\alpha};\tau)\, e^{\tilde{\alpha}\tilde{t}},\;\,\text{where}\;\,
\sigma\in\mathbb{R},\: \tilde{\alpha}\in\mathbb{C},\: \tilde{t}\geq 0,\label{171110:1502}
\end{equation}
with the inverse transformation then given~as
\begin{equation}
\hat{\tilde{u}}_1(\tilde{x},\tilde{\alpha};\tau)=\int_0^\infty\! d\tilde{t} \, \tilde{u}(\tilde{t},\tilde{x};\tau)\, e^{-\tilde{\alpha}\tilde{t}}.\label{171110:1503}
\end{equation}
Now, when transforming this (Laplace-mode) representation back to the inertial frame according to the defining transformation \eqref{171110:1537}, we obtain the wanted invertible non-local representation induced by $\mathfrak{f}_4$ \eqref{171110:1215}
\begin{equation}
u(t,x)= \int_{\sigma-i\infty}^{\sigma+i\infty}\frac{d\alpha}{2\pi i}\, \hat{u}_4\Big(\text{\raisebox{0.075cm}{$\textstyle\frac{x}{\sqrt{1+t/\tau}}$}},\alpha;\tau\Big)
(1+t/\tau)^{\alpha/2},\;\;\;\:\sigma\in\mathbb{R},\: \alpha\in\mathbb{C},\: t\geq 0,\label{171110:2010}
\end{equation}
with its inverse
\begin{equation}
\hat{u}_4(x,\alpha;\tau)=\int_0^\infty \frac{dt}{2\tau(1+t/\tau)}u\Big(t,x{\textstyle\sqrt{1+t/\tau}}\Big)(1+t/\tau)^{-\alpha/2},\label{171110:2011}
\end{equation}
and with the combined result that the spectrum $\alpha$ of the associated eigenvalue problem \eqref{171110:1216} is complex and continuous, thus answering our question we stated above.
Also, as it was the case for the regulating parameter $\lambda$ of $\mathfrak{f}_3$ in the previous section, it should be noted again that the integral evaluation of a representation as \eqref{171110:2010} cannot depend on any regulating parameter, here $\tau$, simply because the general solution $u$ on its left-hand side does not depend on it, which ultimately is based on the fact that the originally formulated Cauchy problem \eqref{151113:1225} is independent to $\tau$.

The next two subsections will now focus on the proposed aim to fully solve the Cauchy problem \eqref{151113:1225} with the invariant-solution approach of the given symmetry algebra $\mathfrak{f}_4$ \eqref{171110:1215}. This will be done in two different ways: Once directly in the originally formulated inertial frame, and once in a changed reference frame defined by the specific non-symmetry transformation \eqref{171110:1537}.

\subsubsection{Direct solution procedure within the originally formulated inertial frame\label{S241}}

Unfortunately the $\mathfrak{f}_4$-induced non-local representation \eqref{171110:2010}-\eqref{171110:2011} is not suitable for to solve the given Cauchy problem \eqref{151113:1225}, due to that the initial condition cannot be implemented into the general solution in a straightforward way. Hence, within the inertial frame, we thus fall back to the invariant single-mode solution \eqref{171110:1217} of $\mathfrak{f}_4$, which, when inserted into \eqref{151113:1225}, reduces the PDE of the Cauchy problem to the following second-order ODE
\begin{equation}
2\tau\partial^2_\rho u_4(\rho,\alpha;\tau)+\rho\partial_\rho u_4(\rho,\alpha;\tau)-\alpha u_4(\rho,\alpha;\tau)=0.\label{171111:1730}
\end{equation}
When trying to solve this equation for its general solution, it is crucial to recognize two important properties of this ODE: Firstly, it represents an own linear eigenvalue problem that is independent of the underlying  linear eigenvalue problem \eqref{171110:1216} induced by the invariant surface condition of $\mathfrak{f}_4$. Although \eqref{171111:1730} is based on the continuous second-order operator $L:=2\tau\partial^2_\rho+\rho\partial_\rho$ and \eqref{171110:1216} on the structurally different first-order operator $2\tau X_1+X_4=2(t+\tau)\partial_t+x\partial_x$, both eigenvalue problems refer to the {\it same} eigenvalue $\alpha$. Hence, the current eigenvalue spectrum of the symmetry operator $2\tau X_1+X_4$, being continuous and complex $(\alpha\in\mathbb{C})$, will now be restricted by the additional operator $L$, emerging from the governing equation \eqref{151113:1225} that admits this symmetry~$2\tau X_1+X_4$. Secondly, the ODE \eqref{171111:1730} is not just any eigenvalue problem, but one of a very specific type, namely that of a
particular Sturm-Liouville problem. Since \eqref{171111:1730} can be equivalently rewritten as
\begin{equation}
L \, u_4(\rho,\alpha;\tau) = \alpha\, u_4(\rho,\alpha;\tau),\;\;\;\text{with}\;\; L=2\tau e^{-\frac{\rho^2}{4\tau}}\partial_\rho
\Big(e^{\frac{\rho^2}{4\tau}}\partial_\rho\Big),\label{171111:1854}
\end{equation}
that means, in terms of the weighted self-adjoint Sturm-Liouville operator $L=\text{\raisebox{0.045cm}{$\frac{1}{w}(-\frac{d}{d\rho}\, p\, \frac{d}{d\rho}+q)$}}$,
where $q=0$ and $2\tau w= -p= \text{\raisebox{-0.025cm}{$e^{\rho^2/4\tau}$}}$ being defined on the unbounded interval $\rho\in (-\infty,\infty)$, the type of Sturm-Liouville problem \eqref{171111:1854} is classified as a singular one (see e.g. \cite{Zettl05}), where, for example, the spectrum can have a discrete and a continuous part. For our purposes, however, it is already sufficient to only extract those eigenfunctions from the general solution of \eqref{171111:1854} which on the lowest degree of complexity offer the following properties: Completeness and orthogonality, or any inverse relation to be able to implement the initial condition and the property to satisfy the boundary conditions of a sufficiently fast solution decay at both time and space infinity. Indeed, such a complete set of regular solutions can be extracted from equation \eqref{171111:1854} if the eigenvalues $\alpha$ are restricted to be real, integer, negative and non-zero, i.e., if $\alpha=-(n+1)$, $n\in\mathbb{N}_0$. The associated eigenfunctions are then given by
\begin{equation}
u_4(\rho,\alpha;\tau)=c(\alpha;\tau)\frac{d^{-(\alpha+1)}}{d\rho^{-(\alpha+1)}}
e^{-\frac{\rho^2}{4\tau}},\;\;\;\alpha=-(n+1),
\; n\in\mathbb{N}_0,
\end{equation}
which can be rewritten in terms of Hermite polynomials
\begin{align}
u_4(\rho,\alpha;\tau)&
=c(\alpha;\tau)\frac{d^{-(\alpha+1)}}{d\rho^{-(\alpha+1)}}
e^{-\frac{\rho^2}{4\tau}}=c(\alpha;\tau)e^{-\frac{\rho^2}{4\tau}}(-1)^{-2(\alpha+1)}
e^{\frac{\rho^2}{4\tau}}\frac{d^{-(\alpha+1)}}{d\rho^{-(\alpha+1)}}
e^{-\frac{\rho^2}{4\tau}}\nonumber\\[0.5em]
& =
c(\alpha;\tau)e^{-\frac{\rho^2}{4\tau}}(-1)^{-(\alpha+1)}(4\tau)^{-\frac{\alpha+1}{2}}
H_{-(\alpha+1)}\!\left({\textstyle\frac{\rho}{\sqrt{4\tau}}}\right)\nonumber\\[0.5em]
&=\hat{c}(\alpha;\tau)e^{-\frac{\rho^2}{4\tau}}
H_{-(\alpha+1)}\!\left({\textstyle\frac{\rho}{\sqrt{4\tau}}}\right),
\end{align}
where $\hat{c}$ is some arbitrary integration function being independent of the ODE's \eqref{171111:1854} defining variable~$\rho$.
Hence, due to the linear superposition principle, the general invariant solution of the governing PDE \eqref{151113:1225} is thus obtained by summing the
particular result \eqref{171110:1217} over all (discrete) eigenvalues $\alpha=-(n+1)$, $n\in\mathbb{N}_0$:
\begin{equation}
u(t,x)=e^{-\frac{\rho^2}{4\tau}}\sum_{n=0}^\infty
\frac{\hat{c}_n(\tau)}{(1+t/\tau)^{\frac{n+1}{2}}}
H_{n}\!\left({\textstyle\frac{\rho}{\sqrt{4\tau}}}\right),\quad\;\text{where}\;\;\;
\rho=\frac{x}{\sqrt{1+t/\tau}}.\label{151112:0941}
\end{equation}
The still arbitrary integration constants $\hat{c}_n(\tau):=\hat{c}(-(n+1);\tau)$ are now
fixed through the given initial condition $u_0$ of the governing Cauchy
problem \eqref{151113:1225}: In making use of the orthogonality
relation for the Hermite polynomials
\begin{equation}
\int_{-\infty}^\infty d\sigma\, e^{-\sigma^{2}}
H_{n}(\sigma)H_{m}(\sigma)=\sqrt{\pi}\,2^{n}\, n!\,\delta_{nm},
\end{equation}
and evaluating \eqref{151112:0941} at $t=0$, will give the
expansion coefficients $\hat{c}_n$ as
\begin{equation}
\hat{c}_n(\tau) = \frac{1}{\sqrt{4\pi\tau}\,2^{n}\,
n!}\int_{-\infty}^\infty dx^\prime
H_{n}\!\left({\textstyle\frac{x^\prime}{\sqrt{4\tau}}}\right)u_0(x^\prime).
\label{151112:0942}
\end{equation}
To make a connection to the general solution \eqref{151113:1713},
we follow \cite{Yaremko14}, in which we first have to spatially
differentiate the following auxiliary (Fourier-transform) relation
\begin{equation}
\frac{1}{2\pi}\int_{-\infty}^\infty dk\, e^{-\tau k^2(1+t/\tau)}\,
e^{ikx}=\frac{1}{\sqrt{4\pi\tau (1+t/\tau)}}\,
e^{-\frac{x^2}{4\tau (1+t/\tau)}},
\end{equation}
$n$-times in order to be then rewritten as
\begin{multline}
\frac{1}{2\pi}\int_{-\infty}^\infty dk\, (ik)^n\, e^{-\tau
k^2(1+t/\tau)}\, e^{ikx}
=\frac{1}{\sqrt{\pi}}(4\tau)^{-\frac{1}{2}}(1+t/\tau)^{-\frac{1}{2}}
\frac{d^n}{dx^n}\, e^{-\frac{x^2}{4\tau
(1+t/\tau)}}\\[0.5em]
=\frac{1}{\sqrt{\pi}}\,(4\tau)^{-\frac{n+1}{2}}(1+t/\tau)^{-\frac{n+1}{2}}\,
 e^{-\frac{x^2}{4\tau
(1+t/\tau)}}\, (-1)^n\,
H_n\!\left({\textstyle\frac{x}{\sqrt{4\tau(1+t/\tau)}}}\right).\;\;
\end{multline}
Hence, we obtain the relation
\begin{equation}
e^{-\frac{\rho^2}{4\tau}}H_n\!\left({\textstyle\frac{\rho}{\sqrt{4\tau}}}\right)=
\frac{(-1)^n}{2\sqrt{\pi}}\,
(4\tau)^{\frac{n+1}{2}}(1+t/\tau)^{\frac{n+1}{2}}\int_{-\infty}^\infty
dk\, (ik)^n\, e^{-\tau k^2(1+t/\tau)}\, e^{ikx},\label{171116:1620}
\end{equation}
which is to be inserted into \eqref{151112:0941}. Collecting then
all terms over which is being summed, including those in
\eqref{151112:0942}, and rewriting them as
\begin{multline}
\sum_{n=0}^\infty
H_{n}\!\left({\textstyle\frac{x^\prime}{\sqrt{4\tau}}}\right)
(4\tau)^{\frac{n+1}{2}}(1+t/\tau)^{\frac{n+1}{2}}(1+t/\tau)^{-\frac{n+1}{2}}\frac{(-1)^n\,
(ik)^n}{2^n\, n!}\\[0.5em]
=(4\tau)^{\frac{1}{2}}\sum_{n=0}^\infty
H_{n}\!\left({\textstyle\frac{x^\prime}{\sqrt{4\tau}}}\right)
\frac{\left(-\frac{1}{2}\, ik\,
(4\tau)^{\frac{1}{2}}\right)^n}{n!},\quad \label{151112:0959}
\end{multline}
we can make use of the following generating function for
the Hermite polynomials
\begin{equation}
e^{2\varphi\zeta-\varphi^2}=\sum_{n=0}^\infty
H_n(\zeta)\frac{\varphi^n}{n!},
\end{equation}
to then formulate \eqref{151112:0959} as
\begin{equation}
\sum_{n=0}^\infty
H_{n}\!\left({\textstyle\frac{x^\prime}{\sqrt{4\tau}}}\right)
(4\tau)^{\frac{n+1}{2}}\frac{(-1)^n\, (ik)^n}{2^n\, n!}=
(4\tau)^{\frac{1}{2}}e^{-ikx^\prime+\tau k^2}.\label{171116:1634}
\end{equation}
Hence, solution \eqref{151112:0941} can be ultimately written as
the general solution \eqref{151113:1713}
\begin{align}
u(t,x) & =
\frac{(4\tau)^{\frac{1}{2}}}{2\sqrt{\pi}\sqrt{4\pi\tau}}
\int_{-\infty}^\infty dx^\prime\, u_0(x^\prime)
\int_{-\infty}^\infty dk\, e^{-\tau k^2(1+t/\tau)-ikx^\prime+\tau
k^2}\, e^{ikx}\nonumber\\[0.5em]
& = \frac{1}{\sqrt{4\pi t}}\int_{-\infty}^\infty dx^\prime
u_0(x^\prime)e^{-\frac{(x-x^\prime)^2}{4t}}. \label{151112:1000}
\end{align}
Remarkable here again is to see how the regularization parameter
$\tau$, which is needed in the single-mode solution \eqref{171110:1217} to ensure a regular
result at the initial point $t=0$, disappears and becomes not relevant anymore in its integral representation
\eqref{151112:1000} when considering all modes collectively.

\subsubsection{Solution procedure within a new reference frame defined by a non-symmetry transformation\label{S242}}

In this approach we exploit the fact that the symmetry algebra $\mathfrak{f}_4$ is physically redundant to~$\mathfrak{f}_1$. Based on the already discussed result \eqref{171110:1322} in Appendix \ref{SB}, there exists an infinite family of different reference frames to which $\mathfrak{f}_4$ is redundant to $\mathfrak{f}_1$. As considered before in \eqref{171110:1537}, let us again choose that change to a new frame which has the lowest degree of mathematical complexity
\begin{equation}
\mathsf{T}\!:\;\;\;\; \tilde{t}={\textstyle\frac{1}{2}}\tau\ln(1+t/\tau),\quad\;\;
\tilde{x}=\frac{x}{\sqrt{1+t/\tau}},\quad\;\;
\tilde{u}=u,\label{171112:1051}
\end{equation}
where again it is to note that the transformed variables $(\tilde{t},\tilde{x},\tilde{u})$ carry the same physical dimension as the original untransformed ones $(t,x,u)$, respectively. Since \eqref{171112:1051} is not a symmetry transformation, the Cauchy problem \eqref{151113:1225} will thus transform into the new frame as the following mathematical formulation
\begin{equation}
\tau\partial _{\tilde{t}} \tilde{u}(\tilde{t},\tilde{x};\tau)-\tilde{x}\partial_{\tilde{x}}\tilde{u}(\tilde{t},\tilde{x};\tau)-2\tau\partial_{\tilde{x}}^2 \tilde{u}(\tilde{t},\tilde{x};\tau)=0,\;\;\;
\tilde{u}(\tilde{t},\tilde{x};\tau)\big|_{\tilde{t}=0}=u_0(\tilde{x}),\;\;\;
(\tilde{t},\tilde{x})\in\mathbb{R}^+_0\times\mathbb{R}.\;\label{171112:1054}
\end{equation}
Although mathematically different it still describes the same physical mechanism of a 1-D diffusion process, however, just in a new physical frame of reference. By construction, the considered symmetry algebra $\mathfrak{f}_4$ now reduces in this new frame to $\mathfrak{f}_1$ where the solution process has already been established in Section \ref{S21}. Indeed, since
\begin{equation}
\mathfrak{f}_4=\{2\tau X_1+X_4+\alpha X_3\}\overset{\mathsf{T}}{\underset{\eqref{171110:1045}}=}\{\tilde{X}_1+\tilde{\alpha}\tilde{X}_3\}=\tilde{\mathfrak{f}}_1,\;\;\text{where $\,\tilde{\alpha}=\alpha/\tau$},
\label{171113:1704}
\end{equation}
the general solution of \eqref{171112:1054} can be represented as the temporal 1-D Laplace transform\footnote[2]{Important to remark here is that in the inertial frame a real discrete subset of the spectrum $\alpha\in\mathbb{C}$ has been used, in particular the subset $\alpha\in\mathbb{Z}^-\!\subset\mathbb{C}$ was used, to fully represent the general solution \eqref{151112:0941}, while with the ansatz \eqref{171112:1230} in the new frame no such restriction need to be made yet --- at this stage of the solution process the full continuous and complex spectrum of $\alpha=\tau\tilde{\alpha}\in\mathbb{C}$ will be assumed. The final result, however (see further in the text), shows that this spectrum need at most only to be restricted to $\text{Re}(\tilde{\alpha})<0$, i.e. to $\sigma<0$, when aiming to represent the general solution of~\eqref{171112:1054} as the Laplace integral \eqref{171112:1230}.} \eqref{171205:1837}
\begin{equation}
\tilde{u}(\tilde{t},\tilde{x};\tau)= \int_{\sigma-i\infty}^{\sigma+i\infty}\frac{d\tilde{\alpha}}{2\pi i}\, \hat{\tilde{u}}_1(\tilde{x},\tilde{\alpha};\tau)\, e^{\tilde{\alpha}\tilde{t}},\;\;
\sigma\in\mathbb{R},\: \tilde{\alpha}\in\mathbb{C},\: \tilde{t}\geq 0,\label{171112:1230}
\end{equation}
with its inverse
\begin{equation}
\hat{\tilde{u}}_1(\tilde{x},\tilde{\alpha};\tau)=\int_0^\infty\! d\tilde{t} \, \tilde{u}(\tilde{t},\tilde{x},\tilde{\alpha};\tau)\, e^{-\tilde{\alpha}\tilde{t}}.\label{171112:1231}
\end{equation}
Hence, since the symmetry algebra $\mathfrak{f}_4$ formulated in the inertial frame induces the same defining Laplace-modes as $\mathfrak{f}_1$ in the new frame connected by $\mathsf{T}$ \eqref{171112:1051}, the symmetry algebra $\mathfrak{f}_4$ is physically redundant to~$\mathfrak{f}_1$. In other words, no new physical modes are generated by the symmetry algebra~$\mathfrak{f}_4$ than those already established by~$\mathfrak{f}_1$. As it was derived above in \eqref{171110:1433}, the shape $\tilde{u}_1$ of the $\mathfrak{f}_1$-induced Laplace-mode $e^{-\tilde{\alpha}\tilde{t}}$ is related to the shape $u_4$ of the $\mathfrak{f}_4$-induced mode $(1+t/\tau)^{\alpha/2}$ simply by:
$u_4(\rho,\alpha;\tau)= \tilde{u}_1(\tilde{x},\tilde{\alpha};\tau)/\tau$, or, equally, by $\hat{u}_4(\rho,\alpha;\tau)=\hat{\tilde{u}}_1(\tilde{x},\tilde{\alpha};\tau)/\tau$, since $\hat{\tilde{u}}_1\!=\!2\pi i\tilde{u}_1$.\linebreak[4]
In particular, when transforming the governing PDE-system \eqref{171112:1054} according to the Laplace transform \eqref{171112:1231}, as it was done in Section \ref{S21} in \eqref{151130:2331}, we yield for the shape $\hat{\tilde{u}}_1$ of the $\mathfrak{f}_1$-induced Laplace mode $e^{-\tilde{\alpha}\tilde{t}}$ the following invariantly reduced ODE-system
\begin{equation}
-\tau u_0(\tilde{x})+\tau\tilde{\alpha}\hat{\tilde{u}}_1(\tilde{x},\tilde{\alpha};\tau)-\tilde{x}\partial_{\tilde{x}}\hat{\tilde{u}}_1(\tilde{x},\tilde{\alpha};\tau)
-2\tau\partial^2_{\tilde{x}}\hat{\tilde{u}}_1(\tilde{x},\tilde{\alpha};\tau)=0,
\label{171112:1218}
\end{equation}
which explicitly involves the initial condition
$u_0$ of the Cauchy problem \eqref{151113:1225}, where we assumed that the solution $\tilde{u}$ temporally decays faster asymptotically than $e^{\text{Re}(\tilde{\alpha})\tilde{t}}$. The general
solution of this ODE \eqref{171112:1218} is given (already in symmetrized form) as\footnote[2]{In \eqref{171112:1659}, the $H(\nu,z)$ are the Hermite functions defined to satisfy the differential equation $w''\!-2zw'\!+2\nu w\!=\!0$, for any $\nu\in\mathbb{C}$, which reduce to the Hermite polynomials if $\nu=n\in\mathbb{N}$, while $\hyf(a,b,z)$ is Kummer's confluent hypergeometric function defined to satisfy the differential equation
$zw''+(b-z)w'-aw=0$, for any $a,b\in\mathbb{C}.$}
\begin{align}
\hat{\tilde{u}}_1(\tilde{x},\tilde{\alpha};\tau) &= C_1(\tilde{\alpha};\tau)e^{-\tilde{\eta}^2}\!H(-2\tilde{a},\tilde{\eta})+ C_2(\tilde{\alpha};\tau)e^{-\tilde{\eta}^2}\!\hyf(\tilde{a},\tilde{b},\tilde{\eta}^2)
\label{171112:1659}\\[0.2em]
&\hspace{0.45cm} +\int_{-\infty}^{\tilde{x}}\frac{d\tilde{x}^\prime u_0(\tilde{x}^\prime)}{\Omega(\tilde{a},\tilde{b},\tilde{x}^\prime;\tau)}e^{-\tilde{\eta}^2+\tilde{\eta}^{\prime 2}}
\!\Big( H(-2\tilde{a},\tilde{\eta})\,\hyf(\tilde{a},\tilde{b},\tilde{\eta}^{\prime 2})\!-\!H(-2\tilde{a},\tilde{\eta}^\prime)\,\hyf(\tilde{a},\tilde{b},\tilde{\eta}^{2})\Big)\nonumber\\[0.2em]
&\hspace{0.45cm} +\int_{\tilde{x}}^\infty\frac{d\tilde{x}^\prime u_0(\tilde{x}^\prime)}{\Omega(\tilde{a},\tilde{b},\tilde{x}^\prime;\tau)}e^{-\tilde{\eta}^2+\tilde{\eta}^{\prime 2}}
\!\Big( H(-2\tilde{a},\tilde{\eta}^\prime)\,\hyf(\tilde{a},\tilde{b},\tilde{\eta}^{2})\!-\!H(-2\tilde{a},\tilde{\eta})\,\hyf(\tilde{a},\tilde{b},\tilde{\eta}^{\prime 2})\Big),\nonumber
\end{align}
where
\begin{gather*}
\Omega(\tilde{a},\tilde{b},\tilde{x}^\prime;\tau)=\frac{4\tilde{a}}{\tau}\Big(2\sqrt{\tau}H(-2\tilde{a}-1,\tilde{\eta}^\prime)\,\hyf(\tilde{a},\tilde{b},\tilde{\eta}^{\prime 2})
+\tilde{x}^\prime H(-2\tilde{a},\tilde{\eta}^\prime)\,\hyf(\tilde{a}+1,\tilde{b}+1,\tilde{\eta}^{\prime 2})\Big),\\[0.2em]
\tilde{\eta}=\frac{\tilde{x}}{2\sqrt{\tau}},\qquad \tilde{\eta}^\prime=\frac{\tilde{x}^\prime}{2\sqrt{\tau}},\qquad \tilde{a}={\textstyle\frac{1}{2}}(1+\tau\tilde{\alpha}),\qquad \tilde{b}={\textstyle\frac{1}{2}},
\end{gather*}
and where $C_1(\tilde{\alpha};\tau)$ and $C_2(\tilde{\alpha};\tau)$ are two arbitrary
integration functions which can only be determined by putting
spatial boundary conditions. But since the considered Cauchy
problem~\eqref{151113:1225}, or \eqref{171112:1054} within the new frame, is free of any spatial boundary
conditions, except for the very weak restriction that the initial
condition $u_0$ may not increase faster than
$e^{x^2}=e^{\tilde{x}^2}$\footnote[3]{At the initial time $t=0$, the untransformed and transformed spatial variable \eqref{171112:1051} coincide $x=\tilde{x}$.} at space
infinity (see footnote to solution \eqref{151113:1713}), we can
specify $C_1(\tilde{\alpha};\tau)$ and $C_2(\tilde{\alpha};\tau)$ freely. A reasonable specification
is of course to choose these functions such that no intermediate singularity in
the solution \eqref{171112:1659} arises when
$|\tilde{x}|\rightarrow\infty$. When looking at the asymptotic structure of~\eqref{171112:1659} under the condition $\lim_{|\tilde{x}|\to\infty}\hat{\tilde{u}}_1=0$, by using the fact that
$\lim_{|\tilde{\eta}|\to \infty}e^{-\tilde{\eta}^2}\!H(-2\tilde{a},\tilde{\eta})=0$ and $\lim_{|\tilde{\eta}|\to \infty}e^{-\tilde{\eta}^2}\!\hyf(\tilde{a},\tilde{b},\tilde{\eta}^2)=0$, for all $\text{Re}(-2\tilde{a})>-1$
\citep{Lebedev72}, and the fact that in the positive asymptotic regime $(\tilde{\eta}>0)\rightarrow\infty$ the former functions decay exponentially faster than the latter ones, then the integration functions can only be chosen as
\begin{equation}
\left.
\begin{aligned}
C_1(\tilde{\alpha};\tau)&=\int_{-\infty}^{\infty}\frac{d\tilde{x}^\prime u_0(\tilde{x}^\prime)}{\Omega(\tilde{a},\tilde{b},-\tilde{x}^\prime;\tau)}e^{\tilde{\eta}^{\prime 2}}\!\hyf(\tilde{a},\tilde{b},\tilde{\eta}^{\prime 2}),
\\[0.25em]
C_2(\tilde{\alpha};\tau)&=\int_{-\infty}^{\infty}\frac{d\tilde{x}^\prime u_0(\tilde{x}^\prime)}{\Omega(\tilde{a},\tilde{b},\tilde{x}^\prime;\tau)}e^{\tilde{\eta}^{\prime 2}}\!H(-2\tilde{a},\tilde{\eta}^\prime),
\end{aligned}
~~~\right\}\; \text{for all}\;\tilde{a}\in\mathbb{C},\:\text{Re}(-2\tilde{a})>-1.\label{171114:0847}
\end{equation}

\pagebreak[4]
\noindent If we assume for convenience $\tau>0$, then the restriction $\text{Re}(-2\tilde{a})>-1$ implies $\text{Re}(\tilde{\alpha})<0$, which again implies the restriction $\sigma<0$ in the Laplace integral \eqref{171112:1230}. Hence, we may only\linebreak consider the left complex half-plane of the symmetry-induced spectrum $\tilde{\alpha}$ \eqref{171113:1704} when aiming to represent the general solution of the governing PDE-system \eqref{171112:1054} as the temporal Laplace integral \eqref{171112:1230}. It is clear that this integral representation, when based on the solution \eqref{171112:1659}, is not analytically tractable anymore. Even its numerical evaluation turns out to be challenging and not straightforward, and hence it is desirable to find an alternative approach to proceed analytically. As before in the inertial frame developed in the foregoing Section \ref{S241}, we don't make the straight ansatz via the general representation \eqref{171112:1230}, but rather make a step back and consider its defining single-mode representation again:
\begin{equation}
\tilde{u}(\tilde{t},\tilde{x};\tau)=\tilde{u}_1(\tilde{x},\tilde{\alpha};\tau)\, e^{\tilde{\alpha}\tilde{t}},\label{171115:0845}
\end{equation}
which, when inserted into the governing equation \eqref{171112:1054}, reduces the PDE in the new frame exactly into the same Sturm-Liouville-ODE \eqref{171111:1730} as in the inertial frame
\begin{equation}
2\tau\partial^2_{\tilde{x}}\tilde{u}_1(\tilde{x},\tilde{\alpha};\tau)+\tilde{x}\partial_{\tilde{x}}\tilde{u}_1(\tilde{x},\tilde{\alpha};\tau)-\tau\tilde{\alpha}\tilde{u}_1(\tilde{x},\tilde{\alpha};\tau)=0.
\label{171114:1018}
\end{equation}
In other words, the shape $u_4$ of the $\mathfrak{f}_4$-induced modes in the inertial frame satisfies the same mathematical equation as the shape $\tilde{u}_1$ of the $\mathfrak{f}_1$-induced Laplace-modes in the new frame. On~the one side this result is to be expected since these two shapes, as derived in \eqref{171110:1433}, only differ by~a constant: $\tilde{u}_1=\tau u_4$, but on the other side this is a remarkable result when considering here the fact that the connection $\mathsf{T}$~\eqref{171112:1051} is {\it not} a symmetry transformation. Ultimately, this combined result \eqref{171115:0845}-\eqref{171114:1018} underpins once more the statement that the symmetry algebra~$\mathfrak{f}_4$ is physically redundant to $\mathfrak{f}_1$, in the very same manner as it was the case for $\mathfrak{f}_3$ to $\mathfrak{f}_2$ already discussed before in Section \ref{S233}. Hence, no new physical modes are generated by the symmetry algebra $\mathfrak{f}_4$ than those already established by $\mathfrak{f}_1$.

Now, in order to obtain the general solution of \eqref{171114:1018} and thus of the underlying Cauchy problem \eqref{171112:1054} in the new frame, we just have to make use of this shape-equivalence and follow the solution procedure as it was outlined before for the inertial frame in Section \ref{S241}, with the corresponding result\footnote[2]{Note that in the new frame the entire single-mode representation \eqref{171115:0845} differs mathematically of course from the corresponding single-mode representation \eqref{171110:1217} in the inertial frame --- a natural difference, which also would exist if the connecting frame-transformation would be a symmetry of the system's governing equation (see e.g. the connecting relation \eqref{171031:1101} induced by the symmetry transformation \eqref{171029:1739}, and then compare to the connecting relation \eqref{171110:1433} induced by the currently considered non-symmetry transformation \eqref{171110:1537}). That means, although the mathematical expressions of the intermediate results themselves in the new frame will differ from those in the inertial frame, the solution process itself, however, is nevertheless equivalent in both frames --- nothing new is to be expected here when solving for the general solution, simply because the shapes $\tilde{u}_1$ and $u_4$ defining the eigenfunctions \eqref{171115:0845} and \eqref{171110:1217} in each frame, respectively, only differ by a constant, $\tilde{u}_1=\tau u_4$, which for a non-symmetry transformation is still a remarkable result by itself.}
\begin{equation}
\tilde{u}(\tilde{t},\tilde{x};\tau)=\sum_{n=0}^\infty \tilde{u}_1(\tilde{x},\tilde{\alpha}_n;\tau) e^{-(n+1)\tilde{t}/\tau},\quad\; \tilde{\alpha}_n=-(n+1)/\tau,\; n\in\mathbb{N}_0,\label{171119:1208}
\end{equation}
where
\begin{equation*}
\tilde{u}_1(\tilde{x},\tilde{\alpha}_n;\tau)=\hat{\tilde{c}}_n(\tau)e^{-\frac{\tilde{x}^2}{4\tau}}H_{-(\tau\tilde{\alpha}_n+1)}\!\left({\textstyle\frac{\tilde{x}}{\sqrt{4\tau}}}\right),
\;\:\text{with}\;\: \hat{\tilde{c}}_n(\tau) = \frac{1}{\sqrt{4\pi\tau}\,2^{n}\,
n!}\int_{-\infty}^\infty d\tilde{x}^\prime
H_{n}\!\left({\textstyle\frac{\tilde{x}^\prime}{\sqrt{4\tau}}}\right)u_0(\tilde{x}^\prime).
\end{equation*}
Then, by using the intermediate result \eqref{171116:1620} in terms of the new-frame variables
\begin{equation}
e^{-\frac{\tilde{x}^2}{4\tau}}H_n\!\left({\textstyle\frac{\tilde{x}}{\sqrt{4\tau}}}\right)=
\frac{(-1)^n}{2\sqrt{\pi}}\,
(4\tau)^{\frac{n+1}{2}}(1+\tilde{t}/\tau)^{\frac{n+1}{2}}\int_{-\infty}^\infty
d\tilde{k}\, (i\tilde{k})^n\, e^{-\tau \tilde{k}^2(1+\tilde{t}/\tau)}\, e^{i\tilde{k}\tilde{x}\sqrt{1+\tilde{t}/\tau}},\label{171116:1623}
\end{equation}
and subsequently the intermediate result \eqref{171116:1634}, when adapted to the Laplace-mode $e^{\tilde{\alpha}_n\tilde{t}}$,
\begin{multline}
\sum_{n=0}^\infty
H_{n}\!\left({\textstyle\frac{\tilde{x}^\prime}{\sqrt{4\tau}}}\right)
(4\tau)^{\frac{n+1}{2}}(1+\tilde{t}/\tau)^{\frac{n+1}{2}}e^{-(n+1)\tilde{t}/\tau}
\frac{(-1)^n\, (i\tilde{k})^n}{2^n\, n!}\\[0.25em]
=(4\tau)^{\frac{1}{2}}(1+\tilde{t}/\tau)^{\frac{1}{2}}e^{-\tilde{t}/\tau}e^{-i\tilde{k}\tilde{x}^\prime e^{-\tilde{t}/\tau}\sqrt{1+\tilde{t}/\tau}+\tau \tilde{k}^2
e^{-2\tilde{t}/\tau}(1+\tilde{t}/\tau)},\quad\;\;\label{171116:1635}
\end{multline}
one directly obtains the general solution of the Cauchy problem \eqref{171112:1054} in the new frame:
\begin{align}
\tilde{u}(\tilde{t},\tilde{x};\tau)&=\frac{(4\tau)^{\frac{1}{2}}(1+\tilde{t}/\tau)^{\frac{1}{2}}e^{-\tilde{t}/\tau}}{2\sqrt{\pi}\sqrt{4\pi\tau}}
\!\int_{-\infty}^\infty\! d\tilde{x}^\prime u_0(\tilde{x}^\prime)\!\int_{-\infty}^\infty\! d\tilde{k}e^{-\tau\tilde{k}^2(1-e^{-2\tilde{t}/\tau})(1+\tilde{t}/\tau)
+i\tilde{k}(\tilde{x}-\tilde{x}^\prime e^{-\tilde{t}/\tau})\sqrt{1+\tilde{t}/\tau}}\nonumber\\[0.5em]
&=\frac{1}{{\textstyle\sqrt{\smash[b]{4\pi\tau(e^{2\tilde{t}/\tau}-1)}}}}\int_{-\infty}^\infty d\tilde{x}^\prime u_0(\tilde{x}^\prime)
e^{-\frac{(e^{\tilde{t}/\tau}\tilde{x}-\tilde{x}^\prime)^2}{4\tau(e^{2\tilde{t}/\tau}-1)}},
\end{align}
which, when transformed back to the inertial frame according to the defining transformation rule \eqref{171112:1051}, turns into the
familiar inertial frame result \eqref{151113:1713} of the Cauchy problem \eqref{151113:1225}:
\begin{equation}
u(t,x)=\frac{1}{\sqrt{4\pi t}}\int_{-\infty}^\infty dx^\prime
u_0(x^\prime)e^{-\frac{(x-x^\prime)^2}{4t}}.\label{171116:1747}
\end{equation}

\vspace{0.25em}
\subsubsection{Short r\'esum\'e of the $\mathfrak{f}_4$-induced solution approach  \label{S243}}

This Section \ref{S24} has shown in detail that the invariant solution \eqref{171110:1217} induced by the symmetry algebra $\mathfrak{f}_4$ did not reveal any new insight into the Cauchy problem \eqref{151113:1225}, simply because $\mathfrak{f}_4$ is physically redundant to the structurally simpler symmetry algebra $\mathfrak{f}_1$. Hence, no new physical modes are generated by the symmetry algebra $\mathfrak{f}_4$ through its invariant solution~\eqref{171110:1217} than those already established by $\mathfrak{f}_1$, being just the temporal Laplace modes~\eqref{171026:1323}. We have seen that within $\mathfrak{f}_4$ there were two alternative solution approaches to generate the general solution of the Cauchy problem: One was based on the continuous part of the symmetry's eigenvalue spectrum~\eqref{171112:1230} and the other one on its discrete part \eqref{171119:1208}.
In particular for the continuous solution approach, Section \ref{S242} showed that it is only accessible if $\mathfrak{f}_4$ is connected to $\mathfrak{f}_1$, i.e, if the redundancy of the symmetry algebra $\mathfrak{f}_4$ to $\mathfrak{f}_1$ is explicitly revealed. Only then a correct mathematical guideline or a correct physical interpretation is provided --- for example, the inverse relation~\eqref{171110:2011} and the general solution \eqref{171112:1659} for the shapes of the continuous $\mathfrak{f}_4$-modes were only constructible with the help of the Laplace modes \eqref{171110:1502}-\eqref{171110:1503} of $\mathfrak{f}_1$.

Moreover, in contrast to $\mathfrak{f}_1$, the symmetry algebra $\mathfrak{f}_4$ introduces an additional parameter~$\tau$ that in the end even turns out to be irrelevant when considering all of its induced modes collectively. In other words, although $\tau$ is part of the single-mode solution \eqref{171110:1217} with the role as a regulating parameter to avoid at the initial point $t=0$ a singularity, the general solution \eqref{171110:2010}, or \eqref{151112:0941}, describing the full dynamics of the originally formulated physical 1-D diffusion problem \eqref{151113:1225}, however, does not depend on it. A trying, therefore, to interpret $\tau$ as a parameter of physical relevance would be delusive again, since this parameter is nothing else than just a mathematical regulator on which the physics may not depend.

Hence, only the two common symmetry algebras $\mathfrak{f}_1$ and $\mathfrak{f}_2$ provide a {\it non}-redundant approach to solve the Cauchy problem \eqref{151113:1225}. The former reveals a temporal insight (via Laplace modes) and the latter a spatial insight (via Fourier modes) into the dynamical process of the 1-D diffusion problem. All other sophisticated symmetry algebras are physically redundant to them, either to $\mathfrak{f}_1$ or to $\mathfrak{f}_2$, simply because the stated physical problem \eqref{151113:1225} is described by only two independent variables: a time and a 1-D space coordinate.

\newpage
\newgeometry{left=2.49cm,right=2.49cm,top=2.50cm,bottom=1.55cm,headsep=1em}

\subsection{Summary and conclusion on the results obtained so far\label{S25}}

In using the principle of linear superposition, the aim was to
construct the general solution \eqref{151113:1713} of the Cauchy
problem for the linear 1-D diffusion equation \eqref{151113:1225}
with the method of invariant solutions generated from the
symmetries of this equation. The crucial aspect to be recognized
in this construction process is that the scaling symmetry $X_3$
\eqref{171026:1546}, in conjunction with any other
symmetries $X_{i\neq 3}$ \eqref{171026:1546}, induces a
linear eigenvalue problem when solving the
condition \eqref{171027:1234} for invariant functions. Depending
on whether the spectrum of the operator is discrete or continuous,
the general solution can then be obtained by summing or
integrating the invariant eigenfunctions over all possible
eigenvalues. For each different symmetry we get different
eigenfunctions, so-called modes with an individual shape factor, where in each case a new mode can give a new or complementary insight
into the physical mechanism of the dynamical process considered. For the 1-D diffusion problem \eqref{151113:1225}, however, it has been shown that only two symmetry algebras provide a {\it non}-redundant approach to solve this problem,
namely only the two common 1-D subalgebras $\mathfrak{f}_1$ and $\mathfrak{f}_2$ \eqref{171027:1229}.  All other symmetry algebras are physically redundant to them, either to $\mathfrak{f}_1$ or to $\mathfrak{f}_2$, where the former reveals a temporal insight (via Laplace modes) and the latter a spatial insight (via Fourier modes) into the dynamical process of the 1-D diffusion problem. It is clear that these two insights are complementary to each other and thus sufficient to give a complete picture of the problem.

In addition, redundant symmetry algebras also can have the feature that next to the physical eigenvalues the induced modes also possess several regularization parameters, which have an effect only on
the single-modes but not in the collective when considering {\it
all} modes. As we have independently seen in this section for the redundant symmetry algebras $\mathfrak{f}_3$ and $\mathfrak{f}_4$, while the freely selectable
regularization parameters $\lambda$ and $\tau$ only have an effect on
the single-modes \eqref{171027:2244} and~\eqref{171110:1217},
respectively, these effects will just cancel as soon as one looks
at the full dynamics \eqref{171028:0947}, \eqref{171110:2010} or \eqref{151112:0941} of
all modes collectively. In all cases, the resultant {\it full}
field~$u(t,x)$ is independent of the free parameters
$\lambda$ and $\tau$. Of course, the full field $u(t,x)$ in the original
coordinates will also not depend on the eigenvalue parameters
$\alpha$, but in contrast to $\lambda$ and  $\tau$ these parameters
are not freely selectable when considering all modes. Instead, they (the $\alpha$'s)
act as complementary coordinates to the original coordinates in
physical space. Hence, caution has to be exercised when
interpreting $\lambda$ and $\tau$ as parameters of physical
relevance, since they are nothing else than regularization
parameters on which the physics may not depend.

To close this section, a formal conclusion will be given. It is clear that the above specific conclusion for the 1-D diffusion problem  \eqref{151113:1225} can be fully generalized to arbitrary dimensions and to arbitrary linear equations.

\subsubsection{Formal conclusion for the general case\label{S251}}

Given is a dynamical system $\Omega$ formulated as a set of mathematical equations $M$ describing or modelling a physical process $P$. Formally, these three entities are related by
\begin{equation}
M \cup I = \Omega \sim P,
\end{equation}
where $I$ stands for any specified initial and boundary conditions defining the input of $M$ to form a unique $\Omega$, while the relation symbol $\sim$ stands for the fact that if $\Omega$ undergoes a change in reference frame $\Omega\rightarrow \tilde{\Omega}$ it still describes the same physical process $P$, however, now in the mathematical formulation $\tilde{\Omega}$, which in general is different to $\Omega$.

Assume that $M$ admits a group of symmetry transformations $\mathsf{S}$ organized into $n$ different $d$-dimensional subalgebras $\mathfrak{h}_k$, $1\leq k\leq n$. Each symmetry algebra $\mathfrak{h}_k$ induces an invariant function $f_k$, that, if it constitutes an eigenfunction\footnote[2]{If $M$ is a linear system and each $\mathfrak{h}_k$ is chosen such that its invariant-function condition forms a linear eigenvalue problem, then the invariant function corresponds to an eigensolution of this condition.} $f_k=f_k^\alpha$ to the eigenvalue $\alpha$ of the invariant condition, consists of a defining mode\footnote[3]{A mode is defined as that minimal part of the eigenfunction which solves the eigenvalue problem.} $m_k^\alpha$ and an associated shape factor $c_k^\alpha=f_k/m_k^\alpha$.\footnote[9]{The shape factor is defined as the remainder of the invariant function $f_k$ and its defining mode $m_k^\alpha$. The expression $f_k/m_k^\alpha$ denotes this remainder and is in general not to be identified as the usual numerical quotient.}

\restoregeometry

\newgeometry{left=2.49cm,right=2.49cm,top=2.50cm,bottom=1.50cm,headsep=1em}

\noindent Let's consider any two arbitrarily chosen subalgebras  $\mathfrak{h}_1$ and $\mathfrak{h}_2$ of the same dimension, and some (invertible) variable transformation $\mathsf{T}$ that induces a change in reference frame, that is, a transformation that only leads to a change in point of view, leaving the underlying physics of the considered system thus unchanged, i.e., $\!\mathsf{T}\!:P\rightarrow \tilde{P}=P$.

Obviously, such a transformation also preserves the symmetry properties of the associated mathematical equations: Because since $\mathfrak{h}_1$ and $\mathfrak{h}_2$ are symmetry algebras of $M$, so are $\mathsf{T}(\mathfrak{h}_1)=\tilde{\mathfrak{h}}_1$ and $\mathsf{T}(\mathfrak{h}_2)=\tilde{\mathfrak{h}}_2$ symmetry algebras of $\mathsf{T}(M)=\tilde{M}$.
Now, if these three chosen entities $\mathfrak{h}_1$, $\mathfrak{h}_2$ and~$\mathsf{T}$ are structured such~that
\begin{itemize}[leftmargin=5mm]
\item[$\boldsymbol{-}$]
if the $\mathsf{T}$-transformed subalgebra $\mathfrak{h}_2$ reduces to $\mathfrak{h}_1$ while $\Omega$ stays invariant, i.e., $\mathsf{T}(\mathfrak{h}_2)=\tilde{\mathfrak{h}}_2\;\hat{\equiv}\;\mathfrak{h}_1$\footnote[2]{Note that the relation $\mathsf{T}(\mathfrak{h}_2)=\tilde{\mathfrak{h}}_2\;\hat{\equiv}\; \mathfrak{h}_1$ can also be equivalently written as $\mathfrak{h}_2\overset{\mathsf{T}}{=}\tilde{\mathfrak{h}}_1$, in accordance with the notation used in \eqref{171120:1439}, \eqref{171104:1644}, \eqref{171120:1433} and \eqref{171113:1704}. The former notation expresses the fact that when the object of transformation of $\mathfrak{h}_2$, namely $\mathsf{T}(\mathfrak{h}_2)$, is denoted or defined as $\tilde{\mathfrak{h}}_2$, then it can be identified as $\mathfrak{h}_1$, while the latter notation expresses the fact that when the transformation relation $\mathsf{T}$ is applied explicitly on the symmetry algebra~$\mathfrak{h}_2$, it then, as a direct result, turns equally into $\tilde{\mathfrak{h}}_1$. Although both notations express the same fact of reduction, they should not be mixed, otherwise wrong conclusion would be the result.} and
$P\sim \tilde{\Omega}=\mathsf{T}(\Omega)=\mathsf{T}(M\cup I)=\tilde{M}\cup\tilde{I}=M\cup I=\Omega$, then $\mathsf{T}\in \mathsf{S}\cap\{\mathsf{S}\: |\: \mathsf{S}(I)=I\}$\footnote[3]{Recall again that the symbol $I$ collectively stands for an arbitrary but fixed (initial-boundary) condition to be imposed on the governing equations $M$. Hence, if the relation $\mathsf{S}(I)=I$ applies, it should be read as an invariance of a certain specified condition $I$ and not as an invariance of all possible conditions which could be imposed on~$M$.} is a full symmetry transformation of
$\Omega\sim P$, and hence $\mathfrak{h}_2$ is said to be {\it fully equivalent} to $\mathfrak{h}_1$.\linebreak[4] In this case the modes induced by $\mathfrak{h}_2$ can be expressed by those of~$\mathfrak{h}_1$, where both modes, when combined with their shape factors, not only are solutions of the same equation set $M$, but also where they represent the very same solution of $M$ due to the same input $I$ in both instances. In other words, when changing the frame of reference by $\mathsf{T}$, the physical process~$P$ is represented by the same mathematical solution in both frames.
\item[$\boldsymbol{-}$]
if the $\mathsf{T}$-transformed subalgebra $\mathfrak{h}_2$ reduces to $\mathfrak{h}_1$ while only $M$ stays invariant, i.e., $\mathsf{T}\in\mathsf{S}$, $\mathsf{T}(\mathfrak{h}_2)=\tilde{\mathfrak{h}}_2\;\hat{\equiv}\; \mathfrak{h}_1$ and $P\sim \tilde{\Omega}=\mathsf{T}(\Omega)=\mathsf{T}(M\cup I)=\tilde{M}\cup\tilde{I}=M\cup \tilde{I}\neq M\cup I$, then $\mathfrak{h}_2$ is said to be only {\it mathematically equivalent} to $\mathfrak{h}_1$. In this case the modes induced by $\mathfrak{h}_2$ can again be expressed by those of~$\mathfrak{h}_1$, where both modes, when including their shape factors, are again solutions of the same equation set $M$, but now representing two different solutions of~$M$, one based on the input $I$ and the other on $\tilde{I}\neq I$. In other words, when changing the frame of reference by~$\mathsf{T}$, the physical process $P$ is represented in each frame by a different solution, resulting, however, from the same set of mathematical equations $M$ with only a different input $I$.
\item[$\boldsymbol{-}$]
if the $\mathsf{T}$-transformed subalgebra $\mathfrak{h}_2$ reduces to $\mathfrak{h}_1$ while $M$ does not stay invariant, i.e., $\mathsf{T}\not\in\mathsf{S}$, $\mathsf{T}(\mathfrak{h}_2)=\tilde{\mathfrak{h}}_2\;\hat{\equiv}\; \mathfrak{h}_1$ and $P\sim \tilde{\Omega}=\mathsf{T}(\Omega)=\mathsf{T}(M\cup I)=\tilde{M}\cup\tilde{I}\neq M\cup \tilde{I}$, then $\mathfrak{h}_2$ is said to be {\it physically redundant} to $\mathfrak{h}_1$, since the modes induced by $\mathfrak{h}_2$ can still be expressed by those of~$\mathfrak{h}_1$, although they are, when combined with their shape factors,\footnote[9]{Worthwhile to note here is that depending on the structure of $\mathsf{T}$, the $M$-reduced equations for the shape factors can turn out to be the same (up to a constant dimensional factor) in both frames, which is a remarkable result when recognizing here that $\mathsf{T}$ is a non-symmetry transformation. In this section, this was the case for the algebra $\mathfrak{f}_3$ when reduced by a non-symmetry-$\mathsf{T}$ to $\mathfrak{f}_2$ (compare \eqref{171107:1852} with \eqref{171103:1628}), as well as for $\mathfrak{f}_4$ when reduced to $\mathfrak{f}_1$ (compare \eqref{171114:1018} with \eqref{171111:1730}), and in the next section it will be the case for $\mathfrak{g}_3$ when reduced to $\mathfrak{g}_1$ or $\mathfrak{g}_2$.} solutions to a different set of mathematical equations $M$ and $\tilde{M}$, respectively. In other words, when changing the frame of reference by~$\mathsf{T}$, the physical process $P$ is represented in each frame by a different solution emerging from a different mathematically formulated set of equations, either from $M$ or from~$\tilde{M}$. Here, in this case, an optimal frame can be defined, as the frame in which $M$ takes the most simplest mathematical structure to depict or to describe the physical process~$P$; any other frame showing a more complicated mathematical structure to describe $P$ can then be regarded as physically redundant.
\end{itemize}
It is clear that full equivalence between two subalgebras as defined above is the strongest relation, and physical redundance the weakest. Hence, full equivalence between two subalgebras implies its mathematical equivalence, which again implies a physical redundancy, but, of course, not vice versa when taking each step backwards again.

\restoregeometry

\section[Application to an example from hydrodynamical stability analysis]{Application to an example from hydrodynamical stability analysis\footnote[2]{The mathematical derivations in this section will not be carried out in such great technical detail as it has been done for the example in the previous Section \ref{S2}. Following the general conclusion described in Section \ref{S251}, only the key results will be presented here, in particular, since also the general solution in physical space of the considered example \eqref{170714:2028} can now only be solved numerically and not analytically anymore; see Appendix \ref{SC}.}\label{S3}}

The case example examined in this section is taken from the recent study by \cite{Hau17}, considering the linearized gas-dynamical equations for a 2-D unbounded homentropic linear shear flow. The first thing to note in \cite{Hau17} is that the governing equations
\begin{equation}
\left.
\begin{aligned}
&\left(\frac{\partial}{\partial t}+Ay\frac{\partial}{\partial x}\right)\frac{\rho}{\rho_0}+\frac{\partial u}{\partial x}+\frac{\partial v}{\partial y}=0,\\
&\left(\frac{\partial}{\partial t}+Ay\frac{\partial}{\partial x}\right)u+Av=-c_s^2\frac{1}{\rho_0}\frac{\partial \rho}{\partial x},\\
&\hspace*{0.55cm}\left(\frac{\partial}{\partial t}+Ay\frac{\partial}{\partial x}\right)v=-c_s^2\frac{1}{\rho_0}\frac{\partial \rho}{\partial y},
\end{aligned}
~~~~~\right \}
\label{170714:2028}
\end{equation}
and its admitted set of Lie-point symmetries
\begin{equation}
X_1=\frac{\partial}{\partial x},\quad X_2=\frac{\partial}{\partial t},\quad X_3=At\frac{\partial}{\partial x}+\frac{\partial}{\partial y},\quad
X_4=u\frac{\partial}{\partial u}+v\frac{\partial}{\partial v}+\rho\frac{\partial}{\partial \rho},\label{170724:1405}
\end{equation}
as presented by Eqs.$\,$[1a-c] and Eq.$\,$[4] in \cite{Hau17}, respectively,
are formulated in an inertial frame of reference. The second thing to note is that the scaling symmetry $X_4$ in the dependent variables is solely due to the
linear character of the governing equations \eqref{170714:2028}. This symmetry, along with the linear superposition principle, is again the key symmetry
to induce a non-local change in representation in the solution space of the dependent variables.

\subsection{Subalgebra $\mathfrak{g}_1$ --- Modal approach\label{S31}}

As a starting point, let us first consider the construction of the invariant solution to the subalgebra $\mathfrak{g}_1$, as it was also done in Sec.$\,$[IV.A] in \cite{Hau17}.
This 2-D subalgebra is spanned by the basis
\begin{equation}
Y_1=X_1+\alpha X_4, \qquad Y_2=X_2+\beta X_4,
\end{equation}
which, due to the occurrence of $X_4$, inherently turns the corresponding system of invariant surface conditions
$Q_1^{\{u,v,\rho\}}=0$ and $Q_2^{\{u,v,\rho\}}=0$, for the solution variables $u$, $v$ and $\rho$, into the following overdetermined, but uncoupled 2-D system
of a linear eigenvalue problem
\begin{equation}
A_1\psi^{\{u,v,\rho\}}=\alpha \psi^{\{u,v,\rho\}},\qquad  A_2\psi^{\{u,v,\rho\}}=\beta \psi^{\{u,v,\rho\}},\label{170714:2126}
\end{equation}
where the group parameters $\alpha$ and $\beta$ are then identified as the eigenvalues to the continuous operators $A_1=\partial_x$ and $A_2=\partial_t$, respectively.
The most general solution of system \eqref{170714:2126} is a continuous spectrum in the complex plane ($\alpha\in\mathbb{C}$, $\beta\in\mathbb{C}$), associated to the complete set
of eigenvectors
\begin{equation}
\psi^{\{u,v,\rho\}}=\psi^{(\text{m})\{u,v,\rho\}} e^{\alpha x+\beta t},\label{170714:2152}
\end{equation}
where the $\psi^{(\text{m})\{u,v,\rho\}}=\psi^{(\text{m})\{u,v,\rho\}}(\alpha,\beta,y)$ are arbitrary integration functions, being the shape factors of the defining modes $ e^{\alpha x+\beta t}$ to the eigenvalues $\alpha$ and $\beta$. For some specified value of these two independent eigenvalues,
the invariant solution \eqref{170714:2152} can be regarded as a particular solution ansatz for the PDE system \eqref{170714:2028},
which in turn will reduce it to an ODE system in the remaining physical variable $y$. By making use of the linear superposition principle, a more general solution
ansatz can be obtained by summing over all eigenvalues (or modes) in the complex plane
\begin{equation}
\psi^{\{u,v,\rho\}}(t,x,y)=\int_{\mathbb{C}^2}d\alpha\, d\beta\,\psi^{(\text{m})\{u,v,\rho\}}(\alpha,\beta,y)\cdot e^{\alpha x+\beta t}.\label{170714:2250}
\end{equation}
Now, when choosing $\alpha=ik$, with $k\in\mathbb{R}$, and the contour of the $\beta$-integration along a straight line at a position $\gamma\in\mathbb{R}$ in the direction of the imaginary axis, such that
it includes all singularities of the integrand when closing the contour around infinity (e.g. by a half circle), one obtains the effect that \eqref{170714:2250} turns into an
{\it invertible} integral relation. Hence, under these conditions, the invariant solution \eqref{170714:2152} to the 2-D subalgebra $\mathfrak{g}_1$ induces the following non-local change of
representation in the dependent solution variables\footnote[2]{The redefined function $\hat{\psi}^{(\text{m})\{u,v,\rho\}}$ in \eqref{170714:2337}-\eqref{170714:2338} is related to the
original function $\psi^{(\text{m})\{u,v,\rho\}}$ in \eqref{170714:2250} by: $\hat{\psi}^{(\text{m})\{u,v,\rho\}}(k,\beta,y)\equiv 2\pi i^2\psi^{(\text{m})\{u,v,\rho\}}(ik,\beta,y)$.}
\begin{equation}
\!\!\psi^{\{u,v,\rho\}}(t,x,y)= \int_{-\infty}^\infty dk \int_{\gamma-i\infty}^{\gamma+i\infty}\frac{d\beta}{2\pi i}\, \hat{\psi}^{(\text{m})\{u,v,\rho\}}(k,\beta,y)\, e^{ik x+\beta t},\;\,
k,\gamma\in\mathbb{R};\, \beta\in\mathbb{C};\, t\geq 0,\label{170714:2337}
\end{equation}
which constitutes the Laplace-Fourier transform of the field $\psi^{\{u,v,\rho\}}$ in physical $(t,x)$-space into its complementary field $\hat{\psi}^{(\text{m})\{u,v,\rho\}}$
in $(\beta,k)$-space, with the inverse transformation then given~as
\begin{equation}
\hat{\psi}^{(\text{m})\{u,v,\rho\}}(k,\beta,y)=\int_0^\infty\! dt \int_{-\infty}^\infty \frac{dx}{2\pi}\, \psi^{\{u,v,\rho\}}(t,x,y)\, e^{-ik x-\beta t}.\label{170714:2338}
\end{equation}
Note that the $\alpha$-integration in \eqref{170714:2250} has to be treated differently than the $\beta$-integration in order to obtain an invertible integral relation. The reason is
that the complementary variable to $\beta$, namely the physical time coordinate $t$, is only defined in the half-domain $0\leq t< \infty$, in contrast to the complementary variable of $\alpha$,
the spatial coordinate $x$, which for unbounded flow in the $x$-direction is defined over the whole domain $-\infty< x< \infty$. Further note that the invariant solution \eqref{170714:2152} formally differs
to Eq.$\,$[20] in \cite{Hau17} in the choice of the complementary variable to $t$, but effectively are equal, since Eq.$\,$[20] contains the inessential parameter $a_2$ that can either be set to one or absorbed into $\beta$. In summary,
the subalgebra $\mathfrak{g}_1$ induces the local (single-mode) representation \eqref{170714:2152} as well as the non-local (general) representation \eqref{170714:2337} of the modal-approach concept in linear stability analysis\label{171202:1438}\footnote[3]{It is clear that if a $\mathfrak{g}_1$-approach to linear stability analysis is chosen, it has to include the analysis of pseudo-spectra \citep{Trefethen05} in order to offer a complete modal description.} (see e.g. \cite{Schmid01}).

\subsection{Subalgebra $\mathfrak{g}_2$ --- Kelvin approach\label{S32}}

Let us now look at the consequences of the 2-D subalgebra $\mathfrak{g}_2$, as analyzed in Sec.$\,$[IV.B] in \cite{Hau17}. This algebra is spanned by the
basis (the inessential parameter $a_3$ is put to one)
\begin{equation}
Y_1=X_1+\alpha X_4, \qquad Y_3=X_3+\beta X_4,\label{170730:1042}
\end{equation}
which gives the invariant surface conditions as the following (overdetermined, but uncoupled) linear eigenvalue problem
\begin{equation}
\big(\partial_x \big)\psi^{\{u,v,\rho\}}=\alpha \psi^{\{u,v,\rho\}},\qquad
\big(At\partial_x+\partial_y\big)\psi^{\{u,v,\rho\}}=\beta \psi^{\{u,v,\rho\}}.\label{170715:1012}
\end{equation}
However, this system of equations is not formulated in the optimal frame of reference. As already mentioned in the beginning of this section, system \eqref{170715:1012}
is formulated within an inertial frame, but, as will be shown now, is not the optimal frame to solve this eigenvalue problem for the associated invariant solution. The optimal frame is
obtained when transforming system \eqref{170715:1012} according to the following (invertible) coordinate transformation
\begin{equation}
\mathsf{K}\!:\quad\; \tilde{t}=t,\quad\; \tilde{x}=x-Ayt,\quad\; \tilde{y}=y,\quad\; \tilde{\psi}^{\{u,v,\rho\}}=\psi^{\{u,v,\rho\}},\label{170715:1051}
\end{equation}
which is best known as the Kelvin transformation. Since the derivatives under \eqref{170715:1051} transform~as
\begin{equation}
\partial_t=\partial_{\tilde{t}}-A\tilde{y}\partial_{\tilde{x}},\quad\;
\partial_x=\partial_{\tilde{x}},\quad\;
\partial_y=\partial_{\tilde{y}}-A\tilde{t}\partial_{\tilde{x}},\label{170715:1406}
\end{equation}
system \eqref{170715:1012} simplifies to
\begin{equation}
\big(\partial_{\tilde{x}}\big)\tilde{\psi}^{\{u,v,\rho\}}=\alpha \tilde{\psi}^{\{u,v,\rho\}},\qquad
\big(\partial_{\tilde{y}}\big)\tilde{\psi}^{\{u,v,\rho\}}=\beta \tilde{\psi}^{\{u,v,\rho\}}.\label{170715:1105}
\end{equation}
As for the algebra $\mathfrak{g}_1$ given in \eqref{170714:2126}, the general solution of \eqref{170715:1105} is again a continuous spectrum of eigenvalues
in the complex plane ($\alpha\in\mathbb{C}$, $\beta\in\mathbb{C}$), however, now associated to the complete set
of eigenvectors complementary to both unbounded spatial variables $\tilde{x}$ and $\tilde{y}$
\begin{equation}
\tilde{\psi}^{\{u,v,\rho\}}=\tilde{\psi}^{(\text{k})\{u,v,\rho\}} e^{\alpha \tilde{x}+\beta \tilde{y}},\label{170715:1119}
\end{equation}
where $\tilde{\psi}^{(\text{k})\{u,v,\rho\}}=\tilde{\psi}^{(\text{k})\{u,v,\rho\}}(\alpha,\beta,\tilde{t})$ are again arbitrary integration functions, but now being the shape factors of the defining (Kelvin) modes $ e^{\alpha \tilde{x}+\beta \tilde{y}}$ induced by $\mathfrak{g}_2$ in the new and optimal Kelvin frame. Following the same construction procedure as outlined in \eqref{170714:2250}-\eqref{170714:2338}, but now by choosing $\alpha=ik_{\tilde{x}}$
and $\beta=ik_{\tilde{y}}$, the invariant solution \eqref{170715:1119}
induces the following non-local change of representation in the dependent solution variables\footnote[2]{The redefined function $\hat{\tilde{\psi}}^{(\text{k})\{u,v,\rho\}}$ in \eqref{170715:1138}-\eqref{171122:1116} is related to the original function $\tilde{\psi}^{(\text{k})\{u,v,\rho\}}$ \eqref{170715:1119} by: $\hat{\tilde{\psi}}^{(\text{k})\{u,v,\rho\}}(\tilde{t},k_{\tilde{x}},k_{\tilde{y}})\equiv i^2\cdot \tilde{\psi}^{(\text{k})\{u,v,\rho\}}(ik_{\tilde{x}},ik_{\tilde{y}},\tilde{t})$.}
\begin{equation}
\tilde{\psi}^{\{u,v,\rho\}}(\tilde{t},\tilde{x},\tilde{y})
= \int_{-\infty}^\infty dk_{\tilde{x}}dk_{\tilde{y}}\, \hat{\tilde{\psi}}^{(\text{k})\{u,v,\rho\}}(\tilde{t},k_{\tilde{x}},k_{\tilde{y}})
\, e^{i(k_{\tilde{x}} \tilde{x}+k_{\tilde{y}} \tilde{y})},\;\:
k_{\tilde{x}},k_{\tilde{y}}\in\mathbb{R},\label{170715:1138}
\end{equation}
with its inverse
\begin{equation}
\hat{\tilde{\psi}}^{(\text{k})\{u,v,\rho\}}(\tilde{t},k_{\tilde{x}},k_{\tilde{y}})
= \int_{-\infty}^\infty \frac{d\tilde{x}}{2\pi}\frac{d\tilde{y}}{2\pi}\, \tilde{\psi}^{\{u,v,\rho\}}(\tilde{t},\tilde{x},\tilde{y})
\, e^{-i(k_{\tilde{x}} \tilde{x}+k_{\tilde{y}} \tilde{y})},\label{171122:1116}
\end{equation}
which constitutes the classical 2-D Fourier-transform of the physical fields $\tilde{\psi}^{\{u,v,\rho\}}$ in the two spatial (unbounded) variables
$\tilde{x}$ and $\tilde{y}$.

Hence, instead of formulating the dynamics in the non-optimal inertial (untilded) frame as done in \cite{Hau17}, the change into the co-moving
accelerated Kelvin (tilded) frame~\eqref{170715:1051} is optimal, since it reduces the complexity of the solutions to a minimum, avoiding the irrelevant temporal shift in the solution
of the inertial frame as presented in Eq.$\,$[21] in \cite{Hau17}. This temporal shift, which leads to a drift in the complementary $k$-space (see e.g. the discussion
to Fig.$\,$[5] in \cite{Hau15}) is only an artefact of describing the dynamics in an inertial frame. In the optimal co-moving frame no such drift exists. Hence, this drift, discussed and interpreted
e.g. in \cite{Hau15}, is a relative, frame-dependent phenomenon, which induces a superimposed motion that is not relevant to understand the dynamics of sound generation by vortex disturbances as
explained in \cite{Hau15}. In fact, this non-relevant superposed motion even distracts and obscures the understanding of the internal dynamical processes at constant shear rate.
Physically relevant in this system is only the existence and occurrence of a critical time due to being a frame {\it independent} system~property. For more details on this issue, please also see the discussion on pure wave propagation in linear shear flow in Appendix \ref{SC}.

Four important things are to be noted here, regarding the optimal frame for $\mathfrak{g}_2$: (i) The coordinates ($\tilde{t},\tilde{x},\tilde{y}$) of the optimal frame, as defined in \eqref{170715:1051},
carry the same physical dimensions as the corresponding coordinates ($t,x,y$) in the inertial frame, respectively. Hence, the eigenvalues $\alpha$ and $\beta$ in \eqref{170715:1105} retain their physical dimensions
in the optimal frame as originally defined in the inertial frame \eqref{170715:1012}. In both frames they carry the same dimension of inverse length: $[\alpha]=[\beta]=1/L$; in contrast, of course, to the eigenvalues $\alpha$ and $\beta$ of the $\mathfrak{g}_1$-algebra \eqref{170714:2126}, where they carry different dimensions, namely that of inverse length and inverse time: $[\alpha]=1/L$ and $[\beta]=1/T$. For this very reason, the modes of the  $\mathfrak{g}_2$-algebra are fundamentally different to the modes of the $\mathfrak{g}_1$-algebra, both in the inertial as well as in the optimal frame. In other words, no frame of reference exists which can transform the
modes of the $\mathfrak{g}_2$-algebra into those of the $\mathfrak{g}_1$-algebra and vice versa. Hence, we face {\it physically} two different approaches to stability analysis for unbounded linear shear flow (here in 2-D):\linebreak A temporal-spatial $(T$-$L)$ mode approach given by the eigensystem \eqref{170714:2126}, also known as the modal or normal-mode\footnote[2]{Using the notion ``modal analysis" or ``normal-mode analysis" for the $\mathfrak{g}_1$-approach, please see again the remark of the second footnote on p.$\,$\pageref{171202:1438}.} approach, and a double-spatial $(L^2)$ mode approach given by the complementary eigensystem \eqref{170715:1105}, also known as the ``non-modal"\footnote[3]{Referring to the $\mathfrak{g}_2$-approach as a ``non-modal analysis", as done in \cite{Hau17} and as first mentioned in \cite{Chagelishvili94,Chagelishvili97}, please see again the key footnote on p.$\,$\pageref{Kelvin}.} or Kelvin mode approach.

(ii) Fixing the optimal frame only on the basis of the eigenvalue problem \eqref{170715:1012}, namely
to reduce it to \eqref{170715:1105}, is not unique. For example, the following invertible and more general transformation
\begin{equation}
\mathsf{K}_\mathsf{t}\!:\quad\;\tilde{t}=t,\quad\; \tilde{x}=x-At\big(y+f_2(t)\big)+f_1(t),\quad\; \tilde{y}=y+f_2(t),\quad\; \tilde{\psi}^{\{u,v,\rho\}}=\psi^{\{u,v,\rho\}},\label{170718:1625}
\end{equation}
where $f_1$ and $f_2$ are arbitrary functions of time, will do the same job as the more simple and basic Kelvin transformation $\mathsf{K}$ \eqref{170715:1051}. In fact, transformation $\mathsf{K}_\mathsf{t}$ \eqref{170718:1625} can be regarded as a generalized~Kelvin transformation since it additionally allows for arbitrary linear accelerations.\footnote[9]{For example, instead of the classical Kelvin ansatz $\psi^{\{u,v,\rho\}}(x,y,t)=\psi^{(\text{k})\{u,v,\rho\}}(t)e^{\alpha x+(\beta^\prime-A\alpha t)y}$,
as presented by Eq.$\,$[21] in \cite{Hau17} for the non-optimal inertial frame, one can use the more general Kelvin ansatz
$\psi^{\{u,v,\rho\}}(x,y,t)=\psi^{(\text{k}_{\text{t}})\{u,v,\rho\}}(t)e^{\alpha (x-At(y+f_2(t))+f_1(t))+\beta^\prime (y+f_2(t))}$, based on \eqref{170718:1625},
in order to reduce the governing system of PDE's \eqref{170714:2028} into a system of ODE's. But careful,
as explained in the main text, this extended ansatz does {\it not} lead to ``new" or ``more general" modes. The functions or shape factors $\psi^{(\text{k}_\text{t})\{u,v,\rho\}}$ are just the consequence of writing the Kelvin modes induced by
\eqref{170715:1105} into a mathematically different (non-classical) representation within a non-optimal frame of reference. In particular, they are even trivially related to the classical ansatz functions by
$\psi^{(\text{k})\{u,v,\rho\}}(t)=\psi^{(\text{k}_\text{t})\{u,v,\rho\}}(t)e^{\alpha(f_1(t)-Atf_2(t))+\beta^\prime f_2(t)}$.} However, the disadvantage of this more general transformation is that it will unnecessarily
turn the governing equations \eqref{170714:2028} into a more complicated system of equations than the basic Kelvin transformation \eqref{170715:1051} will do. In other words, eliminating the free
parameters, namely by choosing $f_1=f_2=0$ to get the basic Kelvin transformation \eqref{170715:1051}, already suffices to solve the problem. Hence, for $f_1,f_2\neq 0$,
\eqref{170718:1625} does not qualify as the defining transformation for an optimal system to the particular eigenvalue problem \eqref{170715:1012}, simply because it is not efficient enough --- it simply features too many free parameters, which only complicates the situation beyond necessity. To find the optimal frame, the real aim is to determine a transformation which manages with a minimum amount of free parameters, true to the principle of Occam's~razor.

(iii) The defining transformation $\mathsf{K}$ \eqref{170715:1051} is {\it not} a symmetry transformation of the
governing system of equations~\eqref{170714:2028}. Because when applying this transformation onto \eqref{170714:2028}, it will be transformed from a spatially into a temporally inhomogeneous PDE system.

(iv)\label{171124:1112} The defining~transformation $\mathsf{K}$ \eqref{170715:1051} also does not constitute a coordinate transformation in the tensorial sense. In other words,
the Kelvin transformation \eqref{170715:1051}, as it stands, does not relate to a
change of a true {\it physical} reference frame, since the components $u$ and $v$ of the velocity vector field are transforming as scalars and not as components of a vector.
However, this can be straightforwardly accomplished by extending transformation \eqref{170715:1051}
with its associated tensor transformation rule for the velocity vector field in an accelerated frame (within Newtonian mechanics)
\vspace{-0.25em}
\begin{equation}
\tilde{u}^i=\frac{\partial\tilde{x}^i}{\partial t}+ \frac{\partial\tilde{x}^i}{\partial x^j}u^j,\quad\; i,j=1,2,\label{170715:1541}
\end{equation}
where the indices $1$ and $2$ stand for the coordinates $x$ and $y$, respectively. Note that $j$ in the above expression is a summation index. Hence, the
corresponding transformation relating to a {\it physical} change of frame according to \eqref{170715:1051} then reads (the additional `\hspace{0.5mm}*\hspace{0.5mm}'-symbol on the transformed variables
indicates that the transformation now refers to a physical frame)
\begin{equation}
\mathsf{K}^*\!:\quad\;\tilde{t}^*=t,\quad\; \tilde{x}^*=x-Ayt,\quad\; \tilde{y}^*=y,\quad\; \tilde{u}^*=u-Atv-Ay,
\quad\; \tilde{v}^*=v,\quad\; \tilde{\rho}^*=\rho,\label{170621:2253}
\end{equation}
which, in order to maintain the simple structure of the solutions \eqref{170715:1119} and \eqref{170715:1138} when applied, can be split and processed in a two step transformation procedure:
First by transforming {\it only} the coordinates according to \eqref{170715:1051} in proceeding to construct the simple solutions \eqref{170715:1119} and \eqref{170715:1138} in the optimal frame as shown above,
and then to correspondingly transform
{\it only} the fields according to the second consecutive transformation
\begin{equation}
\tilde{t}^*=\tilde{t},\quad\; \tilde{x}^*=\tilde{x},\quad\; \tilde{y}^*=\tilde{y},\quad\;
\tilde{u}^*=\tilde{u}-A\tilde{t}\tilde{v}-A\tilde{y},
\quad\; \tilde{v}^*=\tilde{v},\quad\; \tilde{\rho}^*=\tilde{\rho},\label{170715:1526}
\end{equation}
to turn the optimal frame into a physical frame of reference. Indeed, when both independent transformations \eqref{170715:1526} and \eqref{170715:1051} are linked together, one obviously obtains
back the single physical transformation \eqref{170621:2253}. For reasons of simplicity, however, we will consider
coordinate transformations in the following only in their reduced, non-tensorial form, as in \eqref{170715:1051}, since their extension to the physical form through the rule
\eqref{170715:1541} will not provide any new information to the transformation process itself, simply because a coordinate transformation, by definition, is dictated by the coordinates and not by the fields.

In summary, the subalgebra $\mathfrak{g}_2$ induces the local (single-mode) representation \eqref{170715:1119} as well as the non-local (general) representation \eqref{170715:1138} of the Kelvin-mode-approach concept in linear stability analysis for unbounded constant shear rate (see e.g. \cite{Chagelishvili94,Chagelishvili97}). However, in \cite{Chagelishvili94,Chagelishvili97}, this Kelvin-mode approach is also coined  as a ``non-modal" approach, which, as explained in the single footnote on p.$\,$\pageref{Kelvin}, can be misleading and not to be confused with the proper non-modal approach as originally defined e.g. in \cite{Schmid01} and \cite{Schmid07}. --- To see the $\mathfrak{g}_2$-approach in action, in particular as how a general solution of \eqref{170714:2028} for pure wave propagation can be constructed, please see the analysis done in Appendix \ref{SC}.

\subsection{Subalgebra $\mathfrak{g}_3$ --- Redundant approach\label{S33}}

The third and last 2-D subalgebra considered in \cite{Hau17} is $\mathfrak{g}_3$. For this algebra it is claimed that it presents ``a generalized approach" to $\mathfrak{g}_1$ and $\mathfrak{g}_2$, that it
``unites both -- the non-modal (Kelvin mode) and the modal -- approaches'' and that thus ``the class of solutions is
much wider than the well-known Kelvin and modal solutions" since the general invariant solution to $\mathfrak{g}_3$ ``contains two free parameters" where the ``Kelvin mode
and modal ansatz functions ... appear to be limiting cases of
two different representations of the generalized ansatz function". On closer inspection, however, as we mean to show now, this overall claim cannot be confirmed. Correct is that the {\it classical} Kelvin mode and modal solutions of $\mathfrak{g}_2$ and $\mathfrak{g}_1$ appear as limiting cases to the general solution of $\mathfrak{g}_3$, but for the non-limiting case, i.e.,
for non-zero and finite values of both ``free parameters" it is incorrect that this generalized solution will lead to a wider class of solutions than the already established Kelvin and modal solutions.
In fact, in the optimal frame of reference, $\mathfrak{g}_3$ is {\it physically} redundant to either $\mathfrak{g}_2$ or $\mathfrak{g}_1$, depending on how the ``two free parameters" are arranged. Hence, besides the Kelvin and modal solutions, no ``new" or ``more general modes" for linear shear flow exist.

The 2-D subalgebra $\mathfrak{g}_3$, as analyzed in Sec.~[IV.C] in \cite{Hau17}, is spanned by the~basis
\begin{equation}
Y_1=X_1+\alpha X_4, \qquad Y_4=a_2X_2+a_3X_3+\beta X_4,
\end{equation}
which gives the invariant surface conditions as the following (overdetermined, but uncoupled) linear eigenvalue problem
\begin{equation}
\big(\partial_x \big)\psi^{\{u,v,\rho\}}=\alpha \psi^{\{u,v,\rho\}},\qquad
\Big(a_2\partial_t +a_3(At\partial_x+\partial_y)\Big)\psi^{\{u,v,\rho\}}=\beta \psi^{\{u,v,\rho\}},\label{170716:0933}
\end{equation}
where $a_2$ and $a_3$ are two free parameters of the problem, carrying a unique physical dimension only for the ratio $[a_3/a_2]=L/T$, or inversely $[a_2/a_3]=T/L$,
that of a velocity or inverse velocity, respectively.\footnote[2]{The reason why the physical dimensions of the parameters $a_2$ and $a_3$ are not unique, but only their ratio, is that
always one of the two parameters is inessential. This is also pointed out in Sec.~[IV.C] in \cite{Hau17}.}
Indeed, in the limiting case $(a_2=1,a_3=0)$ the above eigenvalue problem \eqref{170716:0933} reduces
to the one of $\mathfrak{g}_1$ \eqref{170714:2126}, and in the case  $(a_2=0,a_3=1)$ to the one of $\mathfrak{g}_2$ \eqref{170715:1012}. However, for the non-limiting case, i.e.,
for $a_2\neq 0$ and $a_3\neq0$, the question now is, does the associated eigenvalue problem to $\mathfrak{g}_3$ \eqref{170716:0933} lead to a wider range of solutions than the
already established solutions of the eigenvalue problems associated to $\mathfrak{g}_1$ \eqref{170714:2126} and $\mathfrak{g}_2$ \eqref{170715:1012}? The answer is clearly no, and can be
straightforwardly seen by just changing the frame of reference! The issue here again is that the system of equations~\eqref{170716:0933} are not formulated in the optimal frame. Changing
the frame either by\footnote[2]{Note that the two coordinate transformations $\mathsf{T}^{(\mathsf{m})}$~\eqref{170716:1207} and $\mathsf{T}^{(\mathsf{k})}$~\eqref{170716:1209} are fully (globally) invertible, without restrictions. Also note that in
both these transformations the new transformed (tilded) variables carry the same physical dimensions as the untransformed (untilded) variables.}
\begin{equation}
\hspace{-2.1cm}\mathsf{T}^{(\mathsf{m})}\!:\quad\; \tilde{t}=t,\quad\; \tilde{x}=x-\frac{a_3}{2a_2}At^2,\quad\; \tilde{y}=y-\frac{a_3}{a_2}t,\quad\; \tilde{\psi}^{\{u,v,\rho\}}=\psi^{\{u,v,\rho\}},\label{170716:1207}
\end{equation}
which will let the derivatives transform as
\begin{equation}
\partial_t=\partial_{\tilde{t}}-\frac{a_3}{a_2}A\tilde{t}\partial_{\tilde{x}}-\frac{a_3}{a_2}\partial_{\tilde{y}},\quad\;
\partial_x=\partial_{\tilde{x}},\quad\;
\partial_y=\partial_{\tilde{y}},\label{170716:1208}
\end{equation}
or, alternatively, by
\begin{equation}
\mathsf{T}^{(\mathsf{k})}\!:\quad\;\tilde{t}=t-\frac{a_2}{a_3}y,\quad\; \tilde{x}=x-Ayt+\frac{a_2}{2a_3}Ay^2,\quad\; \tilde{y}=y,\quad\; \tilde{\psi}^{\{u,v,\rho\}}=\psi^{\{u,v,\rho\}},\label{170716:1209}
\end{equation}
which then will let the derivatives transform as
\begin{equation}
\partial_t=\partial_{\tilde{t}}-A\tilde{y}\partial_{\tilde{x}},\quad\;
\partial_x=\partial_{\tilde{x}},\quad\;
\partial_y=\partial_{\tilde{y}}-A\tilde{t}\partial_{\tilde{x}}
-\frac{a_2}{a_3}\partial_{\tilde{t}},\label{170716:1210}
\end{equation}
will in both cases reduce the ``generalized" eigenvalue problem \eqref{170716:0933}, {\it respectively}, either completely to that of the temporal-spatial ($T$-$L$) mode approach of the $\mathfrak{g}_1$-algebra \eqref{170714:2126} (up to the inessential parameter $a_2$, which can be absorbed into~$\beta$)
\begin{equation}
\big(\partial_{\tilde{x}} \big)\tilde{\psi}^{\{u,v,\rho\}}=\alpha \tilde{\psi}^{\{u,v,\rho\}},\qquad
\big(a_2\partial_{\tilde{t}}\big)\tilde{\psi}^{\{u,v,\rho\}}=\beta \tilde{\psi}^{\{u,v,\rho\}},\label{170716:1225}
\end{equation}
or alternatively to that of the double-spatial ($L^2$) mode approach of the $\mathfrak{g}_2$-algebra in its optimal frame \eqref{170715:1105} (up to the inessential parameter $a_3$
that alternatively can be absorbed again into~$\beta$)
\begin{equation}
\big(\partial_{\tilde{x}} \big)\tilde{\psi}^{\{u,v,\rho\}}=\alpha \tilde{\psi}^{\{u,v,\rho\}},\qquad
\big(a_3\partial_{\tilde{y}}\big)\tilde{\psi}^{\{u,v,\rho\}}=\beta \tilde{\psi}^{\{u,v,\rho\}}.\label{170716:1231}
\end{equation}
Hence, no new or more general stability modes exist than those already established by the two complementary algebras $\mathfrak{g}_1$ \eqref{170714:2152}-\eqref{170714:2338}
and $\mathfrak{g}_2$ \eqref{170715:1119}-\eqref{171122:1116}. In other words, the modes of the invariant solutions induced by the algebra $\mathfrak{g}_3$ turn out to be physically
redundant to those of either $\mathfrak{g}_1$ or $\mathfrak{g}_2$ when written in the optimal frame \eqref{170716:1207} or \eqref{170716:1209}, respectively. The complicated invariant solutions
presented in Eqs.$\,$[22-23] in \cite{Hau17} are just an artefact of describing the dynamics in a mathematically more intricate (non-classical) representation formulated within the non-optimal inertial frame. The additional (superimposed) motion that will be induced by these
solutions is irrelevant to the understanding of the internal dynamical processes at constant shear rate. In other words, the modes of the complementary algebras $\mathfrak{g}_1$ and $\mathfrak{g}_2$
already suffice to fully understand the physics behind unbounded linear shear flow (when extending the modal eigenvalue analysis to also include pseudo-spectra \citep{Trefethen05}). The algebra $\mathfrak{g}_3$ is not needed, since it only forms an unnecessary mathematical sophistication providing no new physical insight.

This issue of redundance or irrelevance we already face in the two publications of \cite{Nold15} and \cite{Nold13}, preceding the current publication of \cite{Hau17}.
For~example, in Sec.~[VI] in \cite{Nold13} ``new invariant modes" in incompressible linear\hfill flow\hfill are\hfill proclaimed.\hfill But,\hfill again,\hfill this\hfill cannot\hfill be\hfill confirmed,\hfill because\hfill when\hfill writing\hfill these

\newgeometry{left=2.49cm,right=2.49cm,top=2.50cm,bottom=1.75cm,headsep=1em}

\noindent ``new invariant modes" in the optimal frame, they simply reduce to the normal temporal-spatial ($T$-$L$) modes as discussed in Sec.~[IV]\footnote[2]{For the normal-mode analysis done in Sec.~[IV] in \cite{Nold13}, the parameter $a_1$ is not relevant and can be put to zero, as can be convincingly seen in the final result Eq.$\,$[21] for the invariant solution. The~reason simply is that the operator $\partial_x$ in Eq.$\,$[14], necessary for the first reduction, is redundant to the operator $\partial_\xi$ in Eq.$\,$[19] for the second reduction. In fact, they are even equal: $\partial_x=(\partial \xi/\partial x)\partial_\xi+(\partial y/\partial x)\partial_y=\partial_\xi$.} in \cite{Nold13}, or as discussed herein by relation~\eqref{170714:2126}. This can be straightforwardly validated: The invariant solution Eq.$\,$[43] in \cite{Nold13} is based on the combined linear eigenvalue problem
\begin{equation}
\big(b_1\partial_x\big)\psi=b_2\psi,\qquad\; \Big(a_1\partial_x+\partial_t+a_4\big(At\partial_x+\partial_y\big)\Big)\psi=a_3\psi,\label{170718:2008}
\end{equation}
associated to the 2-D subalgebra spanned by the basis $X^{(\text{I})}$ Eq.$\,$[36] and $\tilde{X}^{(\text{I})}$ Eq.$\,$[41], where the latter symmetry in the original coordinates (before the first reduction)
is given by
\begin{equation}
\tilde{X}^{(\text{I})}=b_1\partial_x+b_2\psi\partial_\psi=b_1\left(\frac{\partial \bar{x}}{\partial x}\partial_{\bar{x}}+\frac{\partial \bar{y}}{\partial x}\partial_{\bar{y}}\right)
+b_2 f(\bar{x},\bar{y})g(t)\partial_{f(\bar{x},\bar{y})g(t)}=b_1\partial_{\bar{x}}+b_2 f\partial_f,\label{170718:1950}
\end{equation}
where in the last step we have used the fact that the time coordinate $t$ is not a variable anymore in the first reduced system of Eq.$\,$[40], so that the product function
$g(t)=e^{a_3(t+\frac{a_1}{Aa_4})}$ from Eq.$\,$[38] cancels in the last step of \eqref{170718:1950}. Indeed, when substituting the final result Eq.$\,$[43] for the invariant solution into the
defining system \eqref{170718:2008},
it will satisfy this. Now, the optimal frame for system \eqref{170718:2008} is given by the invertible transformation (when choosing the inessential parameter $b_1$ as dimensionless)
\begin{equation}
\tilde{t}=t,\quad\; \tilde{x}=\frac{1}{b_1}\left(x-a_1t-\frac{1}{2}a_4At^2\right),\quad\; \tilde{y}=y-a_4t,\quad\; \tilde{\psi}=\psi,
\end{equation}
which will reduce \eqref{170718:2008} back to the eigenvalue problem of the normal modes \eqref{170714:2126}
\begin{equation}
\big(\partial_{\tilde{x}}\big)\tilde{\psi}=b_2\tilde{\psi},\qquad\; \big(\partial_{\tilde{t}}\big)\tilde{\psi}=a_3\tilde{\psi}.\label{170718:2136}
\end{equation}
Hence, no new invariant modes for incompressible linear shear flow exist as claimed in \cite{Nold13}. This misleading and thus incorrect claim is again repeated in the subsequent publication \cite{Nold15},
however now, particularly for the inviscid case, the analysis is based on an extended symmetry transformation, which serves in \cite{Hau17} as an outlook for future work to find
in linear shear flow even ``more invariant solutions in terms of optimal systems of subalgebras". From the present study it is clear that such an endeavour should not be pursued, as it only would be insignificant
offering no new physical insight.

Coming back to \eqref{170718:2136} in \cite{Nold13}, it is clear that for unbounded linear shear flow in the optimal frame, the reduction induced by \eqref{170718:2136} does not lead to the classical Orr-Sommerfeld equation Eq.$\,$[23], but to its mathematically more sophisticated representation Eq.$\,$[46]. Yet, the latter equation is {\it physically} redundant to the former one, since both representations entail the same physical information, however, only represented in a mathematically different manner.\footnote[3]{Please recall here again the general conclusion made in Sec.$\,$\ref{S251}.} The cause for this physical redundance is again the fact that both representations are based on the same temporal-spatial ($T$-$L$) mode approach.

The construction of the optimal frame of reference for the extended symmetry transformation considered in \cite{Nold15} is more difficult, but nevertheless possible to achieve. A detailed
analysis will be presented in a follow-up study to this one. Therein the redundancy in the invariant solutions also among different base flows will be investigated --- in particular the statement
that the symmetry invariant solutions for certain base flows different from linear shear give rise to modes of ``algebraic growth and decay", will be revealed as a relative and not as an absolute statement.\footnote[9]{It can be shown that those invariant solutions in \cite{Nold15} leading to the ``new" modes of algebraic growth and decay, all fall back to the classical modes of exponential growth and decay when only written in the optimal frame of reference.}
This follow-up study will be a definite generalization
to the present study, where herein the redundancy only within a fixed base flow, namely that of an unbounded linear shear flow, has been investigated, with the current result that next to modal and Kelvin mode solutions no new modes exist, neither in the compressible nor in the incompressible case.

\newgeometry{left=2.49cm,right=2.49cm,top=2.50cm,bottom=2.20cm,headsep=1em}

\section{Final remarks and points for correction in Hau~\emph{et al.}~(2017)\label{S4}}

The analysis and the resulting physical interpretations given in Sec.$\,$[V.C3] and Sec.$\,$[V.D] in \cite{Hau17} on the generalized $\mathfrak{g}_3$-algebra, culminating in the result of Fig.$\,$[1], is, as was
shown in the previous section, only a consequence of describing the dynamics in the non-optimal inertial frame of reference. For example, the result that ``increasing the
parameter $a_2$ leads to an increase of localization in the shearwise direction, i.e., a collection of Kelvin modes with different amplitudes along the $k_y$-axis" is a relative statement and only an artefact
of the inertial frame~used. Because when transforming the dynamics into the optimal frame of reference, no such property exists. Particularly when transforming into the optimal frame given by $\mathsf{T}^{(\mathsf{k})}$~\eqref{170716:1209}, the critical parameter $a_2$, necessary for the conclusions of Sec.$\,$[V.C3] and Sec.$\,$[V.D] in \cite{Hau17}, is then even not part anymore of the associated eigenvalue problem~\eqref{170716:1231}: Instead of operating with the complicated invariant solution Eq.$\,$[22], or Eq.$\,$[43] for $t=0$ in the non-optimal inertial frame,
\begin{equation}
\psi(x,y,t=0)=\psi_0^{(\text{nk})}e^{ik_{x_0}x+i\bar{\beta}y+i\frac{1}{2}Ak_{x_0}\frac{a_2}{a_3}y^2},\label{170720:1817}
\end{equation}
as painstakingly done in Sec.$\,$[V.C3] in \cite{Hau17}, all information is already carried by the more simple solution of \eqref{170716:1231}
in the optimal frame of reference,
\begin{equation}
\psi(x,y,t=0)=\tilde{\psi}(\tilde{x},\tilde{y},\tilde{t})\big|_{t=0}=\tilde{\psi}_0^{(\text{nk})}e^{\alpha \tilde{x}+\frac{\beta}{a_3}\tilde{y}}
=\psi_0^{(\text{nk})}e^{ik_{x_0}\tilde{x}+i\bar{\beta}\tilde{y}},\label{170720:1422}
\end{equation}
where $\tilde{\psi}_0^{(\text{nk})}$, and this is important, is exactly equal to~$\psi_0^{(\text{nk})}$, thus being free of any ``increase of localization" in \eqref{170720:1422}.
The reason for $\tilde{\psi}_0^{(\text{nk})}=\psi_0^{(\text{nk})}$ is that not only the fields transform
invariantly under \eqref{170716:1209}, but that also, by construction, transformation \eqref{170716:1209}, along with the associated solution of \eqref{170716:1231}, will transform
the original underlying equations \eqref{170714:2028} exactly into the same system of Eqs.$\,$[37-39] as in \cite{Hau17}, where, for zero potential vorticity, $\psi_0^{(\text{nk})}$ is its general solution
(which in Sec.$\,$[V.C3] was assumed ``to be independent of $y$ for simplicity").

Hence, the advanced dynamical properties proclaimed in \cite{Hau17}, particularly in Sec.$\,$[V.C3] and in Fig.$\,$[1], are irrelevant
for the understanding of the internal dynamics of linear shear flow processes, simply because they can be transformed away when going into the optimal frame of reference
--- let alone the fact that Fig.$\,$[1] itself is also misleading, as shown~below.

This study closes by pointing out a few technical mistakes and inconsistencies that occurred in the course of constructing Fig.$\,$[1] in \cite{Hau17}.

\subsection{The ``localization" issue\label{S41}}

The discussion and conclusion of the ``increase of localization"-result in Sec.$\,$[V.C3] is based on the assumption that $\tilde{a}_2=Ak_{x_0}a_2\sim a_2$ is a complex number with positive imaginary part. However, apart from the fact that $a_2$ itself is physically ill-defined (see Sec.$\,$\ref{S42}), this assumption of being a complex number is based on an arbitrary choice and is not the result of a mathematical constraint, as misleadingly claimed in \cite{Hau17} after Eq.$\,$[44]. Because, the applied\linebreak 2-D Fourier transform Eq.$\,$[44] on the single mode representation Eq.$\,$[43] is also well-defined for $a_2\in\mathbb{R}$,\footnote[2]{A Fourier transform is also well-defined (in a distributional sense) for non-integrable but bounded functions, e.g.~as for non-decaying but bounded wave functions. The Fourier transform of a bounded function is again bounded, while for unbounded functions the Fourier transform is not well-defined.} and not only if $a_2\in\mathbb{C}$ for $\text{Im}(a_2)\sim\text{Im}(\tilde{a}_2)=\tilde{a}_2^i>0$. Furthermore, the latter choice, as used in \cite{Hau17}, is even stated inaccurately, because to obtain the wanted result of a weighted Gaussian distribution, it is not sufficient to only demand $\tilde{a}_2^i>0$. The sign of the parameter $a_3$ (if chosen real as in Fig.$\,$[1]) is also relevant, thus restricting in the end three parameters in Eq.$\,$[44] in order to yield a Gaussian distribution, namely the imaginary part of $a_2\in \mathbb{C}$ as well as $a_3\in\mathbb{R}$ and $k_{x_0}\in\mathbb{R}$, such that they collectively amount to the constraint $\tilde{a}_2^i/a_3=Ak_{x_0}a_2^i/a_3>0$. However, this restriction
leads to an inevitable inconsistency when examining it more closely. Because, when completely going back to physical position space in trying to construct the corresponding {\it general} and {\it invertible} spectral representation from the single mode representation Eq.$\,$[43] (without the $y$-independence and $t=0$ restriction as assumed in Sec.$\,$[V.C3])
\begin{equation}
\psi(t,x,y)=\psi^{(\text{nk})}(t-{\textstyle\frac{a_2}{a_3}}y, k_{x_0},\bar{\beta})e^{ik_{x_0}(x-Ayt+\frac{a_2}{2a_3}Ay^2)+i\bar{\beta}y},\;\; k_{x_0},\bar{\beta}\in\mathbb{R},\label{170811:1328}
\end{equation}
it is the restriction on $k_{x_0}$ (based on the chosen values of $a_2$ and $a_3$) which makes this inconsistency. To explicitly see it, one has to recognize that the real-valued parameters $k_{x_0}$ and $\bar{\beta}$ are the relevant eigenvalues (or modes) to sum up in order to obtain the {\it general} and {\it invertible} representation of the physical field $\psi(x,y,t)\in\mathbb{R}$ in position space as follows:\footnote[2]{Note that the physical field $\psi(t,x,y)$ on the left-hand side of \eqref{170811:1437} does {\it not} depend on the parameters $a_2$ and $a_3$. In other words, its integral representation on the right-hand side evaluates such that it is independent of $a_2$ and $a_3$. The reason simply is that the original underlying dynamical equations \eqref{170714:2028} for the physical fields $\psi\,\hat{=}\,\psi^{\{u,v,\rho\}}(t,x,y)$ do not contain these parameters, or, metaphorically speaking, do not know of their existence. They only come about through the spectral representation and thus only have their existence in spectral space. Going into physical position space, however, they just average out and thus completely loose their relevance. Hence, the two parameters $a_2$ and $a_3$, or better, its ratio $a_2/a_3$ can thus be regarded at most only as a regulating parameter within spectral space. This result is consistent and to be compared with the conclusions given in Sec.$\,$\ref{S234}, Sec.$\,$\ref{S243}, and the summary in Sec.$\,$\ref{S25}.}
\begin{equation}
\psi(t,x,y)=\int_{-\infty}^\infty dk_{x_0}d\bar{\beta}\, \psi^{(\text{nk})}\big(t-{\textstyle\frac{a_2}{a_3}}y,k_{x_0},\bar{\beta}\big)e^{ik_{x_0}(x-Ayt+\frac{a_2}{2a_3}Ay^2)+i\bar{\beta}y}.\label{170811:1437}
\end{equation}
That the parameters $k_{x_0}$ and $\bar{\beta}$ are indeed the relevant modes to sum up can be easily verified from the single mode expression \eqref{170720:1422} in the optimal frame, where they just arise as normal Fourier modes --- and by transforming this normal (classical) Fourier transform according to the defining transformation \eqref{170716:1209} will then give the (non-optimal) inertial frame result~\eqref{170811:1437}.

Hence, since $k_{x_0}$ is the complementary real-valued Fourier variable to the unrestricted real-valued spatial variable $\tilde{x}$ \eqref{170716:1209}, $k_{x_0}$ must be unrestricted too, independent of the choice of
the parameters $a_2$ and $a_3$, otherwise the invertibility of \eqref{170811:1437} cannot be guaranteed. Restricting $k_{x_0}$ will thus obviously lead to an inconsistency. This inconsistency, however, can only be solved if $a_2\in\mathbb{R}$, i.e., if the imaginary part of $a_2$ is zero, as only this natural choice will not restrict~$k_{x_0}$. But $\text{Im}(a_2)\sim\text{Im}(\tilde{a}_2)=\tilde{a}_2^i=0$ will lead to a completely different result in Eq.$\,$[44] in \cite{Hau17}: Instead of a weighted Gaussian one obtains a pure wave mode distribution, which simply allows for a different conclusion than the misleading one given in Sec.$\,$[V.C3] and then misleadingly also shown in Fig.$\,$[1].

To note here is that the Fourier transform applied in Eq.$\,$[44] (or in Eq.$\,$[E1]) in \cite{Hau17} is not to be confused with the Fourier transform \eqref{170811:1437} given above. The former only transforms
the single particular invariant solution Eq.$\,$[43] itself (the so-called single mode), while the latter sums over all modes, thus giving the correct Fourier (spectral) representation~\eqref{170811:1437} in the inertial frame. In contrast again to the former spectral integral transform Eq.$\,$[44], where its physical relevance is even questionable in so far as the integrand, given by Eq.$\,$[43], is already the (local) spectral representation for physical field $\psi(x,y,t=0)$ in terms of the modes referring to $k_{x_0}$ and $\bar{\beta}$. It does not need to be transformed again into a second spectral representation, as done by the authors by introducing in Eq.$\,$[44] two new additional and independent spectral variables $k_x$ and $k_y$. That the (non-optimal) inertial frame representation \eqref{170811:1437} really represents a 2-D Fourier transform formulated, however, only in a mathematical intricate form, can only be mathematically as well as physically fully understood when going into the corresponding optimal frame. The (unique) inverse of \eqref{170811:1437} is thus given by
\begin{equation}
\psi^{(\text{nk})}\big(t,k_{x_0},\bar{\beta}\big)=\int_{-\infty}^\infty\frac{dx}{2\pi}\frac{dy}{2\pi}\,
\psi\big(t+{\textstyle\frac{a_2}{a_3}}y,x,y\big)e^{-ik_{x_0}(x-Ayt-\frac{a_2}{2a_3}Ay^2)-i\bar{\beta}y}
.\label{170720:2144}
\end{equation}
Finally note that the choice of the physical dimension $[\tilde{a}_2]=1/L$, as given just before Eq.$\,$[44] in \cite{Hau17}, is arbitrary as well. In fact, the physical dimension of $\tilde{a}_2$ is given correctly by $[\tilde{a}_2]=[a_3]/L^2$, thus depending on the dimension of $a_3$, but which itself cannot be uniquely determined. This issue will be discussed next.

\newgeometry{left=2.49cm,right=2.49cm,top=2.50cm,bottom=1.95cm,headsep=1em}

\subsection{The ``unification" issue\label{S42}}

From relation Eq.$\,$[50], the conclusion is drawn, that since $\omega\in\mathbb{C}$ then $a_2\in\mathbb{C}$. This conclusion, however, is invalid, because there is no restriction
that $\beta$ must be real. In general, $\beta$ is complex and thus $a_2$ can be real to also satisfy Eq.$\,$[50] for $\omega\in\mathbb{C}$. Despite this wrong conclusion, Eq.$\,$[50] contains another,
more fundamental mistake: For $a_2\neq 0$ {\it and} $a_3\neq 0$, the constructed ratio $\frac{\beta}{a_3+a_2c_s}$ is physically meaningless. The same is true for the last two relations in Eq.$\,$[48]. The reason is
that the parameters $a_2$, $a_3$ and $\beta$ themselves are not physically defined when denoted separately. Only their ratios, namely $a_3/a_2$, $\beta/a_2$ and $\beta/a_3$, are physically defined. For its absolute values $a_2$, $a_3$ and $\beta$, however, no physical dimension can be determined, since in the solution process for Eq.$\,$[48], which itself is the result of combining Eq.$\,$[22] (first line of Eq.$\,$[48] for $\psi=u$ and $\alpha=ik_x$) and Eq.$\,$[47] (second line of Eq.$\,$[48]), one of these three parameters can always be regarded as inessential (as also already mentioned in the previous section). In contrast, of course, to their ratios, which are physically well-defined, since they have the unique physical dimensions $[a_3/a_2]=L/T$, $[\beta/a_2]=1/T$ and $[\beta/a_3]=1/L$. Hence, the ``general solution (gm)" as presented in the last line of Eq.$\,$[48] is physically meaningless, due to exposing the non-physical absolute parameters. For a physical relation, only their ratios should enter. But this immediately provokes a decision problem: Should the absolute parameters be made relative to $a_2$ {\it or} $a_3$? To make them relative to $a_2$ {\it and} $a_3$ simultaneously is, of course, not possible. So, one has to decide: {\it either} relative to $a_2$, then $\beta$ turns into the physical variable $[\beta/a_2]=[\omega]=1/T$ and one exclusively deals with the modal approach (since the limit $a_2\rightarrow 0$ is then not possible to perform anymore), {\it or} relative to $a_3$, then $\beta$ turns into the physical variable $[\beta/a_3]=[k_y]=1/L$ and one exclusively deals with the Kelvin-mode approach (since then the limit $a_3\rightarrow 0$ is not possible to perform anymore).

The general idea in Eqs.$\,$[48-50], namely to construct a single function which can blend between the modal and non-modal (Kelvin mode) solutions, is an endeavour which is unrealistic from the outset. The modal
and non-modal approaches are {\it complementary} to each other and cannot be realized simultaneously in the sense of a blending function as physically incorrect and misleadingly given by Eq.$\,$[50]; there is no approach which is $x\%$ modal and $(100\!-\!x)\%$ non-modal! --- except, of course, for the limiting cases $x=0$ or $x=100$.

Finally to note here is that several intermediate steps leading to the misleading result Eq.$\,$[50] contain misprints: In the defining equation Eq.$\,$[39] for the function Eq.$\,$[47], ``$a_2$" in the second summand of Eq.$\,$[39] must be replaced by $ic_s^2a_2$, and ``$c_s u^{\text{(nk)}}$" in the last factor on the left-hand side by $c_s^2 u^{\text{(nk)}}$, while in the second summand in Eq.$\,$[47] ``$k_y(t^\prime)\mathcal{A}^{(\text{nk})}$" by $\bar{\beta}(t^\prime)\mathcal{B}^{(\text{nk})}$. Lastly in Eq.$\,$[48] ``$c_s$" has to be replaced by $-c_s$, which then also effects and changes the expressions in Eq.$\,$[49-50]. However, all these misprints, when corrected, have no positive effect on the physical incorrectness of Eqs.$\,$[48-50] as explained above, since this mistake is of methodological nature.

\subsection{The ``asymptotic-decay" issue\label{S43}}

In Appendix [E] it is claimed that the integration constants $\mathcal{A}^{(\text{nk})}$ and $\mathcal{B}^{(\text{nk})}$ have to be chosen as given in Eq.$\,$[E2] in order
to obtain asymptotic decay for the perturbation fields as $y\rightarrow\pm \infty$.\linebreak[4] However, by closer inspection, this claim is not correct. At the example of the initial streamwise velocity field $u_0^{(\text{nk})}$ from Eq.$\,$[47] (exactly the same field as used in Appendix [E] to derive this specification Eq.$\,$[E2]), it can be shown that for $a_2\in\mathbb{C}$, $a_3\in\mathbb{R}$ and $\bar{\beta}\in\mathbb{R}$, as chosen in Fig.$\,$[1], this field is always exponentially decaying, independent of how the constants $\mathcal{A}^{(\text{nk})}$ and $\mathcal{B}^{(\text{nk})}$ are chosen. In other words, these constants have no effect on the asymptotic behaviour of the considered solution Eq.$\,$[47]. Taking exactly the parameters as chosen in Fig.$\,$[1b], the shown initial perturbation field for the streamwise velocity has the following explicit structure~(Eqs.$\,$[22]~\&~[47])
\begin{equation}
\left.
\begin{aligned}
u(x,y,0)&=u_0^{(\text{nk})}e^{ix-\frac{1}{2}(1-i)y^2+5iy}\\
&=e^{ix-\frac{1}{2}(1+i)y^2-5y}\Big[\mathcal{A}^{(\text{nk})}
\cdot\!\!\phantom{.}_1\hspace{-0.25mm}F_1\!\left({\textstyle \frac{1-i}{4},\frac{1}{2},(\frac{2}{5}-\frac{i}{5})}({\scriptstyle 5+(1+i)y})^{\scriptscriptstyle 2}\right)\\[-0.25em]
&\hspace{3.0cm} -\,\mathcal{B}^{(\text{nk})}\big({\textstyle 5+(1+i)y}\big)
\cdot\!\!\phantom{.}_1\hspace{-0.25mm}F_1\!\left({\textstyle \frac{3-i}{4},\frac{3}{2},(\frac{2}{5}-\frac{i}{5})}({\scriptstyle 5+(1+i)y})^{\scriptscriptstyle 2}\right) \Big],
\end{aligned}
~~~ \right \}
\label{170721:1606}
\end{equation}

\restoregeometry

\noindent which in the asymptotic limit $|y|\rightarrow\infty$ behaves as \citep{Abramowitz65}
\begin{equation}
u(x,y,0)\underset{y\rightarrow\pm\infty}{=}\mathcal{A}^{(\text{nk})} e^{-\frac{1}{10}y^2}|y|^{-\frac{1}{2}}f_1(x,y)\mp\mathcal{B}^{(\text{nk})} e^{-\frac{1}{10}y^2}|y|^{-\frac{1}{2}}f_2(x,y),
\end{equation}
where $f_1$ and $f_2$ for all $x\in\mathbb{R}$ are bounded functions  as $y\rightarrow\pm\infty$. Hence, independent of how the constants $\mathcal{A}^{(\text{nk})}$ and $\mathcal{B}^{(\text{nk})}$ are chosen, the
initial streamwise perturbation field \eqref{170721:1606}, as shown in Fig.$\,$[1b], always decays exponentially on both ends of the $y$-domain.

To note is that with the specification of the integration constants as given in Eq.$\,$[E2], the specific single mode $u(x,y,0)$ \eqref{170721:1606} is a complex-valued function (see the remark again of the first footnote on p.$\,$\pageref{171202:2041}), hence, Fig.$\,$[1b] must either show the real or imaginary part of that field. Which part is actually shown is not clear, because, upon reproducing this figure, the result is that the contours of the real part only slightly differ from those of the imaginary part. However, in both cases the contours are tilted towards the negative $x$-direction, and not towards the positive $x$-direction as shown. This can be controlled by the parameter $\bar{\beta}$, depending on its sign, i.e., Fig.$\,$[1b] can thus only refer to $\bar{\beta}=-5$, and not to $\bar{\beta}=5$ as given, which is maybe the problem of another misprint in \cite{Hau17}.

\appendix

\section{Equivalence and redundancy of subalgebra $\mathfrak{f}_3$ to $\mathfrak{f}_2$\label{SA}}

Based on the formal result \eqref{171026:1908}, this section explicitly demonstrates or validates that the 1-D\linebreak[4] subalgebra $\mathfrak{f}_3=\{\lambda X_2+X_5+\alpha X_3\}$ and its associated invariant function is mathematically (group-theoretically) equivalent to that of $\mathfrak{f}_2=\{X_2+\alpha X_3\}$. According to \eqref{171026:1907}, the simplest connecting symmetry transformation $g$ between these two subalgebras is given by $\mathsf{S}_\mathsf{1}\circ\mathsf{S}_\mathsf{6}$ \eqref{171026:1545} when the group parameters are chosen as in \eqref{171026:1908}:
\begin{equation}
\mathsf{S}_\mathsf{1}\circ\mathsf{S}_\mathsf{6}\!:\;\;\; \tilde{t}={\textstyle \frac{t+\varepsilon_1}{1-4(t+\varepsilon_1)\varepsilon_6}},\;\;\; \tilde{x}={\textstyle \frac{x}{1-4(t+\varepsilon_1)\varepsilon_6}},\;\;\;\tilde{u}
={\scriptstyle \sqrt{1-4(t+\varepsilon_1)\varepsilon_6}}\, e^{-\frac{\varepsilon_6 x^2}{1- 4(t+\varepsilon_1)\varepsilon_6}}\, u,\label{171029:1739}
\end{equation}
with $\varepsilon_1=\frac{1}{2}(\lambda-\gamma)$ and $\varepsilon_6=-\frac{1}{2\gamma}$, where $\gamma=1$ in time units, since $[\gamma]=[t]$. Obviously, since the transformation~\eqref{171029:1739} forms a combined additive Lie-group, its inverse is given by\footnote[2]{To note in \eqref{171029:1741} is that $(\mathsf{S}_\mathsf{1}\circ\mathsf{S}_\mathsf{6})^{-1}=\mathsf{S}_\mathsf{6}^{-1}\circ\mathsf{S}_\mathsf{1}^{-1}$.}
\begin{equation}
(\mathsf{S}_\mathsf{1}\circ\mathsf{S}_\mathsf{6})^{-1}\!:\;\;\; t={\textstyle \frac{\tilde{t}}{1+4\tilde{t}\vphantom{A^A}\varepsilon_6}}-\varepsilon_1,\;\;\; x={\textstyle \frac{\tilde{x}}{1+4\tilde{t}\vphantom{A^A}\varepsilon_6}},\;\;\; u
={\scriptstyle \sqrt{1+4\tilde{t}\varepsilon_6}}\, e^{\frac{\varepsilon_6 \tilde{x}^2}{1+ 4\tilde{t}\vphantom{A^a}\varepsilon_6}}\, \tilde{u}.\hspace{2.925cm}\label{171029:1741}
\end{equation}
Now, when transforming the single basis element of $\mathfrak{f}_3$ according to \eqref{171029:1739}, one obtains
\begin{align}
\lambda X_2+X_5+\alpha X_3 &= (2t+\lambda)\partial_x +(\alpha-x) u\partial_u\nonumber\\[0.25em]
&=(2t+\lambda)\left(\frac{\partial\tilde{t}}{\partial x}\partial_{\tilde{t}}+\frac{\partial\tilde{x}}{\partial x}\partial_{\tilde{x}}+\frac{\partial\tilde{u}}{\partial x}\partial_{\tilde{u}}\right)
+(\alpha-x)u\left(\frac{\partial\tilde{t}}{\partial u}\partial_{\tilde{t}}+\frac{\partial\tilde{x}}{\partial u}\partial_{\tilde{x}}+\frac{\partial\tilde{u}}{\partial u}\partial_{\tilde{u}}\right)\nonumber\\[0.25em]
&=\left((2t+\lambda)\frac{\partial\tilde{x}}{\partial x}\right)\partial_{\tilde{x}}
+\left((2t+\lambda)\frac{\partial\tilde{u}}{\partial x}+(\alpha-x)u\frac{\partial\tilde{u}}{\partial u}\right)\partial_{\tilde{u}}\nonumber\\[0.25em]
&=\left(\frac{\gamma^2}{\gamma-2\tilde{t}}\cdot\frac{\gamma-2\tilde{t}}{\gamma}\hspace{0.5mm}\right)\partial_{\tilde{x}}
+\left(\frac{\gamma^2}{\gamma-2\tilde{t}}\cdot\frac{\tilde{x}\tilde{u}}{\gamma}
+\frac{\alpha\big(\gamma-2\tilde{t}\hspace{0.5mm}\big)\tilde{u}-\gamma\tilde{x}\tilde{u}}{\gamma-2\tilde{t}}\right)\partial_{\tilde{u}}\nonumber\\[0.25em]
&= \gamma\partial_{\tilde{x}}+\alpha\tilde{u}\partial_{\tilde{u}}\;\;\hat{\equiv}\;\;
\partial_{\tilde{x}}+\tilde{\alpha}\tilde{u}\partial_{\tilde{u}}=\tilde{X}_2+\tilde{\alpha}\tilde{X}_3,\,\;\text{where $\,\tilde{\alpha}=\alpha/\gamma$,}\label{171104:1640}
\end{align}
which constitutes the single basis element of $\mathfrak{f}_2$ in the new transformed coordinates \eqref{171029:1739}. This reduction can also be demonstrated by transforming the invariant solution \eqref{171027:2244} to  $\mathfrak{f}_3$ directly, with the result
\begin{align}
u(t,x)=u_3(t,\alpha;\lambda)\,e^{-\frac{x(x-2\alpha)}{2(\lambda+2t)}}\;\: &\Leftrightarrow\;\:
\tilde{u}(\tilde{t},\tilde{x};\lambda)= \frac{e^{-\frac{\varepsilon_6 \tilde{x}^2}{1+ 4\tilde{t}\vphantom{A^a}\varepsilon_6}}}{{\scriptstyle \sqrt{1+4\tilde{t}\varepsilon_6}}}
u_3\!\left({\textstyle \frac{\tilde{t}}{1+4\tilde{t}\vphantom{A^A}\varepsilon_6}}-\varepsilon_1,\alpha;\lambda\right)
e^{-\frac{x(x-2\alpha)}{2(\lambda+2t)}}\nonumber\\[0.25em]
&\hspace{2.275cm}=\frac{e^{-\frac{\varepsilon_6 \tilde{x}^2}{1+ 4\tilde{t}\vphantom{A^a}\varepsilon_6}}}{{\scriptstyle \sqrt{1+4\tilde{t}\varepsilon_6}}}
u_3\!\left({\textstyle \frac{\tilde{t}}{1+4\tilde{t}\vphantom{A^A}\varepsilon_6}}-\varepsilon_1,\alpha;\lambda\right)
e^{-\frac{\tilde{x}^2}{2(\gamma- 2\tilde{t}\vphantom{A^a})}}e^{\frac{\alpha\tilde{x}}{\gamma}}\nonumber\\[0.25em]
&\hspace{2.275cm}=\frac{u_3{\scriptstyle\left({\textstyle \frac{\tilde{t}}{1+4\tilde{t}\vphantom{A^A}\varepsilon_6}}-\varepsilon_1,\alpha;\lambda\right)}}{{\scriptstyle\sqrt{1+4\tilde{t}\varepsilon_6}}}
\, e^{\frac{\alpha\tilde{x}}{\gamma}}\:\equiv\: {\textstyle\frac{1}{\gamma}}\, \tilde{u}_2(\tilde{t},\tilde{\alpha};\lambda)\, e^{\tilde{\alpha}\tilde{x}},\label{171031:1101}
\end{align}
where in the last line the fact has been used that the prefactor can be identified as an own function $\tilde{u}_2$, depending only on the new-frame time variable $\tilde{t}$ along with the two parameters  $\tilde{\alpha}=\alpha/\gamma$ and $\lambda$. However, note that this last line in \eqref{171031:1101} actually has to be identified by $\tilde{u}_2/\gamma$ and not by $\tilde{u}_2$, which would be dimensionally inconsistent since the modes of $\mathfrak{f}_2$ should remain dimensionally invariant in either frame according to its defining transformation \eqref{171029:1739}, i.e., $[u_2]=[\tilde{u}_2]$. Actually, this single-mode identification in \eqref{171031:1101} has its origin in the corresponding non-local representations \eqref{171028:0947} and \eqref{171027:1929} when considering all modes collectively, as only these representations ultimately determine the physical dimensions of $u_3$ and $u_2$, respectively. The result then is that $[\alpha u_3]=[\tilde{\alpha}\tilde{u}_2]$, or equivalently $[u_3]=[\tilde{u}_2/\gamma]$, which also can be explicitly seen when transforming instead of the local (single-mode) representation \eqref{171027:2244}, as shown in~\eqref{171031:1101}, the associated non-local representation \eqref{171028:0947}
\begin{align}
u(t,x)=\int_{\mathbb{C}}d\alpha\, u_3(t,\alpha;\lambda)\,e^{-\frac{x(x-2\alpha)}{2(\lambda+2t)}} \;\;\: &\Leftrightarrow\;\;\:
\tilde{u}(\tilde{t},\tilde{x};\lambda)=\int_{\mathbb{C}}d\alpha
\frac{u_3{\scriptstyle\left({\textstyle \frac{\tilde{t}}{1+4\tilde{t}\vphantom{A^A}\varepsilon_6}}-\varepsilon_1,\alpha;\lambda\right)}}{{\scriptstyle\sqrt{1+4\tilde{t}\varepsilon_6}}}
\, e^{\frac{\alpha\tilde{x}}{\gamma}}\nonumber\\[0.25em]
&\hspace{2.35cm}=\int_{\mathbb{C}}d\tilde{\alpha}\, \gamma
\frac{u_3{\scriptstyle\left({\textstyle \frac{\tilde{t}}{1+4\tilde{t}\vphantom{A^A}\varepsilon_6}}-\varepsilon_1,\alpha;\lambda\right)}}{{\scriptstyle\sqrt{1+4\tilde{t}\varepsilon_6}}}
\, e^{\tilde{\alpha}\tilde{x}}\hspace{1.25cm}\nonumber\\[0.25em]
&\hspace{2.35cm}\equiv \int_{\mathbb{C}}d\tilde{\alpha}\, \tilde{u}_2(\tilde{t},\tilde{\alpha};\lambda)\, e^{\tilde{\alpha}\tilde{x}}.\label{171106:1255}
\end{align}
Hence, the representations \eqref{171027:2244}-\eqref{171028:0947} induced by $\mathfrak{f}_3$ can be expressed by or reduced to the corresponding ones \eqref{171027:1900}-\eqref{171027:1929} of $\mathfrak{f}_2$. Since \eqref{171031:1101} and \eqref{171106:1255} are based on the  symmetry transformation \eqref{171029:1739} leaving the underlying dynamical system \eqref{151113:1225} (up to the initial condition) invariant, the subalgebra $\mathfrak{f}_3$ is thus said to be mathematically {\it equivalent} to $\mathfrak{f}_2$.

However, the symmetry transformation \eqref{171029:1739} is not the only transformation for which $\mathfrak{f}_3$ can be expressed by or reduced to $\mathfrak{f}_2$. For example, when considering (invertible) coordinate transformations to only invoke a physical change in the reference frame
\begin{equation}
\mathsf{T}\!:\;\;\; \tilde{t}=\tilde{t}(t,x),\;\;\; \tilde{x}=\tilde{x}(t,x),\;\;\; \tilde{u}=\phi_1(t,x)\cdot u+\phi_2(t,x),\label{171101:1426}
\end{equation}
where the scalar field $u$ transforms as a tensorial scalar up to a local re-gauging $\phi$, then an infinite family of frames exist where the inertial frame symmetry algebra $\mathfrak{f}_3$ turns into~$\mathfrak{f}_2$:
\begin{align}
\lambda X_2+X_5+\alpha X_3 &= (2t+\lambda)\partial_x +(\alpha-x) u\partial_u\nonumber\\[0.25em]
&=(2t+\lambda)\left(\frac{\partial\tilde{t}}{\partial x}\partial_{\tilde{t}}+\frac{\partial\tilde{x}}{\partial x}\partial_{\tilde{x}}+\frac{\partial\tilde{u}}{\partial x}\partial_{\tilde{u}}\right)
+(\alpha-x)u\frac{\partial\tilde{u}}{\partial u}\partial_{\tilde{u}}\nonumber\\[0.25em]
&= \xi\partial_{\tilde{x}}+\alpha\tilde{u}\partial_{\tilde{u}}
\;\;\hat{\equiv}\;\;\partial_{\tilde{x}}+\tilde{\alpha}\tilde{u}\partial_{\tilde{u}}=\tilde{X}_2+\tilde{\alpha}\tilde{X}_3,\;\;\text{with}\;\,\tilde{\alpha}=\alpha/\xi,\label{171103:1025}
\end{align}
where the defining differential conditions\footnote[2]{Note that in order for \eqref{171103:1025} to be dimensionally consistent, a dimensional factor $\xi$ in units of time $t$ has to be introduced. However, for this particular operator \eqref{171103:1025} in view of determining invariant functions, this factor turns out to be an inessential parameter as it can be absorbed into the essential parameter $\alpha$, to give $\tilde{\alpha}=\alpha/\xi$ which then has units of inverse length $1/x$.}
\begin{equation}
(2t+\lambda)\frac{\partial\tilde{t}}{\partial x}=0,\qquad (2t+\lambda)\frac{\partial\tilde{x}}{\partial x}=\xi,\qquad (2t+\lambda)\frac{\partial\tilde{u}}{\partial x}+(\alpha-x)u\frac{\partial\tilde{u}}{\partial u}=\alpha\tilde{u},\label{171104:1456}
\end{equation}
lead to the general solution for \eqref{171101:1426}
\begin{equation}
\mathsf{T}\!:\;\;\; \tilde{t}=f_1(t),\;\;\; \tilde{x}={\textstyle \frac{\xi x}{\lambda+2t}}+f_2(t),\;\;\; \tilde{u}=f_3(t)e^{\frac{x^2}{2(\lambda+2t)}}\cdot u+f_4(t)e^{\frac{\alpha x}{\lambda+2t}},\label{171101:1447}
\end{equation}
which, obviously, also includes the symmetry transformation \eqref{171029:1739} as a specific choice\footnote[3]{Choosing $f_1(t)=\frac{\gamma}{2}-\frac{\gamma^2}{2(\lambda+2t)}$, $f_2(t)=0$, $f_3(t)={\scriptstyle \sqrt{\frac{\lambda+2t}{\gamma}}}$, $f_4(t)=0$ and $\xi=\gamma$, will reduce \eqref{171101:1447} to the\linebreak[4] symmetry transformation \eqref{171029:1739}. Note here that $f_1$ carries the physical dimension of time, that $f_3$ is dimensionless, and that $\tilde{x}$ in \eqref{171101:1447} is again a spatial variable carrying the same dimension as $x$.} of the temporally arbitrary but $\mathsf{T}$-invertible integration functions $f_i$, $1\leq i\leq 4$. As before for the symmetry transformation in \eqref{171031:1101}, this reduction from $\mathfrak{f}_3$ to $\mathfrak{f}_2$ can also directly be seen when transforming the associated invariant solution to $\mathfrak{f}_3$ according to
$\mathsf{T}$ \eqref{171101:1447}
\begin{align}
u(t,x)=u_3(t,\alpha;\lambda)\,e^{-\frac{x(x-2\alpha)}{2(\lambda+2t)}}\;\; &\Leftrightarrow\;\;
\tilde{u}(\tilde{t},\tilde{x};\lambda)= f_3(t)e^{\frac{x^2}{2(\lambda+2t)}}\cdot u_3(t,\alpha;\lambda)
e^{-\frac{x(x-2\alpha)}{2(\lambda+2t)}}+f_4(t)e^{\frac{\alpha x}{\lambda+2t}}\hspace{0.45cm}\nonumber\\[0.25em]
&\hspace{2.275cm}= \Big(f_3(t)u_3(t,\alpha;\lambda)+f_4(t)\Big)
e^{\frac{\alpha x}{\lambda+2t}}\nonumber\\[0.25em]
&\hspace{2.275cm}= \Big(f_3(t)u_3(t,\alpha;\lambda)+f_4(t)\Big)
e^{-\frac{\alpha f_2(t)}{\xi}} e^{\frac{\alpha \tilde{x}}{\xi}}\nonumber\\[0.25em]
&\hspace{2.275cm} \equiv\: g(t,\tilde{\alpha};\lambda)\, e^{\tilde{\alpha} \tilde{x}} = g\big((f_1^{-1}(\tilde{t}),\tilde{\alpha};\lambda\big)\, e^{\tilde{\alpha} \tilde{x}}\nonumber\\[0.25em]
&\hspace{2.275cm} \equiv\: {\textstyle\frac{1}{\xi}}\, \tilde{u}_2(\tilde{t},\tilde{\alpha};\lambda)\, e^{\tilde{\alpha}\tilde{x}},\label{171101:0928}
\end{align}
which corresponds to the invariant solution \eqref{171027:1900} induced by $\mathfrak{f}_2$ within the new frame. That the single mode in the last line in~\eqref{171101:0928} has to be identified as $\tilde{u}_2/\xi$ and not simply as $\tilde{u}_2$ can again be explicitly validated when considering all modes collectively, namely by transforming the associated non-local representation \eqref{171028:0947} accordingly. As already said before, it is because only these representations determine the physical dimensions of the complementary fields, with the result that $[u_3]=[\tilde{u}_2/\xi]$ as obtained~from
\begin{align}
\! u(t,x)=\int_{\mathbb{C}}d\alpha\, u_3(t,\alpha;\lambda)\,e^{-\frac{x(x-2\alpha)}{2(\lambda+2t)}} \; &\Leftrightarrow\;
\tilde{u}(\tilde{t},\tilde{x};\lambda)=\int_{\mathbb{C}}d\alpha\,\Big(f_3(t)u_3(t,\alpha;\lambda)+f_4(t)\Big)\,
e^{-\frac{\alpha f_2(t)}{\xi}} e^{\frac{\alpha \tilde{x}}{\xi}}\nonumber\\[0.25em]
&\hspace{2.15cm}=\int_{\mathbb{C}}d\tilde{\alpha}\: \xi\, \Big(f_3(t)u_3(t,\alpha;\lambda)+f_4(t)\Big)\,
e^{-\tilde{\alpha} f_2(t)} e^{\tilde{\alpha} \tilde{x}}\nonumber\\[0.25em]
&\hspace{2.15cm}\equiv\int_{\mathbb{C}}d\tilde{\alpha}\: \xi\, g\big((f_1^{-1}(\tilde{t}),\tilde{\alpha};\lambda\big)\, e^{\tilde{\alpha} \tilde{x}}\nonumber\\[0.25em]
&\hspace{2.15cm}\equiv \int_{\mathbb{C}}d\tilde{\alpha}\, \tilde{u}_2(\tilde{t},\tilde{\alpha};\lambda)\, e^{\tilde{\alpha}\tilde{x}},\label{171106:1337}
\end{align}
which corresponds to the global representation \eqref{171027:1929} induced by $\mathfrak{f}_2$ within the new frame. Hence, since $\mathsf{T}$ \eqref{171101:1447} is in general not a symmetry transformation of the underlying mathematical equation \eqref{151113:1225}, making the subalgebra $\mathfrak{f}_3$ thus mathematically inequivalent to $\mathfrak{f}_2$, it still is physically {\it redundant} to it, simply because (i) when changing the physical frame of reference according to $\mathsf{T}$ \eqref{171101:1447}, the old-frame symmetry algebra $\mathfrak{f}_3$ changes to the new-frame symmetry algebra $\mathfrak{f}_2$,\linebreak[4] and (ii) since from both $\mathsf{T}$-connected frames \eqref{171101:1447} the {\it same} physical process of a 1-D diffusion is described, yet for each frame in a mathematically different way, one can thus always regard one mathematical description as physically redundant over the other one.

\noindent The difference in the mathematical formulation is of course expressed only in the shape~$\tilde{u}_2$ of the defining $\mathfrak{f}_2$-induced mode $e^{\alpha x}$, which in the new frame surely is different to its shape $u_2$~in the inertial frame, i.e.,
$\tilde{u}_2(\tilde{t},\tilde{\alpha};\lambda)\neq u_2(f^{-1}_1(\tilde{t}),\xi\tilde{\alpha})=u_2(t,\alpha)$. In particular, as derived above in~\eqref{171101:0928}, the shape $\tilde{u}_2$ of the $\mathfrak{f}_2$-mode is universally related to the shape $u_3$ of the $\mathfrak{f}_3$-mode, by
\begin{align}
u_3(t,\alpha;\lambda)&=\frac{f_3^{-1}(f_1^{-1}(\tilde{t}))}{\xi}e^{\tilde{\alpha}f_2(f_1^{-1}(\tilde{t}))}\cdot \tilde{u}_2(\tilde{t},\tilde{\alpha};\lambda)
-f_3^{-1}(f_1^{-1}(\tilde{t}))\, f_4(f_1^{-1}(\tilde{t}))\nonumber\\[0.25em]
&\equiv F_1(\tilde{t})\cdot \tilde{u}_2(\tilde{t},\tilde{\alpha};\lambda)+F_2(\tilde{t}),
\end{align}
which holds irrespective of whether the change of reference frame is defined by a symmetry transformation or not.
Hence, since the symmetry algebra $\mathfrak{f}_3$ formulated in the inertial frame induces the same modal approach as $\mathfrak{f}_2$ in any of the new frames defined by \eqref{171101:1447}, no new physical modes are generated by $\mathfrak{f}_3$ than those already established by $\mathfrak{f}_2$.

It is clear that the optimal frame of reference to describe a pure diffusion process is the inertial frame, as mathematically formulated for the 1-D problem by equation \eqref{151113:1225}. A change of frame in this case is only of interest if within the optimal (here inertial) frame a non-optimal symmetry algebra as e.g. $\mathfrak{f}_3$ is chosen to mathematically solve the problem. As shown in Section~\ref{S23}, only the reduction from $\mathfrak{f}_3$ to $\mathfrak{f}_2$, in changing the frame either by $\mathsf{S}_\mathsf{1}\circ\mathsf{S}_\mathsf{6}$ \eqref{171029:1739} or more generally by $\mathsf{T}$ \eqref{171101:1447}, allows us to give the correct mathematical guidelines and the correct physical interpretation when using the non-optimal symmetry algebra~$\mathfrak{f}_3$ within the inertial frame to solve the 1-D diffusion equation~\eqref{151113:1225}.

\section{Redundancy of subalgebra $\mathfrak{f}_4$ to $\mathfrak{f}_1$\label{SB}}

This section demonstrates that although the symmetry algebra $\mathfrak{f}_4$ \eqref{171027:1229} is mathematically (group-theoretically) inequivalent to $\mathfrak{f}_1$, it still is physically redundant to it. By changing the physical reference frame according to the following pure coordinate transformation
\begin{equation}
\mathsf{T}\!:\;\;\; \tilde{t}=\tilde{t}(t,x),\;\;\; \tilde{x}=\tilde{x}(t,x),\;\;\; \tilde{u}=u,\label{171110:1036}
\end{equation}
where, for simplicity, the scalar field $u$ transforms as a true scalar without any local re-gauging, there obviously exists an infinite family of frames where the inertial frame symmetry algebra $\mathfrak{f}_4$ turns into~$\mathfrak{f}_1$:
\begin{align}
2\tau X_1+X_4+\alpha X_3 &= 2(t+\tau)\partial_t +x\partial_x+\alpha u\partial_u\nonumber\\[0.25em]
&=2(t+\tau)\left(\frac{\partial\tilde{t}}{\partial t}\partial_{\tilde{t}}+\frac{\partial\tilde{x}}{\partial t}\partial_{\tilde{x}}\right)
+x\left(\frac{\partial\tilde{t}}{\partial x}\partial_{\tilde{t}}+\frac{\partial\tilde{x}}{\partial x}\partial_{\tilde{x}}\right)
+\alpha \tilde{u}\partial_{\tilde{u}}\nonumber\\[0.25em]
&= \tau\partial_{\tilde{t}}+\alpha\tilde{u}\partial_{\tilde{u}}
\;\;\hat{\equiv}\;\;\partial_{\tilde{t}}+\tilde{\alpha}\tilde{u}\partial_{\tilde{u}}=\tilde{X}_1+\tilde{\alpha}\tilde{X}_3,\;\;\text{with}\;\,\tilde{\alpha}=\alpha/\tau,\label{171110:1045}
\end{align}
where the defining differential conditions
\begin{equation}
2(t+\tau)\frac{\partial\tilde{t}}{\partial t}+x\frac{\partial\tilde{t}}{\partial x}=\tau,\qquad 2(t+\tau)\frac{\partial\tilde{x}}{\partial t}+x\frac{\partial\tilde{x}}{\partial x}=0,\label{171110:1104}
\end{equation}
then lead to the general solution for \eqref{171110:1036}
\begin{equation}
\mathsf{T}\!:\;\;\;\; \tilde{t}={\textstyle\frac{1}{2}}\tau\ln(1+t/\tau)+f_1\Big(\text{\raisebox{0.075cm}{$\textstyle\frac{x}{\sqrt{1+t/\tau}}$}}\Big),\quad\;\;
\tilde{x}=f_2\Big(\text{\raisebox{0.075cm}{$\textstyle\frac{x}{\sqrt{1+t/\tau}}$}}\Big),\quad\;\;
\tilde{u}=u.\label{171110:1322}
\end{equation}

\section{General solution induced by the 2-D subalgebra $\mathfrak{g}_2$\label{SC}}

This section develops a general solution of the governing equations \eqref{170714:2028} (for $\rho_0=1$)
\begin{equation}
\left.
\begin{aligned}
&\left(\frac{\partial}{\partial t}+Ay\frac{\partial}{\partial x}\right)\rho+\frac{\partial u}{\partial x}+\frac{\partial v}{\partial y}=0,\\[0.5em]
&\left(\frac{\partial}{\partial t}+Ay\frac{\partial}{\partial x}\right)u+Av+c_s^2\frac{\partial \rho}{\partial x}=0,\\[0.5em]
&\left(\frac{\partial}{\partial t}+Ay\frac{\partial}{\partial x}\right)v+c_s^2\frac{\partial \rho}{\partial y}=0,
\end{aligned}
~~~~~\right \}
\label{171123:0928}
\end{equation}
when based on the invariant-solution approach of the 2-D symmetry algebra $\mathfrak{g}_2$ \eqref{170730:1042}
\begin{align}
\mathfrak{g}_2&\underset{\phantom{\eqref{170724:1405}}}{=}\{X_1+\alpha X_4,\, X_3+\beta X_4\}\nonumber\\[-0.25em]
&\underset{\eqref{170724:1405}}{=}\{\partial_x+\alpha(u\partial_u+v\partial_v+\rho\partial_\rho),\, At\partial_x+\partial_y+\beta(u\partial_u+v\partial_v+\rho\partial_\rho)\}.\label{171123:0941}
\end{align}
In general, system \eqref{171123:0928} allows for vortex and wave solutions in linear shear flow. For convenience and for analytical simplicity, only the case of {\it pure} wave propagation will be considered, as outlined in
Sec.$\,$[6] in \cite{Hau16}. Therein a particular solution class is considered where \eqref{171123:0928} is turned into a Cauchy problem to satisfy the initial condition
\begin{equation}
\left.
\begin{aligned}
u(t,x,y)\big|_{t=0}& =\int_{-\infty}^\infty dk_x dk_y\, \hat{u}(t,k_x,k_y)\big|_{t=0}\, e^{i(k_x x+k_y y)},\\[0.25em]
v(t,x,y)\big|_{t=0}& =\int_{-\infty}^\infty dk_x dk_y\, \hat{v}(t,k_x,k_y)\big|_{t=0}\, e^{i(k_x x+k_y y)},\label{171123:1509}
\end{aligned}
~~~\right\}
\end{equation}
for the stream- and shearwise velocity in a 2-D Fourier representation. The corresponding complementary initial fields are given as (see Appendix$\,$[D] in \cite{Hau16})
\begin{equation}
\left.
\begin{aligned}
\hat{u}(t,k_x,k_y)\big|_{t=0}=\; &
\left(\frac{c_s k_{x}}{\sqrt{k_{x}^2+k_{y}^2}}-i
\frac{Ak_{y}}{k_{x}^2+k_{y}^2}\right)\Psi^{(+)}(k_x,k_y)\, e^{i\varphi},\\[0.5em]
\hat{v}(t,k_x,k_y)\big|_{t=0}=\; &
\left(\frac{c_s k_{y}}{\sqrt{k_{x}^2+k_{y}^2}}+i
\frac{Ak_{x}}{k_{x}^2+k_{y}^2}\right)\Psi^{(+)}(k_x,k_y)\, e^{i\varphi},\label{171123:1354}
\end{aligned}
~~~\right\}
\end{equation}
where $\Psi^{(+)}(k_x,k_y)$ is the (sufficiently fast decaying) spectral wave-packet
\begin{equation}
\Psi^{(+)}(k_x,k_y)=\frac{1}{2}\Big(G(\vk,\vk_0)+|G(\vk,\vk_0)|+G(\vk,-\vk_0)+|G(\vk,-\vk_0)|\Big),
\label{171123:1258}
\end{equation}
with
\begin{equation}
G(\vk,\vk_0)=\epsilon\arctan\!\left(k_R^2-(k_{x}-k_{x_0})^2-\bigg(k_{y}-\frac{k_{y_0}}{k_{x_0}}k_{x}\bigg)^{\!2}\right).
\label{171123:1259}
\end{equation}
The packet consists of two oval shaped localization areas, whose size\footnote[2]{The argument of the $\arctan$-function in \eqref{171123:1259} is measured relative to a constant wavenumber $k:=\lVert\vk_0^c\rVert$, which has been set to 1; otherwise the expression would be dimensionally inconsistent.} is defined by $k_R$ with centers situated at $(k_{x_0},k_{y_0})$ and $(-k_{x_0},-k_{y_0})$, respectively. The parameter $\epsilon$
defines the initial disturbance amplitude, having the physical dimension of length squared: $[\epsilon]=L^2$. The remaining free parameters in the above condition are set as in Fig.$\,$[6.6] \& [6.7] in \cite{Hau16}: $k_{x_0}=0.15\pi$, $k_{y_0}=0.6\pi$, $k_R=0.075\pi$, $\epsilon=0.1$, $\varphi=55^\circ$, $c_s=1$, and $A=c_s k_{x_0}$ in order to get the (maximum) Mach number which was fixed at $\mathcal{M}=A/(c_s k_{x_0})=1$.\label{171126:0950}

\noindent Although the Cauchy problem \eqref{171123:0928} is a dynamical system for three independent fields $u$, $v$ and $\rho$, only two fields can be chosen arbitrarily at the initial time $t=0$, as e.g. $u$ and $v$ in \eqref{171123:1509}. The third initial condition, here for $\rho$, is automatically fixed by the following conservation law that is being admitted by this system (see Eq.$\,$[24] in \cite{Hau17})
\begin{equation}
\left(\frac{\partial}{\partial t} +A y\frac{\partial}{\partial x}\right)\left[\frac{\partial u}{\partial y}-\frac{\partial v}{\partial x}-A\rho\right]=0,\label{171123:1519}
\end{equation}
stating that {\it any} solution of system \eqref{171123:0928} will identically satisfy equation \eqref{171123:1519}. Hence, when forcing the expression in the square bracket\footnote[2]{The expression in the square bracket in \eqref{171123:1519} is known as the potential vorticity \citep{Hau17}, which, if set to zero, the system will only allow for pure wave propagation.} to zero, the initial condition for the density field is thus determined by \eqref{171123:1509} as (Eq.$\,$[D.21a] in \cite{Hau16})
\begin{equation}
\rho(t,x,y)\big|_{t=0} =\int_{-\infty}^\infty dk_x dk_y\, \hat{\rho}(t,k_x,k_y)\big|_{t=0}\, e^{i(k_x x+k_y y)},\;\;\text{with}\;\; \hat{\rho}\big|_{t=0}=\Psi^{(+)}\, e^{i\varphi}.\label{171123:1630}
\end{equation}
It is clear now that the initial conditions \eqref{171123:1509} and \eqref{171123:1630} uniquely determine the Cauchy problem~\eqref{171123:0928} not only in real physical space, but also in its complementary spectral space, where, after a 2-D Fourier transform in all three fields
\begin{equation}
\psi^{\{u,v,\rho\}}(t,x,y)
= \int_{-\infty}^\infty dk_{x}dk_{y}\, \hat{\psi}^{\{u,v,\rho\}}(t,k_{x},k_{y})
\, e^{i(k_x x+k_y y)},\label{171123:1640}
\end{equation}
the system \eqref{171123:0928} has the form
\begin{equation}
\left.
\begin{aligned}
&\left(\frac{\partial}{\partial t}-Ak_x\frac{\partial}{\partial k_y}\right)\hat{\rho}+ik_x\hat{u}+ik_y\hat{v}=0,\\[0.5em]
&\left(\frac{\partial}{\partial t}-Ak_x\frac{\partial}{\partial k_y}\right)\hat{u}+A\hat{v}+ic_s^2k_x\hat{\rho}=0,\\[0.5em]
&\left(\frac{\partial}{\partial t}-Ak_x\frac{\partial}{\partial k_y}\right)\hat{v}+ic_s^2k_y\hat{\rho}=0,
\end{aligned}
~~~~~\right\}\label{171123:1651}
\end{equation}
by admitting the spectral form of the conservation law \eqref{171123:1519}
\begin{equation}
\left(\frac{\partial}{\partial t}-Ak_x\frac{\partial}{\partial k_y}\right)\Bigg[ik_y\hat{u}-ik_x\hat{v}-A\hat{\rho}\Bigg]=0,\label{171123:1649}
\end{equation}
and being supplemented by the initial conditions \eqref{171123:1354} and \eqref{171123:1630}. Finally, using Lie's theorem in going from infinitesimal to finite group transformations (see e.g. \cite{Bluman96}), the symmetry algebra $\mathfrak{g}_2$ \eqref{171123:0941} can also be transformed into spectral space to take the form
\begin{equation}
\hat{\mathfrak{g}}_2=\{(-ik_x+\alpha)(\hat{u}\partial_{\hat{u}}+\hat{v}\partial_{\hat{v}}+\hat{\rho}\partial_{\hat{\rho}}),\, (-ik_xAt-ik_y+\beta)(\hat{u}\partial_{\hat{u}}+\hat{v}\partial_{\hat{v}}+\hat{\rho}\partial_{\hat{\rho}})\},\label{171123:1758}
\end{equation}
showing that the (time-dependent) coordinate translations in physical space turn into corresponding phase-shifts of the spectral~fields.

However, as explained and discussed in Section \ref{S32}, to solve the equations \eqref{171123:0928}, or equivalently \eqref{171123:1651}, which are all set in the inertial frame, is not optimal, simply because the inertial frame is not the optimal frame to solve these equations. The complexity of the Cauchy problem~\eqref{171123:0928} in constructing a general solution can be significantly reduced when changing into the co-moving (accelerating) Kelvin frame via the transformation $\mathsf{K}$ \eqref{170715:1051}\footnote[3]{Note that the coordinate transformation $\mathsf{K}$ \eqref{171123:1839} does not change to a new {\it physical} frame, since the velocity fields $u$ and $v$ are transforming as scalars and not as components of a vector. For more details on this issue and how to turn  $\mathsf{K}$ \eqref{171123:1839} into a physical transformation, please see the remark (iv) on p.$\,$\pageref{171124:1112}.}
\begin{equation}
\mathsf{K}\!:\quad\; \tilde{t}=t,\quad\; \tilde{x}=x-Ayt,\quad\; \tilde{y}=y,\quad\; \tilde{\psi}^{\{u,v,\rho\}}=\psi^{\{u,v,\rho\}},\label{171123:1839}
\end{equation}
which turns the spatially {\it in}homogeneous equations \eqref{171123:0928} into the following spatially homogeneous (but now temporally inhomogeneous) set of equations
\begin{equation}
\left.
\begin{aligned}
&\frac{\partial\tilde{\rho}}{\partial\tilde{t}}+\frac{\partial\tilde{u}}{\partial\tilde{x}}
+\left(\frac{\partial}{\partial\tilde{y}}-A\tilde{t}\frac{\partial}{\partial\tilde{x}}\right)\tilde{v}=0,\\[0.5em]
&\frac{\partial\tilde{u}}{\partial\tilde{t}}+A\tilde{v}
+c_s^2\frac{\partial\tilde{\rho}}{\partial\tilde{x}}=0,\\[0.5em]
&\frac{\partial\tilde{v}}{\partial\tilde{t}}
+c_s^2\left(\frac{\partial}{\partial\tilde{y}}-A\tilde{t}\frac{\partial}{\partial\tilde{x}}\right)\tilde{\rho}=0,\label{171123:1856}
\end{aligned}
~~~~~\right\}
\end{equation}
and similarly the admitted conservation law \eqref{171123:1519} into a pure temporally conserved relation
\begin{equation}
\frac{\partial}{\partial\tilde{t}}\left[\left(\frac{\partial}{\partial\tilde{y}}-A\tilde{t}\frac{\partial}{\partial\tilde{x}}\right)\tilde{u}
-\frac{\partial\tilde{v}}{\partial\tilde{x}}-A\tilde{\rho}
\right]=0.\label{171123:1855}
\end{equation}
The associated initial conditions \eqref{171123:1509}-\eqref{171123:1259} and \eqref{171123:1630}, however, transform invariantly since at the initial time $t=0$ the Kelvin transformation \eqref{171123:1839} turns into an identity transformation:
\begin{equation}
\tilde{\psi}^{\{u,v,\rho\}}(\tilde{t},\tilde{x},\tilde{y})\big|_{\tilde{t}=0}=\int_{-\infty}^\infty dk_{\tilde{x}} dk_{\tilde{y}}\, \hat{\tilde{\psi}}^{\{u,v,\rho\}}(\tilde{t},k_{\tilde{x}},k_{\tilde{y}})\big|_{\tilde{t}=0}\,
e^{i(k_{\tilde{x}} \tilde{x}+k_{\tilde{y}} \tilde{y})},\label{171123:1924}
\end{equation}
with
\begin{equation}
\left.
\begin{aligned}
\hat{\tilde{u}}(\tilde{t},k_{\tilde{x}},k_{\tilde{y}})\big|_{\tilde{t}=0}\:=\; &
\left(\frac{c_s k_{\tilde{x}}}{\sqrt{k_{\tilde{x}}^2+k_{\tilde{y}}^2}}-i
\frac{Ak_{\tilde{y}}}{k_{\tilde{x}}^2+k_{\tilde{y}}^2}\right)\Psi^{(+)}(k_{\tilde{x}},k_{\tilde{y}})\, e^{i\varphi},\\[0.5em]
\hat{\tilde{v}}(\tilde{t},k_{\tilde{x}},k_{\tilde{y}})\big|_{\tilde{t}=0}\:=\; &
\left(\frac{c_s k_{\tilde{y}}}{\sqrt{k_{\tilde{x}}^2+k_{\tilde{y}}^2}}+i
\frac{Ak_{\tilde{x}}}{k_{\tilde{x}}^2+k_{\tilde{y}}^2}\right)\Psi^{(+)}(k_{\tilde{x}},k_{\tilde{y}})\, e^{i\varphi},\\[0.5em]
\hat{\tilde{\rho}}(\tilde{t},k_{\tilde{x}},k_{\tilde{y}})\big|_{\tilde{t}=0}\:=\;&\:
\Psi^{(+)}(k_{\tilde{x}},k_{\tilde{y}})\, e^{i\varphi},\label{171123:2035}
\end{aligned}
~~~\right\}
\end{equation}
where $\Psi^{(+)}$ is the same function \eqref{171123:1258}-\eqref{171123:1259} as it was defined in the inertial frame. Since the Kelvin transformation \eqref{171123:1839} corresponds in spectral space\footnote[2]{The derivation of the spectral Kelvin transformation $\hat{\mathsf{K}}$ \eqref{171123:2105} is presented in Appendix \ref{SD}.} to
\begin{equation}
\hat{\mathsf{K}}\!:\quad\; \tilde{t}=t,\quad\; k_{\tilde{x}}=k_x,\quad\; k_{\tilde{y}}=k_y+Atk_x,\quad\; \hat{\tilde{\psi}}^{\{u,v,\rho\}}=\hat{\psi}^{\{u,v,\rho\}},\label{171123:2105}
\end{equation}
the transformed spectral equations \eqref{171123:1651} are thus given as
\begin{equation}
\left.
\begin{aligned}
&\frac{\partial\hat{\tilde{\rho}}}{\partial \tilde{t}}+ik_{\tilde{x}}\hat{\tilde{u}}
+i\Big(k_{\tilde{y}}-A\tilde{t}k_{\tilde{x}}\Big)\hat{\tilde{v}}=0,\\[0.5em]
&\frac{\partial\hat{\tilde{u}}}{\partial \tilde{t}}+A\hat{\tilde{v}}+ic_s^2k_{\tilde{x}}\hat{\tilde{\rho}}=0,\\[0.5em]
&\frac{\partial\hat{\tilde{v}}}{\partial \tilde{t}}+ic_s^2\Big(k_{\tilde{y}}-A\tilde{t}k_{\tilde{x}}\Big)\hat{\tilde{\rho}}=0,\label{171123:2132}
\end{aligned}
~~~~~\right\}
\end{equation}
and the admitted conservation law \eqref{171123:1649} as
\begin{equation}
\frac{\partial}{\partial \tilde{t}}\bigg[i\Big(k_{\tilde{y}}-A\tilde{t}k_{\tilde{x}}\Big)\hat{\tilde{u}}
-ik_{\tilde{x}}\hat{\tilde{v}}-A\hat{\tilde{\rho}}\bigg]=0,\label{171123:2131}
\end{equation}
all supplemented by the initial conditions \eqref{171123:2035}. The symmetry algebra $\mathfrak{g}_2$ \eqref{171123:0941} simplifies in the new Kelvin frame in physical space to
\begin{equation}
\tilde{\mathfrak{g}}_2=\{\partial_{\tilde{x}}+\alpha(\tilde{u}\partial_{\tilde{u}}+\tilde{v}\partial_{\tilde{v}}+\tilde{\rho}\partial_{\tilde{\rho}}),\, \partial_{\tilde{y}}+\beta(\tilde{u}\partial_{\tilde{u}}+\tilde{v}\partial_{\tilde{v}}+\tilde{\rho}\partial_{\tilde{\rho}})\},\label{171123:2152}
\end{equation}
and \eqref{171123:1758} in spectral space to
\begin{equation}
\hat{\tilde{\mathfrak{g}}}_2=\{(-ik_{\tilde{x}}+\alpha)(\hat{\tilde{u}}\partial_{\hat{\tilde{u}}}+\hat{\tilde{v}}\partial_{\hat{\tilde{v}}}+\hat{\tilde{\rho}}\partial_{\hat{\tilde{\rho}}}),\, (-ik_{\tilde{y}}+\beta)(\hat{\tilde{u}}\partial_{\hat{\tilde{u}}}+\hat{\tilde{v}}\partial_{\hat{\tilde{v}}}+\hat{\tilde{\rho}}\partial_{\hat{\tilde{\rho}}})\}.\label{171123:2153}
\end{equation}
Recognizing that $\mathsf{K}$ \eqref{171123:1839} is {\it not} a symmetry transformation of the Cauchy problem \eqref{171123:0928}, the obtained result \eqref{171123:2132} is a remarkable result, since a PDE-system got reduced to an ODE-system without applying any symmetry transformation.

It is clear, of course, that the same result \eqref{171123:2132} would have also been obtained when applying the invariant-solution approach of the symmetry algebra~$\mathfrak{g}_2$ \eqref{171123:0941} to the governing PDE-system~\eqref{171123:0928}: Solving the invariant surface condition for this symmetry algebra, one obtains the local (single-mode) representation (Eq.$\,$[21] in \cite{Hau17})
\begin{equation}
\psi^{\{u,v,\rho\}}(t,x,y)=\psi^{(\text{k})\{u,v,\rho\}}(t,\alpha,\beta)e^{\alpha x +(\beta-\alpha At)y},\label{171124:1120}
\end{equation}
as well as the non-local (general) representation when summing over all eigenvalues $\alpha$ and $\beta$
\begin{equation}
\psi^{\{u,v,\rho\}}(t,x,y)=\int_{\mathbb{C}^2} d\alpha d\beta\, \psi^{(\text{k})\{u,v,\rho\}}(t,\alpha,\beta)e^{\alpha x +(\beta-\alpha At)y},\label{171124:1050}
\end{equation}
where at this stage, however, it is not clear yet whether this integral relation can be inverted or not, i.e., whether \eqref{171124:1050} constitutes a true representation or not. Nevertheless, when inserting \eqref{171124:1120} into the governing PDE-system~\eqref{171123:0928}, it will be reduced to the ODE-system \eqref{171123:2132}, where the modal shape factors $\psi^{(\text{k})\{u,v,\rho\}}$ for each field in \eqref{171124:1120} can then be identified as the corresponding complementary Fourier fields \raisebox{-0.025cm}{$\hat{\tilde{\psi}}^{\{u,v,\rho\}}$} \eqref{171123:2105} in the new Kelvin frame:
\begin{equation}
\!\!u^{(\text{k})}(t,\alpha,\beta)\equiv c\cdot\hat{\tilde{u}}(\tilde{t},k_{\tilde{x}},k_{\tilde{y}}),\quad v^{(\text{k})}(t,\alpha,\beta)\equiv c\cdot\hat{\tilde{v}}(\tilde{t},k_{\tilde{x}},k_{\tilde{y}}),\quad \rho^{(\text{k})}(t,\alpha,\beta)\equiv c\cdot\hat{\tilde{\rho}}(\tilde{t},k_{\tilde{x}},k_{\tilde{y}}),\:\,\label{171124:1212}
\end{equation}
with the eigenvalues then being naturally identified as $\alpha=ik_{\tilde{x}}$ and $\beta=ik_{\tilde{y}}$, for $(k_{\tilde{x}},k_{\tilde{y}})\in\mathbb{R}^2$, and the time $t=\tilde{t}$. The parameter $c\in\mathbb{C}$ in \eqref{171124:1212} is an arbitrary constant.

The reason why the shape factors $\psi^{(\text{k})\{u,v,\rho\}}$ of the defining modes $e^{\alpha x +(\beta-\alpha At)y}$ induced by~$\mathfrak{g}_2$ in \eqref{171124:1120} could have been identified as Fourier fields in the new Kelvin frame, as related in~\eqref{171124:1212}, is that the (non-symmetry) Kelvin transformation $\mathsf{K}$ \eqref{171123:1839} reduces the inertial frame symmetry algebra $\mathfrak{g}_2$ \eqref{171123:0941} in the new frame to the simple symmetry algebra $\tilde{\mathfrak{g}}_2$ \eqref{171123:2152}, which induces the well-known local (single-mode) 2-D Fourier representation \eqref{170715:1119}
\begin{equation}
\tilde{\psi}^{\{u,v,\rho\}}(\tilde{t},\tilde{x},\tilde{y})=\tilde{\psi}^{(\text{k})\{u,v,\rho\}}(\tilde{t},\alpha,\beta)e^{\alpha\tilde{x}+\beta\tilde{y}},\label{171124:1352}
\end{equation}
leading to the general 2-D Fourier transform when summing over all eigenvalues $\alpha$ and $\beta$
\begin{align}
\tilde{\psi}^{\{u,v,\rho\}}(\tilde{t},\tilde{x},\tilde{y})&=\int_{\mathbb{C}^2} d\alpha d\beta\, \tilde{\psi}^{(\text{k})\{u,v,\rho\}}(\tilde{t},\alpha,\beta)e^{\alpha\tilde{x}+\beta\tilde{y}}\nonumber\\[0.5em]
&\hspace{-0.275cm}\underset{\eqref{171123:2153}}{=}\int_{-\infty}^\infty dk_{\tilde{x}}dk_{\tilde{y}}\,
\hat{\tilde{\psi}}^{\{u,v,\rho\}}(\tilde{t},k_{\tilde{x}},k_{\tilde{y}})e^{i(k_{\tilde{x}}\tilde{x}+k_{\tilde{y}}\tilde{y})},
\label{171124:1353}
\end{align}
which is identical to the result obtained in \eqref{170715:1138}, since $\hat{\tilde{\psi}}^{\{u,v,\rho\}}=\hat{\tilde{\psi}}^{\text{(k)}\{u,v,\rho\}}$, if the free parameter in \eqref{171124:1212} is chosen as $c=-1$. Note that in the last line of \eqref{171124:1353} the associated spectral symmetry $\hat{\tilde{\mathfrak{g}}}_2$~\eqref{171123:2153} has been used, which naturally dictates that the eigenvalues $\alpha$ and $\beta$ have to be chosen as $ik_{\tilde{x}}$ and $ik_{\tilde{y}}$, in order to guarantee a non-zero invariant solution also in spectral space within the new Kelvin frame. Such a non-zero invariant solution in spectral space cannot be created in the (non-optimal) inertial frame when using its associate symmetry $\hat{\mathfrak{g}}_2$~\eqref{171123:1758}, simply because from the result of the performed symmetry analysis \eqref{170724:1405}, $\beta$ has to be a constant for all times $t$ and thus cannot be identified as a temporal quantity, i.e., $\beta\neq iAtk_x+ik_y$. Hence, the only invariant spectral solution induced by $\hat{\mathfrak{g}}_2$~\eqref{171123:1758} in the inertial frame is the trivial zero solution. Finally note that the key result \eqref{171124:1353} could have also been obtained by directly transforming the inertial frame relation \eqref{171124:1050} into the optimal Kelvin frame via $\mathsf{K}$ \eqref{171123:1839}:
\begin{align}
&\psi^{\{u,v,\rho\}}(t,x,y)=\int_{\mathbb{C}^2} d\alpha d\beta\, \psi^{(\text{k})\{u,v,\rho\}}(t,\alpha,\beta)e^{\alpha x +(\beta-\alpha At)y}\nonumber\\[0.5em]
\Leftrightarrow\quad\; &
\tilde{\psi}^{\{u,v,\rho\}}(\tilde{t},\tilde{x},\tilde{y})=\int_{\mathbb{C}^2} d\alpha d\beta\, \psi^{(\text{k})\{u,v,\rho\}}(t,\alpha,\beta)e^{\alpha (\tilde{x}+A\tilde{t}\tilde{y}) +(\beta-\alpha A\tilde{t})\tilde{y}}\nonumber\\[0.5em]
&\phantom{\tilde{\psi}^{\{u,v,\rho\}}(\tilde{t},\tilde{x},\tilde{y})}\hspace{-0.275cm}\underset{\eqref{171124:1212}}{=}
i^2c\cdot\int_{-\infty}^\infty dk_{\tilde{x}} dk_{\tilde{y}}\,
\hat{\tilde{\psi}}^{\{u,v,\rho\}}(\tilde{t},k_{\tilde{x}},k_{\tilde{y}})e^{i(k_{\tilde{x}}\tilde{x}+k_{\tilde{y}}\tilde{y})},\label{171124:1520}
\end{align}
which, for $c=-1$, then turns into the result \eqref{171124:1353}. This finally also answers the above question on the invertibility of relation \eqref{171124:1050}, which again was only achieved with the insight that \eqref{171124:1050} can be reduced and thus being redundant to the Fourier transform \eqref{171124:1353}.

\subsection{Analytical solution in spectral space\label{SC1}}

The considered Cauchy problem \eqref{171123:0928}-\eqref{171123:1630}, which is set up to only allow for pure wave propagation in linear shear flow, is best solved when changing the frame of reference from the non-optimal inertial frame to the optimal co-moving Kelvin frame. When choosing within this new frame the spectral representation, the considered Cauchy problem \eqref{171123:0928}-\eqref{171123:1630}, being a PDE-system, reduces to the ODE-system \eqref{171123:2132}-\eqref{171123:2131}
\begin{equation}
\left.
\begin{aligned}
&\frac{\partial\hat{\tilde{\rho}}}{\partial \tilde{t}}+ik_{\tilde{x}}\hat{\tilde{u}}
+i\Big(k_{\tilde{y}}-A\tilde{t}k_{\tilde{x}}\Big)\hat{\tilde{v}}=0,\\[0.5em]
&\frac{\partial\hat{\tilde{u}}}{\partial \tilde{t}}+A\hat{\tilde{v}}+ic_s^2k_{\tilde{x}}\hat{\tilde{\rho}}=0,\\[0.5em]
&\frac{\partial\hat{\tilde{v}}}{\partial \tilde{t}}+ic_s^2\Big(k_{\tilde{y}}-A\tilde{t}k_{\tilde{x}}\Big)\hat{\tilde{\rho}}=0,\label{171124:1813}
\end{aligned}
~~~~~\right\}
\end{equation}
with the condition for zero potential vorticity (pure wave propagation)
\begin{equation}
i\Big(k_{\tilde{y}}-A\tilde{t}k_{\tilde{x}}\Big)\hat{\tilde{u}}
-ik_{\tilde{x}}\hat{\tilde{v}}-A\hat{\tilde{\rho}}=0,\label{171124:1814}
\end{equation}
and with the initial conditions \eqref{171123:2035}. This overdetermined ODE-system \eqref{171124:1813}-\eqref{171124:1814} can now be analytically solved in a closed (fully integrated)
form in terms of hypergeometric functions. Its general solution is given as (Eq.$\,$[46] in \cite{Hau17})
\begin{equation}
\left.
\begin{aligned}
\hat{\tilde{u}}=\; &
\bigg[ C_1(k_{\tilde{x}},k_{\tilde{y}})\cdot \hyf\!\left({\textstyle\frac{A+ic_sk_{\tilde{x}}}{4A},\frac{1}{2},\frac{ic_s(k_{\tilde{y}}-A\tilde{t}k_{\tilde{x}})^2}{Ak_{\tilde{x}}}}\right)\\[0.25em]
&\;\; +C_2(k_{\tilde{x}},k_{\tilde{y}})\cdot(k_{\tilde{y}}-A\tilde{t}k_{\tilde{x}})
\cdot \hyf\!\left({\textstyle\frac{3A+ic_sk_{\tilde{x}}}{4A},\frac{3}{2},\frac{ic_s(k_{\tilde{y}}-A\tilde{t}k_{\tilde{x}})^2}{Ak_{\tilde{x}}}}\right)\bigg]
e^{ic_s\big(k_{\tilde{y}}-\frac{1}{2}Ak_{\tilde{x}}\tilde{t}\big)\tilde{t}},\\[0.5em]
\hat{\tilde{v}}=\; & \frac{1}{A^2+c_s^2k_{\tilde{x}}^2}\left(c_s^2k_{\tilde{x}}\big(k_{\tilde{y}}-A\tilde{t}k_{\tilde{x}}\big)\hat{\tilde{u}}
-A\frac{d\hat{\tilde{u}}}{d\tilde{t}}\right),\\[0.5em]
\hat{\tilde{\rho}}=\; & -\frac{i}{A}\left(k_{\tilde{x}}\hat{\tilde{v}}-\big(k_{\tilde{y}}-A\tilde{t}k_{\tilde{x}}\big)\hat{\tilde{u}}\right),
\end{aligned}
~~~~~\right\}\label{171124:1828}
\end{equation}
where $\hyf(a,b,z)$ is Kummer's confluent hypergeometric function, being regular at $z=0$, and where $C_1$ and $C_2$ are abritrary
integration functions depending on the spectral coordinates $(k_{\tilde{x}},k_{\tilde{y}})$. To note is that the general solution only carries two integrations functions
$C_1$ and $C_2$ that can be arbitrarily fixed, i.e., the dynamics of the ODE-system \eqref{171124:1813}-\eqref{171124:1814} is fully determined by only placing two
initial conditions, simply for the reason that this system is overdetermined facing four equations for only three dynamical fields $\hat{\tilde{u}}$, $\hat{\tilde{v}}$ and $\hat{\tilde{\rho}}$.
Hence, inserting any two conditions from the specific initial input \eqref{171123:2035} will uniquely determine the integration functions $C_1$ and~$C_2$. The third and remaining condition from \eqref{171123:2035}
will then, by construction, be automatically consistent to \eqref{171124:1828}.

The complementary solution for each field $\tilde{u}$, $\tilde{v}$ and $\tilde{\rho}$ in real physical space within the Kelvin frame is then given by the 2-D Fourier transform \eqref{171124:1353} of the spectral result \eqref{171124:1828}:
\begin{equation}
\tilde{\psi}^{\{u,v,\rho\}}(\tilde{t},\tilde{x},\tilde{y})=\int_{-\infty}^\infty dk_{\tilde{x}}dk_{\tilde{y}}\,
\hat{\tilde{\psi}}^{\{u,v,\rho\}}(\tilde{t},k_{\tilde{x}},k_{\tilde{y}})e^{i(k_{\tilde{x}}\tilde{x}+k_{\tilde{y}}\tilde{y})}.
\label{171124:2333}
\end{equation}
In the non-optimal inertial frame, however, when transforming according to the defining transformation $\hat{\mathsf{K}}$ \eqref{171123:2105}, the optimal spectral Kelvin solution \eqref{171124:1828} changes to
\begin{equation}
\left.
\begin{aligned}
\hat{u}=\; &
\bigg[ C_1(k_x,k_y+Atk_x)\cdot \hyf\Big({\textstyle\frac{A+ic_sk_x}{4A},\frac{1}{2},\frac{ic_sk_y^2}{Ak_x}}\Big)\\[0.25em]
&\;\; +C_2(k_x,k_y+Atk_x)\cdot k_y
\cdot \hyf\Big({\textstyle\frac{3A+ic_sk_x}{4A},\frac{3}{2},\frac{ic_sk_y^2}{Ak_x}}\Big)\bigg]
e^{ic_s\big(k_y+\frac{1}{2}Ak_x t\big)t},\\[0.5em]
\hat{v}=\; & \frac{1}{A^2+c_s^2k_x^2}\left[c_s^2k_xk_y\hat{u}
-A\left(\frac{\partial}{\partial t}-Ak_x\frac{\partial}{\partial k_y}\right)\hat{u}\right],\\[0.5em]
\hat{\rho}=\; & -\frac{i}{A}\left(k_x\hat{v}-k_y\hat{u}\right),
\label{171124:2139}
\end{aligned}
~~~~~\right\}
\end{equation}
being now a solution which obviously will show a higher dynamical complexity than the optimal solution~\eqref{171124:1828}, particularly due to exhibiting an explicit time dependence in the integration functions $C_1$ and~$C_2$. Indeed, the solution \eqref{171124:2139} satisfies the spectral inertial frame equations \eqref{171123:1651}-\eqref{171123:1649}. It is clear that since the dependent variables transform invariantly under $\hat{\mathsf{K}}$~\eqref{171123:2105}, the full inertial frame solution \eqref{171124:2139} emerges from the Kelvin solution \eqref{171124:1828} by actually just applying the simple relation
\begin{equation}
\hat{\tilde{\psi}}^{\{u,v,\rho\}}(t,k_x,k_y+Atk_x)=\hat{\psi}^{\{u,v,\rho\}}(t,k_x,k_y).\label{171124:2326}
\end{equation}
The complementary solution for each field $u$, $v$ and $\rho$ in real physical space within the inertial frame is then given by the 2-D Fourier transform \eqref{171123:1640} of the spectral result \eqref{171124:2326}:
\begin{equation}
\psi^{\{u,v,\rho\}}(t,x,y)
= \int_{-\infty}^\infty dk_{x}dk_{y}\, \hat{\psi}^{\{u,v,\rho\}}(t,k_{x},k_{y})
\, e^{i(k_x x+k_y y)}.\label{171124:2327}
\end{equation}
Also in the real physical space, it is clear that since the dependent variables transform invariantly under $\mathsf{K}$~\eqref{171123:1839}, the full inertial frame solution \eqref{171124:2327} emerges from the Kelvin solution \eqref{171124:2333} by actually just applying the simple relation
\begin{equation}
\tilde{\psi}^{\{u,v,\rho\}}(t,x-Ayt,y)=\psi^{\{u,v,\rho\}}(t,x,y).
\end{equation}

\subsection{Visualization and discussion of a non-physical solution\label{SC2}}

\begin{figure}[h!]
\begin{subfigure}[t]{.24\textwidth}
\FigureXYLabel{\includegraphics[type=pdf,ext=.pdf,read=.pdf,width=0.96\textwidth]{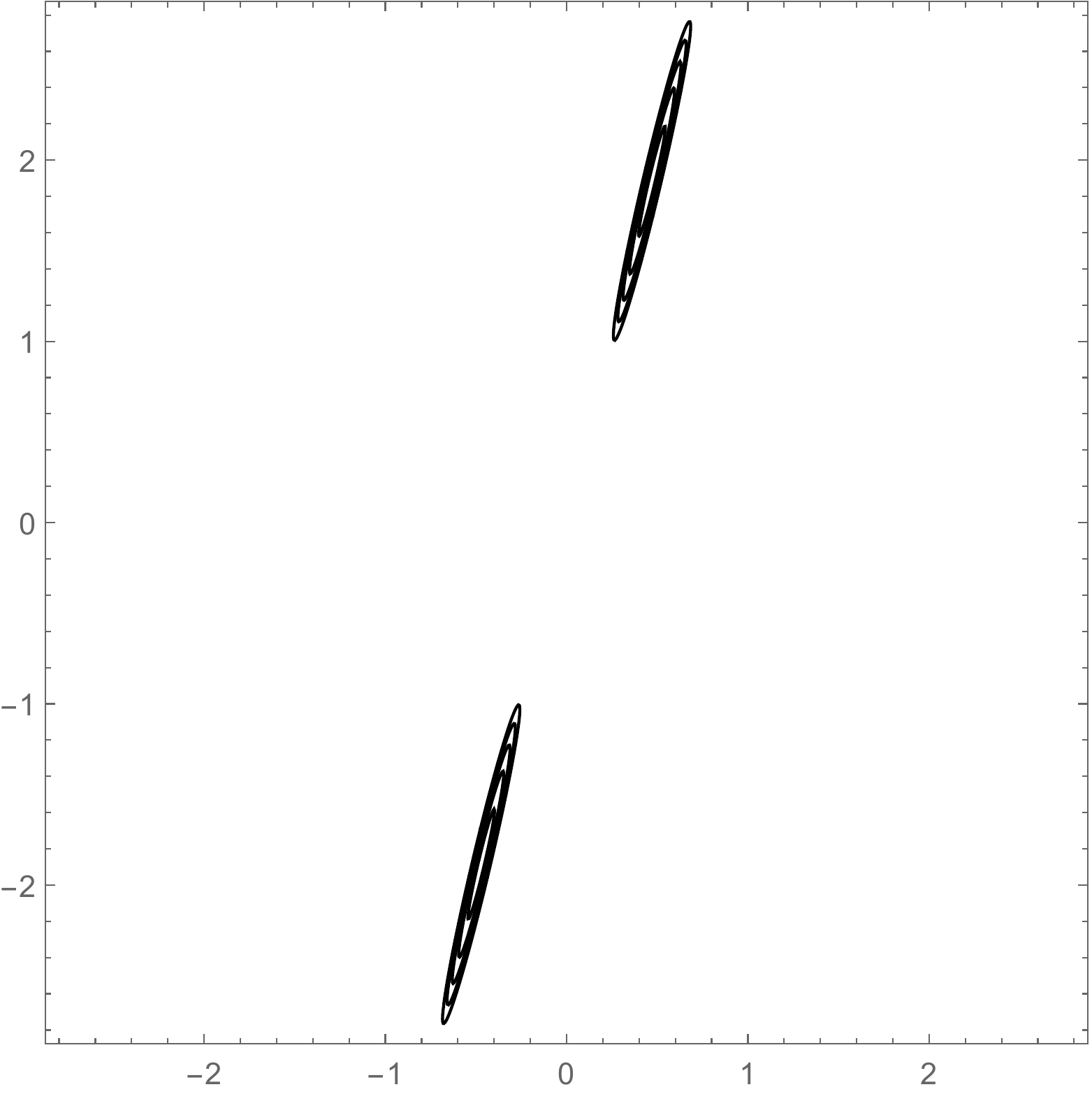}}
{${\scriptstyle\hspace{0.25cm} k_{\tilde{x}}}$}{-1mm}{\begin{rotate}{0}
\end{rotate}}{0mm}
\vspace*{-1.25em}\subcaption*{{\scriptsize (a.1) $\tilde{t}=0$}}\vspace*{1.25em}
\end{subfigure}
\hspace*{-.2cm}
\begin{subfigure}[t]{.24\textwidth}
\FigureXYLabel{\includegraphics[type=pdf,ext=.pdf,read=.pdf,width=0.96\textwidth]{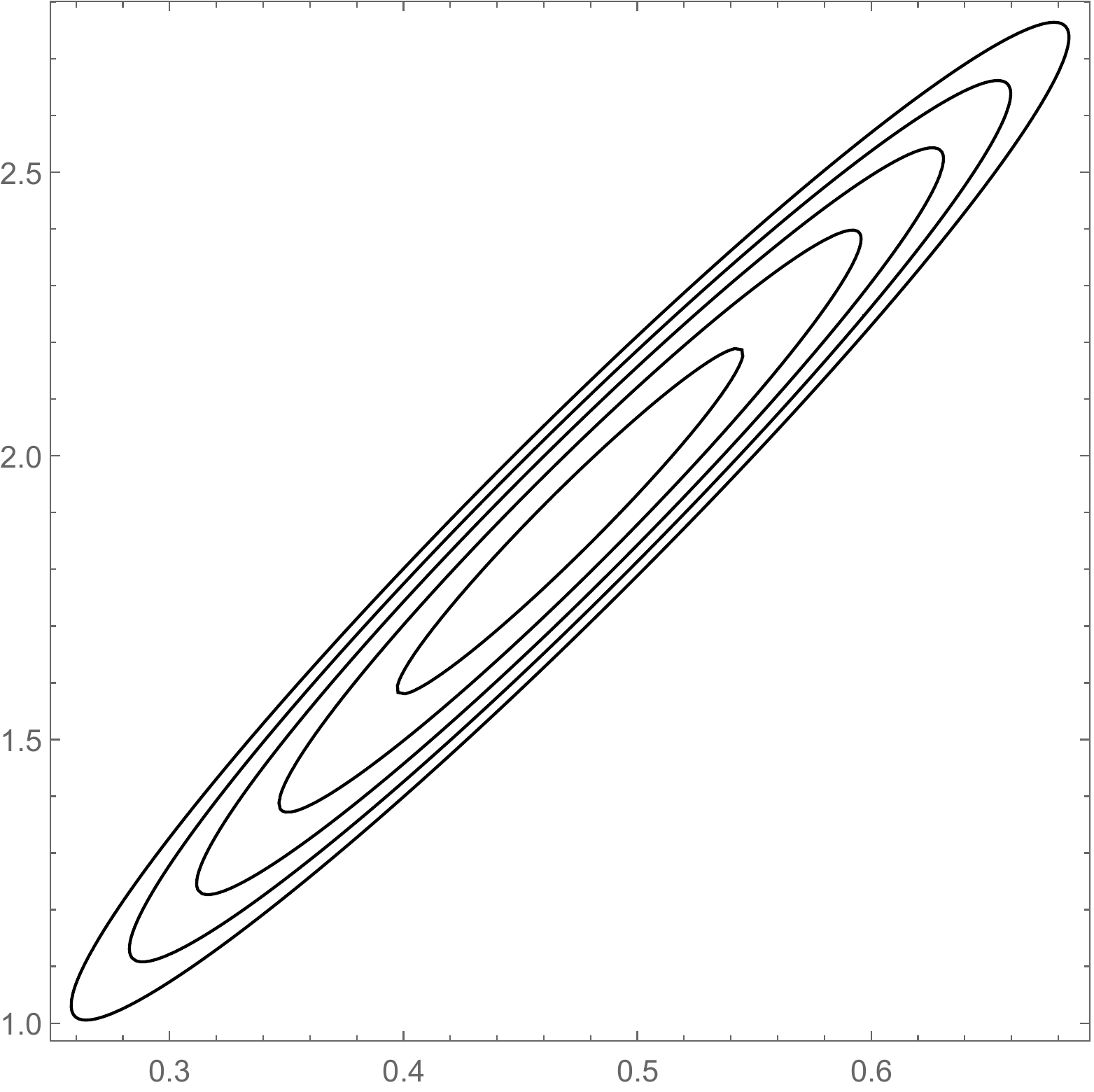}}
{${\scriptstyle\hspace{0.25cm} k_{\tilde{x}}}$}{-1mm}{\begin{rotate}{0}
\end{rotate}}{0mm}
\vspace*{-1.25em}\subcaption*{{\scriptsize (a.2) $\tilde{t}=0$, zoom}}\vspace*{1.25em}
\end{subfigure}
\begin{subfigure}[c]{0.0001\textwidth}
\vspace*{-3.5cm}${\scriptstyle \! k_{\tilde{y}}}$
\end{subfigure}
\begin{subfigure}[t]{.24\textwidth}
\FigureXYLabel{\includegraphics[type=pdf,ext=.pdf,read=.pdf,width=0.96\textwidth]{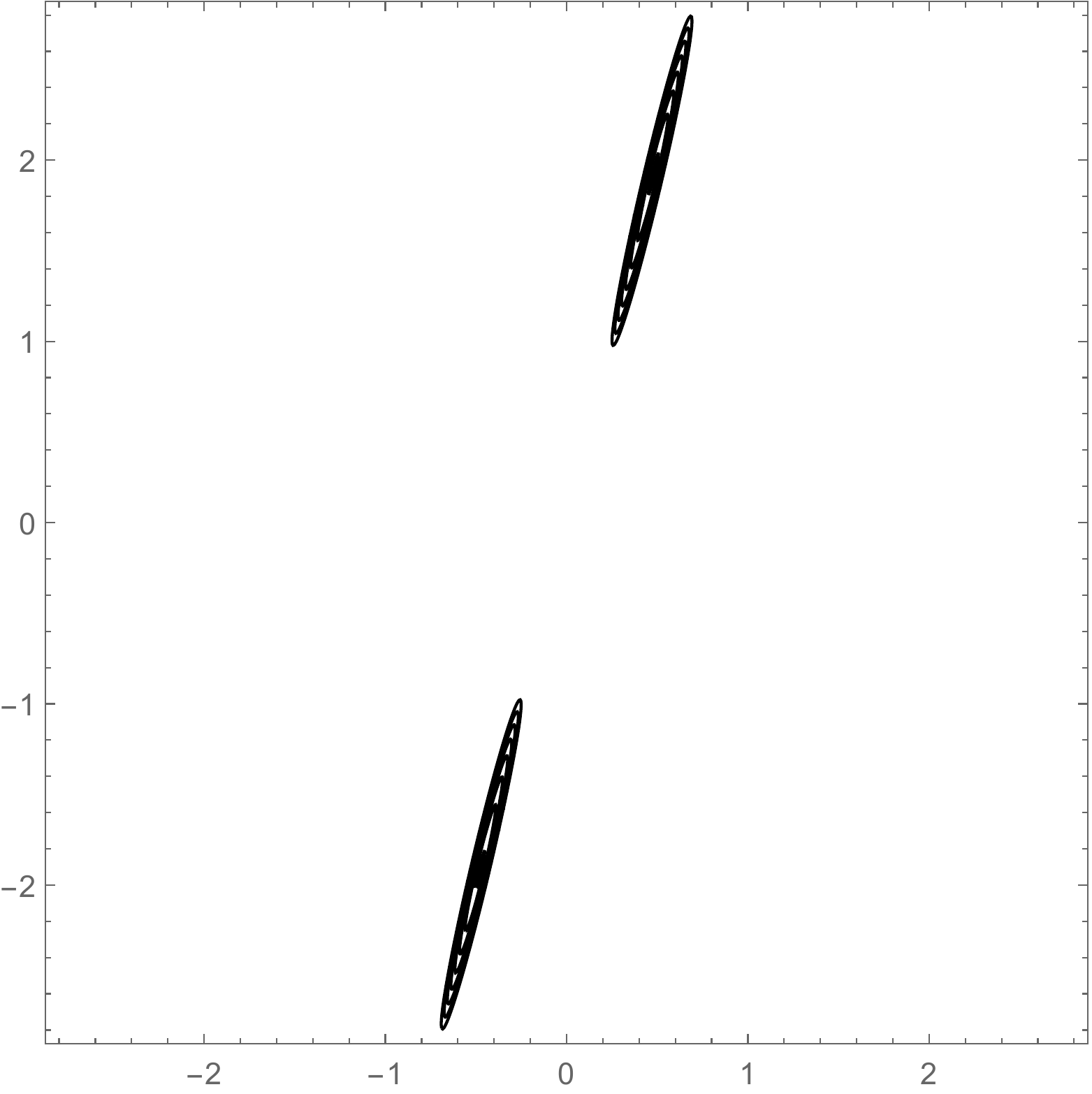}}
{${\scriptstyle\hspace{0.25cm} k_{\tilde{x}}}$}{-1mm}{\begin{rotate}{0}
\end{rotate}}{0mm}
\vspace*{-1.25em}\subcaption*{{\scriptsize (b.1) $\tilde{t}=1/2\tilde{t}^*$}}\vspace*{1.25em}
\end{subfigure}
\hspace*{-.2cm}
\begin{subfigure}[t]{.24\textwidth}
\FigureXYLabel{\includegraphics[type=pdf,ext=.pdf,read=.pdf,width=0.96\textwidth]{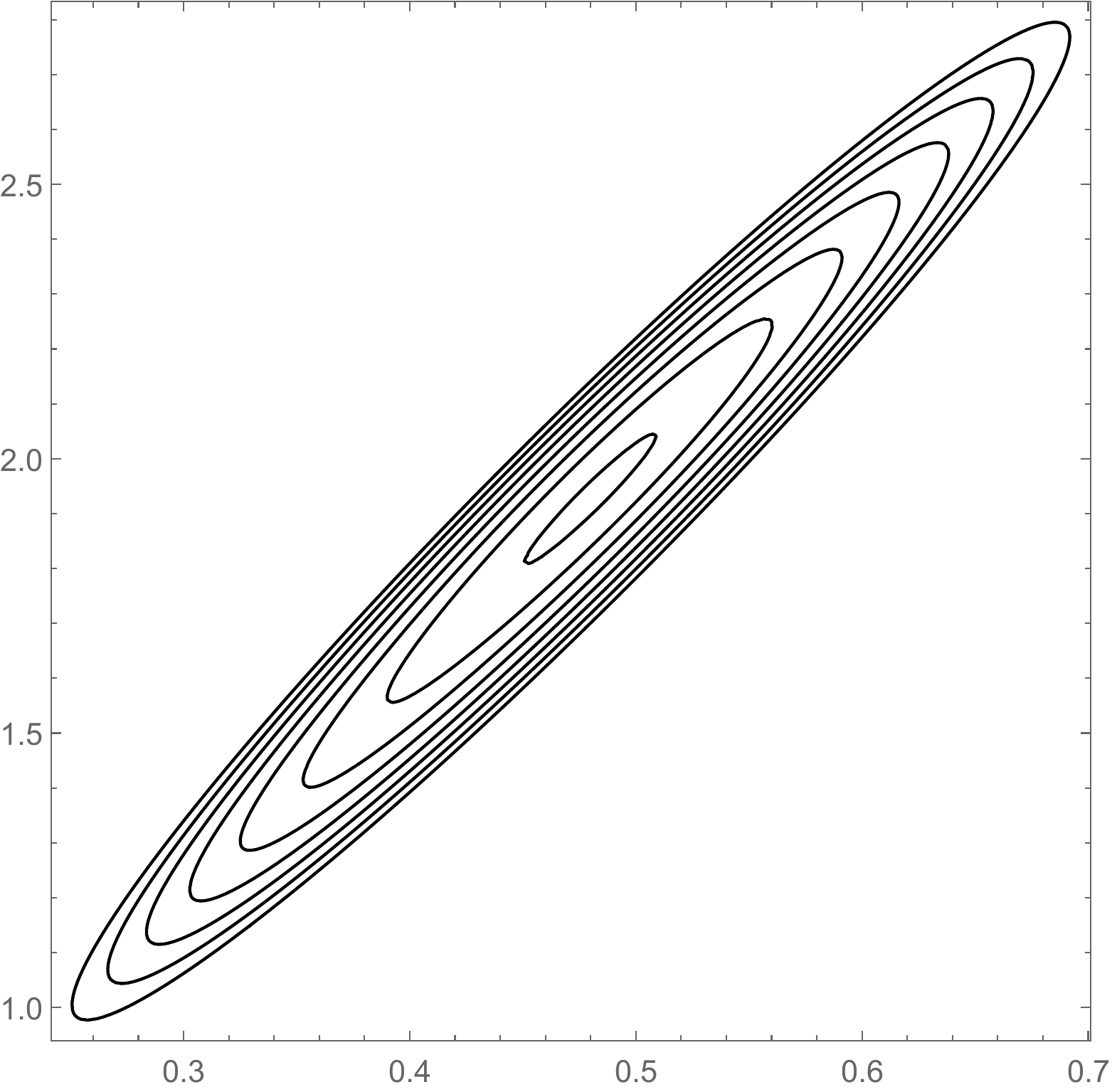}}
{${\scriptstyle\hspace{0.25cm} k_{\tilde{x}}}$}{-1mm}{\begin{rotate}{0}
\end{rotate}}{0mm}
\vspace*{-1.25em}\subcaption*{{\scriptsize (b.2) $\tilde{t}=1/2\tilde{t}^*$, zoom}}\vspace*{1.25em}
\end{subfigure}
\begin{subfigure}[t]{.24\textwidth}
\FigureXYLabel{\includegraphics[type=pdf,ext=.pdf,read=.pdf,width=0.96\textwidth]{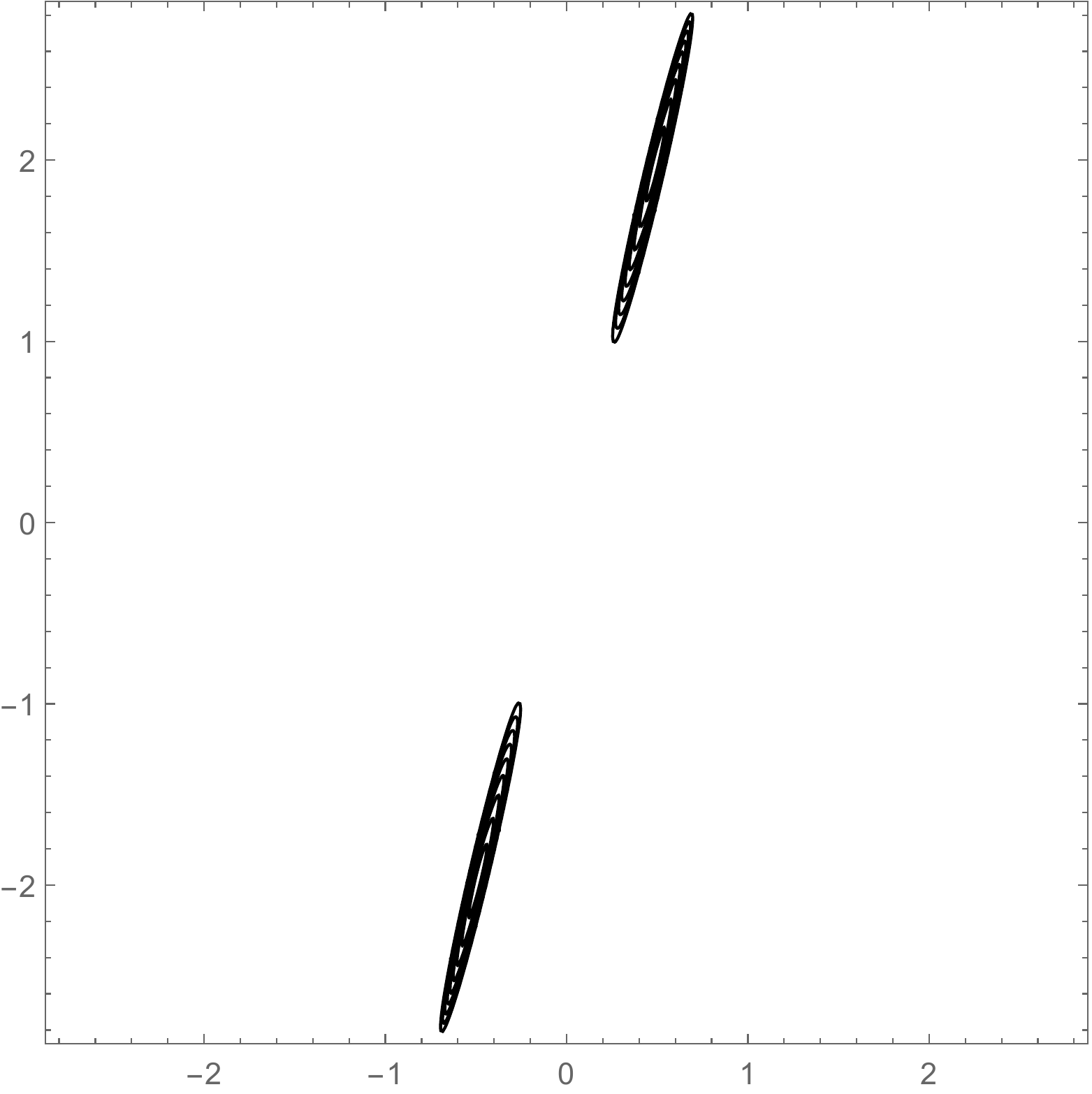}}
{${\scriptstyle\hspace{0.25cm} k_{\tilde{x}}}$}{-1mm}{\begin{rotate}{0}
\end{rotate}}{0mm}
\vspace*{-1.25em}\subcaption*{{\scriptsize (c.1) $\tilde{t}=\tilde{t}^*$}}\vspace*{1.25em}
\end{subfigure}
\hspace*{-.2cm}
\begin{subfigure}[t]{.24\textwidth}
\FigureXYLabel{\includegraphics[type=pdf,ext=.pdf,read=.pdf,width=0.96\textwidth]{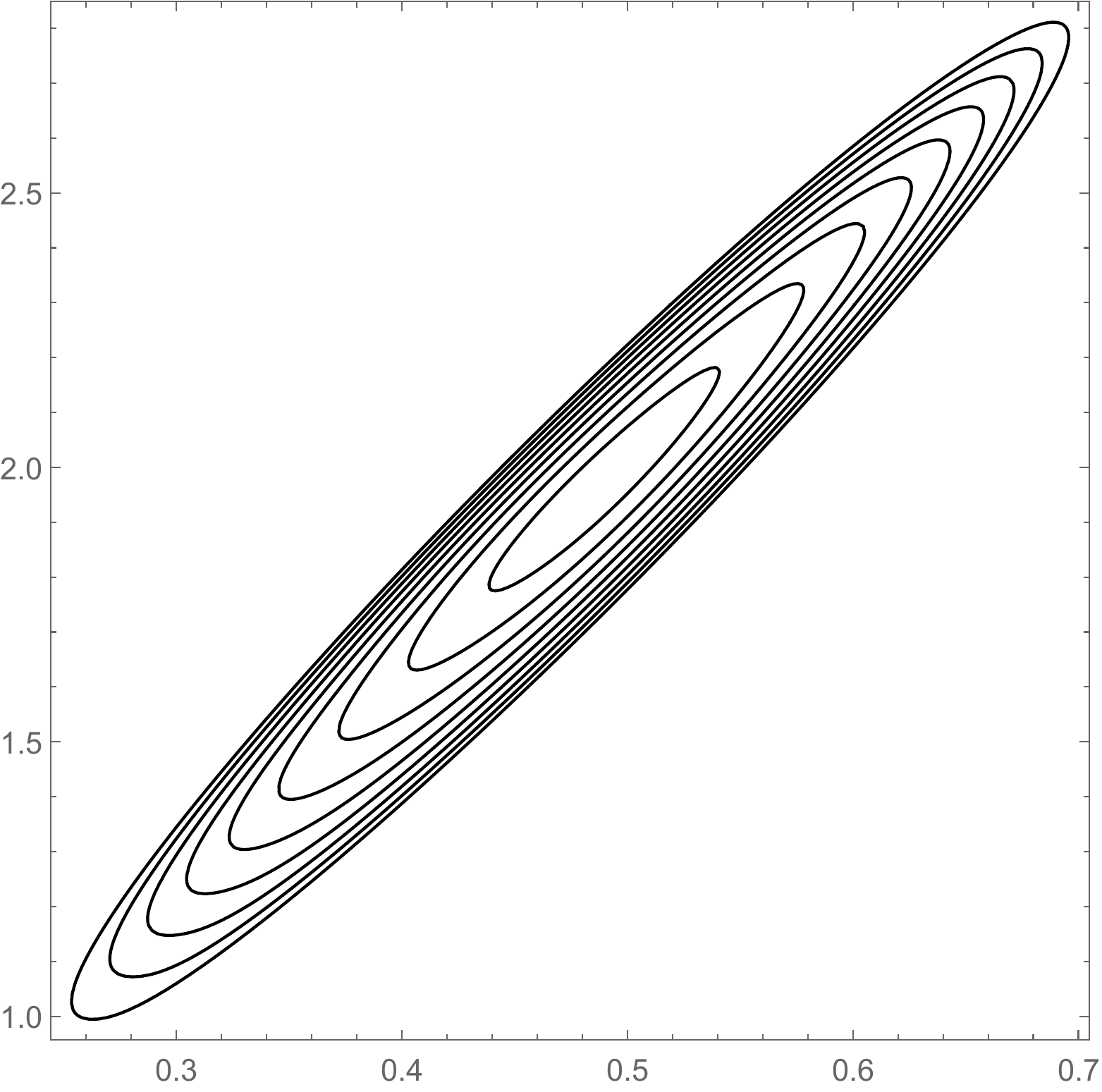}}
{${\scriptstyle\hspace{0.25cm} k_{\tilde{x}}}$}{-1mm}{\begin{rotate}{0}
\end{rotate}}{0mm}
\vspace*{-1.25em}\subcaption*{{\scriptsize (c.2) $\tilde{t}=\tilde{t}^*$, zoom}}\vspace*{1.25em}
\end{subfigure}
\begin{subfigure}[t]{0.0001\textwidth}
\vspace*{-2.0cm}${\scriptstyle \! k_{\tilde{y}}}$
\end{subfigure}
\begin{subfigure}[t]{.24\textwidth}
\FigureXYLabel{\includegraphics[type=pdf,ext=.pdf,read=.pdf,width=0.96\textwidth]{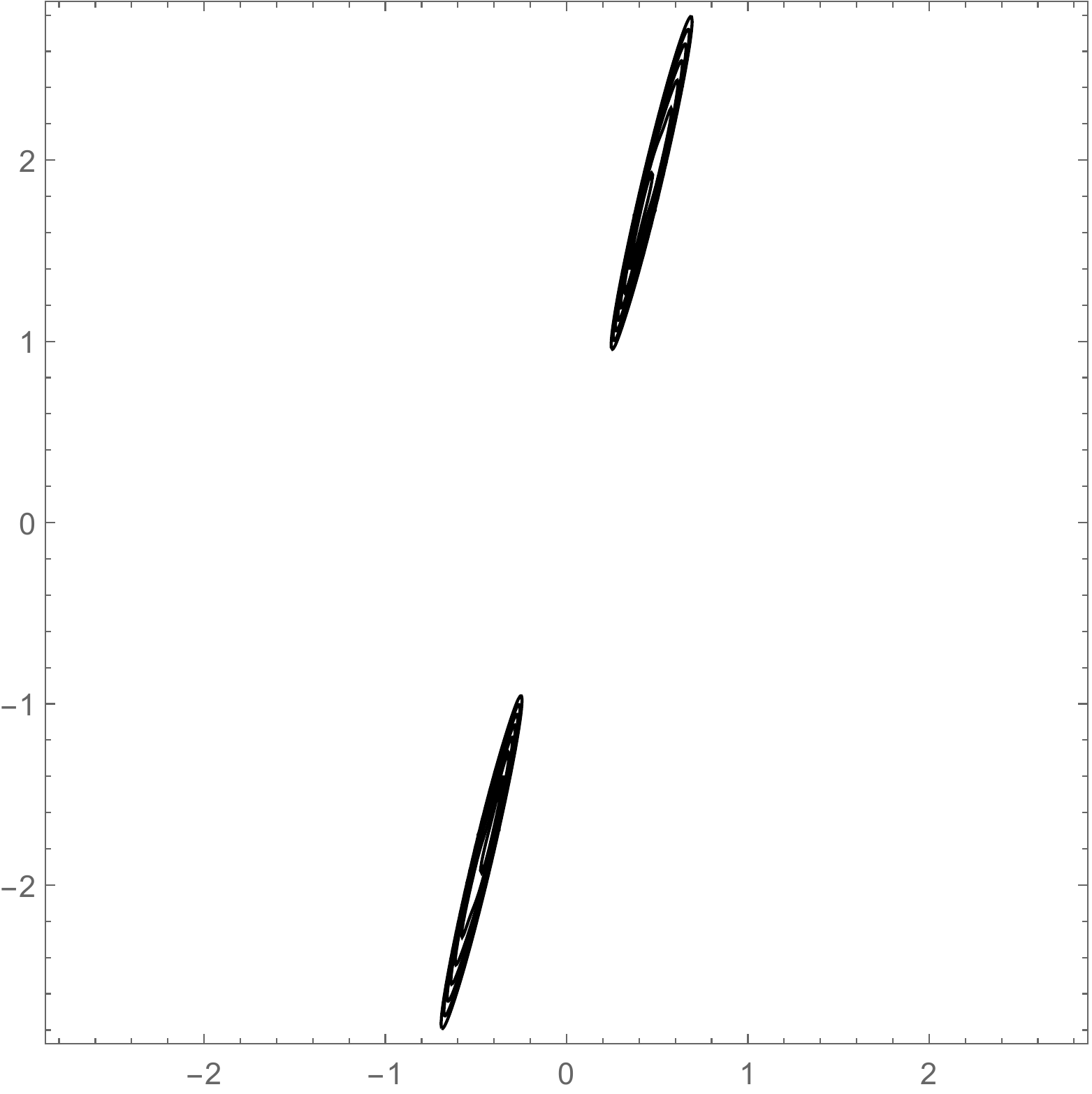}}
{${\scriptstyle\hspace{0.25cm} k_{\tilde{x}}}$}{-1mm}{\begin{rotate}{0}
\end{rotate}}{0mm}
\vspace*{-1.25em}\subcaption*{{\scriptsize (d.1) $\tilde{t}=3/2\tilde{t}^*$}}\vspace*{1.25em}
\end{subfigure}
\hspace*{-.2cm}
\begin{subfigure}[t]{.24\textwidth}
\FigureXYLabel{\includegraphics[type=pdf,ext=.pdf,read=.pdf,width=0.96\textwidth]{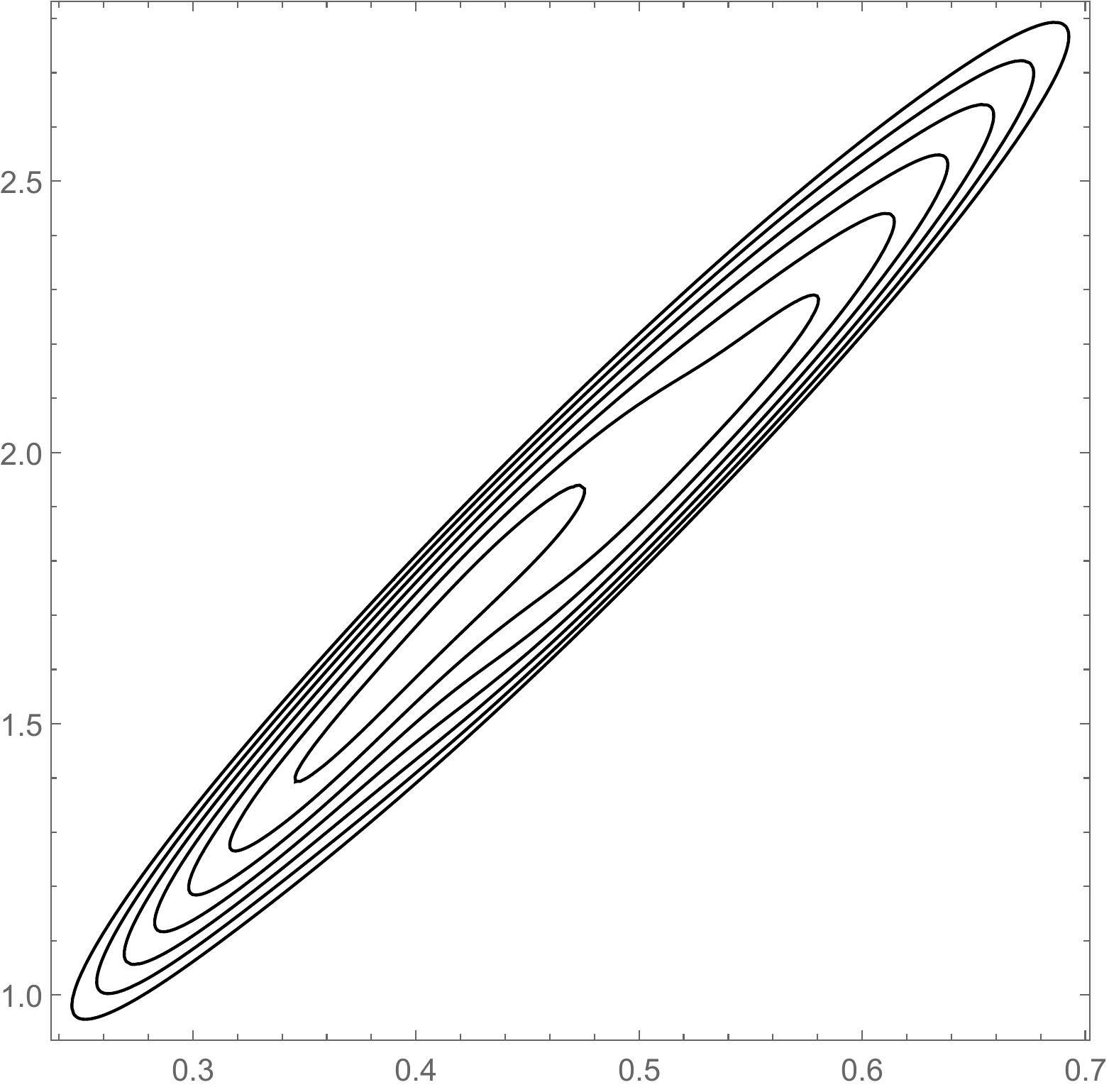}}
{${\scriptstyle\hspace{0.25cm} k_{\tilde{x}}}$}{-1mm}{\begin{rotate}{0}
\end{rotate}}{0mm}
\vspace*{-1.25em}\subcaption*{{\scriptsize (d.2) $\tilde{t}=3/2\tilde{t}^*$, zoom}}\vspace*{1.25em}
\end{subfigure}
\begin{subfigure}[t]{.24\textwidth}
\FigureXYLabel{\includegraphics[type=pdf,ext=.pdf,read=.pdf,width=0.96\textwidth]{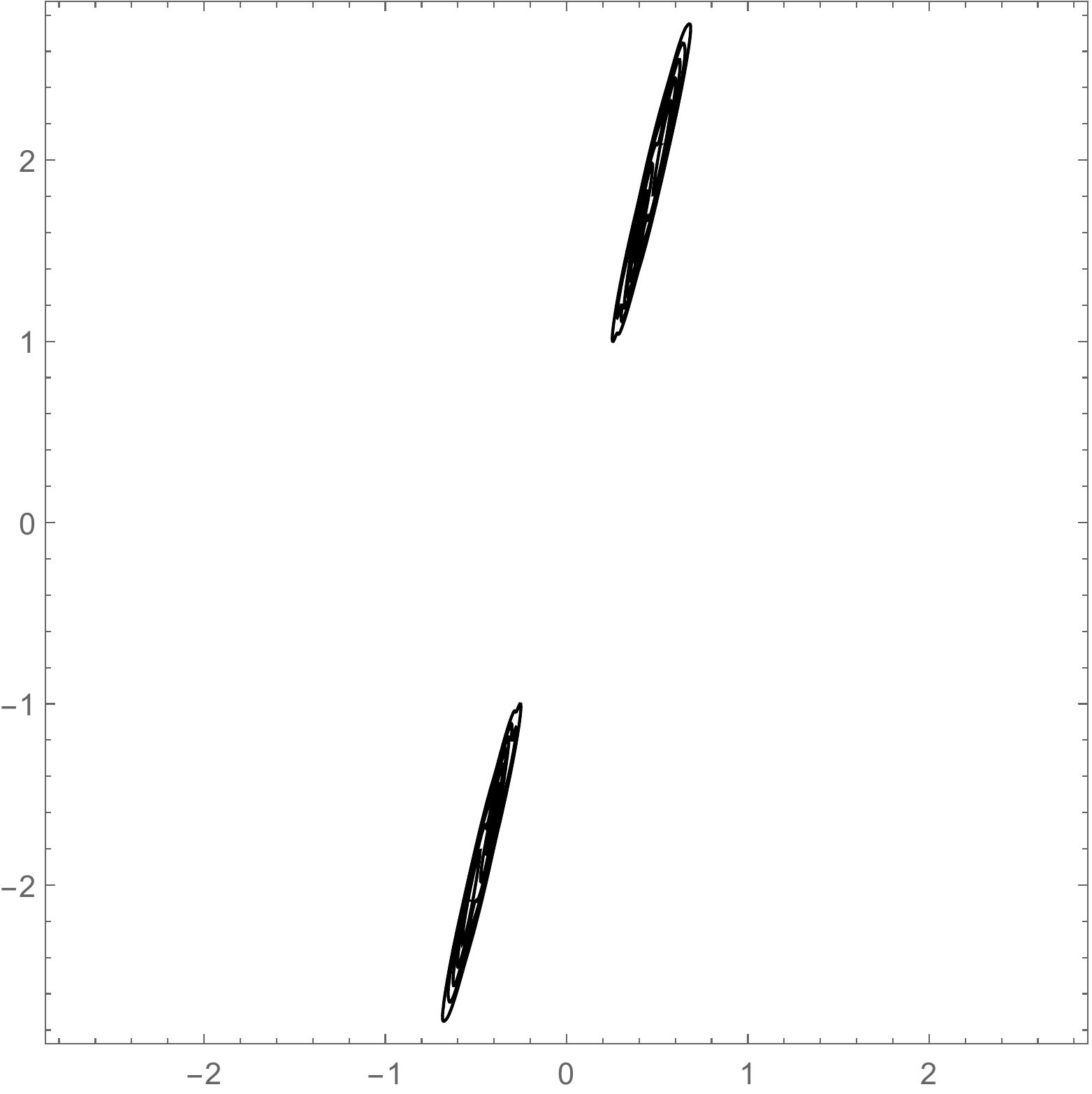}}
{${\scriptstyle\hspace{0.25cm} k_{\tilde{x}}}$}{-1mm}{\begin{rotate}{0}
\end{rotate}}{0mm}
\vspace*{-1.25em}\subcaption*{{\scriptsize (e.1) $\tilde{t}=2\tilde{t}^*$}}\vspace*{1.25em}
\end{subfigure}
\hspace*{-.2cm}
\begin{subfigure}[t]{.24\textwidth}
\FigureXYLabel{\includegraphics[type=pdf,ext=.pdf,read=.pdf,width=0.96\textwidth]{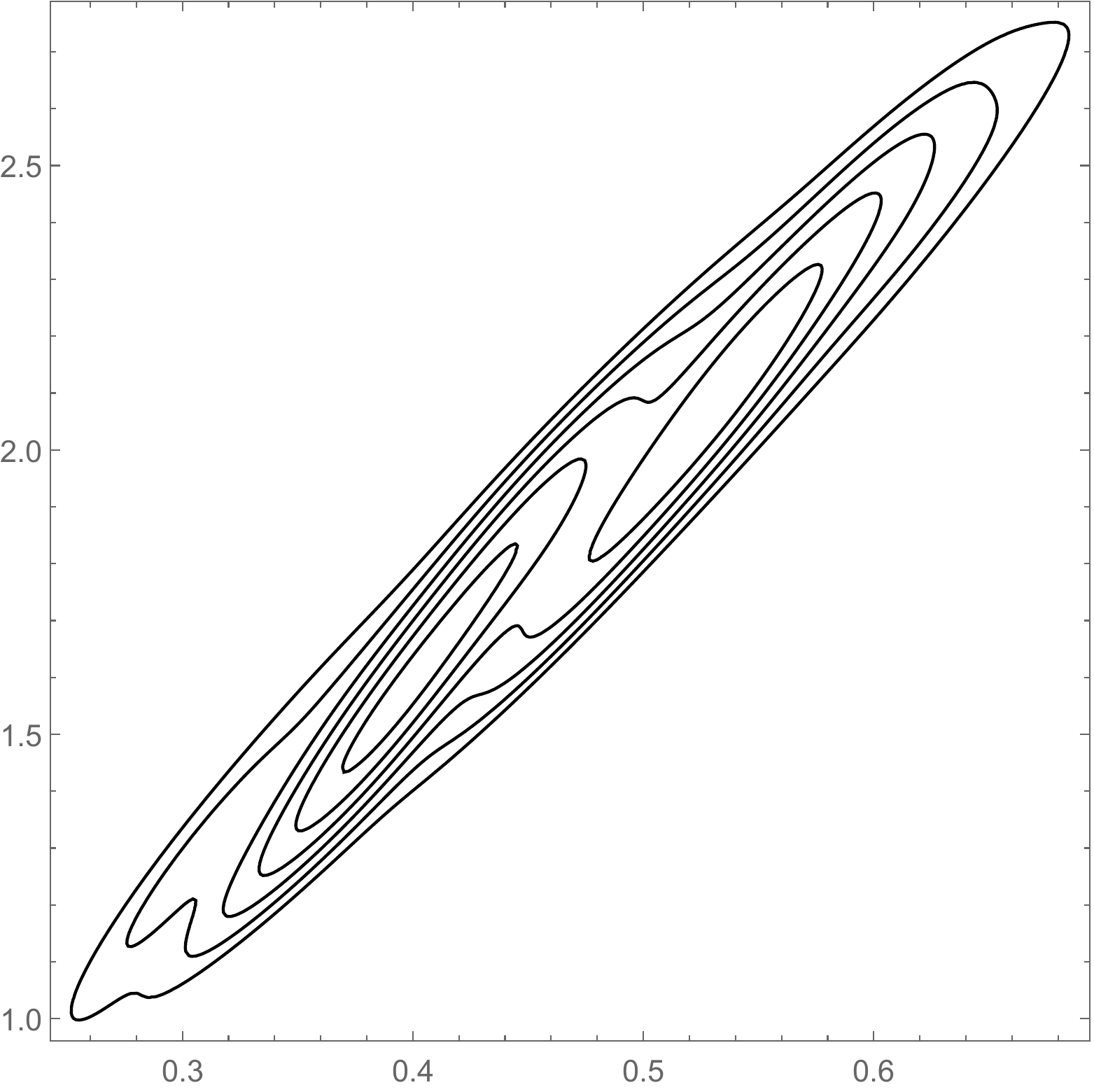}}
{${\scriptstyle\hspace{0.25cm} k_{\tilde{x}}}$}{-1mm}{\begin{rotate}{0}
\end{rotate}}{0mm}
\vspace*{-1.25em}\subcaption*{{\scriptsize (e.2) $\tilde{t}=2\tilde{t}^*$, zoom}}\vspace*{1.25em}
\end{subfigure}
\begin{subfigure}[t]{0.0001\textwidth}
\vspace*{-2.0cm}${\scriptstyle \! k_{\tilde{y}}}$
\end{subfigure}
\begin{subfigure}[t]{.24\textwidth}
\FigureXYLabel{\includegraphics[type=pdf,ext=.pdf,read=.pdf,width=0.96\textwidth]{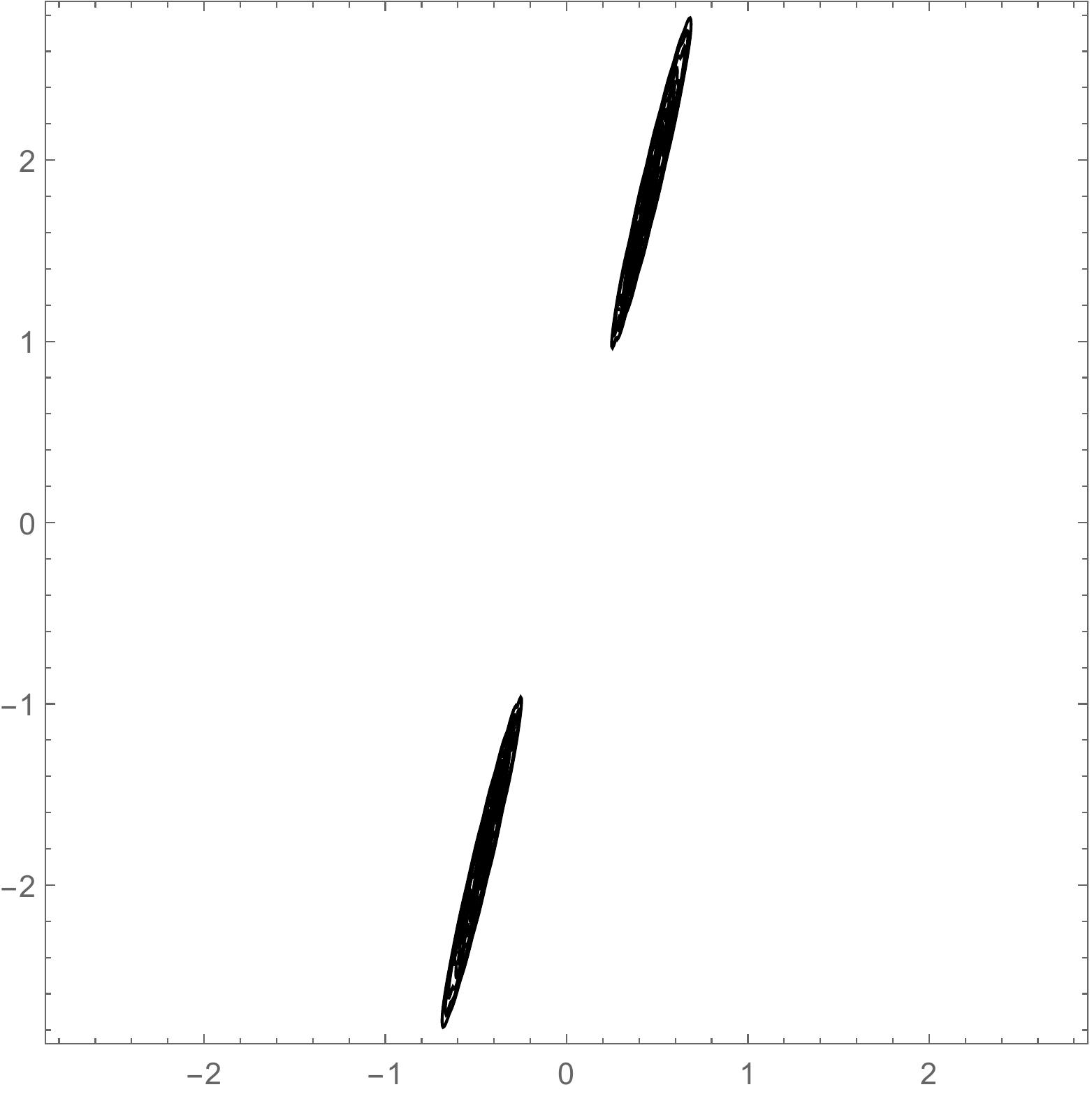}}
{${\scriptstyle\hspace{0.25cm} k_{\tilde{x}}}$}{-1mm}{\begin{rotate}{0}
\end{rotate}}{0mm}
\vspace*{-1.25em}\subcaption*{{\scriptsize (f.1) $\tilde{t}=5/2\tilde{t}^*$}}\vspace*{1.25em}
\end{subfigure}
\hspace*{-.2cm}
\begin{subfigure}[t]{.24\textwidth}
\FigureXYLabel{\includegraphics[type=pdf,ext=.pdf,read=.pdf,width=0.96\textwidth]{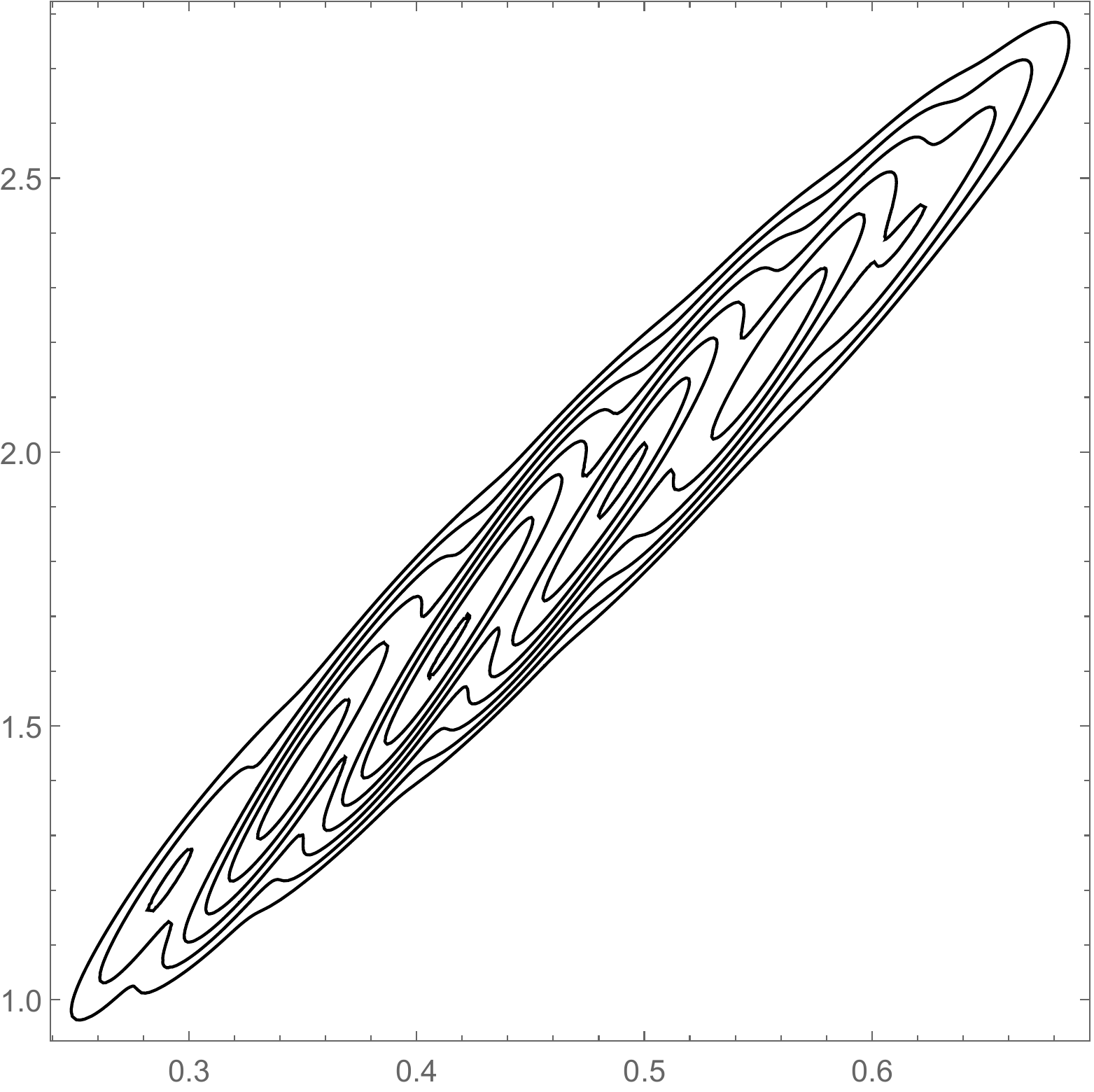}}
{${\scriptstyle\hspace{0.25cm} k_{\tilde{x}}}$}{-1mm}{\begin{rotate}{0}
\end{rotate}}{0mm}
\vspace*{-1.25em}\subcaption*{{\scriptsize (f.2) $\tilde{t}=5/2\tilde{t}^*$, zoom}}\vspace*{1.25em}
\end{subfigure}
\vspace*{-1em}\caption{{\footnotesize Evolution of the Kelvin-frame solution $|\hat{\tilde{\rho}}|$ in the spectral $(k_{\tilde{x}},k_{\tilde{y}})$-plane, according to the ODE-solution
\eqref{171124:1828} with the initial condition \eqref{171123:2035}. For all times $\tilde{t}\geq 0$ the wave packet is stationary and not drifting in $k$-space. Furthermore, the wave packet
even keeps its expansion size for all times, i.e., it is also not diffusing in $k$-space. Inside the wave packet, however, the dynamics is changing. After reaching
the critical time $\tilde{t}^*=t^*=k_{y_0}/(Ak_{x_0})\sim 8.5$, which is defined in the center of the packet
when $k_{\tilde{y}}(\tilde{t}):=k_{\tilde{y}}-A\tilde{t}^* k_{\tilde{x}}=0$, the topological structure of the wave packet changes significantly. Particularly,
the initial oval symmetry in the contour lines along its major axes breaks at $\tilde{t}> \tilde{t}^*$. However, the solution shown here is not truly physical, since the considered initial condition \eqref{171123:2035}, as taken identically from \cite{Hau16}, violates the reality constraint of the Fourier transform, having the effect that the above spectral solution will not lead to a real-valued solution in physical space. The real part of this associated solution in physical space is shown in Fig.$\,$\ref{fig3}.\label{fig1}}}
\end{figure}

\begin{figure}[t!]
\begin{subfigure}[t]{.24\textwidth}
\FigureXYLabel{\includegraphics[type=pdf,ext=.pdf,read=.pdf,width=0.96\textwidth]{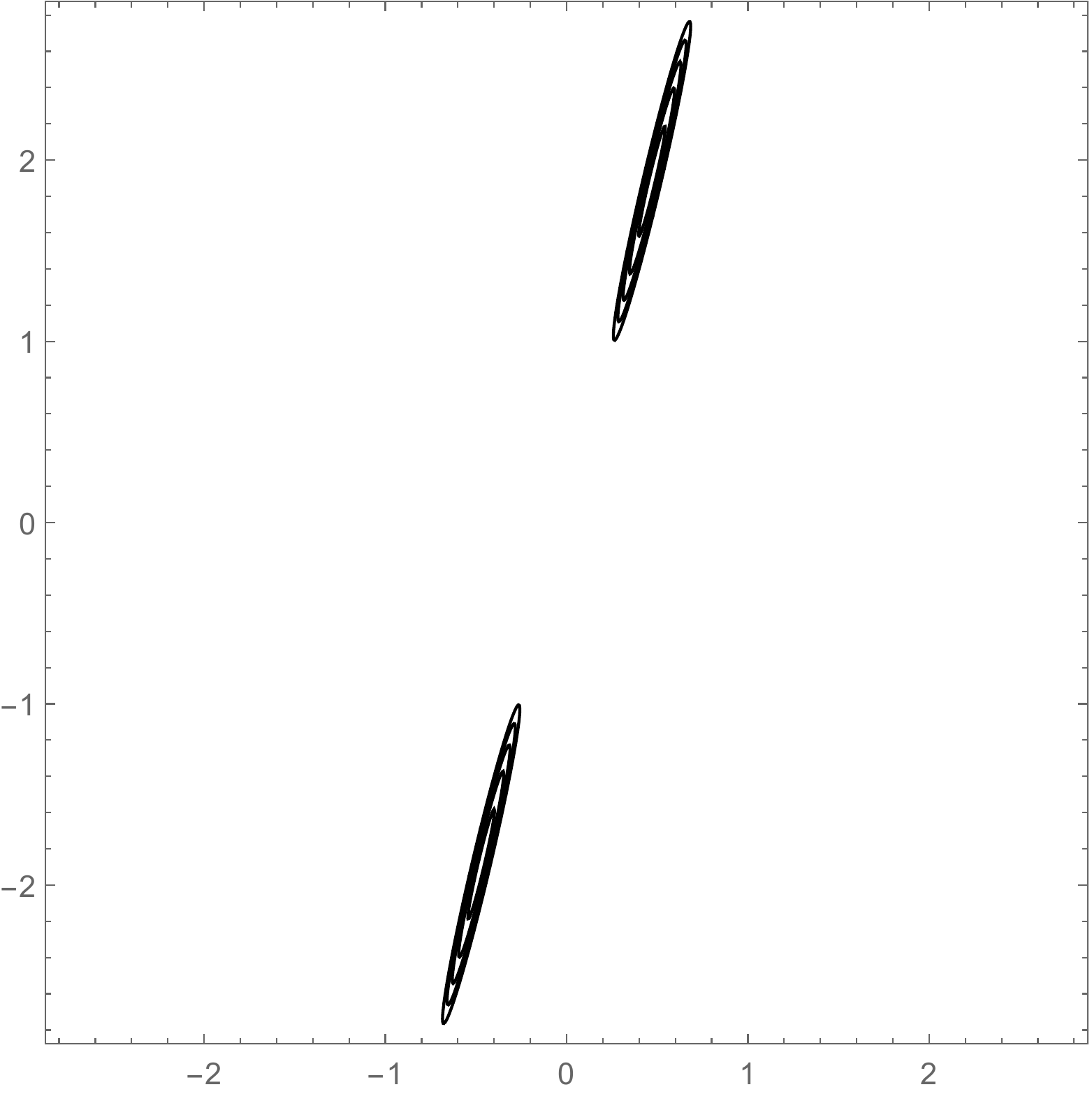}}
{${\scriptstyle\hspace{0.25cm} k_x}$}{-1mm}{\begin{rotate}{0}
\end{rotate}}{0mm}
\vspace*{-1.25em}\subcaption*{{\scriptsize (a.1) $t=0$}}\vspace*{1.25em}
\end{subfigure}
\hspace*{-.2cm}
\begin{subfigure}[t]{.24\textwidth}
\FigureXYLabel{\includegraphics[type=pdf,ext=.pdf,read=.pdf,width=0.96\textwidth]{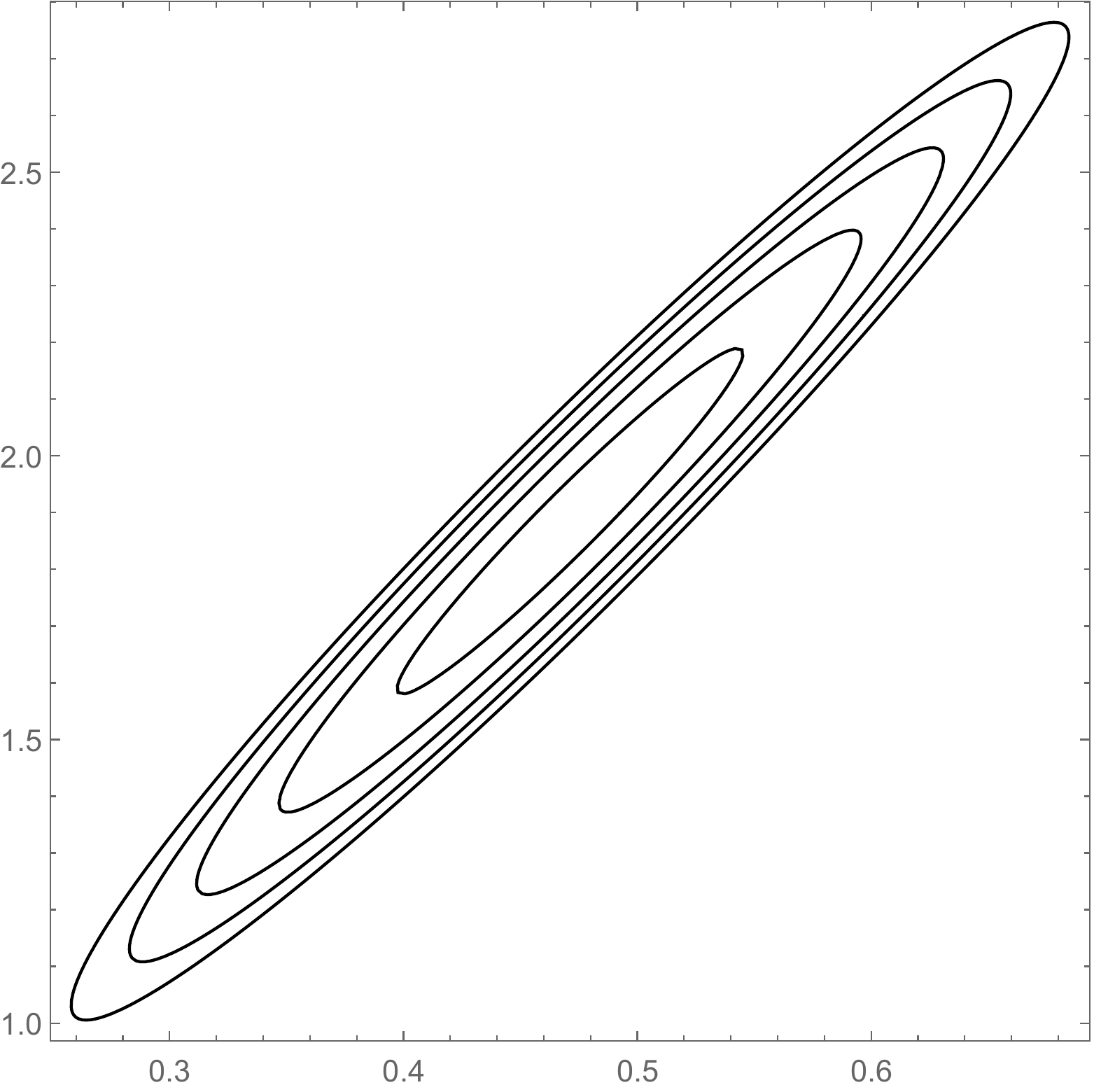}}
{${\scriptstyle\hspace{0.25cm} k_x}$}{-1mm}{\begin{rotate}{0}
\end{rotate}}{0mm}
\vspace*{-1.25em}\subcaption*{{\scriptsize (a.2) $t=0$, zoom}}\vspace*{1.25em}
\end{subfigure}
\begin{subfigure}[c]{0.0001\textwidth}
\vspace*{-3.5cm}${\scriptstyle \! k_y}$
\end{subfigure}
\begin{subfigure}[t]{.24\textwidth}
\FigureXYLabel{\includegraphics[type=pdf,ext=.pdf,read=.pdf,width=0.96\textwidth]{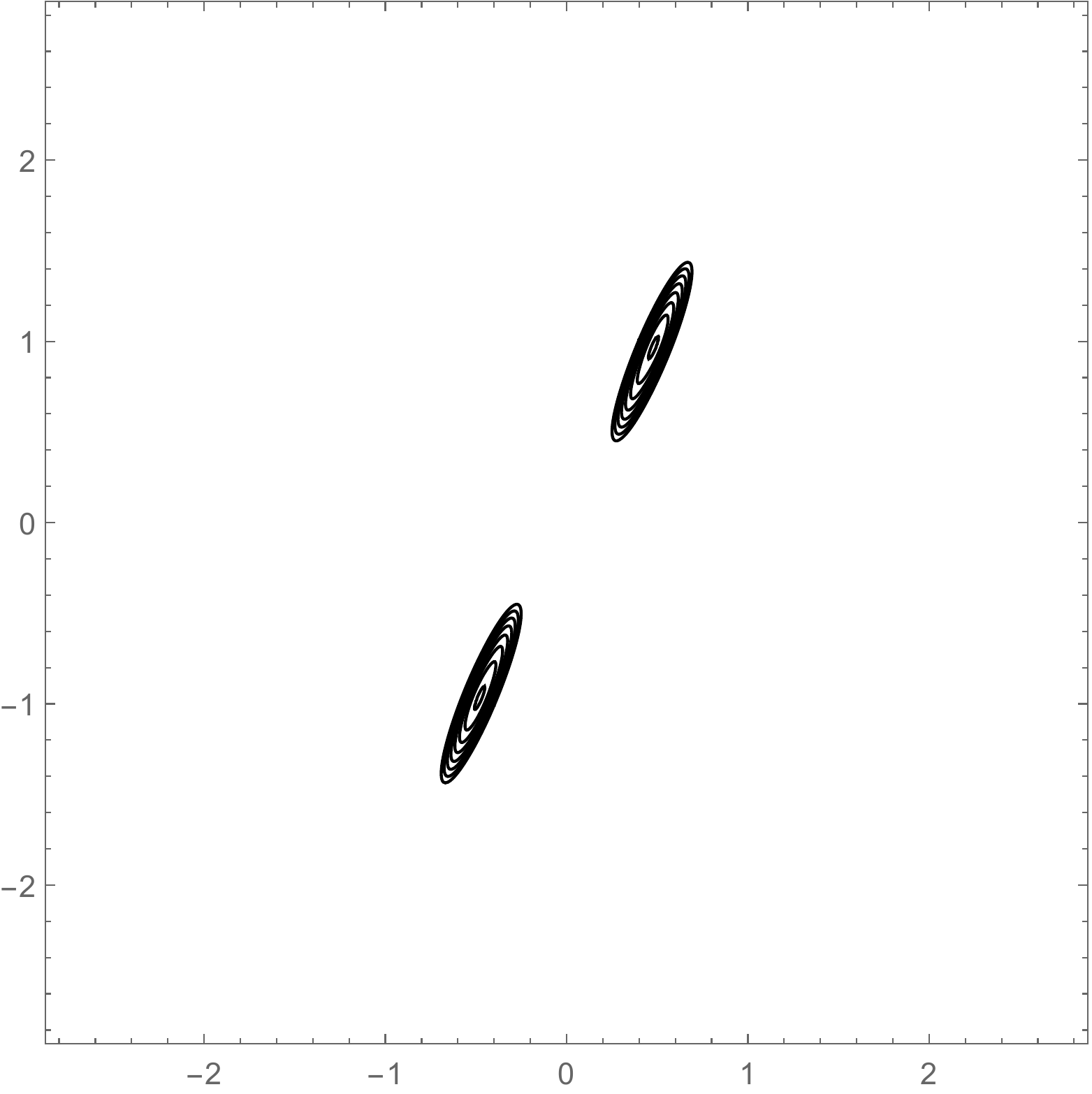}}
{${\scriptstyle\hspace{0.25cm} k_x}$}{-1mm}{\begin{rotate}{0}
\end{rotate}}{0mm}
\vspace*{-1.25em}\subcaption*{{\scriptsize (b.1) $t=1/2t^*$}}\vspace*{1.25em}
\end{subfigure}
\hspace*{-.2cm}
\begin{subfigure}[t]{.24\textwidth}
\FigureXYLabel{\includegraphics[type=pdf,ext=.pdf,read=.pdf,width=0.96\textwidth]{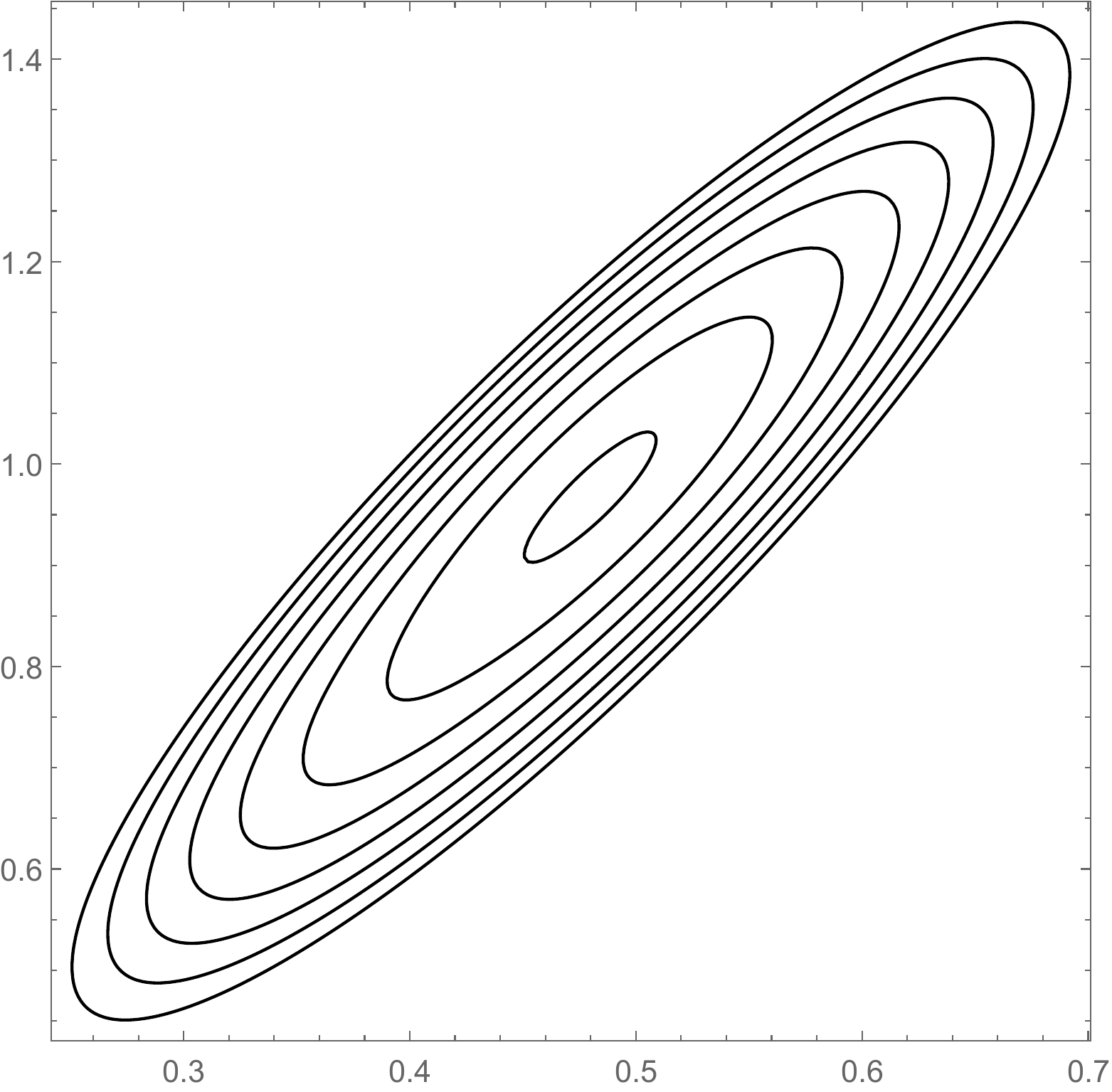}}
{${\scriptstyle\hspace{0.25cm} k_x}$}{-1mm}{\begin{rotate}{0}
\end{rotate}}{0mm}
\vspace*{-1.25em}\subcaption*{{\scriptsize (b.2) $t=1/2t^*$, zoom}}\vspace*{1.25em}
\end{subfigure}
\begin{subfigure}[t]{.24\textwidth}
\FigureXYLabel{\includegraphics[type=pdf,ext=.pdf,read=.pdf,width=0.96\textwidth]{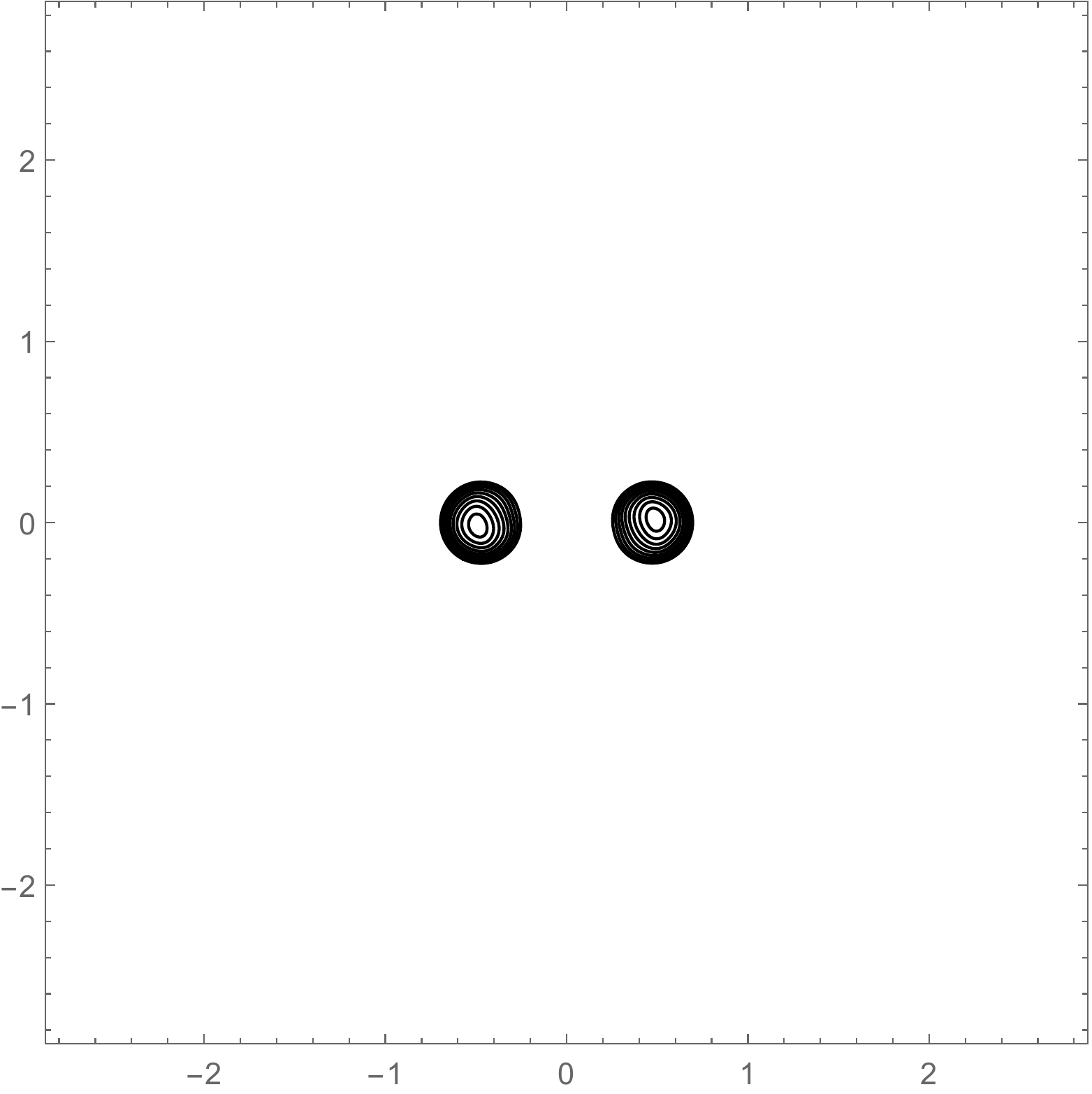}}
{${\scriptstyle\hspace{0.25cm} k_x}$}{-1mm}{\begin{rotate}{0}
\end{rotate}}{0mm}
\vspace*{-1.25em}\subcaption*{{\scriptsize (c.1) $t=t^*$}}\vspace*{1.25em}
\end{subfigure}
\hspace*{-.2cm}
\begin{subfigure}[t]{.24\textwidth}
\FigureXYLabel{\includegraphics[type=pdf,ext=.pdf,read=.pdf,width=0.96\textwidth]{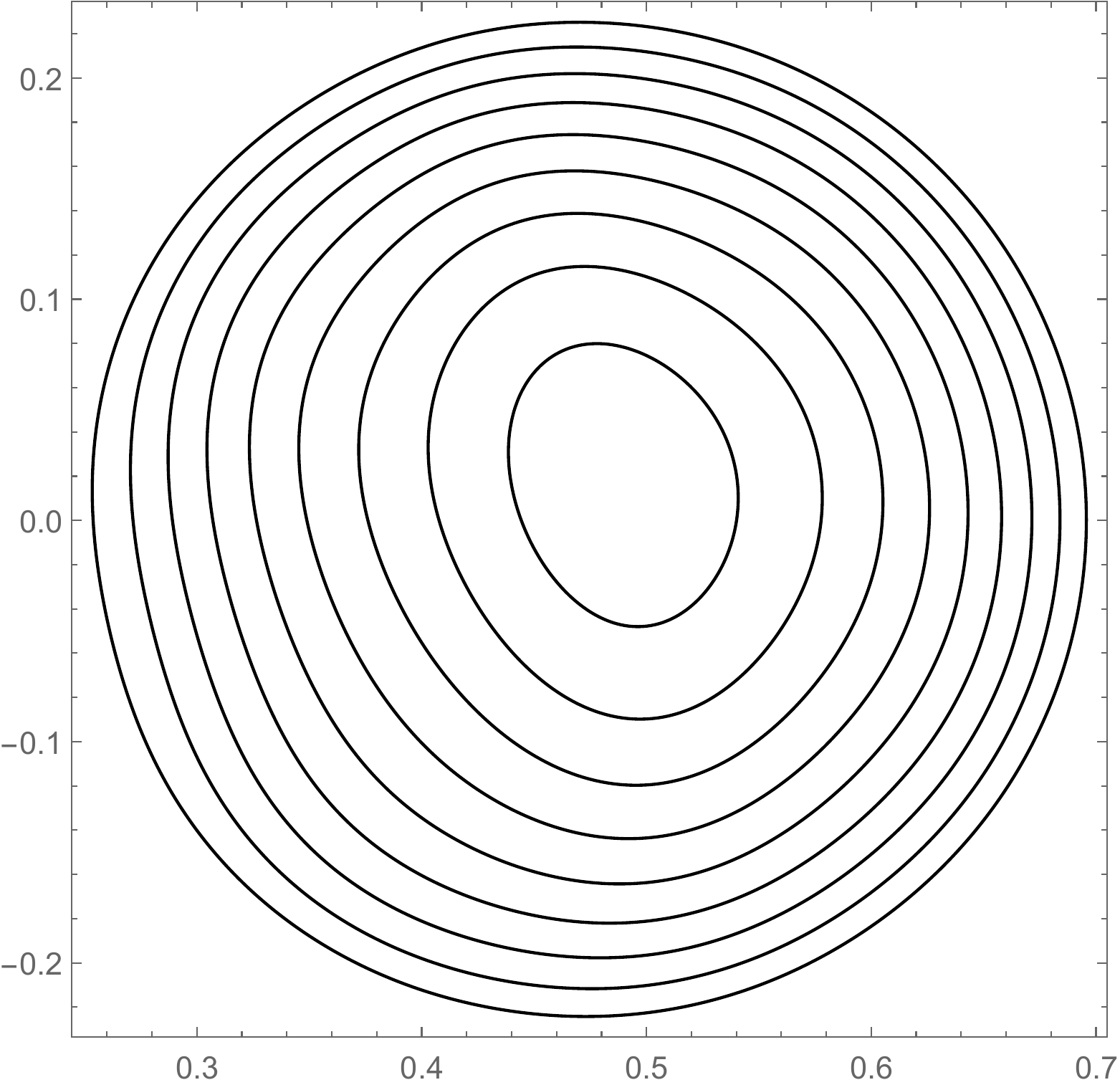}}
{${\scriptstyle\hspace{0.25cm} k_x}$}{-1mm}{\begin{rotate}{0}
\end{rotate}}{0mm}
\vspace*{-1.25em}\subcaption*{{\scriptsize (c.2) $t=t^*$, zoom}}\vspace*{1.25em}
\end{subfigure}
\begin{subfigure}[t]{0.0001\textwidth}
\vspace*{-2.0cm}${\scriptstyle \! k_y}$
\end{subfigure}
\begin{subfigure}[t]{.24\textwidth}
\FigureXYLabel{\includegraphics[type=pdf,ext=.pdf,read=.pdf,width=0.96\textwidth]{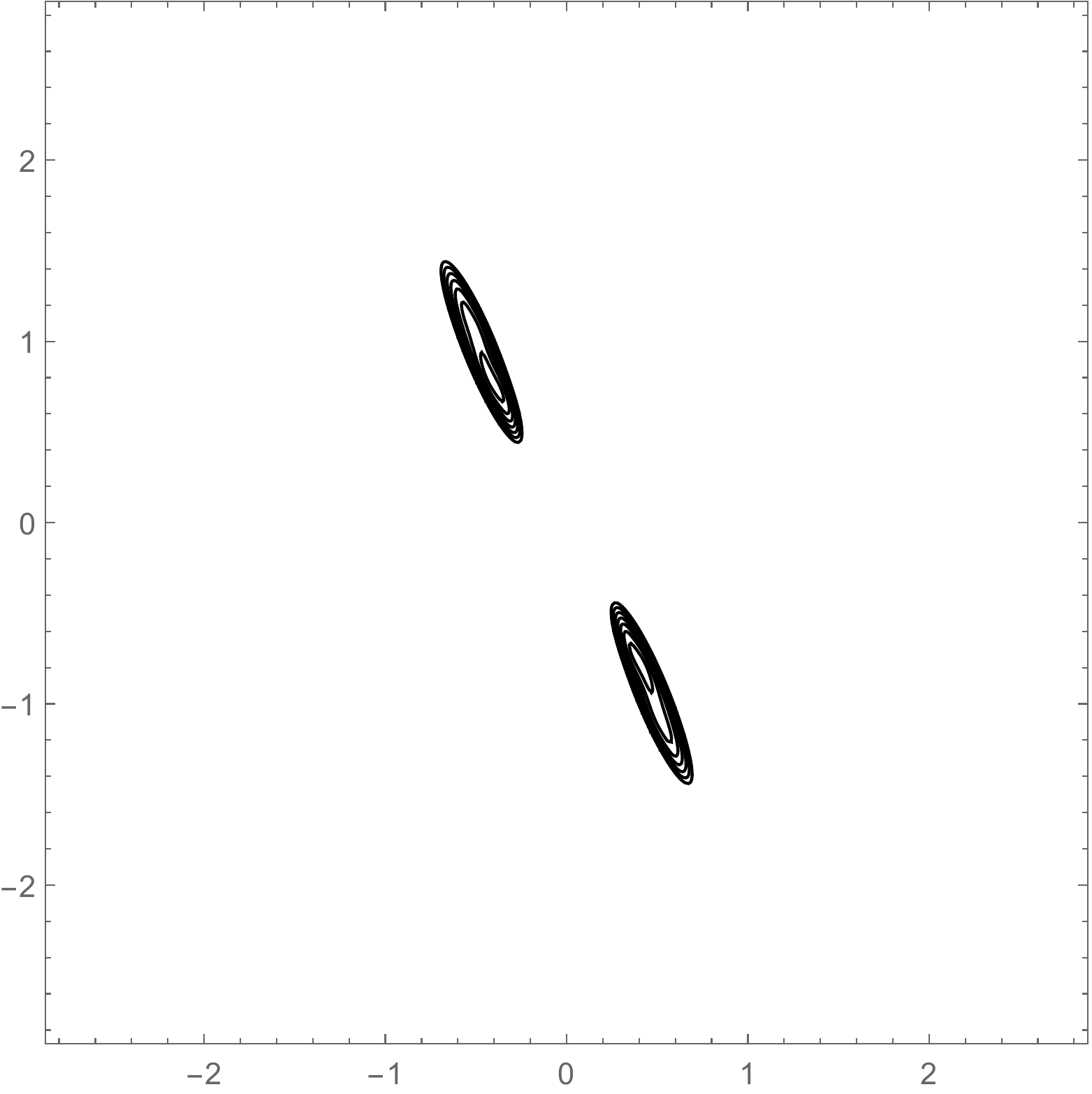}}
{${\scriptstyle\hspace{0.25cm} k_x}$}{-1mm}{\begin{rotate}{0}
\end{rotate}}{0mm}
\vspace*{-1.25em}\subcaption*{{\scriptsize (d.1) $t=3/2t^*$}}\vspace*{1.25em}
\end{subfigure}
\hspace*{-.2cm}
\begin{subfigure}[t]{.24\textwidth}
\FigureXYLabel{\includegraphics[type=pdf,ext=.pdf,read=.pdf,width=0.96\textwidth]{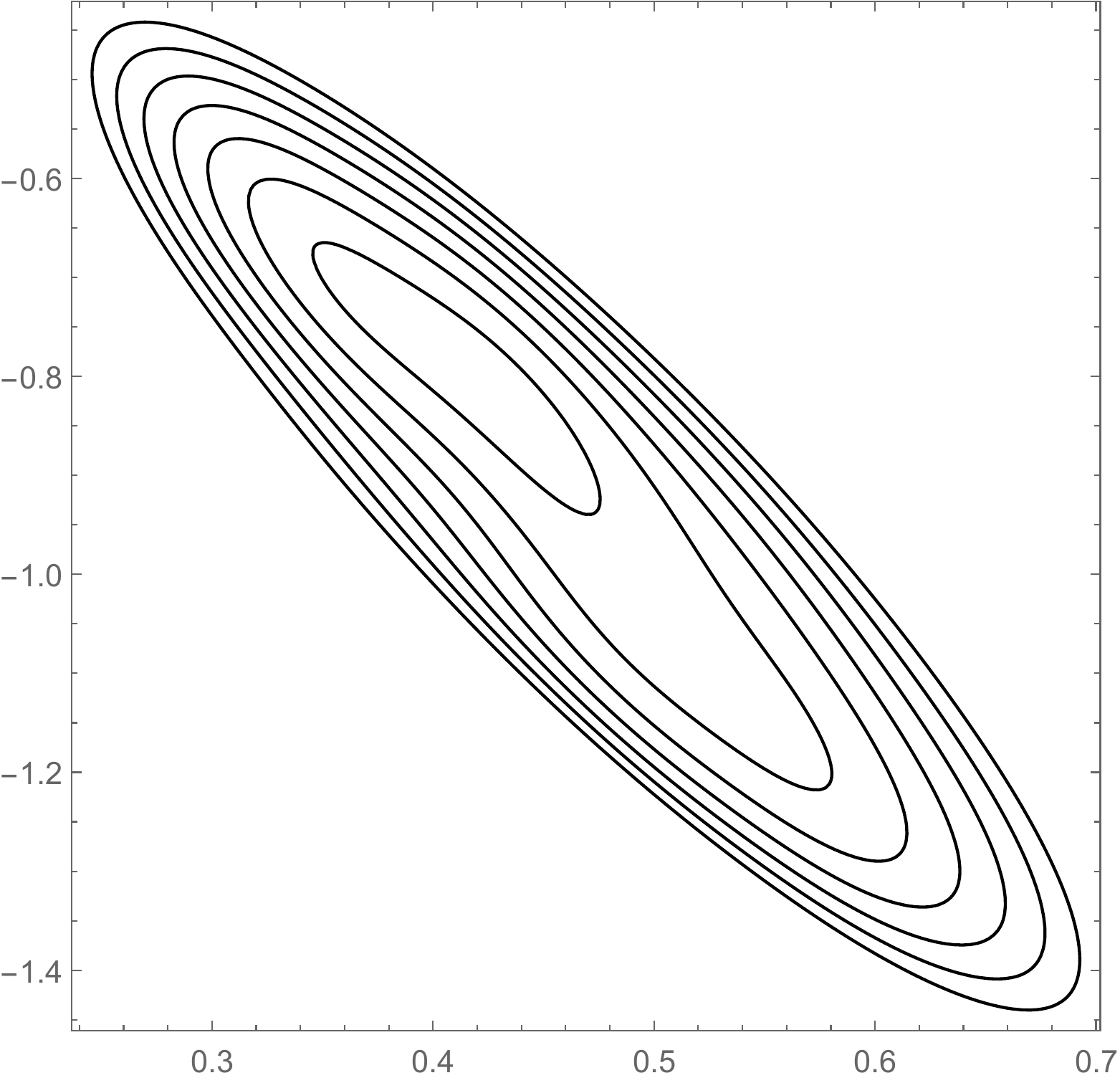}}
{${\scriptstyle\hspace{0.25cm} k_x}$}{-1mm}{\begin{rotate}{0}
\end{rotate}}{0mm}
\vspace*{-1.25em}\subcaption*{{\scriptsize (d.2) $t=3/2t^*$, zoom}}\vspace*{1.25em}
\end{subfigure}
\begin{subfigure}[t]{.24\textwidth}
\FigureXYLabel{\includegraphics[type=pdf,ext=.pdf,read=.pdf,width=0.96\textwidth]{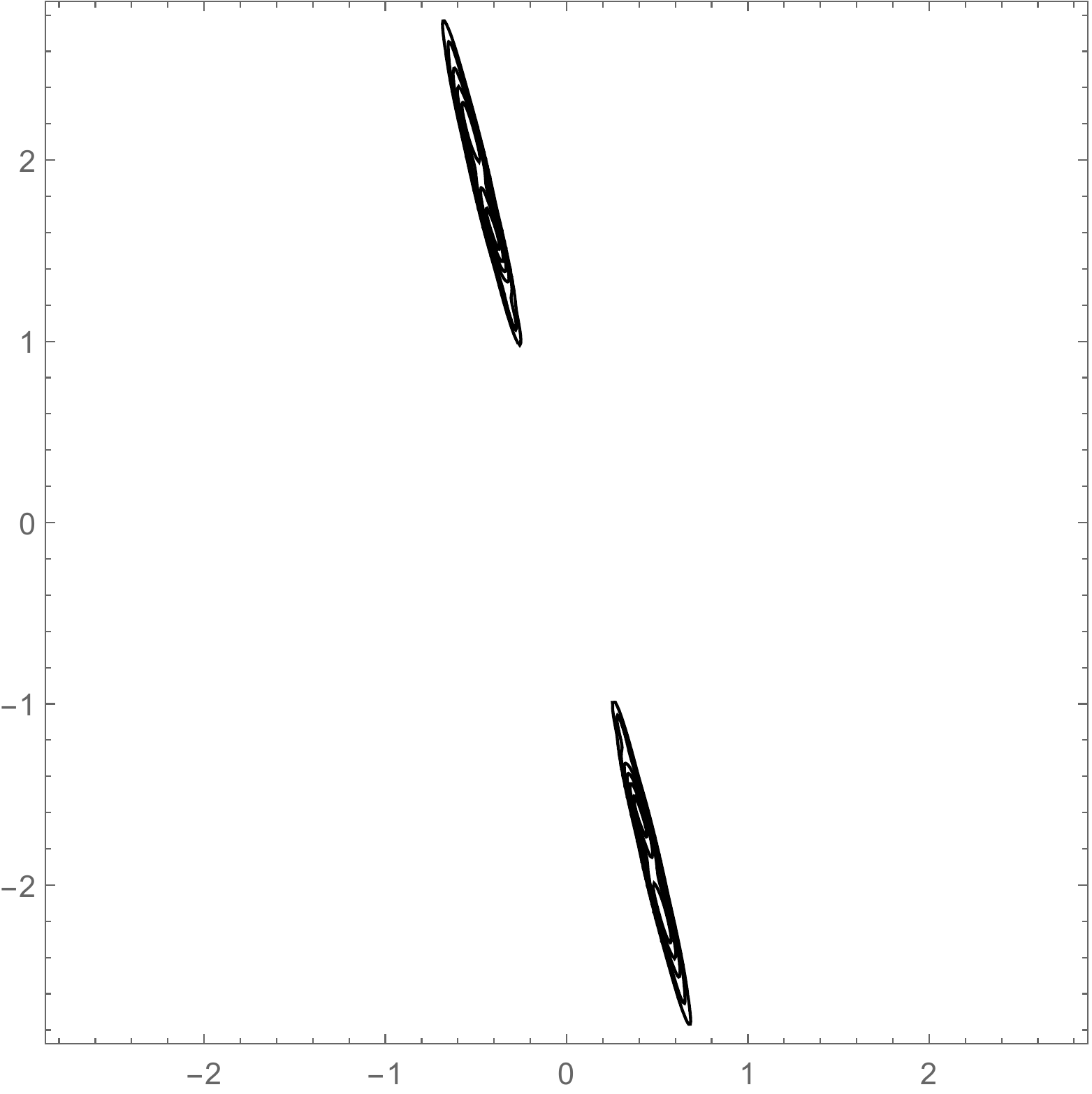}}
{${\scriptstyle\hspace{0.25cm} k_x}$}{-1mm}{\begin{rotate}{0}
\end{rotate}}{0mm}
\vspace*{-1.25em}\subcaption*{{\scriptsize (e.1) $t=2t^*$}}\vspace*{1.25em}
\end{subfigure}
\hspace*{-.2cm}
\begin{subfigure}[t]{.24\textwidth}
\FigureXYLabel{\includegraphics[type=pdf,ext=.pdf,read=.pdf,width=0.96\textwidth]{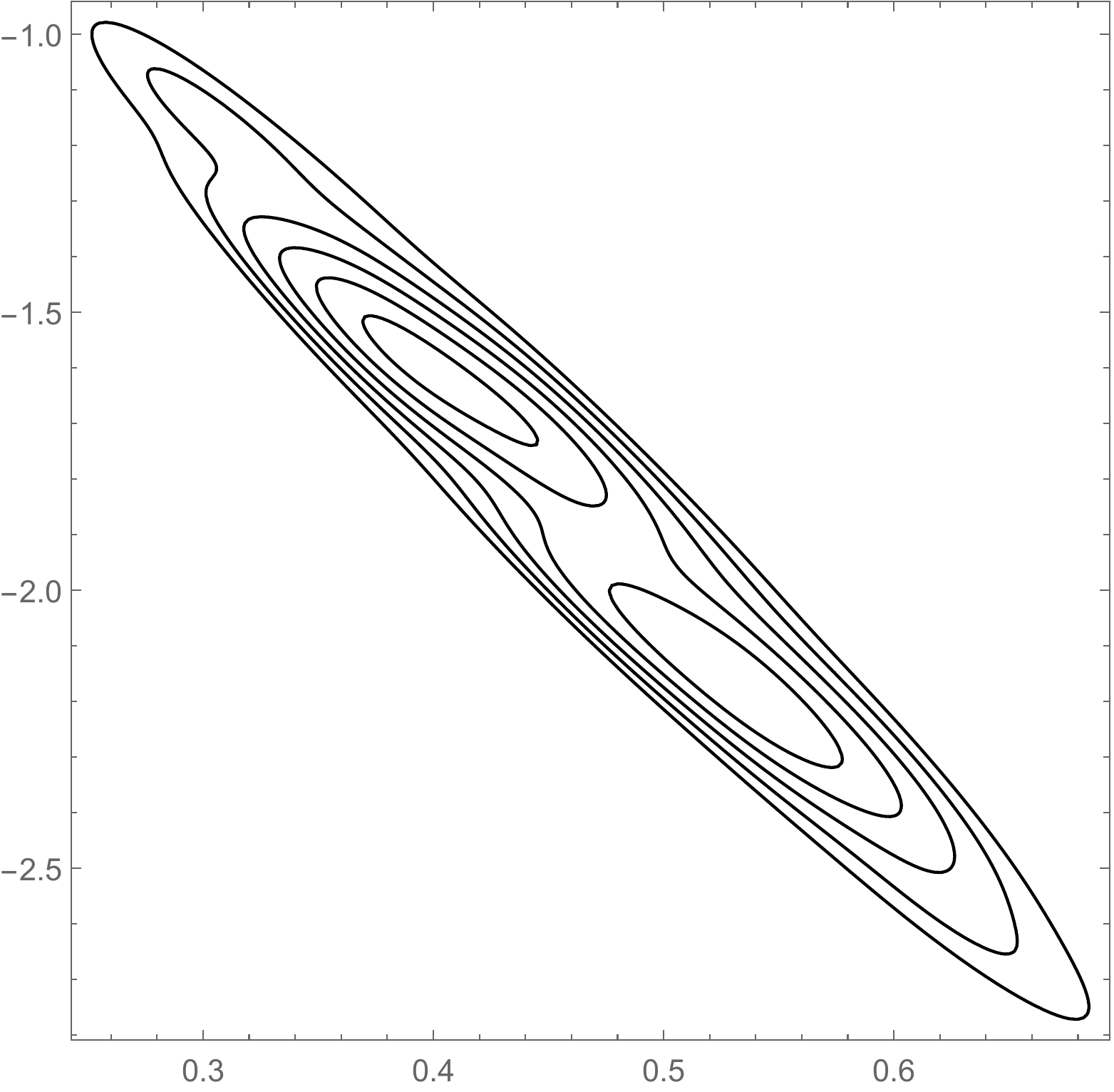}}
{${\scriptstyle\hspace{0.25cm} k_x}$}{-1mm}{\begin{rotate}{0}
\end{rotate}}{0mm}
\vspace*{-1.25em}\subcaption*{{\scriptsize (e.2) $t=2t^*$, zoom}}\vspace*{1.25em}
\end{subfigure}
\begin{subfigure}[t]{0.0001\textwidth}
\vspace*{-2.0cm}${\scriptstyle \! k_y}$
\end{subfigure}
\begin{subfigure}[t]{.24\textwidth}
\FigureXYLabel{\includegraphics[type=pdf,ext=.pdf,read=.pdf,width=0.96\textwidth]{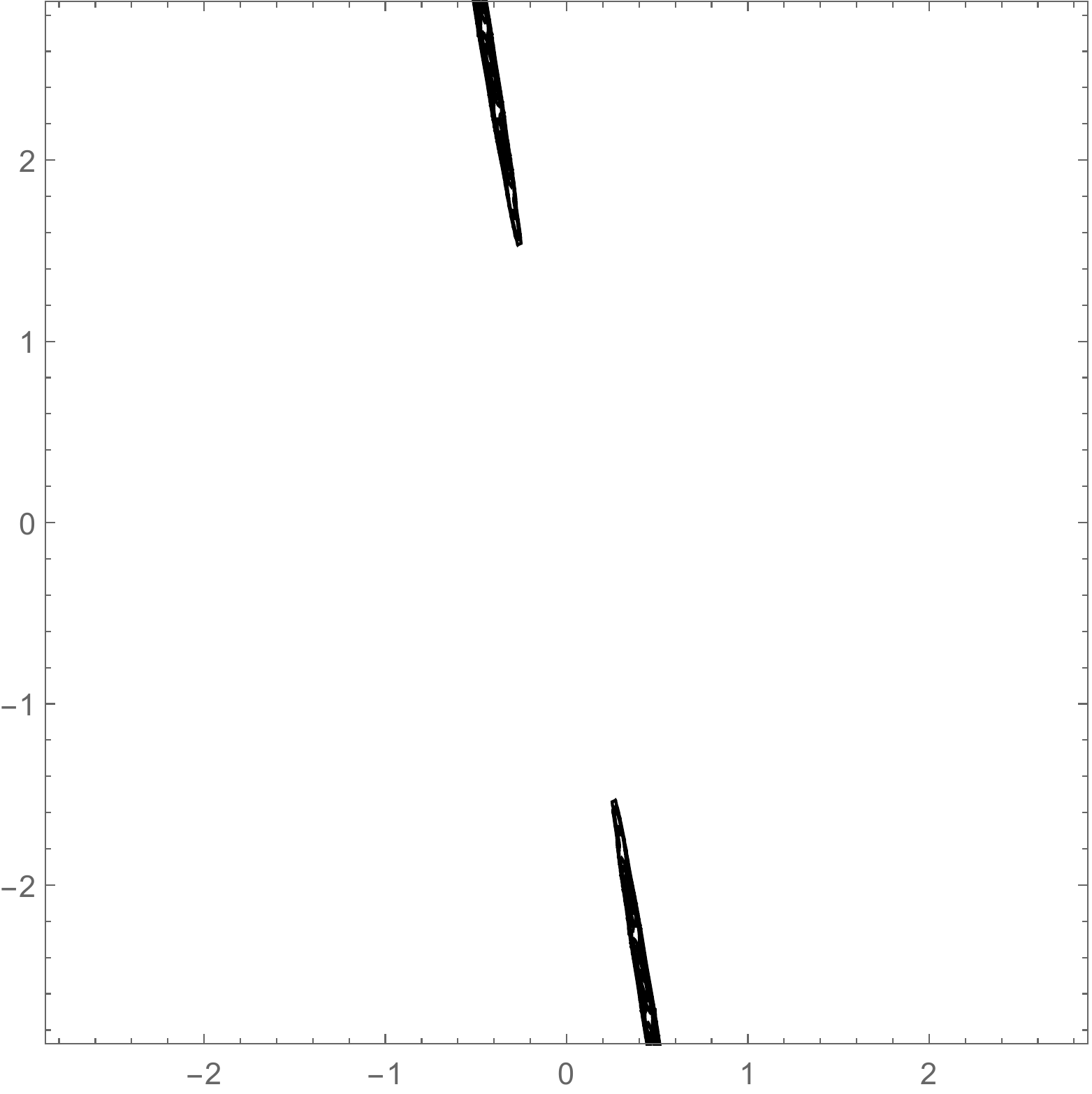}}
{${\scriptstyle\hspace{0.25cm} k_x}$}{-1mm}{\begin{rotate}{0}
\end{rotate}}{0mm}
\vspace*{-1.25em}\subcaption*{{\scriptsize (f.1) $t=5/2t^*$}}\vspace*{1.25em}
\end{subfigure}
\hspace*{-.2cm}
\begin{subfigure}[t]{.24\textwidth}
\FigureXYLabel{\includegraphics[type=pdf,ext=.pdf,read=.pdf,width=0.96\textwidth]{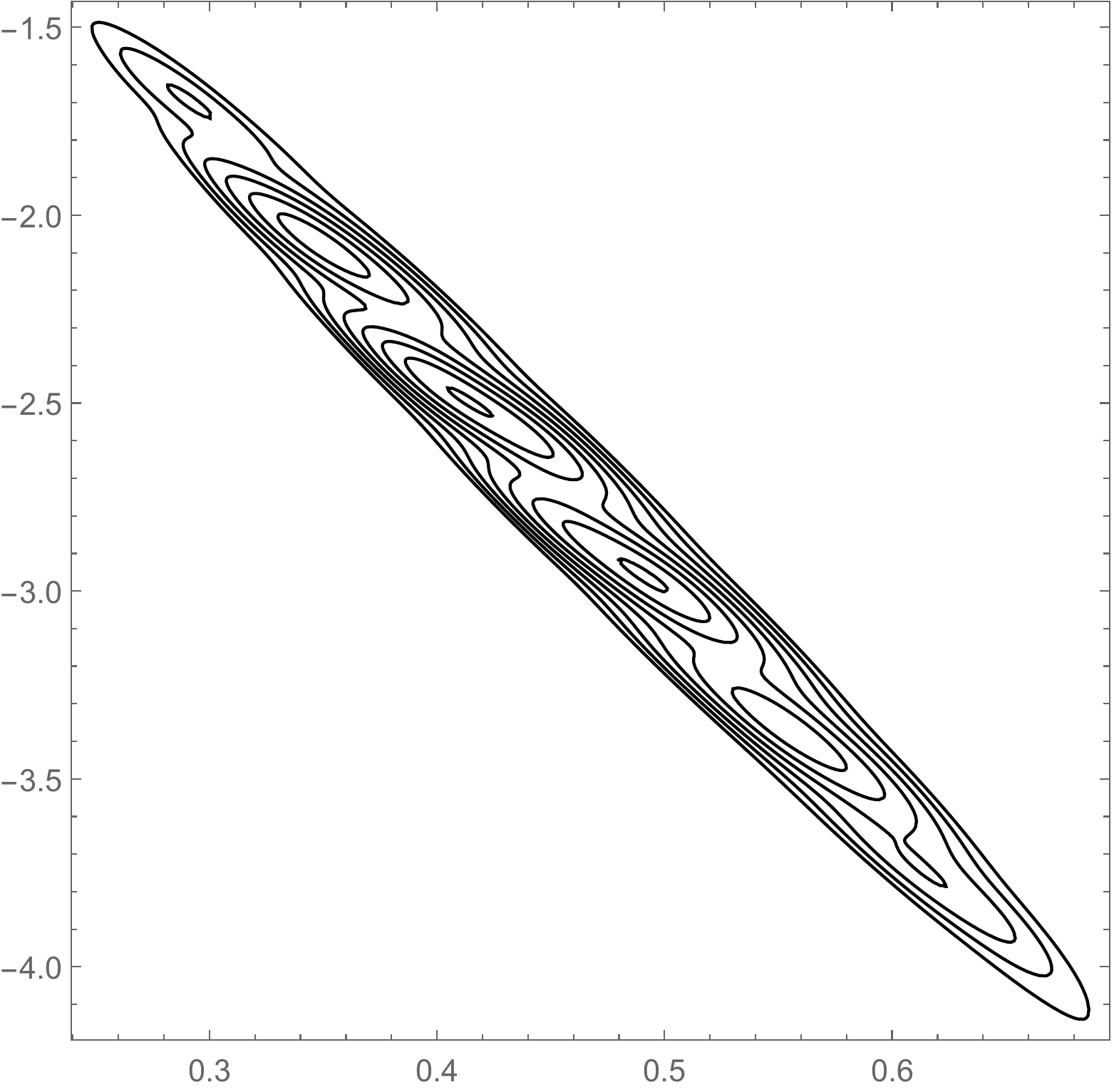}}
{${\scriptstyle\hspace{0.25cm} k_x}$}{-1mm}{\begin{rotate}{0}
\end{rotate}}{0mm}
\vspace*{-1.25em}\subcaption*{{\scriptsize (f.2) $t=5/2t^*$, zoom}}\vspace*{1.25em}
\end{subfigure}
\vspace*{-1em}\caption{{\footnotesize Evolution of the inertial-frame solution $|\hat{\rho}|$ in the spectral $(k_x,k_y)$-plane, according to the PDE-solution
\eqref{171124:2139} with the initial condition \eqref{171123:1354}. This figure matches with Fig.$\,$[6.6] in \cite{Hau16}. By construction, the wave-packet is drifting along
the $k_y$-direction. However, besides this drift the wave packet is also changing its expansion size. Before the critical time
$t^*$ is reached, the packet is shrinking in the $k_y$- and expanding in the $k_x$-direction, while after that point in time, for $t>t^*$, it is expanding in the $k_y$- and shrinking in the
$k_x$-direction. Note that since the time coordinate transforms as an invariant $t=\tilde{t}$ under $\mathsf{K}$ \eqref{171123:1839}, the value of critical time~$t^*$ is the same as
that of $\tilde{t}^*$ in Fig.$\,$\ref{fig1}, that is, $t^*=\tilde{t}^*\sim 8.5$. However here in the inertial frame, the time~$t^*$ expresses itself differently than in the optimal Kelvin frame of Fig.$\,$\ref{fig1}, namely as the time when the packet crosses the $k_x$-axis, since by its definition from Fig.$\,$\ref{fig1} we obtain $0=k_{\tilde{y}}-A\tilde{t}^*k_{\tilde{x}}=k_y+At^*k_x-At^*k_x=k_y$. As already said for Fig.$\,$\ref{fig1}, the solution shown here is also not truly physical, since the considered initial condition~\eqref{171123:1354}, as taken identically from \cite{Hau16}, violates the reality constraint of the Fourier transform, having the effect that the above spectral solution will not lead to a real-valued solution in physical space. The real part of this associated solution in physical space is shown in Fig.$\,$\ref{fig4}.\label{fig2}}}
\end{figure}

\begin{figure}[H]
\begin{subfigure}[t]{.49\textwidth}
\FigureXYLabel{\includegraphics[type=pdf,ext=.pdf,read=.pdf,width=0.85\textwidth]{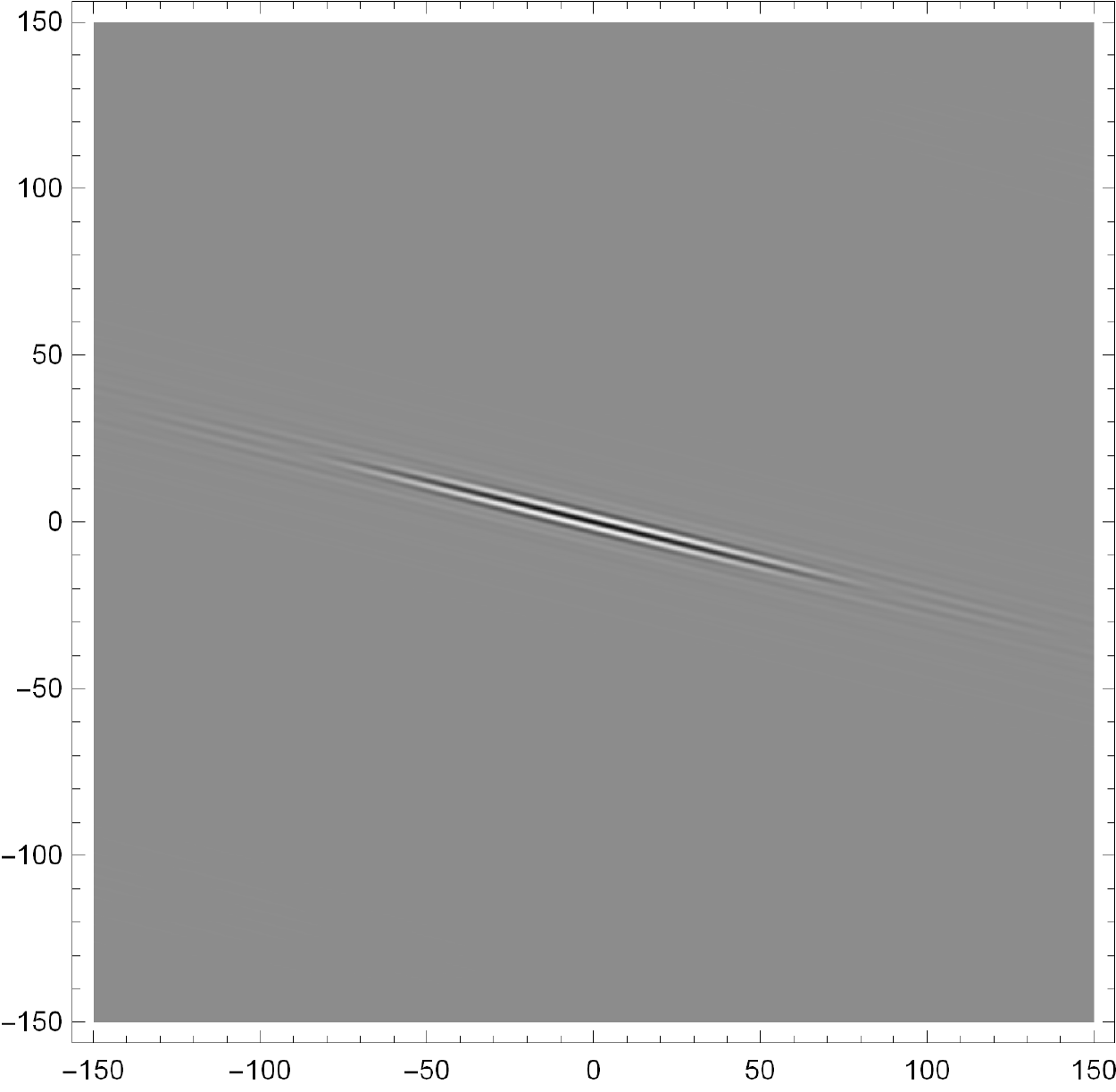}}
{${\scriptstyle\hspace{0.45cm}\tilde{x}}$}{-2mm}{\begin{rotate}{0}
\end{rotate}}{0mm}
\vspace*{-1.25em}\subcaption*{{\scriptsize (a) $\tilde{t}=0$}}\vspace*{0.5em}
\end{subfigure}
\hspace*{.2cm}
\begin{subfigure}[c]{0.0001\textwidth}
\vspace*{-6.5cm}${\scriptstyle \!\! \tilde{y}}$
\end{subfigure}
\begin{subfigure}[t]{.49\textwidth}
\FigureXYLabel{\includegraphics[type=pdf,ext=.pdf,read=.pdf,width=0.85\textwidth]{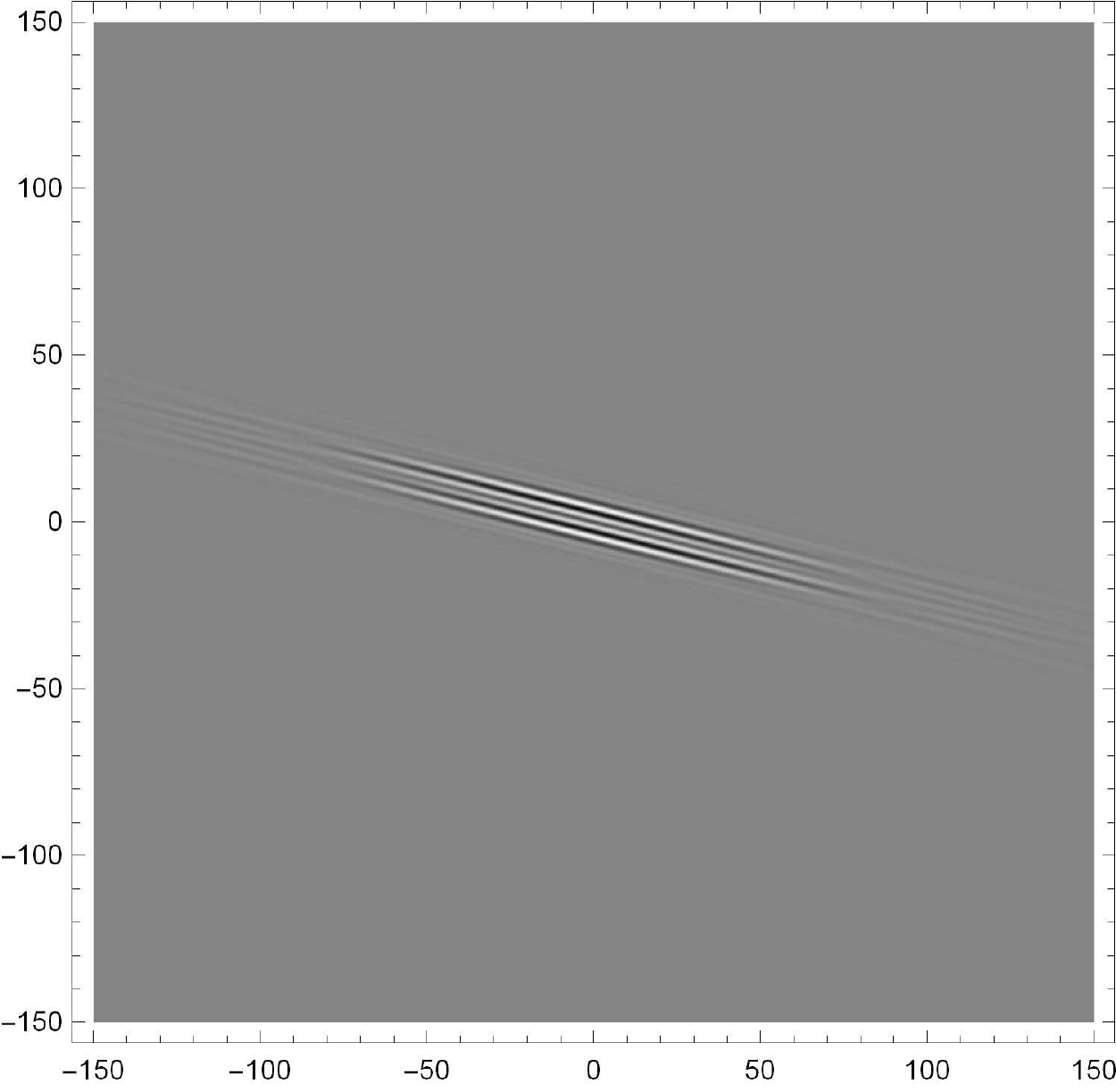}}
{${\scriptstyle\hspace{0.45cm}\tilde{x}}$}{-2mm}{\begin{rotate}{0}
\end{rotate}}{0mm}
\vspace*{-1.25em}\subcaption*{{\scriptsize (b) $\tilde{t}=1/2\tilde{t}^*$}}\vspace*{0.5em}
\end{subfigure}
\begin{subfigure}[t]{.49\textwidth}
\FigureXYLabel{\includegraphics[type=pdf,ext=.pdf,read=.pdf,width=0.85\textwidth]{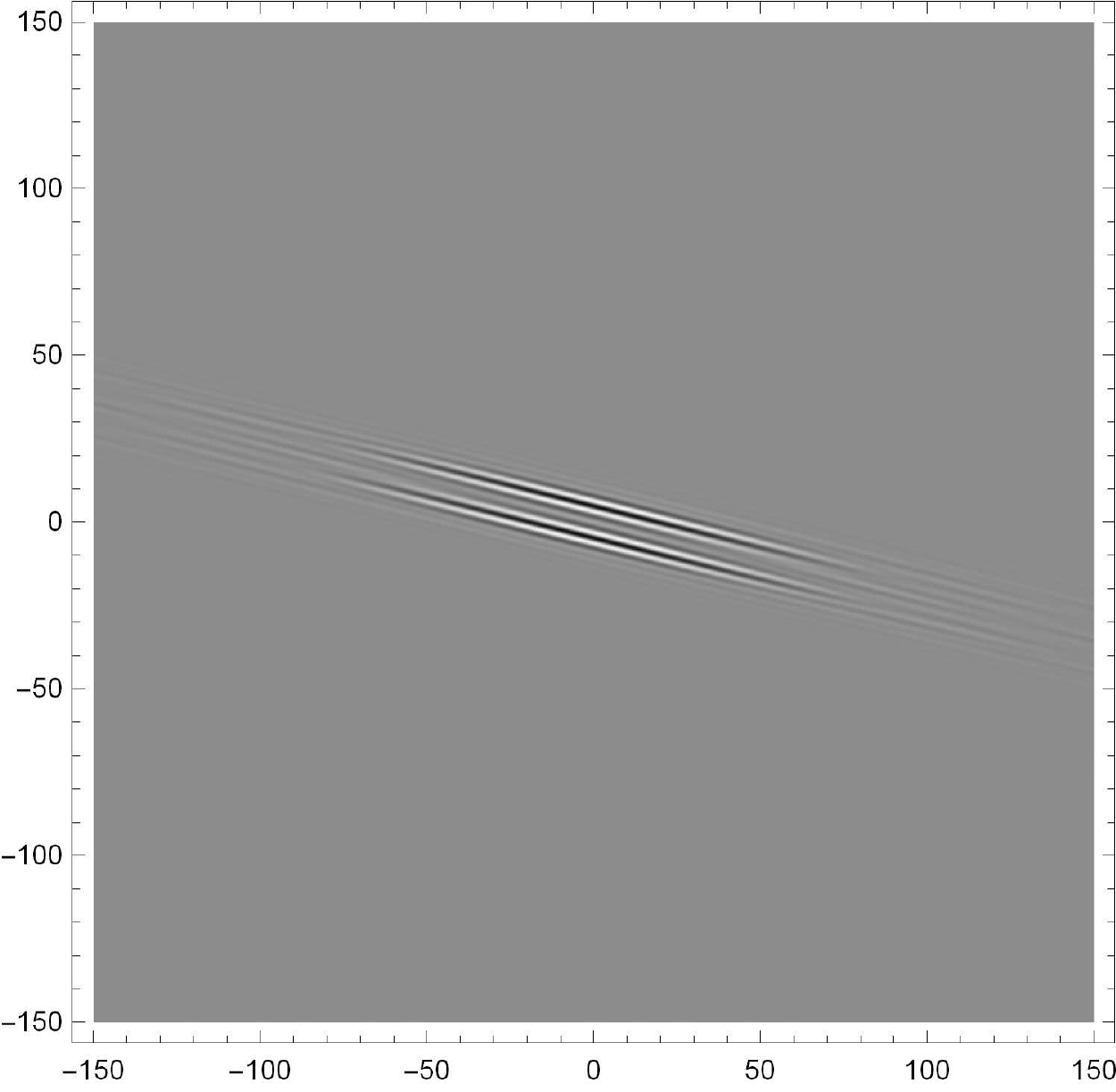}}
{${\scriptstyle\hspace{0.45cm}\tilde{x}}$}{-2mm}{\begin{rotate}{0}
\end{rotate}}{0mm}
\vspace*{-1.25em}\subcaption*{{\scriptsize (c) $\tilde{t}=\tilde{t}^*$}}\vspace*{0.5em}
\end{subfigure}
\hspace*{.2cm}
\begin{subfigure}[c]{0.0001\textwidth}
\vspace*{-6.5cm}${\scriptstyle \!\! \tilde{y}}$
\end{subfigure}
\begin{subfigure}[t]{.49\textwidth}
\FigureXYLabel{\includegraphics[type=pdf,ext=.pdf,read=.pdf,width=0.85\textwidth]{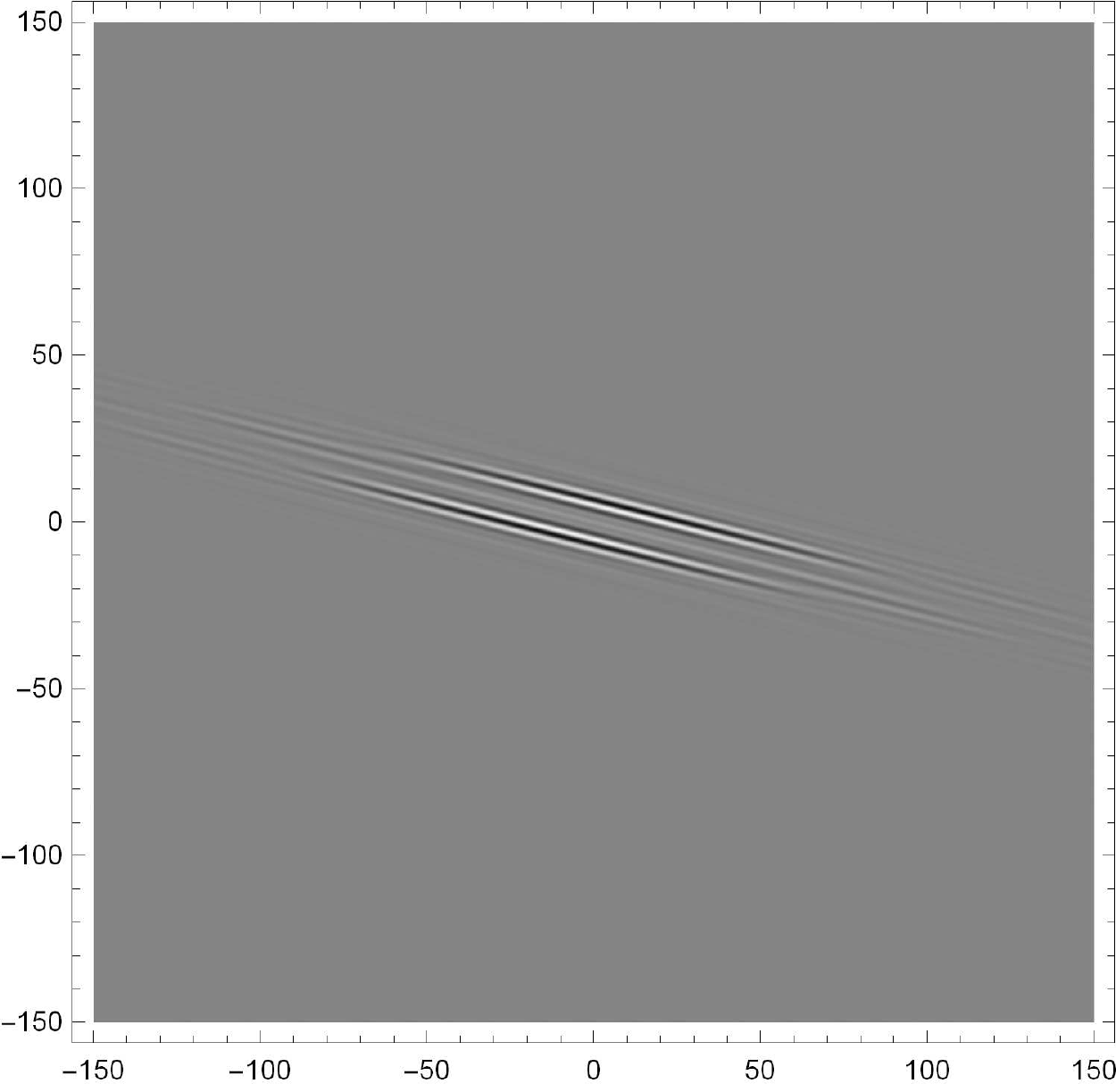}}
{${\scriptstyle\hspace{0.45cm}\tilde{x}}$}{-2mm}{\begin{rotate}{0}
\end{rotate}}{0mm}
\vspace*{-1.25em}\subcaption*{{\scriptsize (d) $\tilde{t}=3/2\tilde{t}^*$}}\vspace*{0.5em}
\end{subfigure}
\begin{subfigure}[t]{.49\textwidth}
\FigureXYLabel{\includegraphics[type=pdf,ext=.pdf,read=.pdf,width=0.85\textwidth]{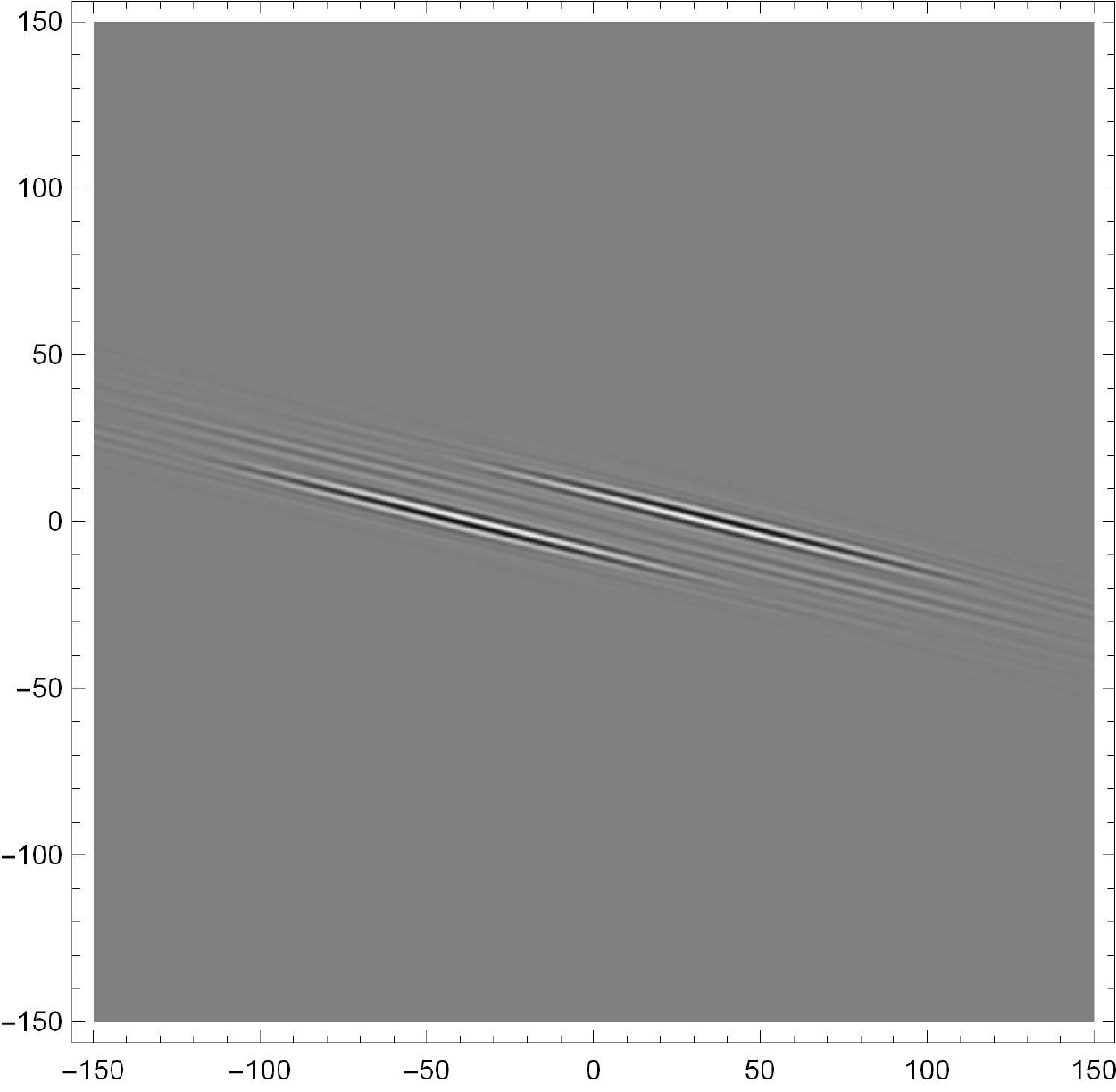}}
{${\scriptstyle\hspace{0.45cm}\tilde{x}}$}{-2mm}{\begin{rotate}{0}
\end{rotate}}{0mm}
\vspace*{-1.25em}\subcaption*{{\scriptsize (e) $\tilde{t}=2\tilde{t}^*$}}\vspace*{0.5em}
\end{subfigure}
\hspace*{.2cm}
\begin{subfigure}[c]{0.0001\textwidth}
\vspace*{-6.5cm}${\scriptstyle \!\! \tilde{y}}$
\end{subfigure}
\begin{subfigure}[t]{.49\textwidth}
\FigureXYLabel{\includegraphics[type=pdf,ext=.pdf,read=.pdf,width=0.85\textwidth]{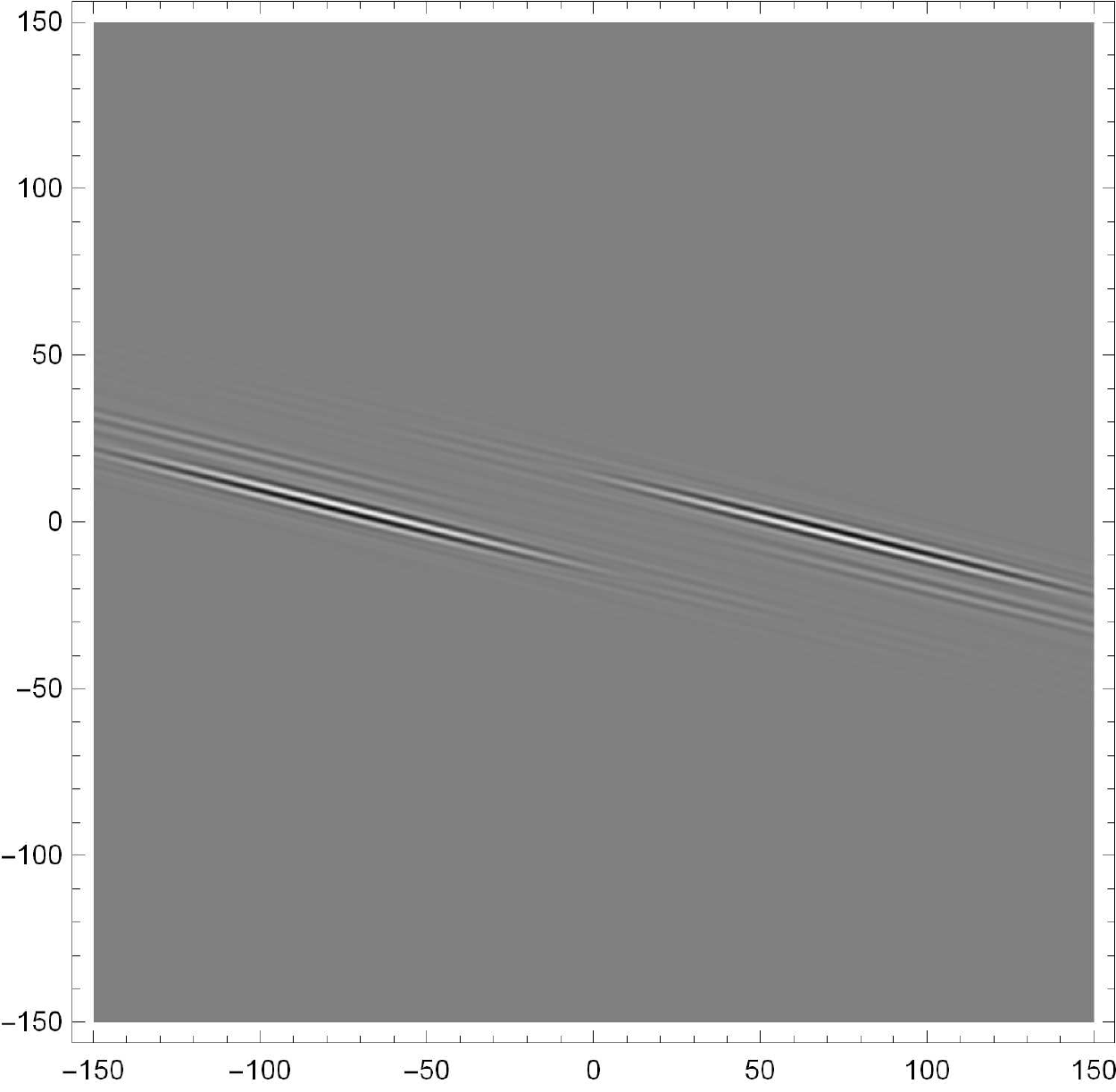}}
{${\scriptstyle\hspace{0.45cm}\tilde{x}}$}{-2mm}{\begin{rotate}{0}
\end{rotate}}{0mm}
\vspace*{-1.25em}\subcaption*{{\scriptsize (f) $\tilde{t}=5/2\tilde{t}^*$}}\vspace*{0.5em}
\end{subfigure}
\vspace*{-1em}\caption{{\footnotesize Evolution in physical space of the density field in the accelerated Kelvin frame,
synchronically corresponding to the evolution of its complementary density field in spectral space as shown in Fig.$\,$\ref{fig1}. Explicitly shown
in (a)-(f) is only $\text{Re}(\tilde{\rho})$, the real part of the density
field in physical space. Due to that the initial condition \eqref{171123:2035} violates the reality constraint of the Fourier transformation,
the density field $\tilde{\rho}$ itself is non-real, i.e. $\text{Im}(\tilde{\rho})\neq 0$, and thus unphysical.\label{fig3}}}
\end{figure}

\begin{figure}[H]
\begin{subfigure}[t]{.49\textwidth}
\FigureXYLabel{\includegraphics[type=pdf,ext=.pdf,read=.pdf,width=0.85\textwidth]{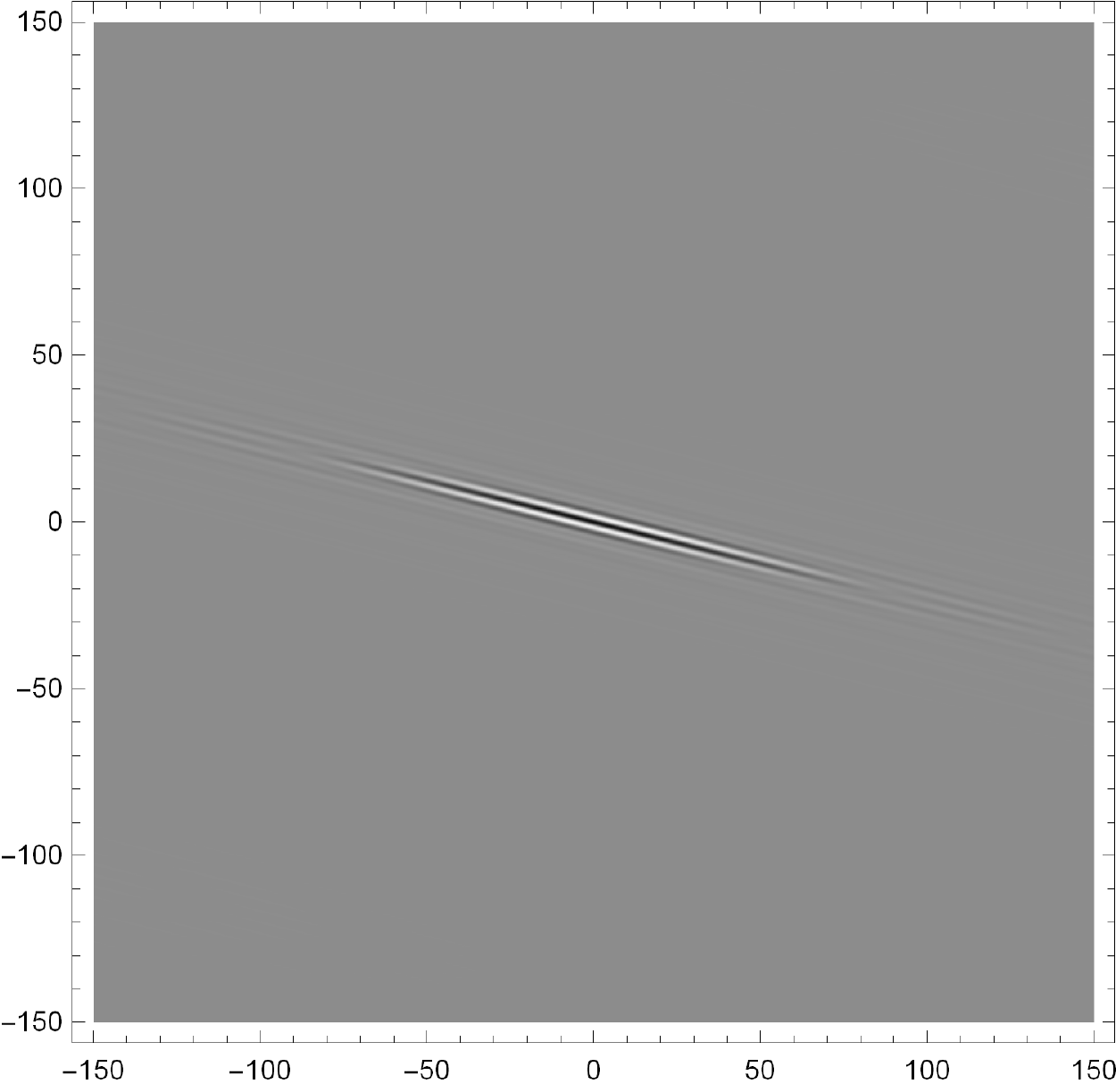}}
{${\scriptstyle\hspace{0.45cm}x}$}{-2mm}{\begin{rotate}{0}
\end{rotate}}{0mm}
\vspace*{-1.25em}\subcaption*{{\scriptsize (a) $t=0$}}\vspace*{0.5em}
\end{subfigure}
\hspace*{.2cm}
\begin{subfigure}[c]{0.0001\textwidth}
\vspace*{-6.5cm}${\scriptstyle \!\! y}$
\end{subfigure}
\begin{subfigure}[t]{.49\textwidth}
\FigureXYLabel{\includegraphics[type=pdf,ext=.pdf,read=.pdf,width=0.85\textwidth]{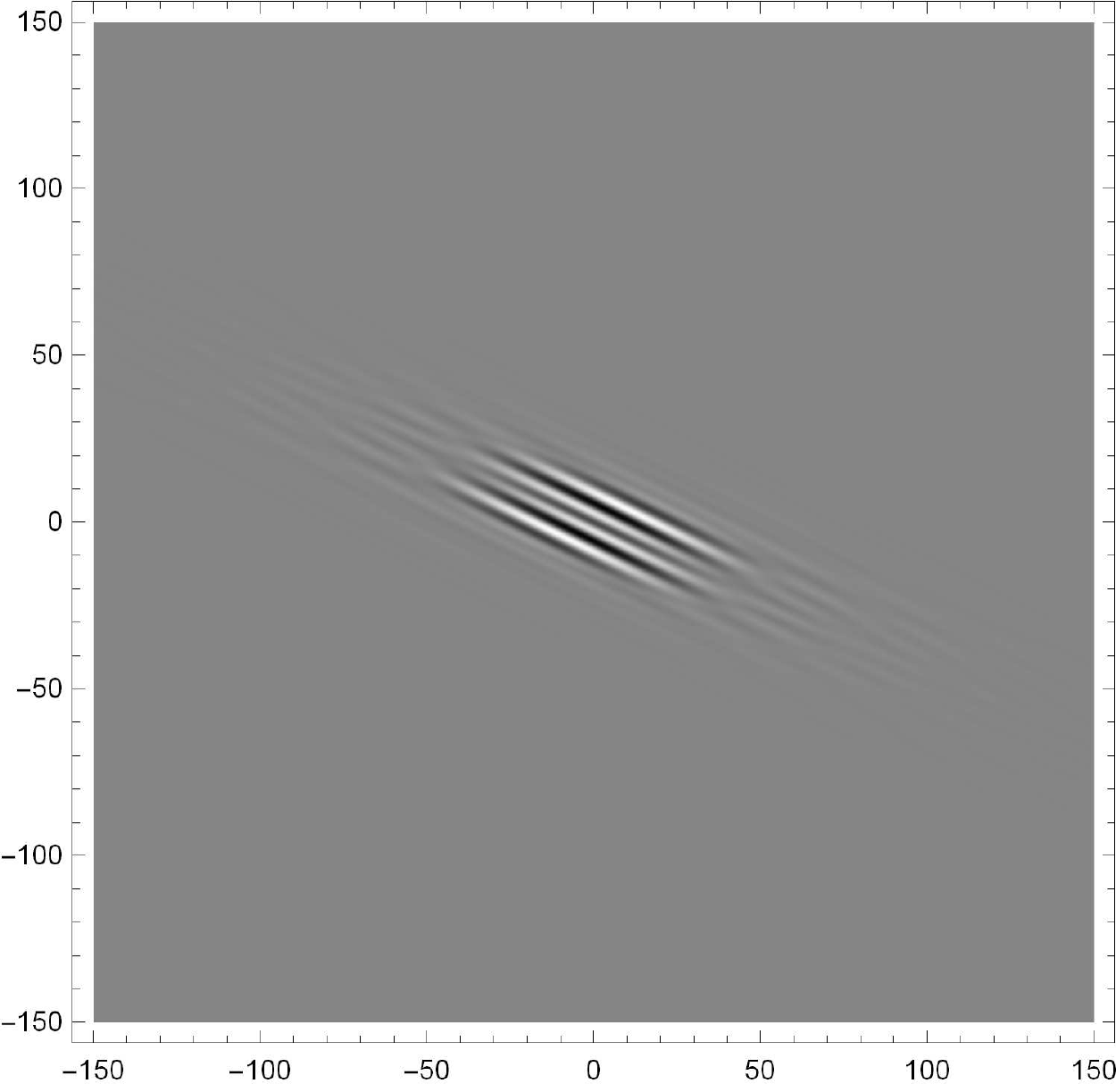}}
{${\scriptstyle\hspace{0.45cm} x}$}{-2mm}{\begin{rotate}{0}
\end{rotate}}{0mm}
\vspace*{-1.25em}\subcaption*{{\scriptsize (b) $t=1/2t^*$}}\vspace*{0.5em}
\end{subfigure}
\begin{subfigure}[t]{.49\textwidth}
\FigureXYLabel{\includegraphics[type=pdf,ext=.pdf,read=.pdf,width=0.85\textwidth]{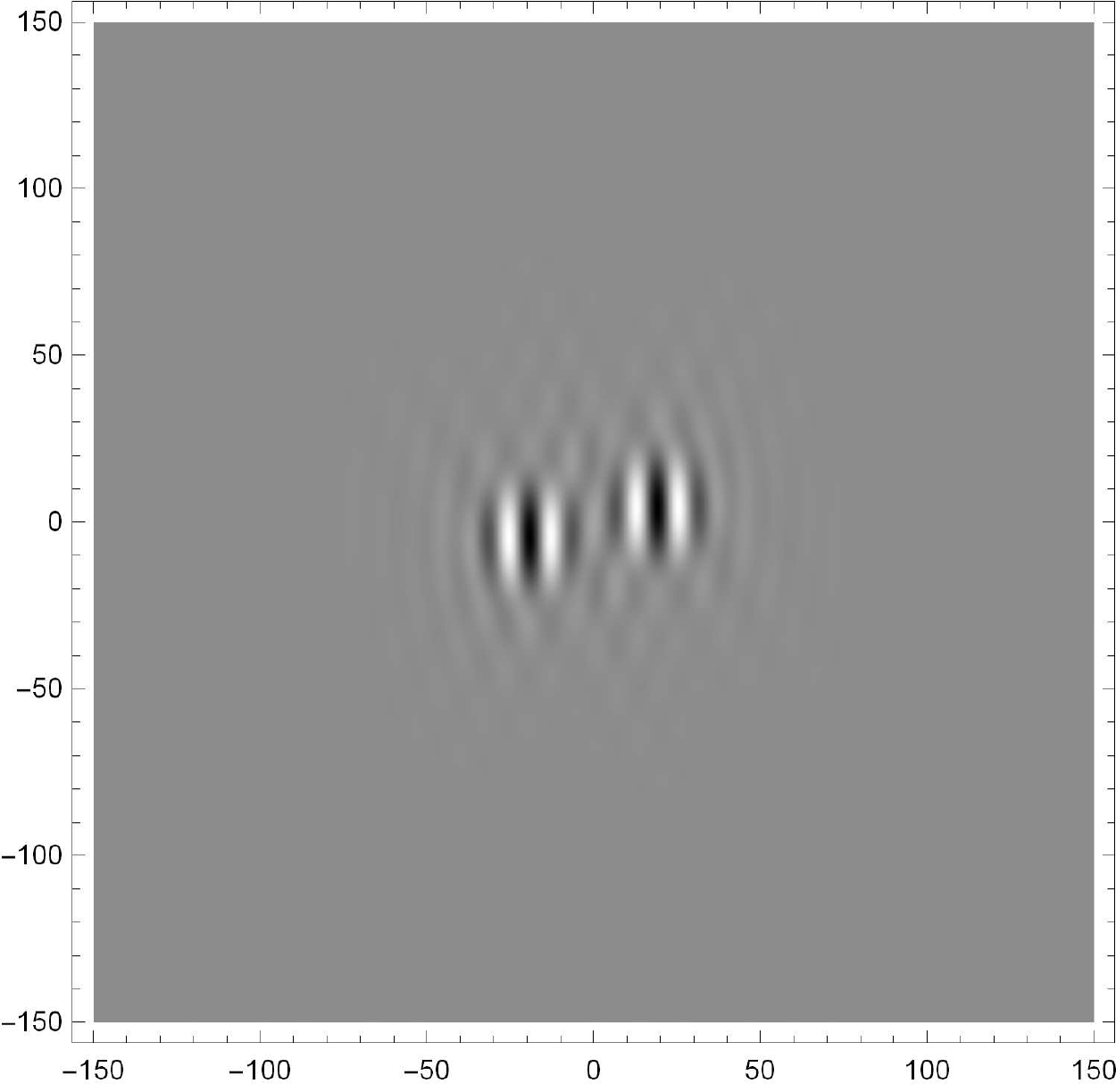}}
{${\scriptstyle\hspace{0.45cm}x}$}{-2mm}{\begin{rotate}{0}
\end{rotate}}{0mm}
\vspace*{-1.25em}\subcaption*{{\scriptsize (c) $t=t^*$}}\vspace*{0.5em}
\end{subfigure}
\hspace*{.2cm}
\begin{subfigure}[c]{0.0001\textwidth}
\vspace*{-6.5cm}${\scriptstyle \!\! y}$
\end{subfigure}
\begin{subfigure}[t]{.49\textwidth}
\FigureXYLabel{\includegraphics[type=pdf,ext=.pdf,read=.pdf,width=0.85\textwidth]{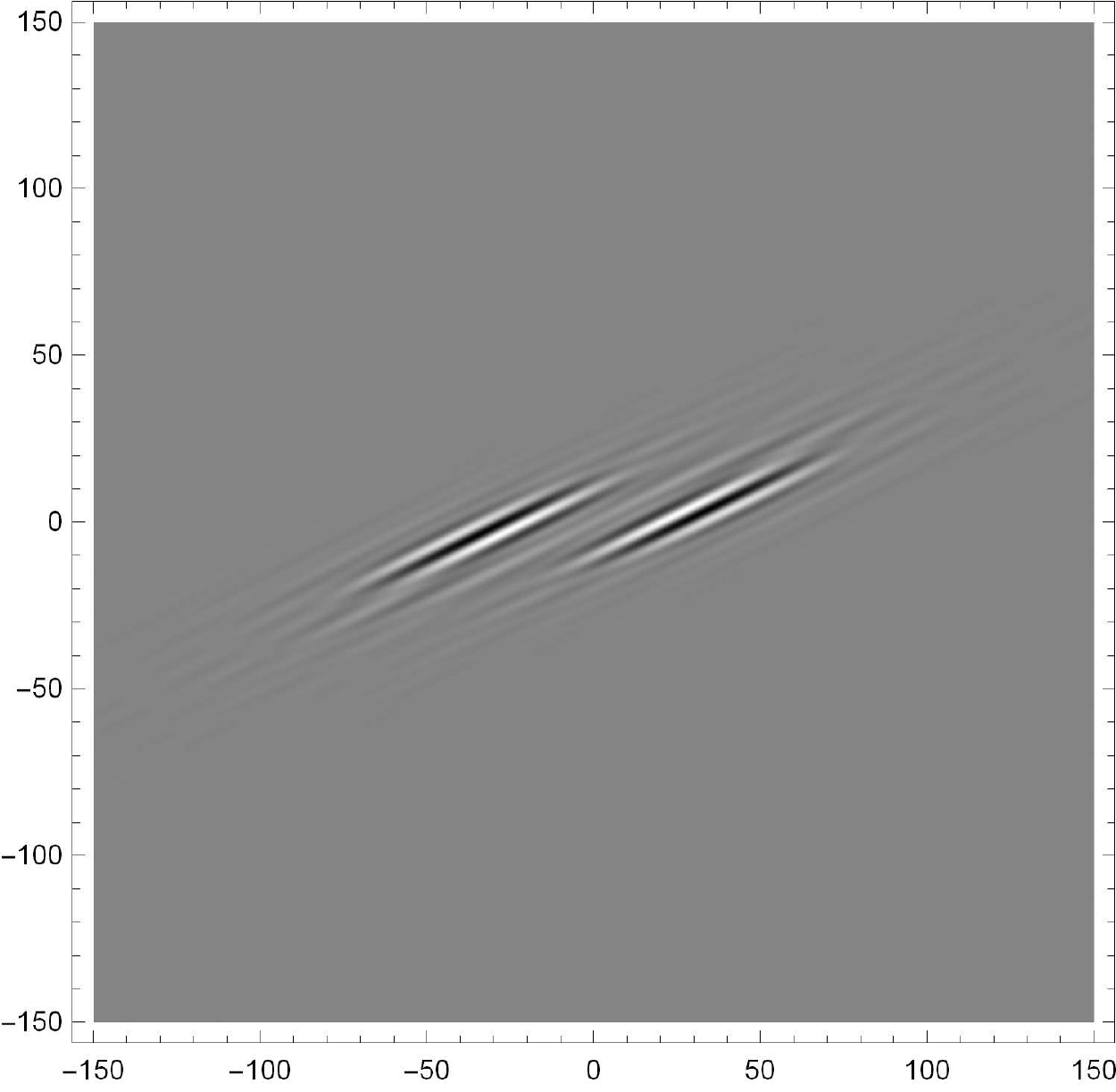}}
{${\scriptstyle\hspace{0.45cm} x}$}{-2mm}{\begin{rotate}{0}
\end{rotate}}{0mm}
\vspace*{-1.25em}\subcaption*{{\scriptsize (d) $t=3/2t^*$}}\vspace*{0.5em}
\end{subfigure}
\begin{subfigure}[t]{.49\textwidth}
\FigureXYLabel{\includegraphics[type=pdf,ext=.pdf,read=.pdf,width=0.85\textwidth]{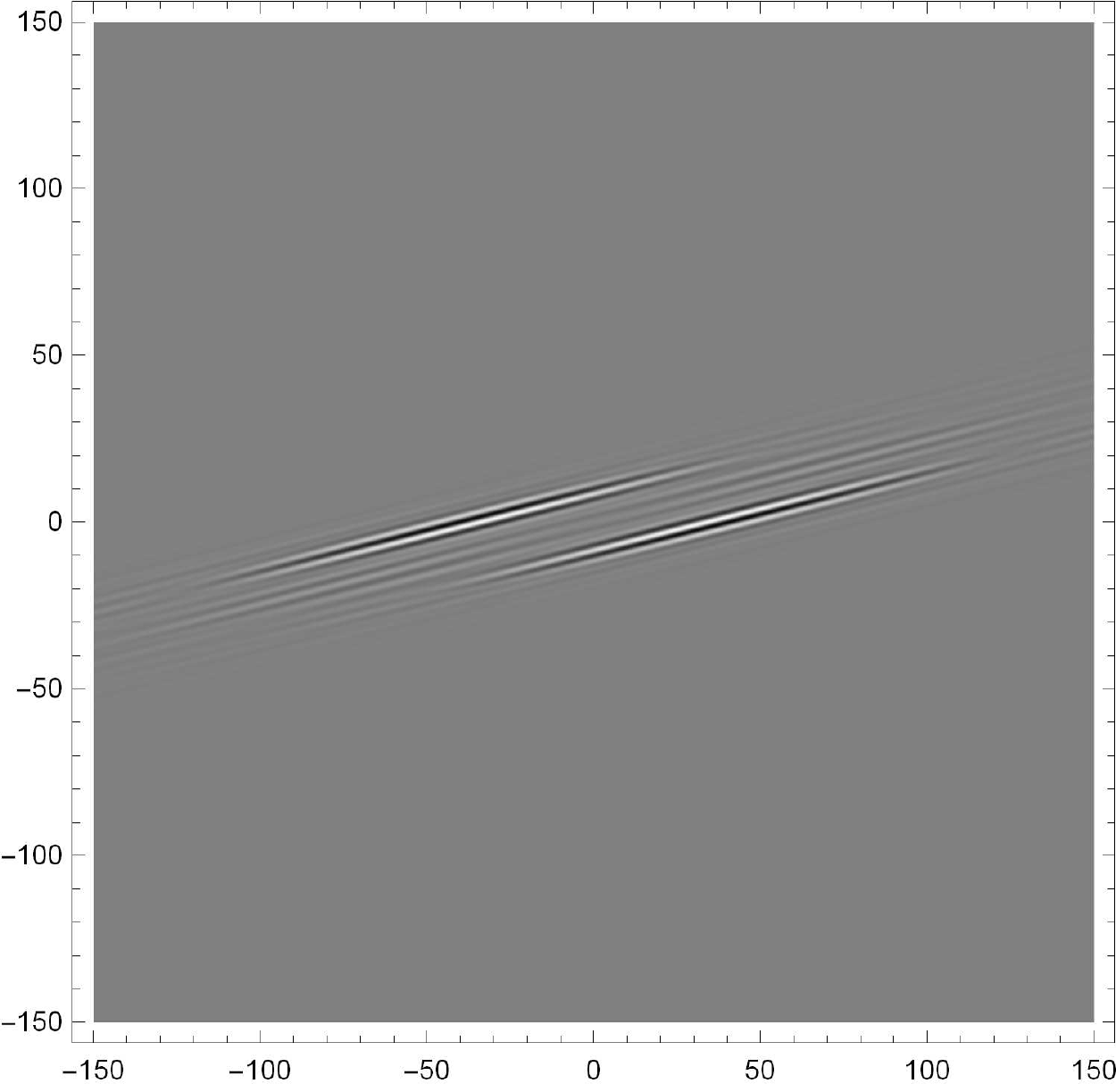}}
{${\scriptstyle\hspace{0.45cm} x}$}{-2mm}{\begin{rotate}{0}
\end{rotate}}{0mm}
\vspace*{-1.25em}\subcaption*{{\scriptsize (e) $t=2t^*$}}\vspace*{0.5em}
\end{subfigure}
\hspace*{.2cm}
\begin{subfigure}[c]{0.0001\textwidth}
\vspace*{-6.5cm}${\scriptstyle \!\! y}$
\end{subfigure}
\begin{subfigure}[t]{.49\textwidth}
\FigureXYLabel{\includegraphics[type=pdf,ext=.pdf,read=.pdf,width=0.85\textwidth]{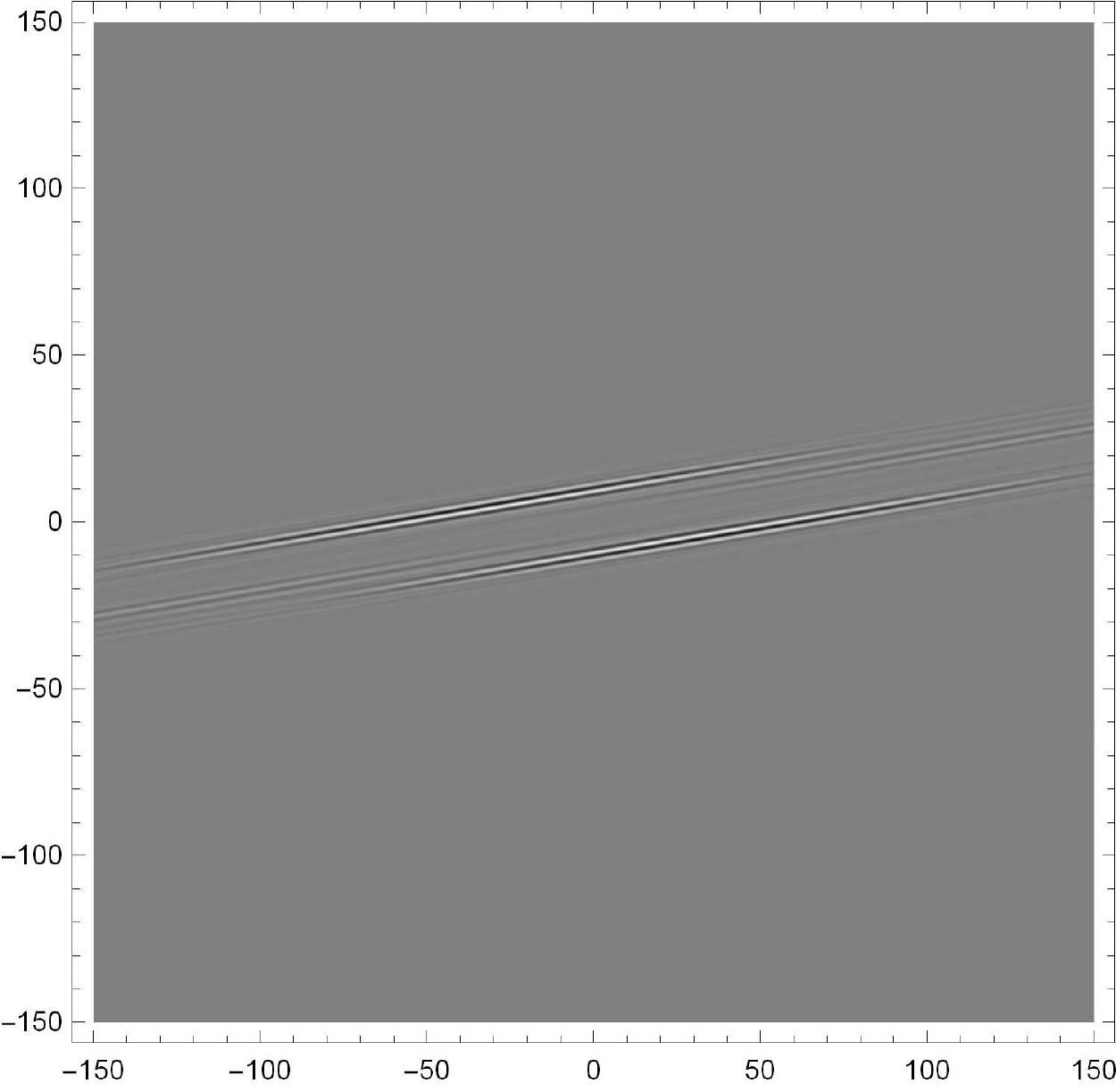}}
{${\scriptstyle\hspace{0.45cm} x}$}{-2mm}{\begin{rotate}{0}
\end{rotate}}{0mm}
\vspace*{-1.25em}\subcaption*{{\scriptsize (f) $t=5/2t^*$}}\vspace*{0.5em}
\end{subfigure}
\vspace*{-1em}\caption{{\footnotesize Evolution in physical space of the density field in the inertial frame,
synchronically corresponding to the evolution of its complementary density field in spectral space as shown in Fig.$\,$\ref{fig2}. Explicitly shown
in (a)-(f) is only $\text{Re}(\rho)$, the real part of the density
field in physical space. Due to that the initial condition \eqref{171123:1354} violates the reality constraint of the Fourier transformation,
the density field $\rho$ itself is non-real, i.e. $\text{Im}(\rho)\neq 0$, and thus unphysical. This figure should match Fig.$\,$[6.7] in \cite{Hau16}, but it does not. For the reason why
its associated Fig.$\,$\ref{fig2} in spectral space matches with Fig.$\,$[6.6] in \cite{Hau16} but not its complementary
picture in physical space, please see the discussion in the text.\label{fig4}}}
\end{figure}

For the initial condition \eqref{171123:1509}-\eqref{171123:1259} as proposed in \cite{Hau16}, Fig.$\,$\ref{fig1} shows the spectral Kelvin frame solution of the absolute-valued density field \eqref{171124:1828}, i.e., $|\hat{\tilde{\rho}}|=\sqrt{\smash[b]{\text{Re}(\hat{\tilde{\rho}})^2+\text{Im}(\hat{\tilde{\rho}})^2}}$, and~Fig.$\,$\ref{fig2} the corresponding spectral solution
$|\hat{\rho}|=\sqrt{\smash[b]{\text{Re}(\hat{\rho})^2+\text{Im}(\hat{\rho})^2}}$ \eqref{171124:2139} in the non-optimal inertial frame, while Fig.$\,$\ref{fig3} and Fig.$\,$\ref{fig4} their associated real-valued solutions $\text{Re}(\tilde{\rho})$ \eqref{171124:2333} and $\text{Re}(\rho)$~\eqref{171124:2327} in physical space, respectively.

Visualized in Fig.$\,$\ref{fig1} \& Fig.$\,$\ref{fig2} are contour plots of $|\hat{\tilde{\rho}}|$ and $|\hat{\rho}|$ for different times $\tilde{t}=t$ relative to the critical time $\tilde{t}^*=t^*$, which, as a frame invariant, is defined as $t^*=k_{y0}/(Ak_{x0})$\footnote[2]{For the specified parameters as they are set on p.$\,$\pageref{171126:0950} for the considered initial condition \eqref{171123:1509}-\eqref{171123:1259}, the explicit value of the critical time is $t^*\sim 8.5$.}, while\linebreak[4] Fig.$\,$\ref{fig3} \& Fig.$\,$\ref{fig4} are density plots of $\text{Re}(\tilde{\rho})$ and $\text{Re}(\rho)$ synchronically corresponding to their complementary spectral fields of Fig.$\,$\ref{fig1} \& Fig.$\,$\ref{fig2}, respectively. The grey shading in these density plots is
coded as follows: White gives the lowest negative value, while black gives the highest positive value. Zero is
then shaded as the grey tone in-between.

Unlike the spectral solutions $\hat{\tilde{\rho}}$ \eqref{171124:1828} and $\hat{\rho}$ \eqref{171124:2139}, the associated physical-space solutions $\tilde{\rho}$ \eqref{171124:2333} and $\rho$ \eqref{171124:2327}
cannot be solved or evaluated in an analytically closed form. Hence, in contrast to the analytical solutions shown in Fig.$\,$\ref{fig1} \& Fig.$\,$\ref{fig2}, only numerical results are shown in Fig.$\,$\ref{fig3} \& Fig.$\,$\ref{fig4}. For that, the defining 2-D Fourier integral has been solved numerically, by first discretizing it and then using the algorithm of a fast Fourier transform (FFT). In physical space a computational box of $(2\times 300)\times (2\times 300)=600\times 600$ with $2^{10}=1024$ sampling points was chosen, which proved to be sufficient enough to present the density field free from numerical artefacts (which inevitably
emerges from the finiteness of the underlying computational box, if it would be chosen too small in its extent). Visualized in Fig.$\,$\ref{fig3} \& Fig.$\,$\ref{fig4} is the density field only in the domain
$(2\times 150)\times (2\times 150)=300\times 300$, half the size of the computational box, in order to compare with the corresponding Fig.$\,$[6.7] in \cite{Hau16}.

To note here is that Fig.$\,$\ref{fig2} should correspond to Fig.$\,$[6.6] and Fig.$\,$\ref{fig4} to Fig.$\,$[6.7] in \cite{Hau16}, since the same initial conditions \eqref{171123:1509}-\eqref{171123:1259} have been employed as in \cite{Hau16}. Although exactly the same conditions and parameters have been used as in \cite{Hau16}, only the former Fig.$\,$[6.6] can be reproduced, the latter Fig.$\,$[6.7], however, not. In particular, that the splitting of the wave packet in physical space only occurs after the critical time $t=t^*$ has been reached, cannot be confirmed. This is surprising, because the initial conditions for the wave packet in\linebreak[4]\cite{Hau16} were exactly constructed to exhibit this particular property of spatial confinement, but in reality does not show it --- the error already lies in the idea that the group velocity of the wave packet in physical space can be controlled by the proposed spectral field decomposition as done in Appendix~[D] in \cite{Hau16}. Since also the lower Mach number case Fig.$\,$[6.5] in\linebreak[4] \cite{Hau16} cannot be reproduced, which should illustrate more vividly the property of wave over-reflection and the existence of a ``critical layer"\footnote[2]{Important to note here is that one of the main objectives in \cite{Hau16} is to disprove the existence of a critical layer as being a layer ``that is responsible for the `absorption of sound' (Campos et al. 1999)" and that its current support is actually based on a misconception that ``might be compared with the prediction of infinite pressure when approaching the so-called sound barrier and the Prandtl-Glauert singularity" [p.$\,$123]. According to \cite{Hau16}, this misconception is due to ``using a non-optimal [$\mathfrak{g}_1$-modal] approach", while when using the ``non-modal" $\mathfrak{g}_2$-approach, it allows to resolve this misconception and ``to comprehend the phenomena of the wave over-reflection and the `critical layer'" [p.$\,$123]. Unfortunately, however, it is \cite{Hau16} who misunderstood the results obtained by \cite{Campos99} regarding the existence of a critical layer for sound propagation in linear shear flow. For details on this issue, please see the discussion at the end of this main section on pp.$\,$\pageref{171201:0725}-\pageref{171205:2313}.} for acoustic propagation in shear flow, the overall conclusions given in \cite{Hau16} to this regard do not match when referring to the corresponding and correct spectral results shown in Fig.$\,$[6.4].

The difference between the optimal Kelvin frame solutions (Fig.$\,$\ref{fig1} \& Fig.$\,$\ref{fig3}) and the non-optimal inertial frame solutions (Fig.$\,$\ref{fig2} \& Fig.$\,$\ref{fig4}), is that the latter solutions show an irrelevant motion which is not needed to understand the dynamics of wave propagation in linear shear flow: The inertial spectral frame solution (Fig.$\,$\ref{fig2}) shows an irrelevant drift of the wave packet along the $k_y$-axis and its associated physical-space solution (Fig.$\,$\ref{fig4}) a corresponding irrelevant clockwise rotation about the origin. This irrelevant motion is absent in the optimal Kelvin frame (Fig.$\,$\ref{fig1} \& Fig.$\,$\ref{fig3}). In particular for the spectral wave-packet (Fig.$\,$\ref{fig1}), not only its position, but also its geometrical structure and its size (spectral extent) stay unchanged as time evolves, which is obviously not the case for the inertial frame (Fig.$\,$\ref{fig2}).

Hence, what only matters to understand the dynamics of the considered system \eqref{171123:0928} is fully decoded in the optimal Kelvin frame, which for pure wave propagation under a high Mach number is shown, for example, in Fig.$\,$\ref{fig3}: A critical time $t^*$ exists, where before this time is reached~($\tilde{t}< t^*$), the wave packet undergoes a split into two equal sized packets moving in opposite directions, while after the critical time has passed ($\tilde{t}> t^*$), the two wave packets start to get sheared apart. Of course, these key dynamical processes also exist and can be seen in the non-optimal inertial frame Fig.$\,$\ref{fig4}, however, they appear less clear.

It is evident that this remarkable difference between the observed dynamics in the inertial and Kelvin frame will also exist when considering wave propagation induced by initial pure vortex mode perturbations as particularly discussed in \cite{Hau15}. Both in spectral as well as in physical space the accelerated Kelvin frame is the optimal frame, since here the internal dynamics of the wave packet is {\it not} superimposed and thus obscured by some extra, non-relevant motion --- only the relevant dynamics is exposed in this frame. The drift in spectral and the rotation in physical space, however, are relative phenomena which can be transformed away, and hence are ultimately irrelevant for the understanding of the internal dynamics of the system.

A small but decisive remark at the end: All solutions shown here so far in Fig.$\,$\ref{fig1}-\ref{fig4} are not truly physical, simply because the considered initial condition given in spectral space \eqref{171123:1354} violates the reality constraint of the Fourier transform \eqref{171123:1509}. This constraint on each of the three spectral~fields\footnote[2]{The *-symbol in \eqref{171127:1622} denotes complex conjugation.}
\begin{equation}
\hat{\psi}^{*\{u,v,\rho\}}(t,k_x,k_y)=\hat{\psi}^{\{u,v,\rho\}}(t,-k_x,-k_y),\;\;\forall t\geq 0,\label{171127:1622}
\end{equation}
which should hold for all times, in particular thus also for the initial time $t=0$, guarantees that when transforming the spectral fields back to original physical space that they are real valued.\linebreak[4] The initial condition \eqref{171123:1354}, however, does not satisfy this constraint --- the problem lies not in the wave-packet $\Psi^{(+)}$ itself, but in the factors attached to it. Hence, when starting the spectral dynamics with
this particular non-real ($\mathbb{C}$) initial condition \eqref{171123:1354}, it will ultimately lead to $\mathbb{C}$-solutions in physical space, which of course is unphysical.

Unfortunately, this crucial constraint \eqref{171127:1622} has not been taken into account by \cite{Hau16}. This marks a further problem in \cite{Hau16}, which is independent of the inconsistency problem mentioned above, namely that the (correct) spectral solutions shown in Fig.$\,$[6.4] \& Fig.$\,$[6.6] do~not match with their associated physical-space solutions in Fig.$\,$[6.5] \& Fig.$\,$[6.7], respectively. Because, when correctly adjusting the initial condition \eqref{171123:1354} to satisfy the reality constraint \eqref{171127:1622}, as will be done in the next section, this mismatch between spectral and physical-space solutions in \cite{Hau16} is not resolved and remains, irrespective of how the initial condition~\eqref{171123:1354} will be adjusted.

%
% Non-physical Solution %%%%%%%%%%%%%%%%%%%%%%%%%%%%%%%%%%%%%%%%%%%%%%%%%%%%%%%%%%%%%%%%%%%%%%%%%%%%%%%%%%%%%%%%%%%%%%%%%%%%%%%
%
\newpage

\newpage

\phantom{x}\newpage

\newpage
%
%%%%%%%%%%%%%%%%%%%%%%%%%%%%%%%%%%%%%%%%%%%%%%%%%%%%%%%%%%%%%%%%%%%%%%%%%%%%%%%%%%%%%%%%%%%%%%%%%%%%%%%%%
%

\subsection{Visualization of a physical solution\label{SC3}}

\begin{figure}[h!]
\begin{subfigure}[t]{.24\textwidth}
\FigureXYLabel{\includegraphics[type=pdf,ext=.pdf,read=.pdf,width=0.96\textwidth]{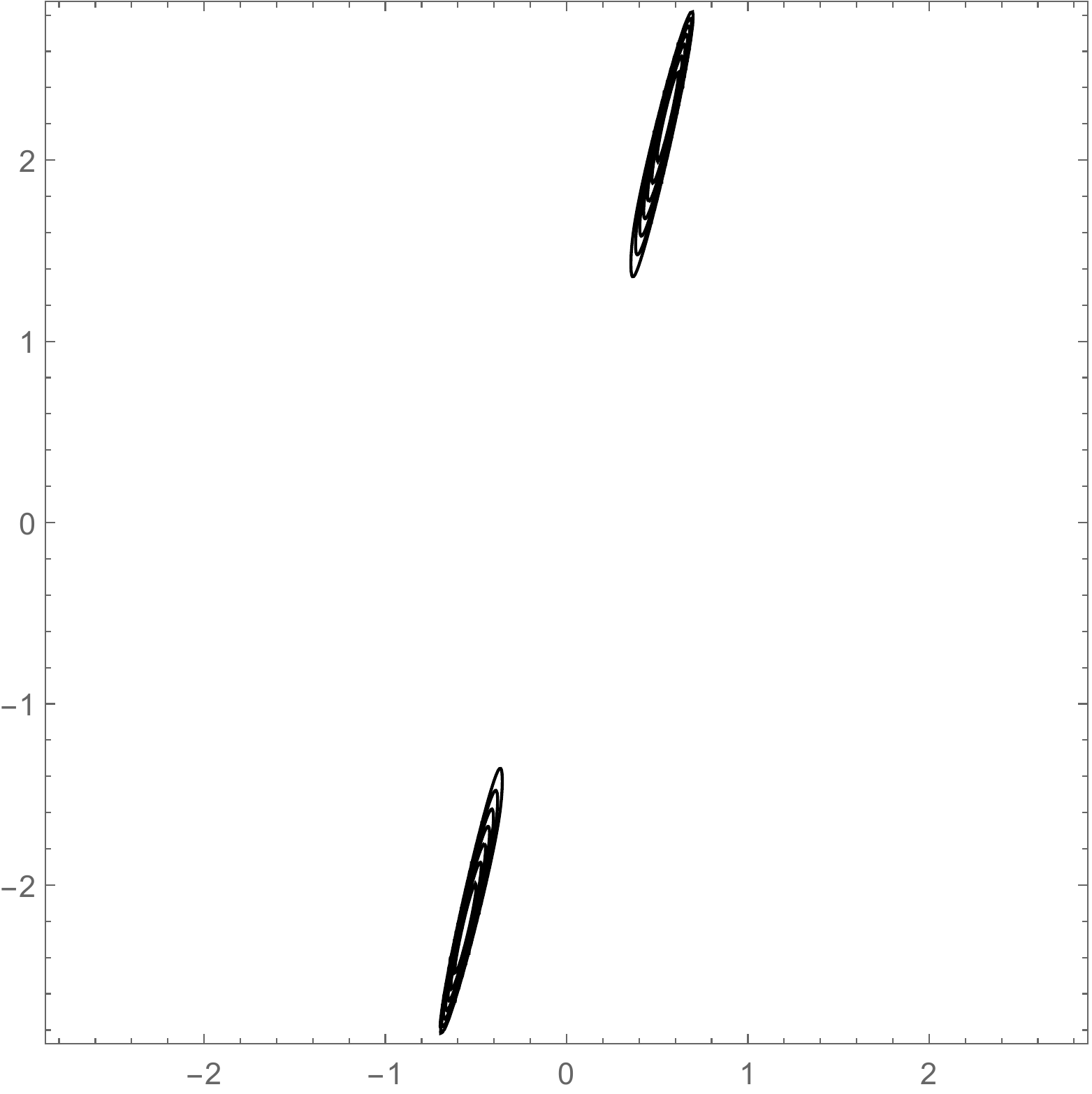}}
{${\scriptstyle\hspace{0.25cm} k_{\tilde{x}}}$}{-1mm}{\begin{rotate}{0}
\end{rotate}}{0mm}
\vspace*{-1.25em}\subcaption*{{\scriptsize (a.1) $\tilde{t}=0$}}\vspace*{1.25em}
\end{subfigure}
\hspace*{-.2cm}
\begin{subfigure}[t]{.24\textwidth}
\FigureXYLabel{\includegraphics[type=pdf,ext=.pdf,read=.pdf,width=0.96\textwidth]{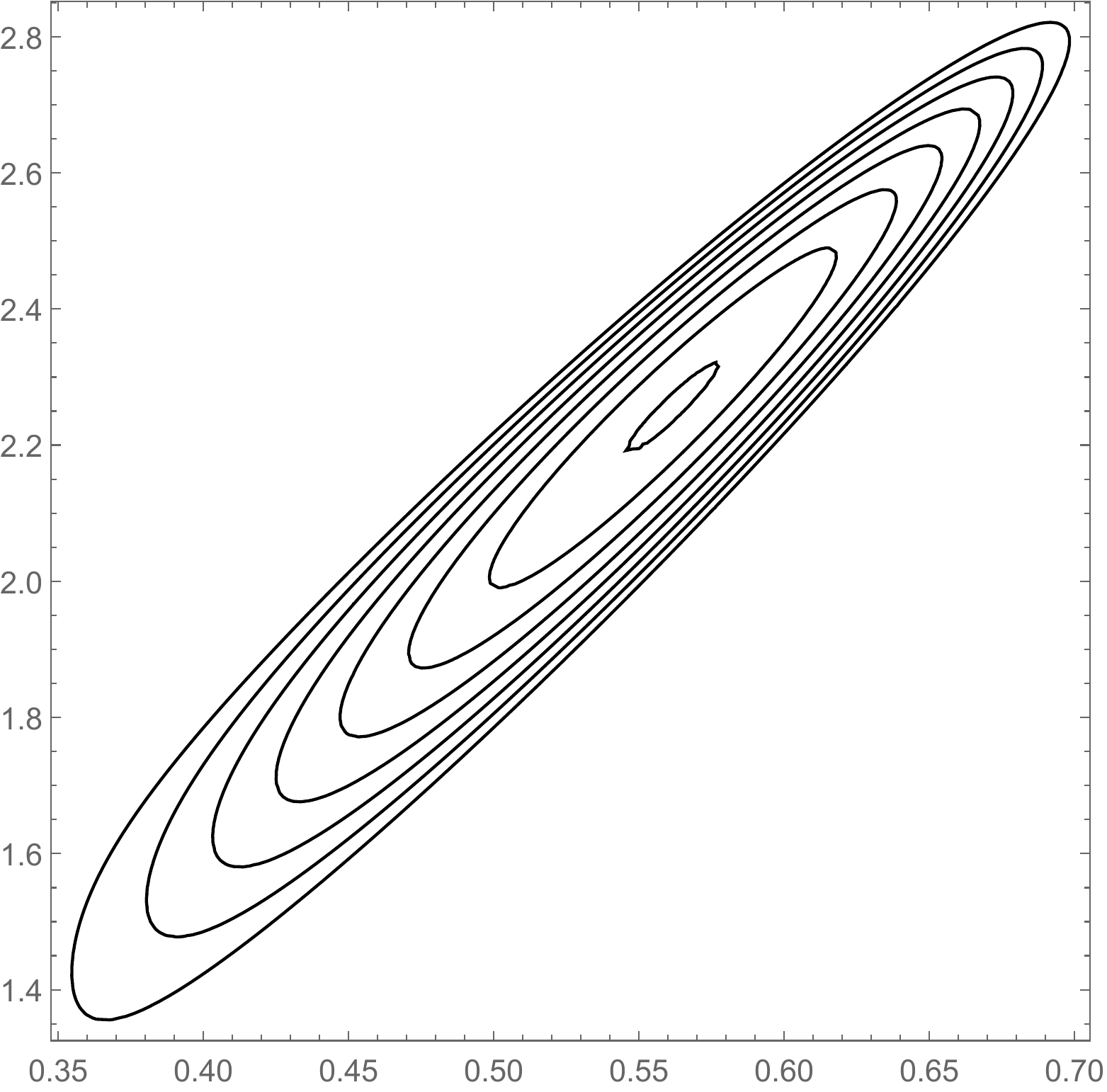}}
{${\scriptstyle\hspace{0.25cm} k_{\tilde{x}}}$}{-1mm}{\begin{rotate}{0}
\end{rotate}}{0mm}
\vspace*{-1.25em}\subcaption*{{\scriptsize (a.2) $\tilde{t}=0$, zoom}}\vspace*{1.25em}
\end{subfigure}
\begin{subfigure}[c]{0.0001\textwidth}
\vspace*{-3.5cm}${\scriptstyle \! k_{\tilde{y}}}$
\end{subfigure}
\begin{subfigure}[t]{.24\textwidth}
\FigureXYLabel{\includegraphics[type=pdf,ext=.pdf,read=.pdf,width=0.96\textwidth]{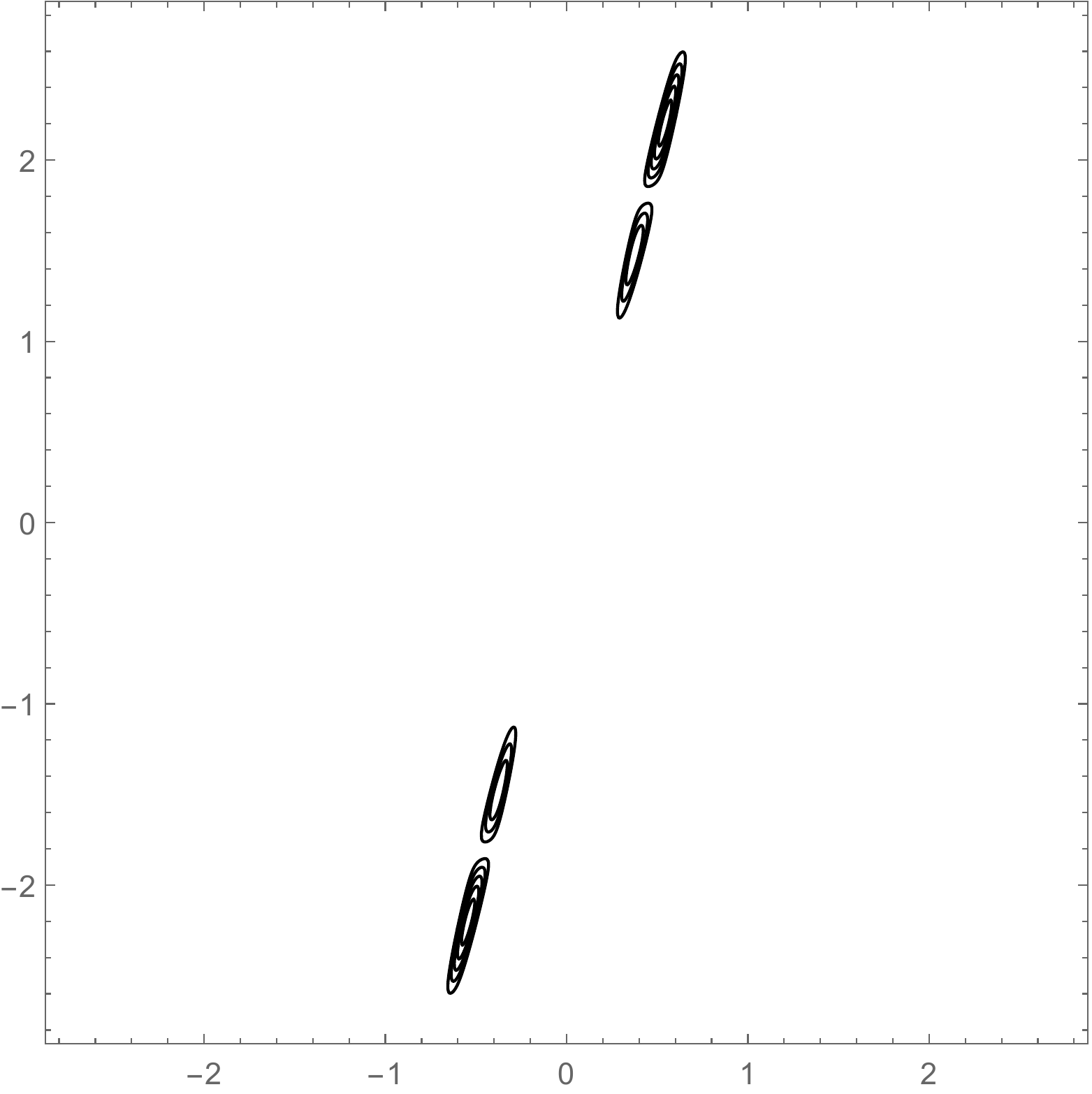}}
{${\scriptstyle\hspace{0.25cm} k_{\tilde{x}}}$}{-1mm}{\begin{rotate}{0}
\end{rotate}}{0mm}
\vspace*{-1.25em}\subcaption*{{\scriptsize (b.1) $\tilde{t}=1/2\tilde{t}^*$}}\vspace*{1.25em}
\end{subfigure}
\hspace*{-.2cm}
\begin{subfigure}[t]{.24\textwidth}
\FigureXYLabel{\includegraphics[type=pdf,ext=.pdf,read=.pdf,width=0.96\textwidth]{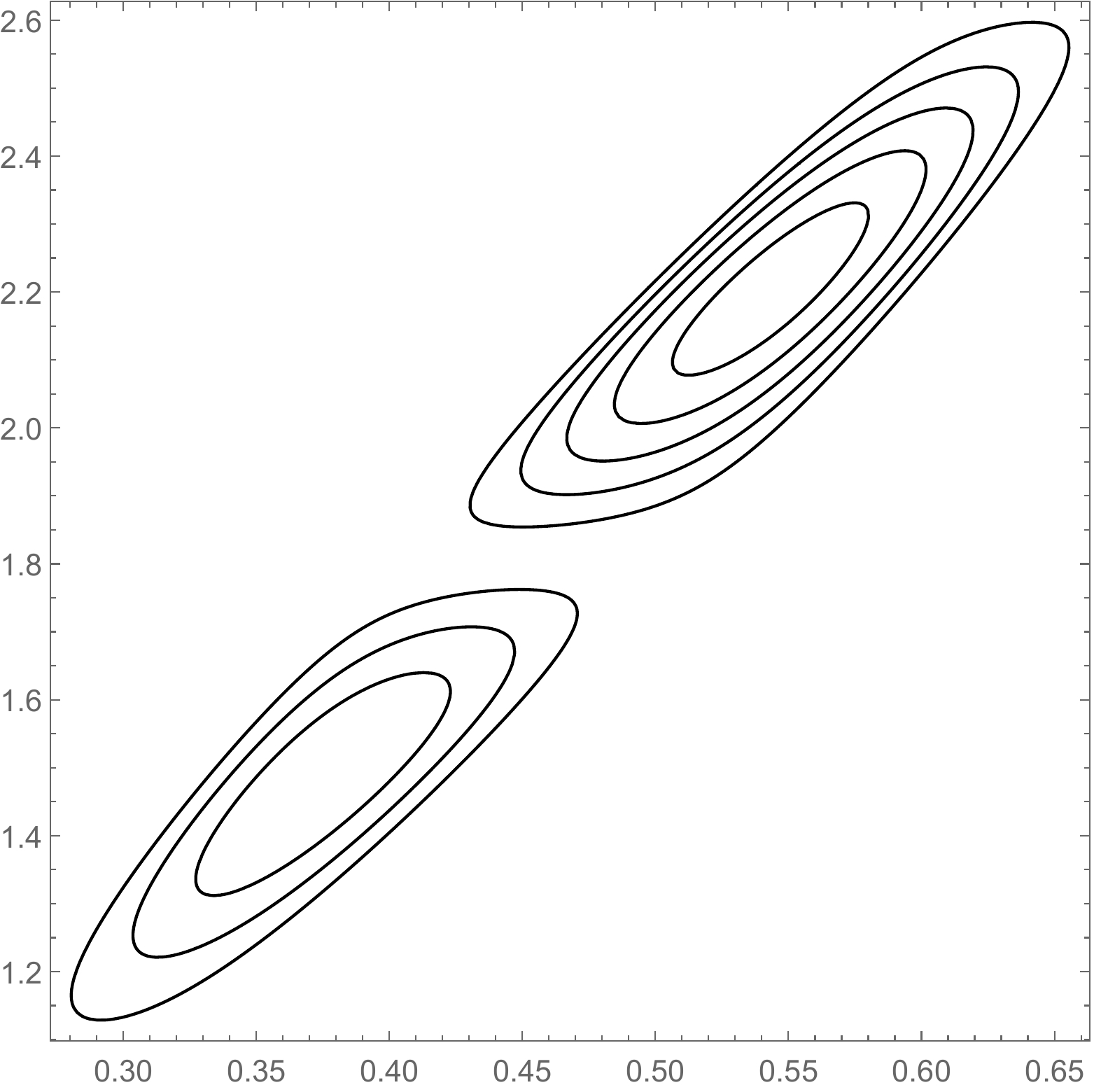}}
{${\scriptstyle\hspace{0.25cm} k_{\tilde{x}}}$}{-1mm}{\begin{rotate}{0}
\end{rotate}}{0mm}
\vspace*{-1.25em}\subcaption*{{\scriptsize (b.2) $\tilde{t}=1/2\tilde{t}^*$, zoom}}\vspace*{1.25em}
\end{subfigure}
\begin{subfigure}[t]{.24\textwidth}
\FigureXYLabel{\includegraphics[type=pdf,ext=.pdf,read=.pdf,width=0.96\textwidth]{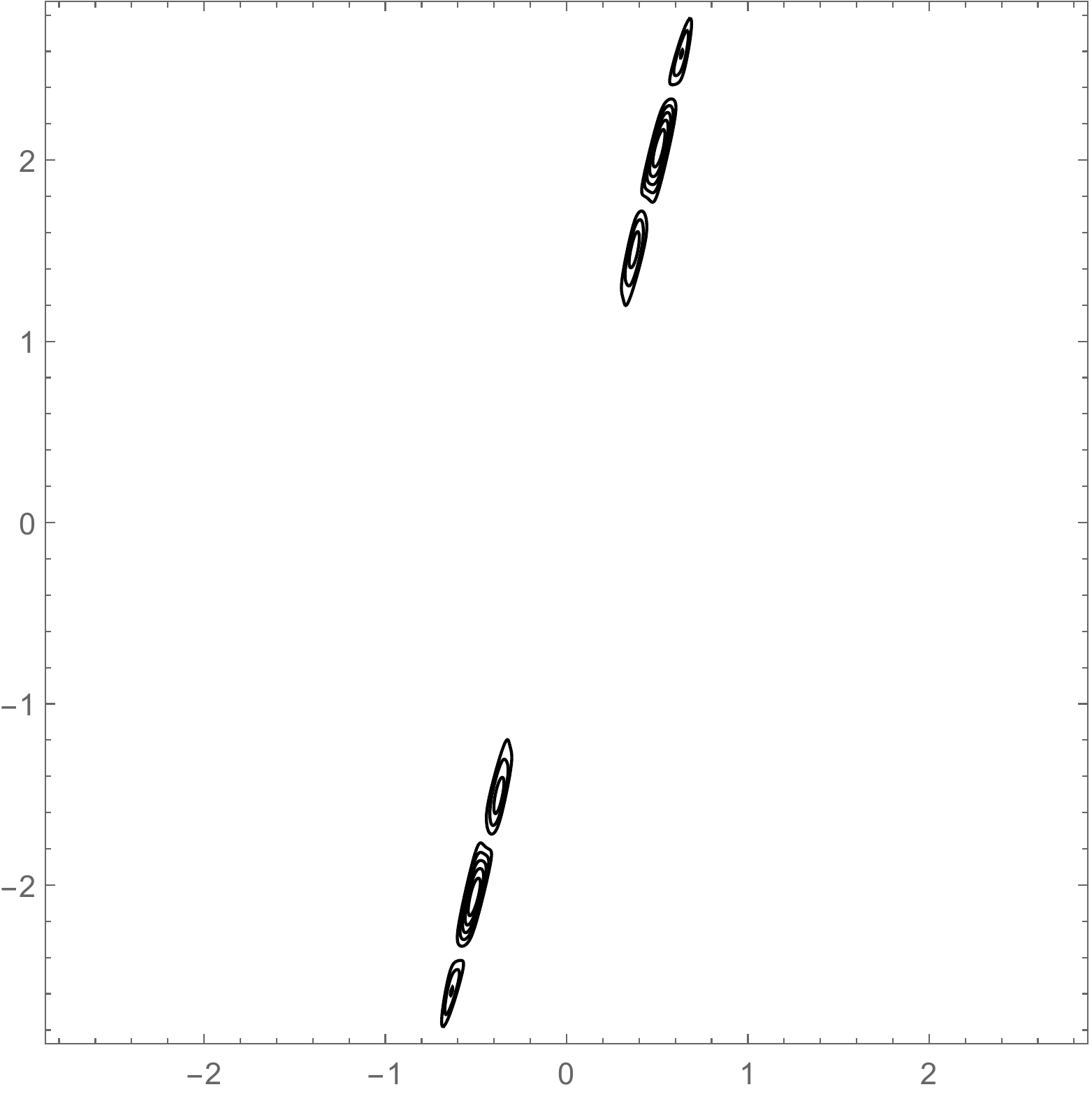}}
{${\scriptstyle\hspace{0.25cm} k_{\tilde{x}}}$}{-1mm}{\begin{rotate}{0}
\end{rotate}}{0mm}
\vspace*{-1.25em}\subcaption*{{\scriptsize (c.1) $\tilde{t}=\tilde{t}^*$}}\vspace*{1.25em}
\end{subfigure}
\hspace*{-.2cm}
\begin{subfigure}[t]{.24\textwidth}
\FigureXYLabel{\includegraphics[type=pdf,ext=.pdf,read=.pdf,width=0.96\textwidth]{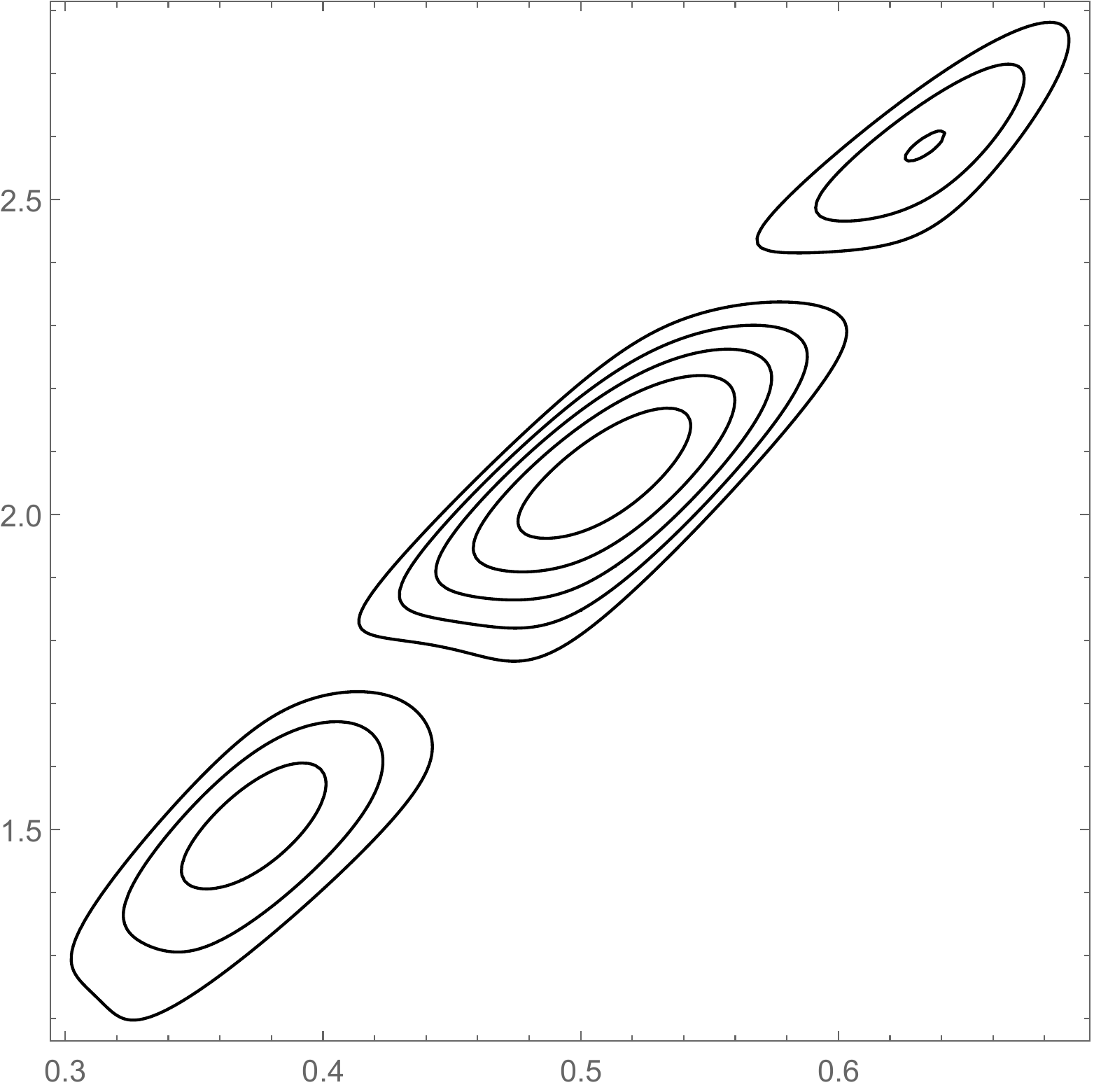}}
{${\scriptstyle\hspace{0.25cm} k_{\tilde{x}}}$}{-1mm}{\begin{rotate}{0}
\end{rotate}}{0mm}
\vspace*{-1.25em}\subcaption*{{\scriptsize (c.2) $\tilde{t}=\tilde{t}^*$, zoom}}\vspace*{1.25em}
\end{subfigure}
\begin{subfigure}[t]{0.0001\textwidth}
\vspace*{-2.0cm}${\scriptstyle \! k_{\tilde{y}}}$
\end{subfigure}
\begin{subfigure}[t]{.24\textwidth}
\FigureXYLabel{\includegraphics[type=pdf,ext=.pdf,read=.pdf,width=0.96\textwidth]{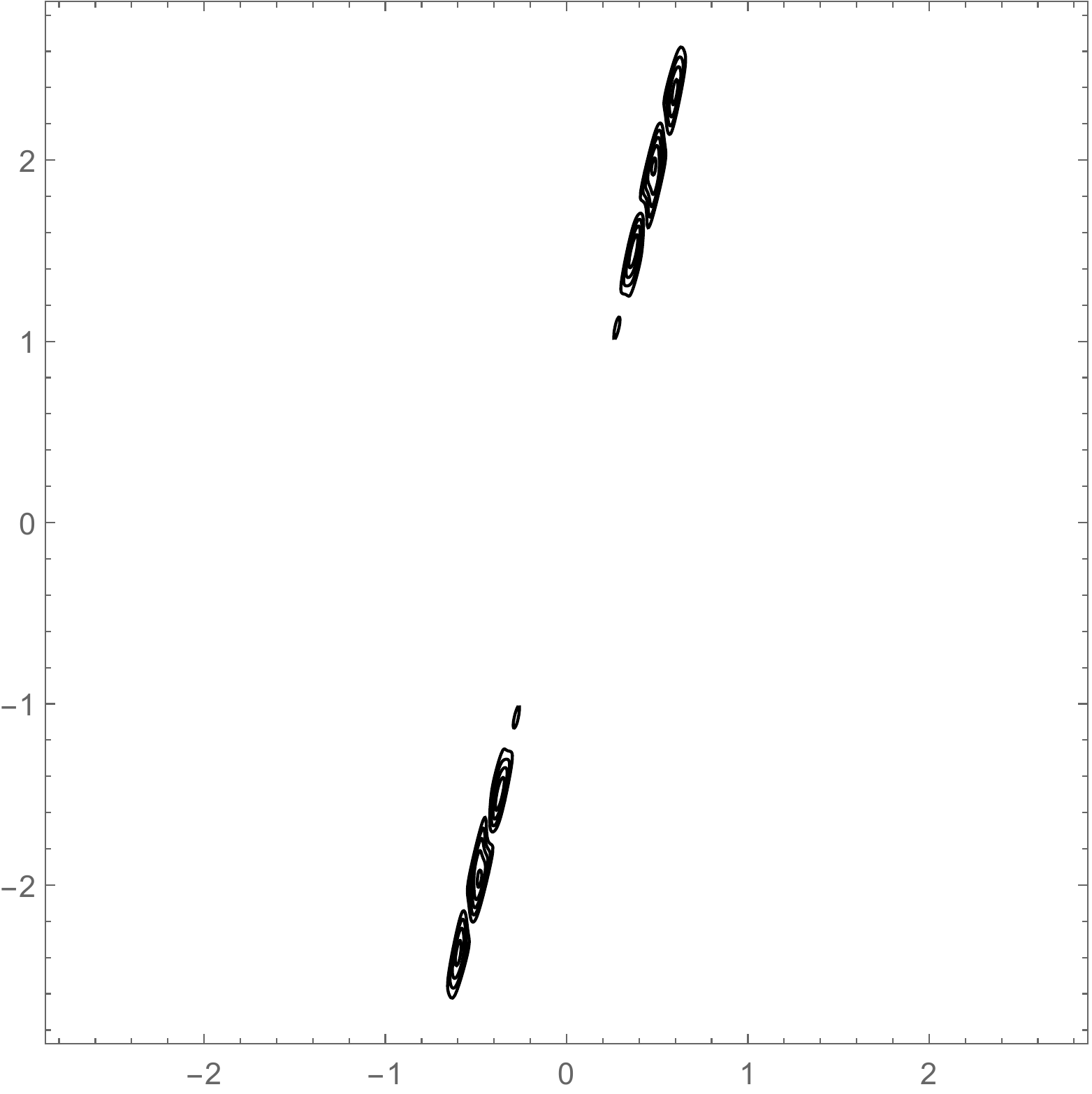}}
{${\scriptstyle\hspace{0.25cm} k_{\tilde{x}}}$}{-1mm}{\begin{rotate}{0}
\end{rotate}}{0mm}
\vspace*{-1.25em}\subcaption*{{\scriptsize (d.1) $\tilde{t}=3/2\tilde{t}^*$}}\vspace*{1.25em}
\end{subfigure}
\hspace*{-.2cm}
\begin{subfigure}[t]{.24\textwidth}
\FigureXYLabel{\includegraphics[type=pdf,ext=.pdf,read=.pdf,width=0.96\textwidth]{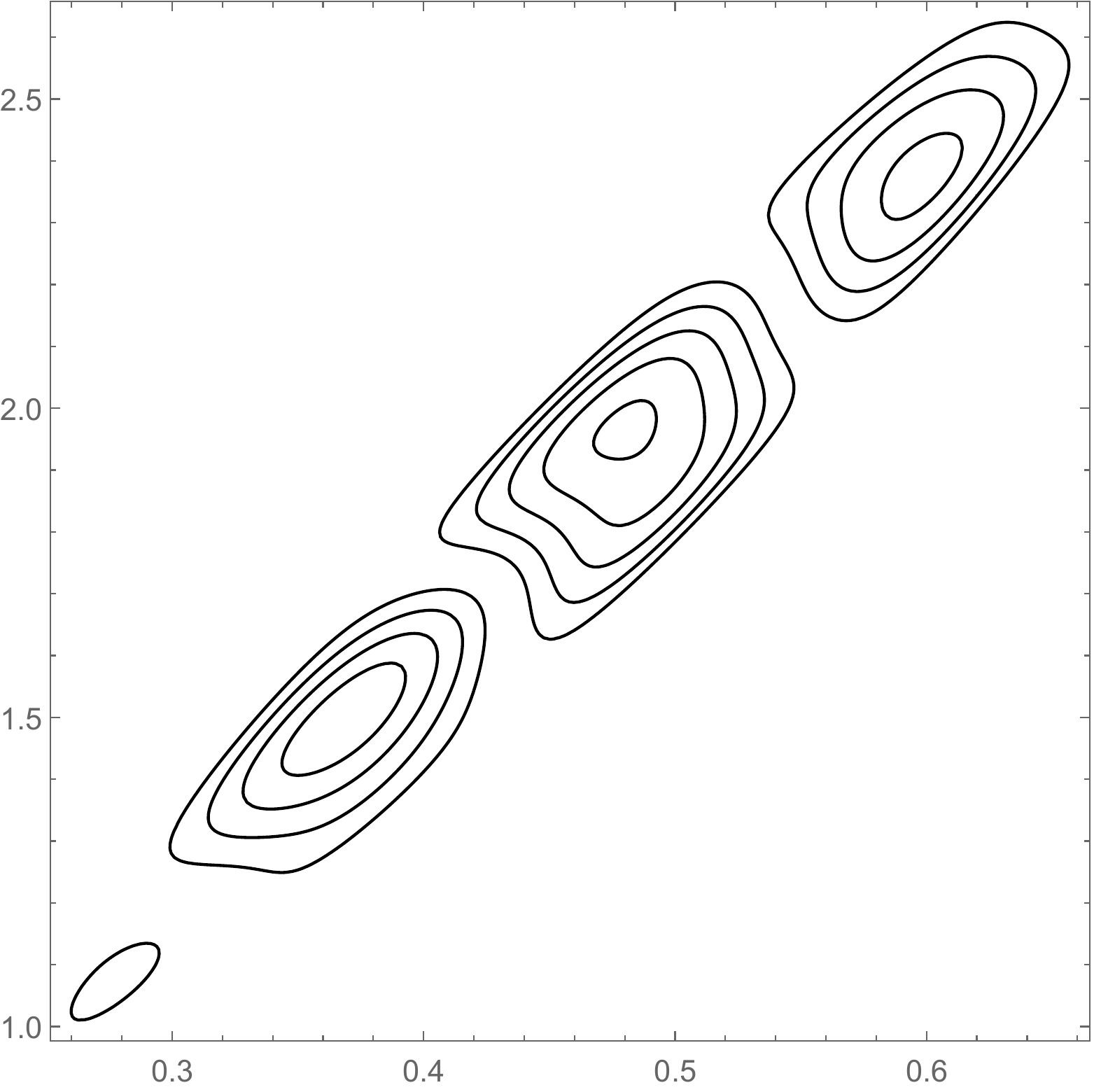}}
{${\scriptstyle\hspace{0.25cm} k_{\tilde{x}}}$}{-1mm}{\begin{rotate}{0}
\end{rotate}}{0mm}
\vspace*{-1.25em}\subcaption*{{\scriptsize (d.2) $\tilde{t}=3/2\tilde{t}^*$, zoom}}\vspace*{1.25em}
\end{subfigure}
\begin{subfigure}[t]{.24\textwidth}
\FigureXYLabel{\includegraphics[type=pdf,ext=.pdf,read=.pdf,width=0.96\textwidth]{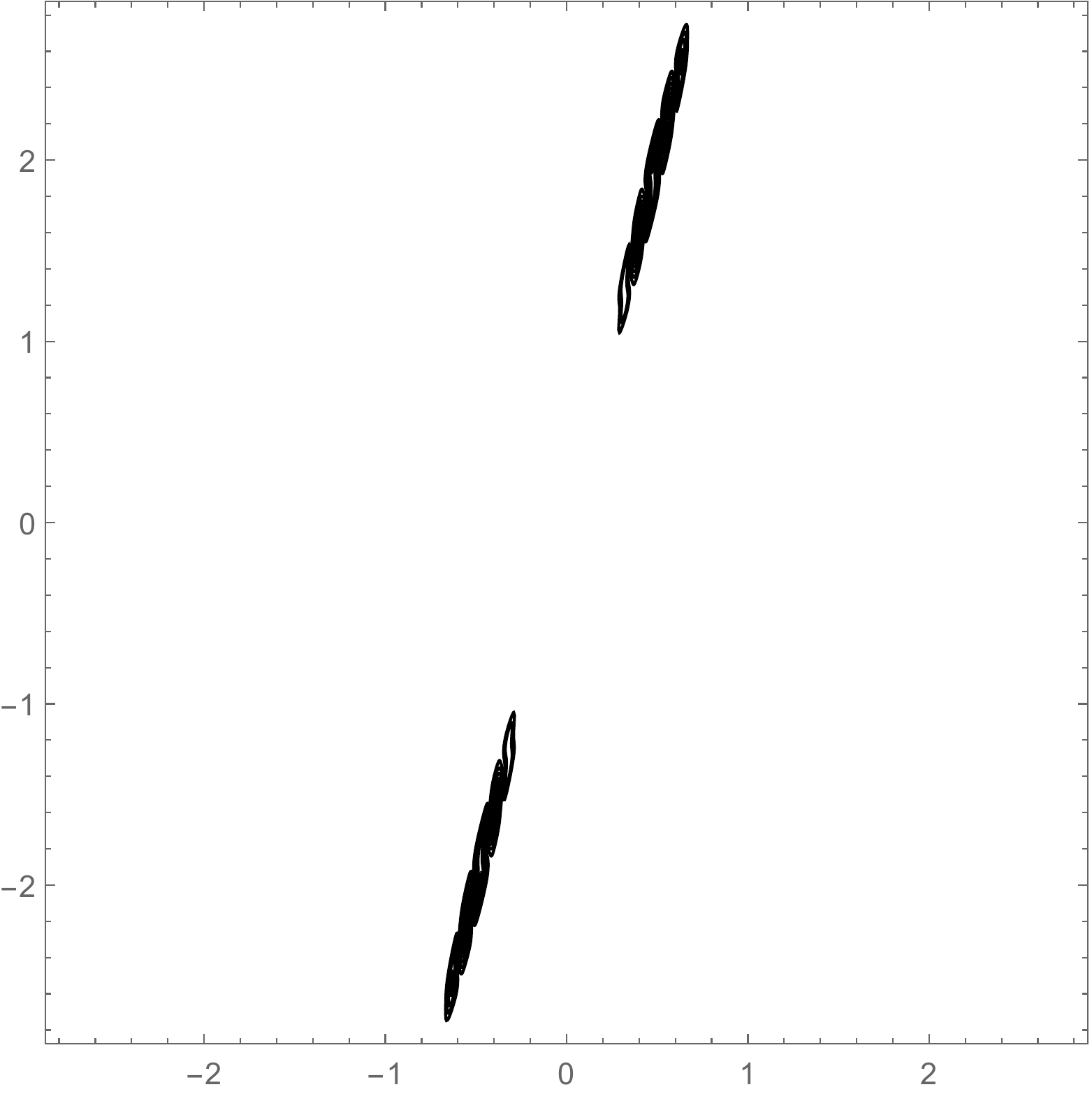}}
{${\scriptstyle\hspace{0.25cm} k_{\tilde{x}}}$}{-1mm}{\begin{rotate}{0}
\end{rotate}}{0mm}
\vspace*{-1.25em}\subcaption*{{\scriptsize (e.1) $\tilde{t}=2\tilde{t}^*$}}\vspace*{1.25em}
\end{subfigure}
\hspace*{-.2cm}
\begin{subfigure}[t]{.24\textwidth}
\FigureXYLabel{\includegraphics[type=pdf,ext=.pdf,read=.pdf,width=0.96\textwidth]{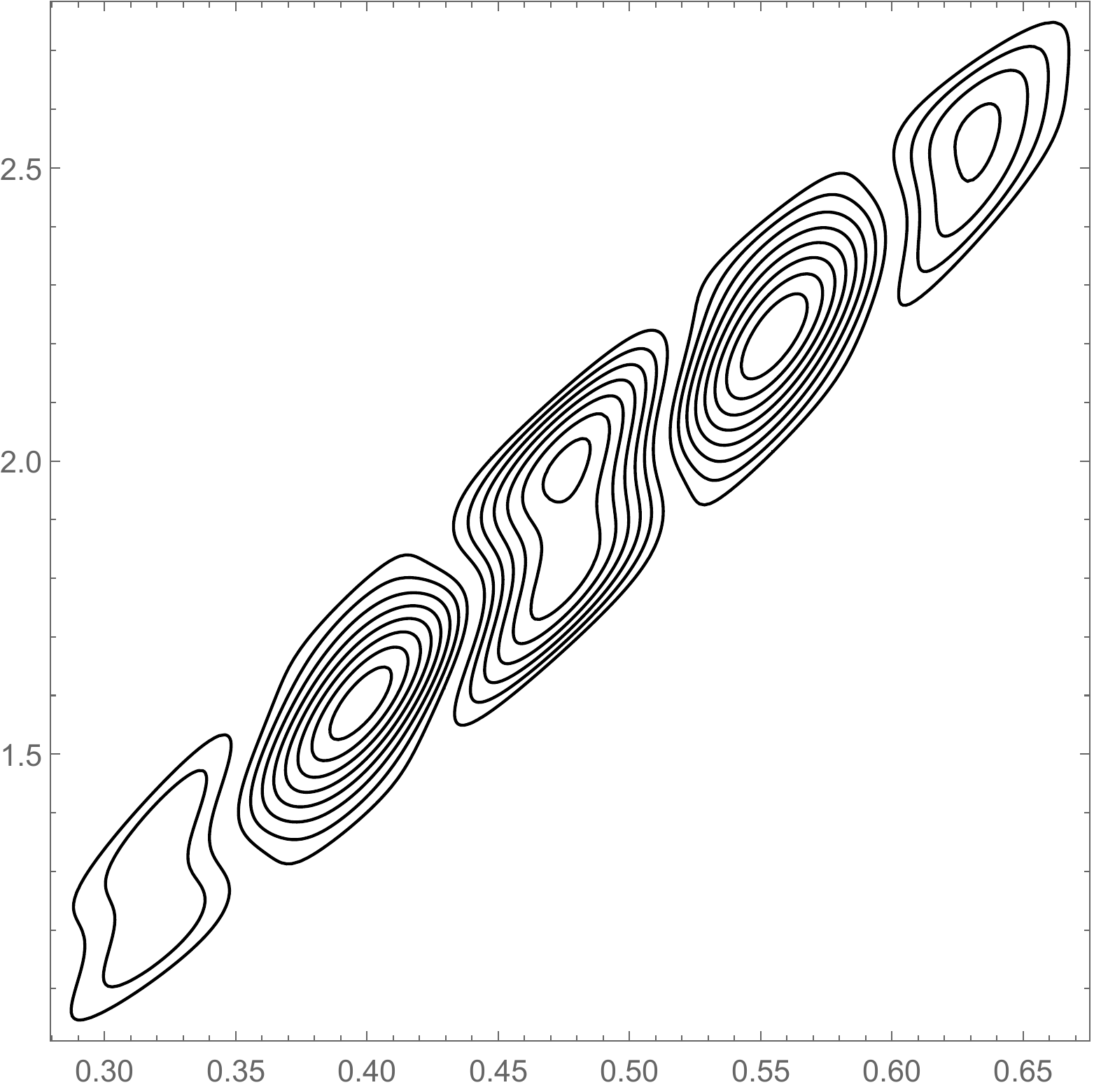}}
{${\scriptstyle\hspace{0.25cm} k_{\tilde{x}}}$}{-1mm}{\begin{rotate}{0}
\end{rotate}}{0mm}
\vspace*{-1.25em}\subcaption*{{\scriptsize (e.2) $\tilde{t}=2\tilde{t}^*$, zoom}}\vspace*{1.25em}
\end{subfigure}
\begin{subfigure}[t]{0.0001\textwidth}
\vspace*{-2.0cm}${\scriptstyle \! k_{\tilde{y}}}$
\end{subfigure}
\begin{subfigure}[t]{.24\textwidth}
\FigureXYLabel{\includegraphics[type=pdf,ext=.pdf,read=.pdf,width=0.96\textwidth]{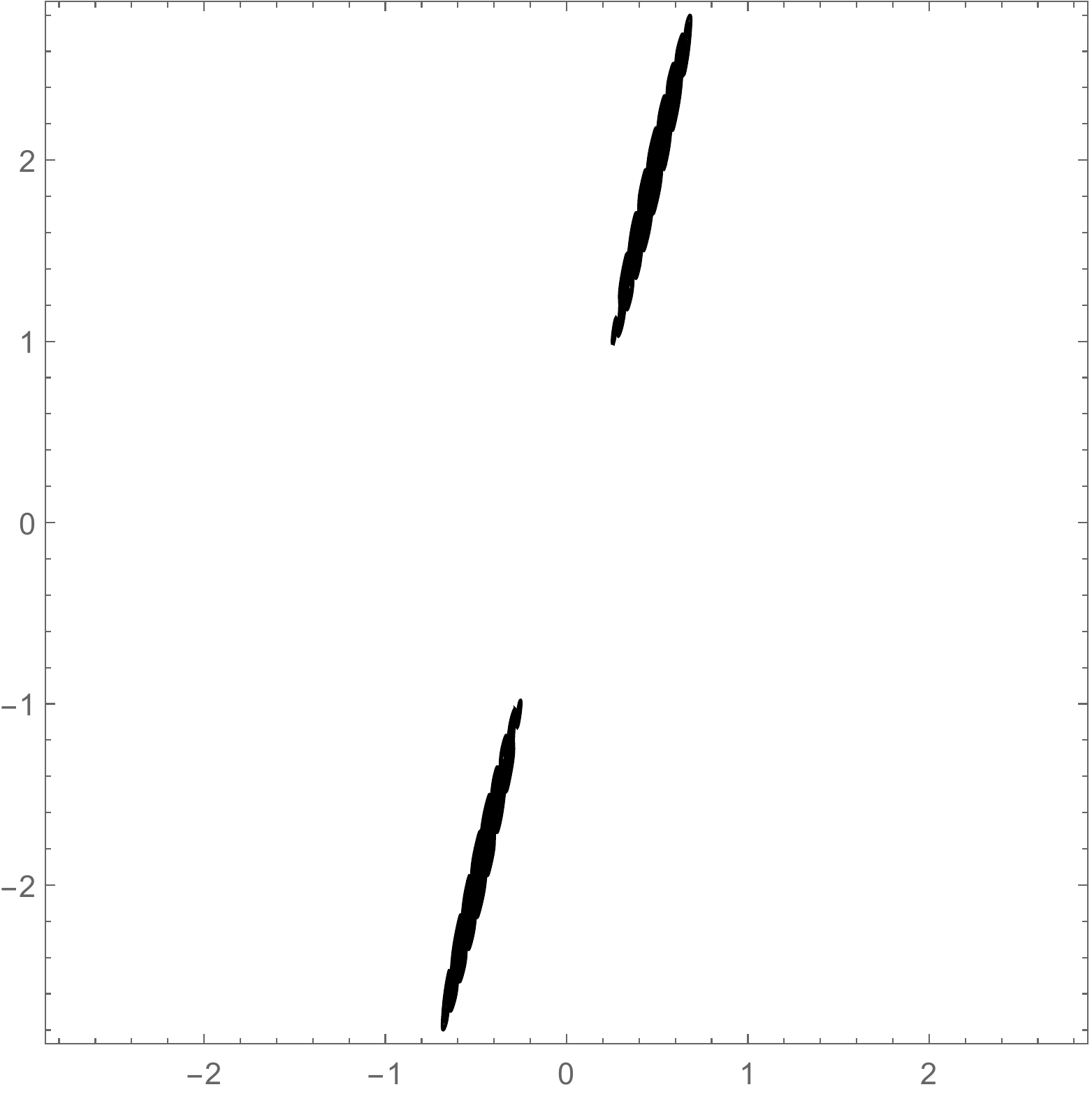}}
{${\scriptstyle\hspace{0.25cm} k_{\tilde{x}}}$}{-1mm}{\begin{rotate}{0}
\end{rotate}}{0mm}
\vspace*{-1.25em}\subcaption*{{\scriptsize (f.1) $\tilde{t}=5/2\tilde{t}^*$}}\vspace*{1.25em}
\end{subfigure}
\hspace*{-.2cm}
\begin{subfigure}[t]{.24\textwidth}
\FigureXYLabel{\includegraphics[type=pdf,ext=.pdf,read=.pdf,width=0.96\textwidth]{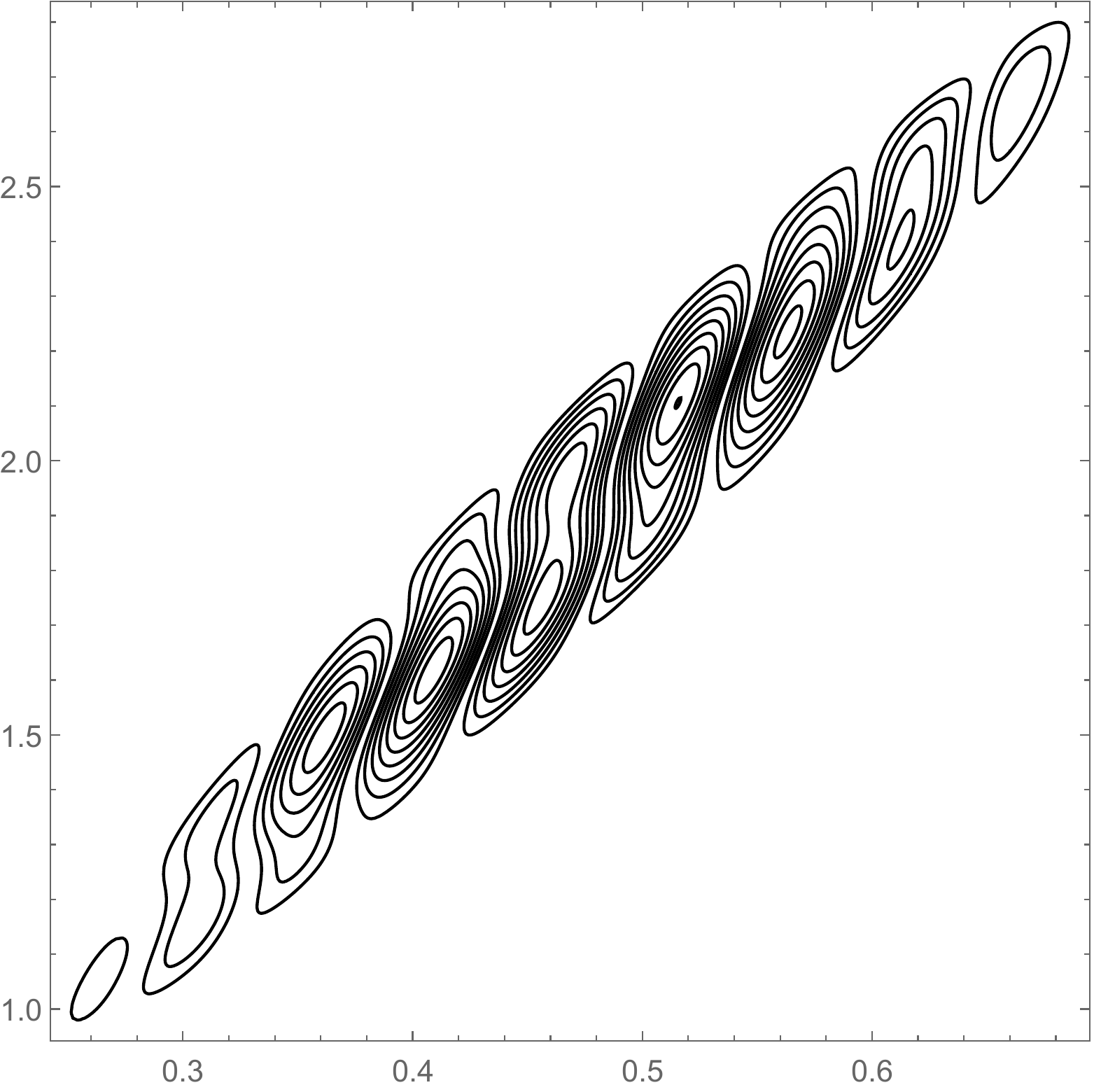}}
{${\scriptstyle\hspace{0.25cm} k_{\tilde{x}}}$}{-1mm}{\begin{rotate}{0}
\end{rotate}}{0mm}
\vspace*{-1.25em}\subcaption*{{\scriptsize (f.2) $\tilde{t}=5/2\tilde{t}^*$, zoom}}\vspace*{1.25em}
\end{subfigure}
\vspace*{-1em}\caption{{\footnotesize Evolution of the physical Kelvin-frame solution $|\hat{\tilde{\rho}}|$ in the spectral $(k_{\tilde{x}},k_{\tilde{y}})$-plane, according to the ODE-solution
\eqref{171124:1828} with the now physical initial condition \eqref{171129:1053}-\eqref{171129:1036}. As before for the non-physical case in Fig.$\,$\ref{fig1}, the wave packet is again not drifting and not diffusing in $k$-space. It is stationary and keeps its expansion size for all times. Inside the wave packet, however, the dynamics is changing again, but distinctively different to the corresponding non-physical case (Fig.$\,$\ref{fig1}): Before the critical time $\tilde{t}^*=t^*=k_{y_0}/(Ak_{x_0})\sim 8.5$ is reached, which again is defined in the center of the packet
when $k_{\tilde{y}}(\tilde{t}):=k_{\tilde{y}}-A\tilde{t}^* k_{\tilde{x}}=0$, no oval symmetry in the contour lines along its major axes is present or maintained. Instead, the initial wave packet rather starts to break apart and continues to break into smaller and smaller ones as time progresses. Hence, the contour lines are no longer connected as it is the case for the non-physical solution (Fig.$\,$\ref{fig1}) resulting from the analysis of \cite{Hau16}.\label{fig5}}}
\end{figure}

\begin{figure}[t!]
\begin{subfigure}[t]{.24\textwidth}
\FigureXYLabel{\includegraphics[type=pdf,ext=.pdf,read=.pdf,width=0.96\textwidth]{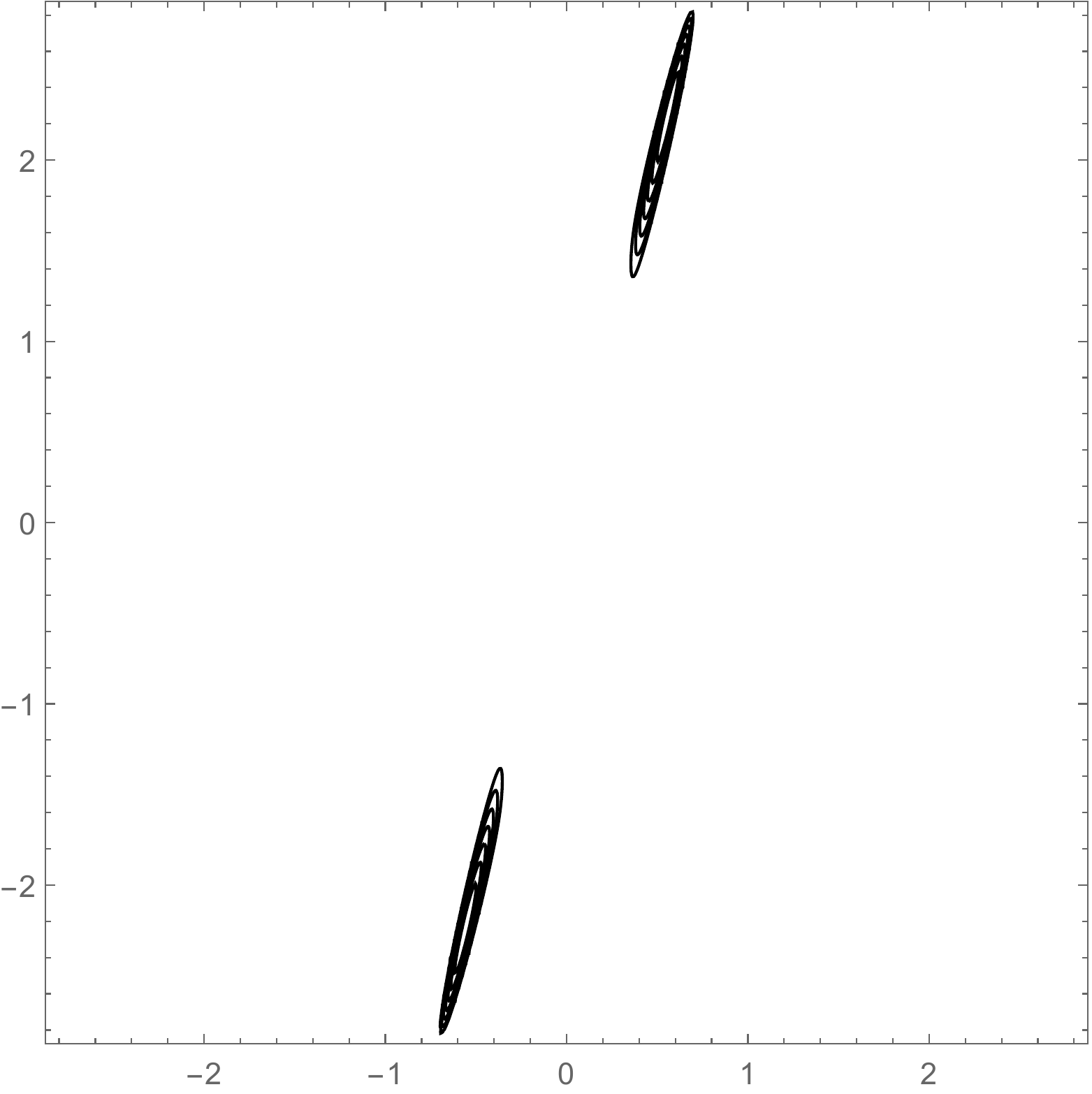}}
{${\scriptstyle\hspace{0.25cm} k_x}$}{-1mm}{\begin{rotate}{0}
\end{rotate}}{0mm}
\vspace*{-1.25em}\subcaption*{{\scriptsize (a.1) $t=0$}}\vspace*{1.25em}
\end{subfigure}
\hspace*{-.2cm}
\begin{subfigure}[t]{.24\textwidth}
\FigureXYLabel{\includegraphics[type=pdf,ext=.pdf,read=.pdf,width=0.96\textwidth]{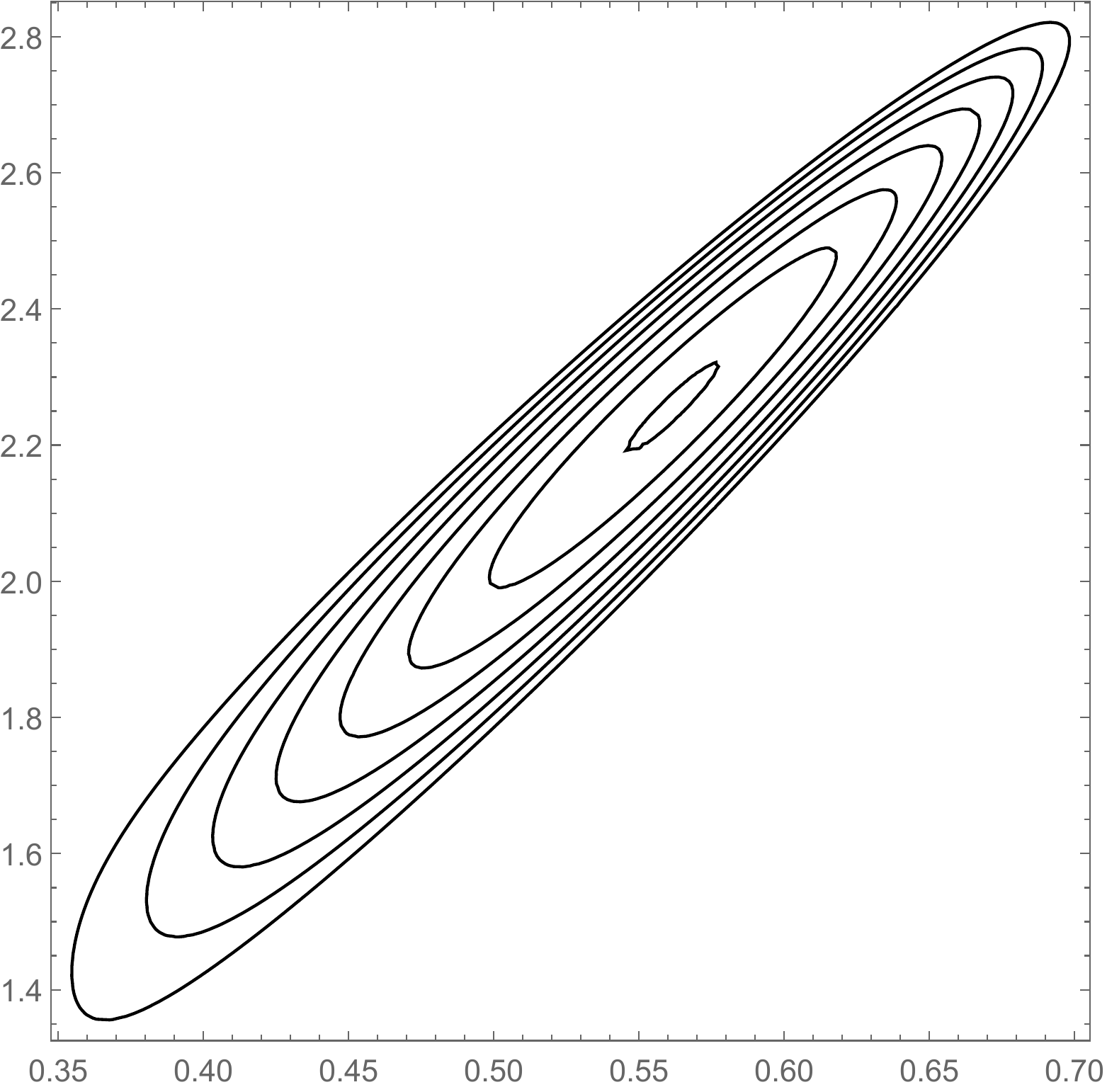}}
{${\scriptstyle\hspace{0.25cm} k_x}$}{-1mm}{\begin{rotate}{0}
\end{rotate}}{0mm}
\vspace*{-1.25em}\subcaption*{{\scriptsize (a.2) $t=0$, zoom}}\vspace*{1.25em}
\end{subfigure}
\begin{subfigure}[c]{0.0001\textwidth}
\vspace*{-3.5cm}${\scriptstyle \! k_y}$
\end{subfigure}
\begin{subfigure}[t]{.24\textwidth}
\FigureXYLabel{\includegraphics[type=pdf,ext=.pdf,read=.pdf,width=0.96\textwidth]{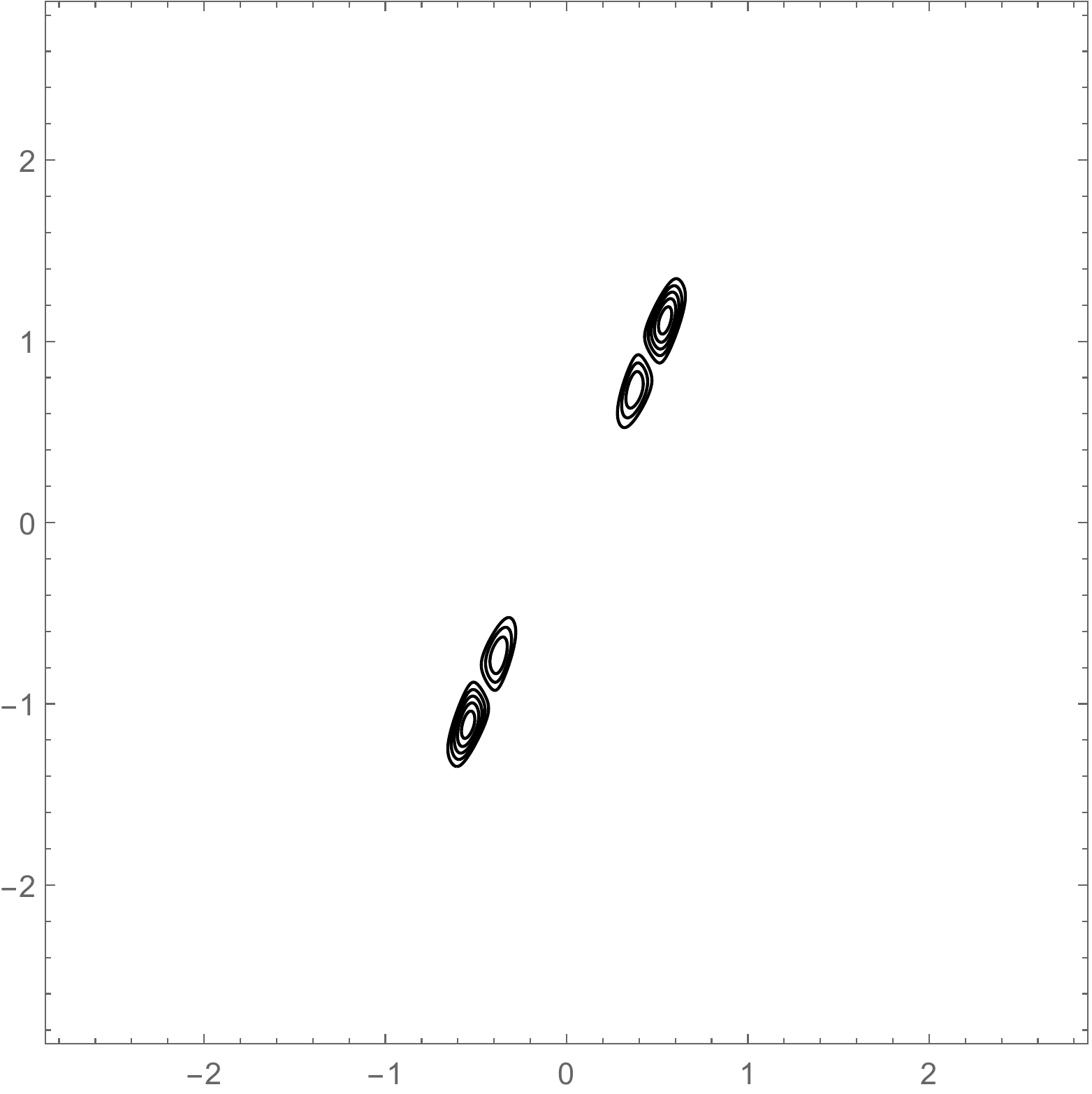}}
{${\scriptstyle\hspace{0.25cm} k_x}$}{-1mm}{\begin{rotate}{0}
\end{rotate}}{0mm}
\vspace*{-1.25em}\subcaption*{{\scriptsize (b.1) $t=1/2t^*$}}\vspace*{1.25em}
\end{subfigure}
\hspace*{-.2cm}
\begin{subfigure}[t]{.24\textwidth}
\FigureXYLabel{\includegraphics[type=pdf,ext=.pdf,read=.pdf,width=0.96\textwidth]{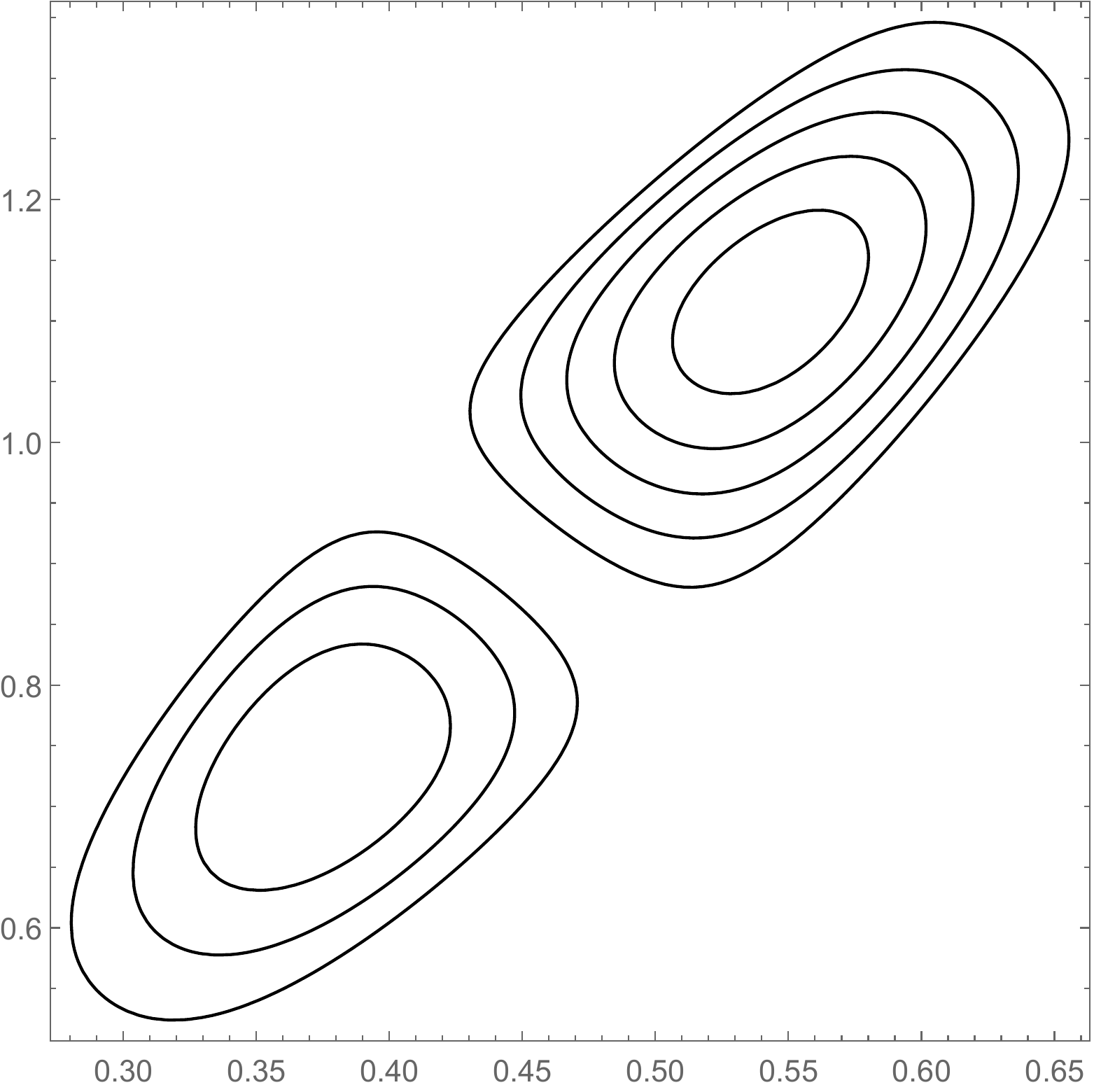}}
{${\scriptstyle\hspace{0.25cm} k_x}$}{-1mm}{\begin{rotate}{0}
\end{rotate}}{0mm}
\vspace*{-1.25em}\subcaption*{{\scriptsize (b.2) $t=1/2t^*$, zoom}}\vspace*{1.25em}
\end{subfigure}
\begin{subfigure}[t]{.24\textwidth}
\FigureXYLabel{\includegraphics[type=pdf,ext=.pdf,read=.pdf,width=0.96\textwidth]{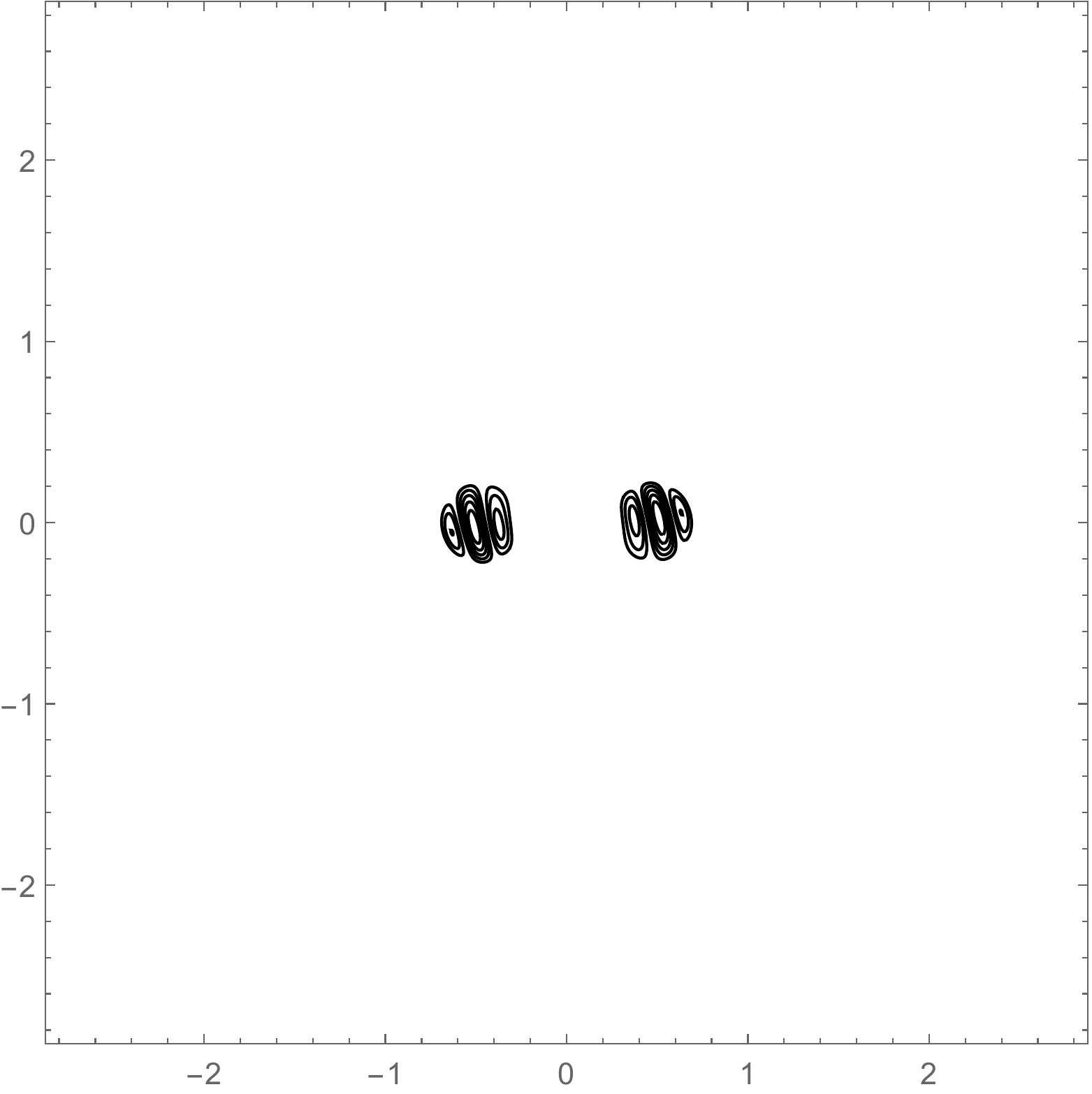}}
{${\scriptstyle\hspace{0.25cm} k_x}$}{-1mm}{\begin{rotate}{0}
\end{rotate}}{0mm}
\vspace*{-1.25em}\subcaption*{{\scriptsize (c.1) $t=t^*$}}\vspace*{1.25em}
\end{subfigure}
\hspace*{-.2cm}
\begin{subfigure}[t]{.24\textwidth}
\FigureXYLabel{\includegraphics[type=pdf,ext=.pdf,read=.pdf,width=0.96\textwidth]{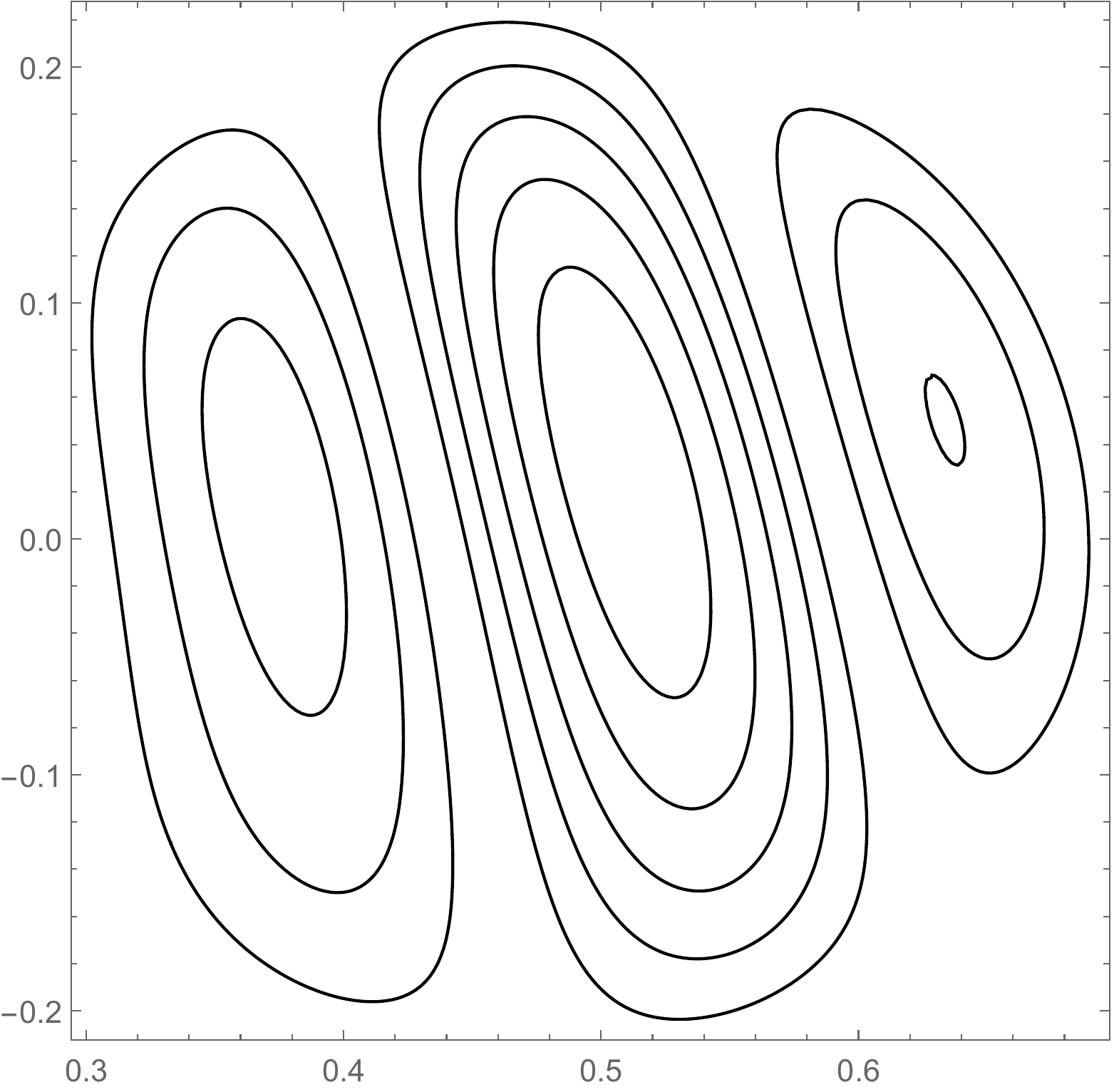}}
{${\scriptstyle\hspace{0.25cm} k_x}$}{-1mm}{\begin{rotate}{0}
\end{rotate}}{0mm}
\vspace*{-1.25em}\subcaption*{{\scriptsize (c.2) $t=t^*$, zoom}}\vspace*{1.25em}
\end{subfigure}
\begin{subfigure}[t]{0.0001\textwidth}
\vspace*{-2.0cm}${\scriptstyle \! k_y}$
\end{subfigure}
\begin{subfigure}[t]{.24\textwidth}
\FigureXYLabel{\includegraphics[type=pdf,ext=.pdf,read=.pdf,width=0.96\textwidth]{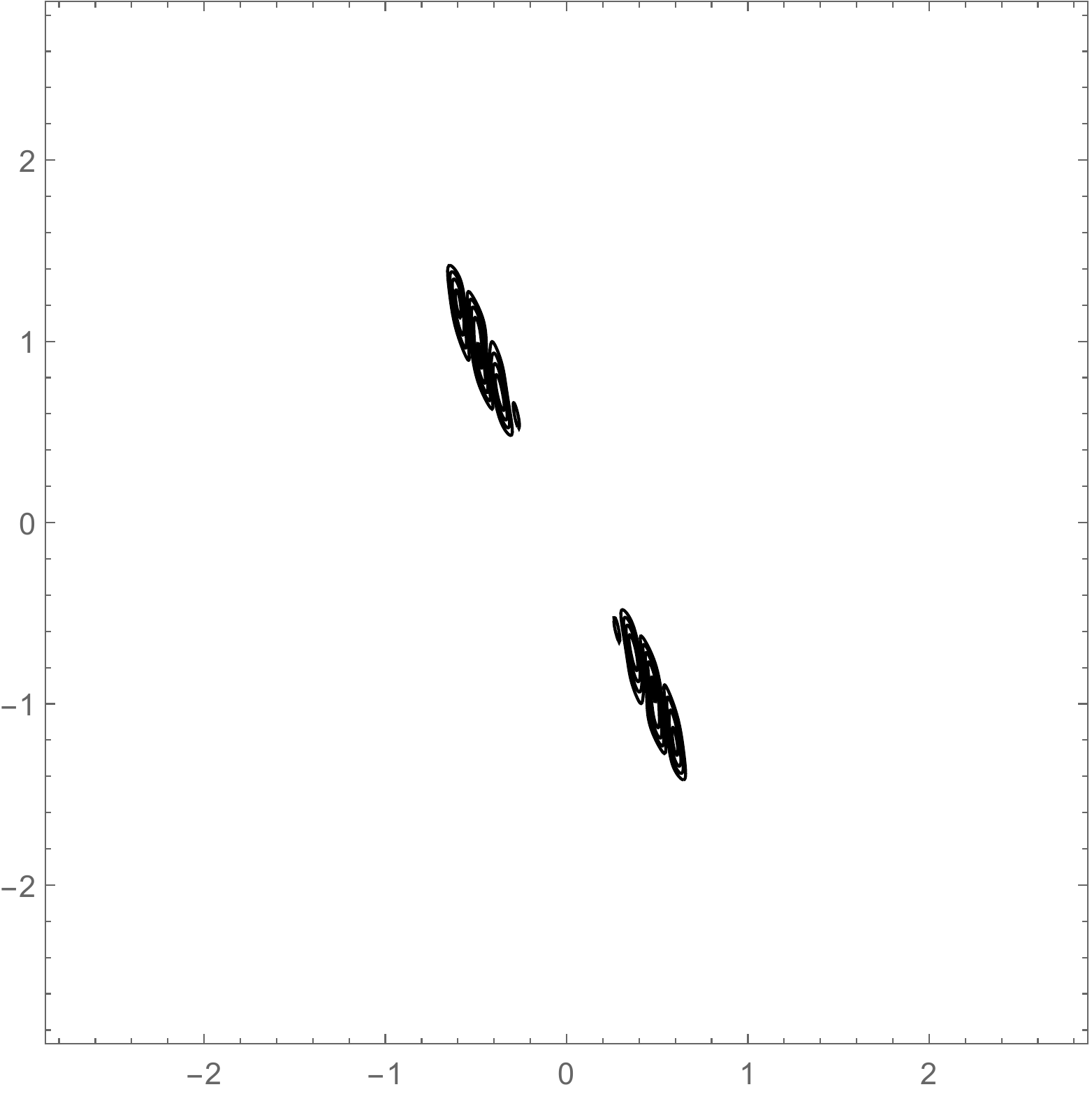}}
{${\scriptstyle\hspace{0.25cm} k_x}$}{-1mm}{\begin{rotate}{0}
\end{rotate}}{0mm}
\vspace*{-1.25em}\subcaption*{{\scriptsize (d.1) $t=3/2t^*$}}\vspace*{1.25em}
\end{subfigure}
\hspace*{-.2cm}
\begin{subfigure}[t]{.24\textwidth}
\FigureXYLabel{\includegraphics[type=pdf,ext=.pdf,read=.pdf,width=0.96\textwidth]{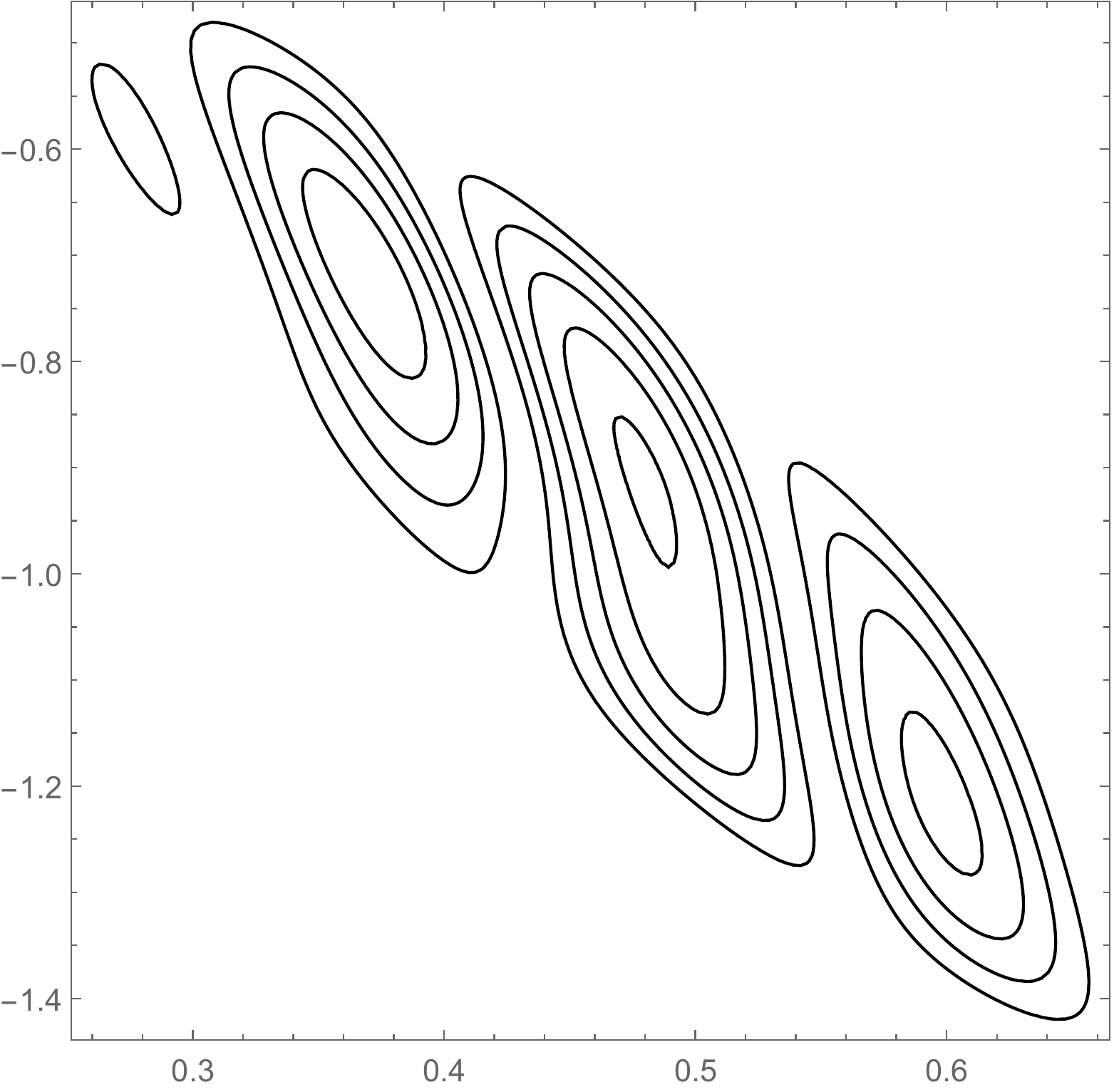}}
{${\scriptstyle\hspace{0.25cm} k_x}$}{-1mm}{\begin{rotate}{0}
\end{rotate}}{0mm}
\vspace*{-1.25em}\subcaption*{{\scriptsize (d.2) $t=3/2t^*$, zoom}}\vspace*{1.25em}
\end{subfigure}
\begin{subfigure}[t]{.24\textwidth}
\FigureXYLabel{\includegraphics[type=pdf,ext=.pdf,read=.pdf,width=0.96\textwidth]{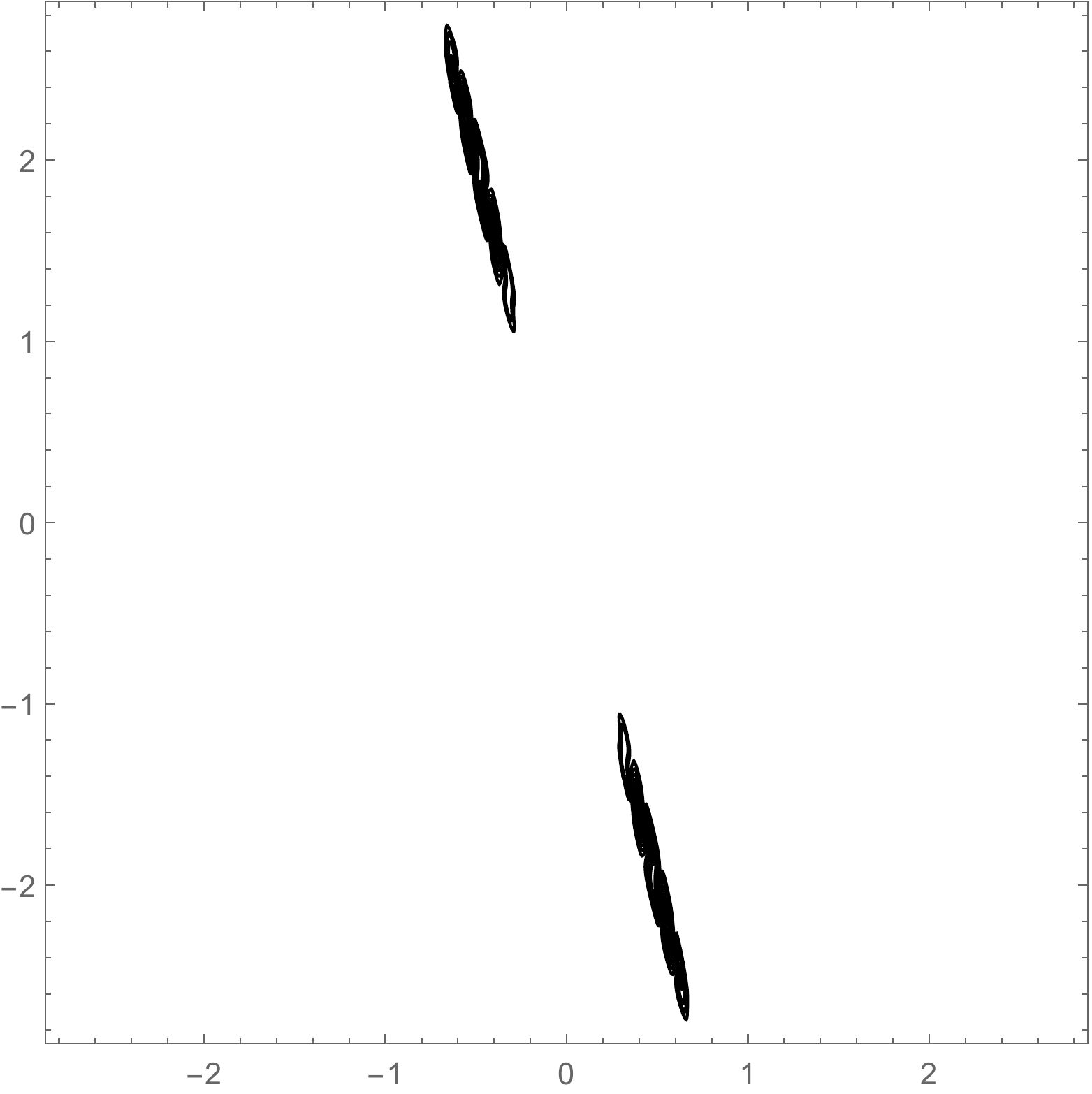}}
{${\scriptstyle\hspace{0.25cm} k_x}$}{-1mm}{\begin{rotate}{0}
\end{rotate}}{0mm}
\vspace*{-1.25em}\subcaption*{{\scriptsize (e.1) $t=2t^*$}}\vspace*{1.25em}
\end{subfigure}
\hspace*{-.2cm}
\begin{subfigure}[t]{.24\textwidth}
\FigureXYLabel{\includegraphics[type=pdf,ext=.pdf,read=.pdf,width=0.96\textwidth]{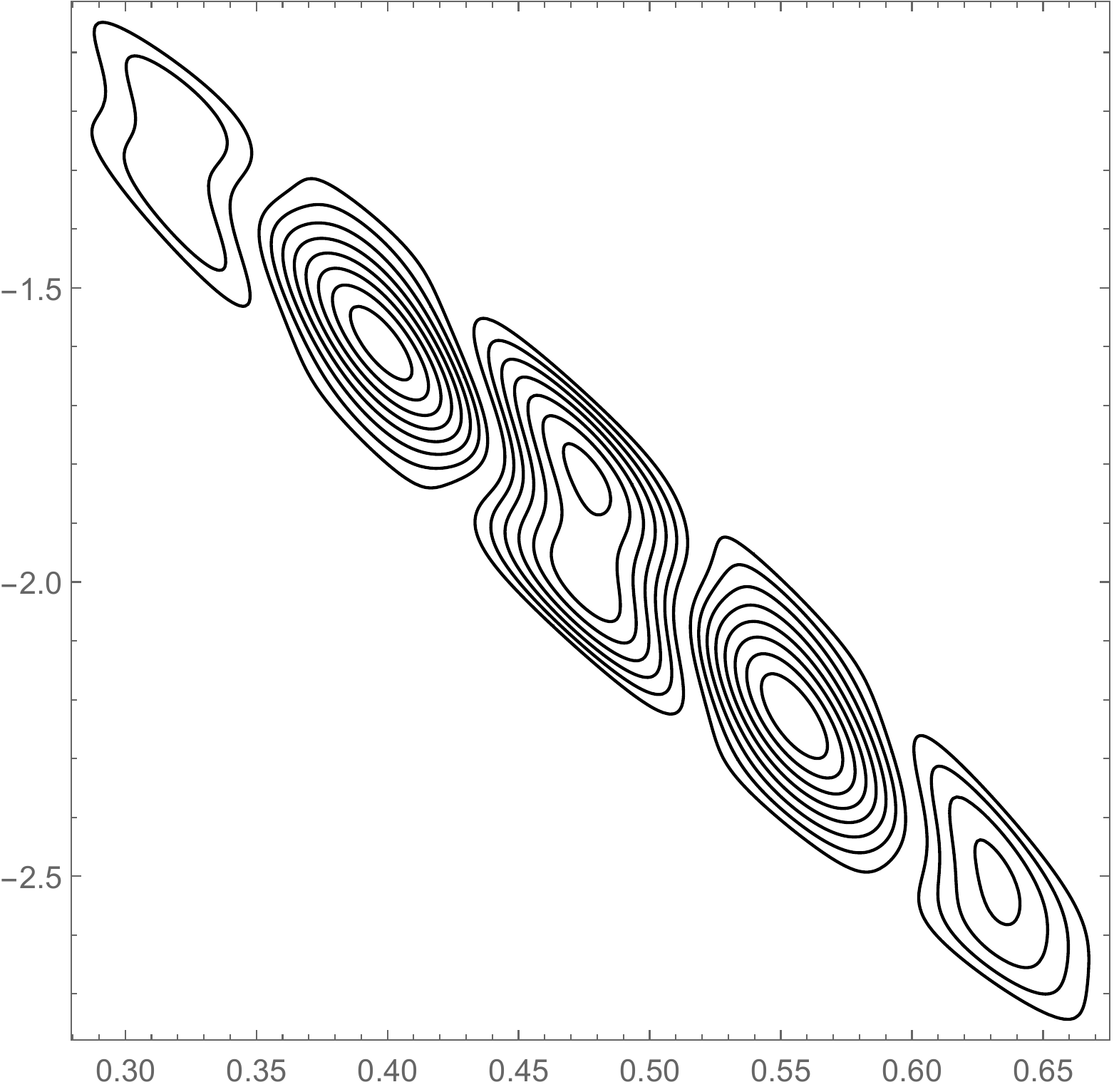}}
{${\scriptstyle\hspace{0.25cm} k_x}$}{-1mm}{\begin{rotate}{0}
\end{rotate}}{0mm}
\vspace*{-1.25em}\subcaption*{{\scriptsize (e.2) $t=2t^*$, zoom}}\vspace*{1.25em}
\end{subfigure}
\begin{subfigure}[t]{0.0001\textwidth}
\vspace*{-2.0cm}${\scriptstyle \! k_y}$
\end{subfigure}
\begin{subfigure}[t]{.24\textwidth}
\FigureXYLabel{\includegraphics[type=pdf,ext=.pdf,read=.pdf,width=0.96\textwidth]{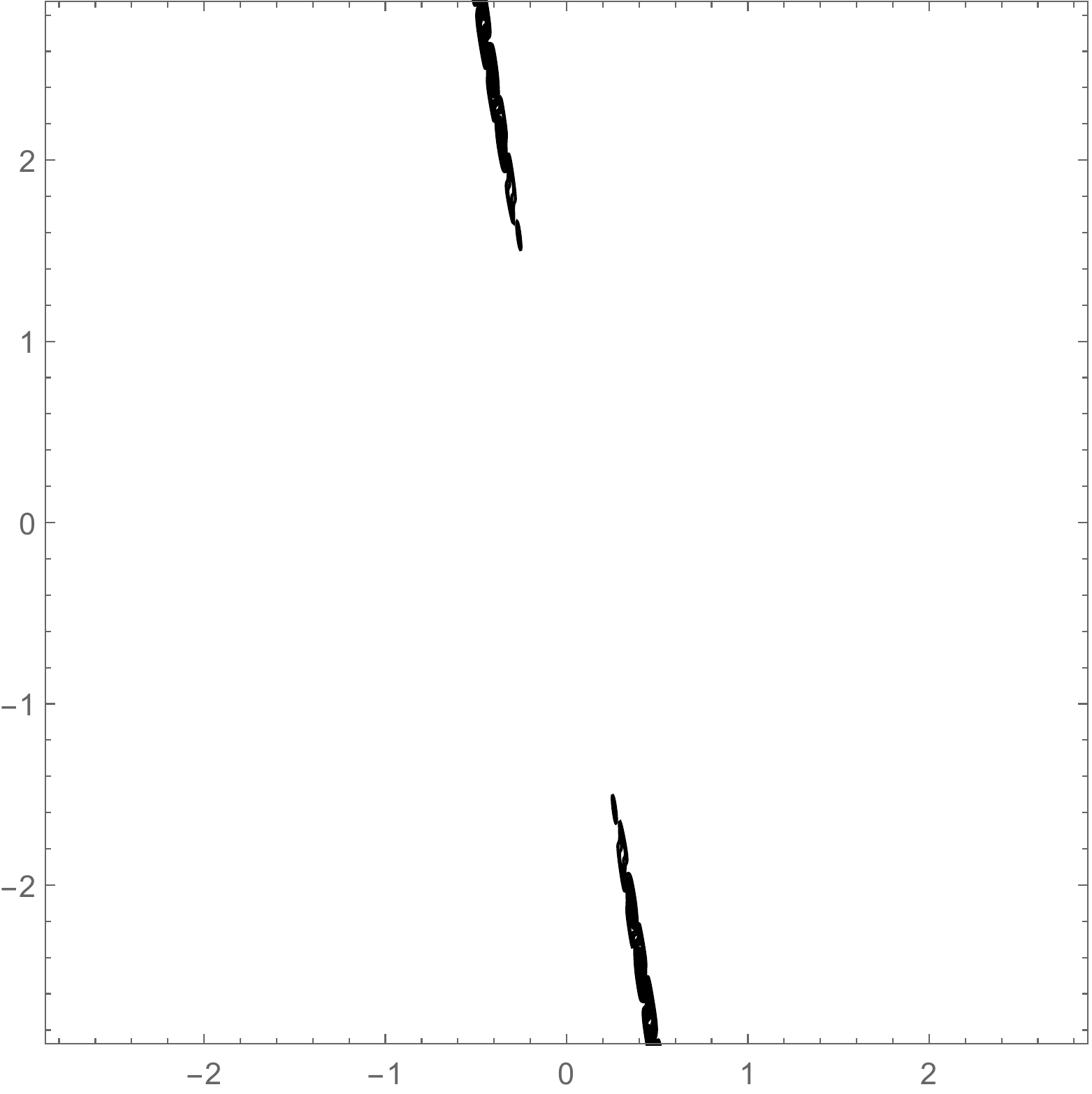}}
{${\scriptstyle\hspace{0.25cm} k_x}$}{-1mm}{\begin{rotate}{0}
\end{rotate}}{0mm}
\vspace*{-1.25em}\subcaption*{{\scriptsize (f.1) $t=5/2t^*$}}\vspace*{1.25em}
\end{subfigure}
\hspace*{-.2cm}
\begin{subfigure}[t]{.24\textwidth}
\FigureXYLabel{\includegraphics[type=pdf,ext=.pdf,read=.pdf,width=0.96\textwidth]{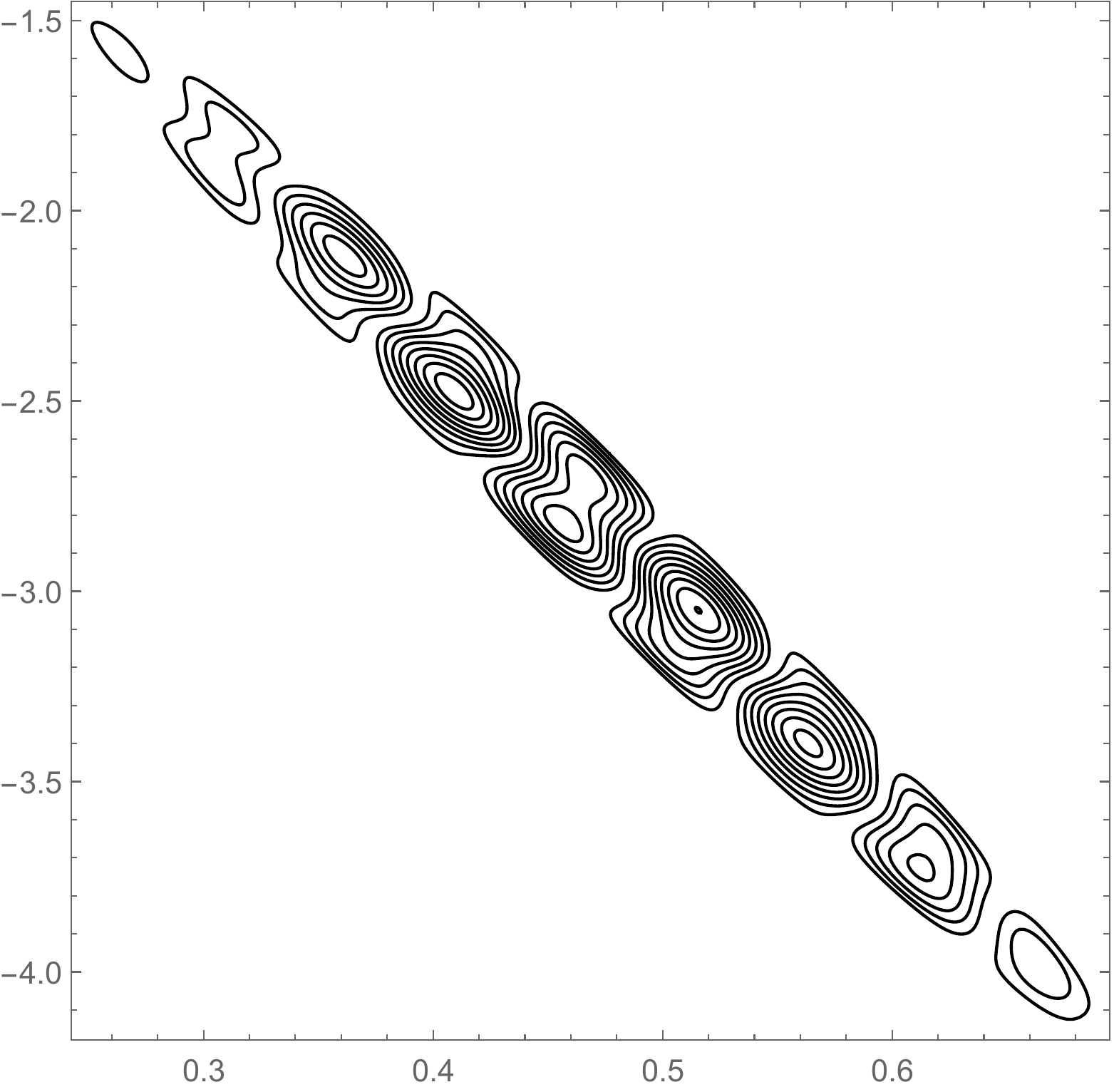}}
{${\scriptstyle\hspace{0.25cm} k_x}$}{-1mm}{\begin{rotate}{0}
\end{rotate}}{0mm}
\vspace*{-1.25em}\subcaption*{{\scriptsize (f.2) $t=5/2t^*$, zoom}}\vspace*{1.25em}
\end{subfigure}
\vspace*{-1em}\caption{{\footnotesize Evolution of the physical inertial-frame solution $|\hat{\rho}|$ in the spectral $(k_x,k_y)$-plane, according to the PDE-solution
\eqref{171124:2139} with the now physical initial condition \eqref{171129:0941}-\eqref{171129:1015}. This figure is the correction to Fig.$\,$[6.6] in \cite{Hau16}, since it now will lead to a real-valued solution in physical space (see Fig.$\,$\ref{fig8}). As before for the non-physical case (Fig.$\,$\ref{fig2}), the wave packet is drifting and changing its expansion size again, however, now in a more intricate fashion:  Before the critical time
$t^*$ is reached, the initial wave packet is already broken into smaller ones, collectively shrinking in the $k_y$- and expanding in the $k_x$-direction, while after that point in time, for $t>t^*$, it continues to break into even smaller and smaller packets as time progresses, all collectively expanding in the $k_y$- and shrinking in the $k_x$-direction. Note again that the time coordinate under $\mathsf{K}$ \eqref{171123:1839} is transforming as an invariant, that is, the critical time in the inertial frame takes exactly the same value $t^*=\tilde{t}^*\sim 8.5$ as in the Kelvin frame of Fig.$\,$\ref{fig5} ---  in the above inertial frame, however, the critical time~$t^*$ expresses itself differently than in the optimal Kelvin frame, namely as the time when the packet crosses the $k_x$-axis, since from its definition in Fig.$\,$\ref{fig5} we obtain $0=k_{\tilde{y}}-A\tilde{t}^*k_{\tilde{x}}=k_y+At^*k_x-At^*k_x=k_y$. Hence, topologically as well as geometrically, this physical solution differs significantly from the corresponding solution Fig.$\,$[6.6] in \cite{Hau16}, which, as explained and discussed in the text, is unphysical.\label{fig6}}}
\end{figure}

\begin{figure}[H]
\begin{subfigure}[t]{.49\textwidth}
\FigureXYLabel{\includegraphics[type=pdf,ext=.pdf,read=.pdf,width=0.85\textwidth]{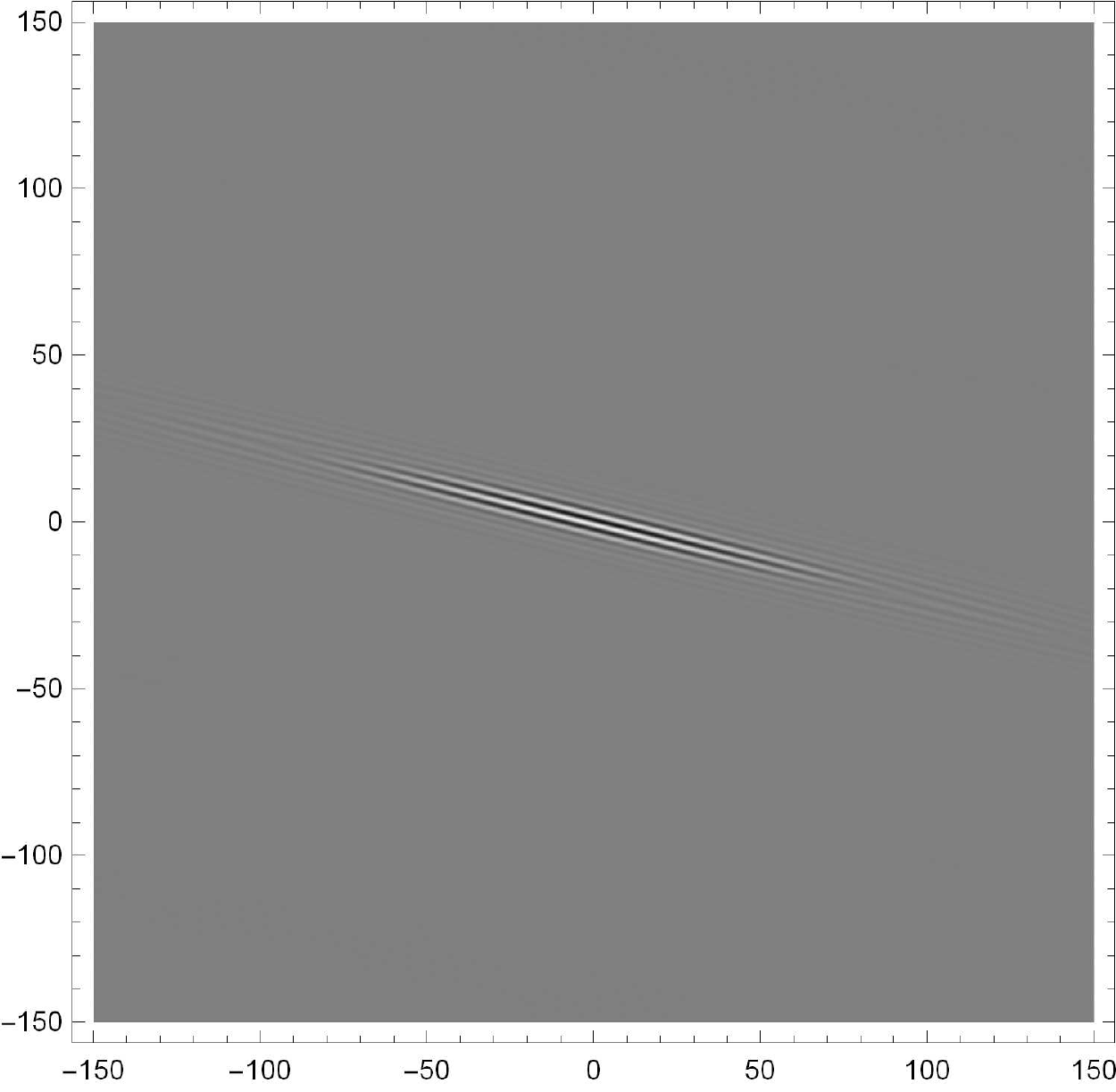}}
{${\scriptstyle\hspace{0.45cm}\tilde{x}}$}{-2mm}{\begin{rotate}{0}
\end{rotate}}{0mm}
\vspace*{-1.25em}\subcaption*{{\scriptsize (a) $\tilde{t}=0$}}\vspace*{0.5em}
\end{subfigure}
\hspace*{.2cm}
\begin{subfigure}[c]{0.0001\textwidth}
\vspace*{-6.5cm}${\scriptstyle \!\! \tilde{y}}$
\end{subfigure}
\begin{subfigure}[t]{.49\textwidth}
\FigureXYLabel{\includegraphics[type=pdf,ext=.pdf,read=.pdf,width=0.85\textwidth]{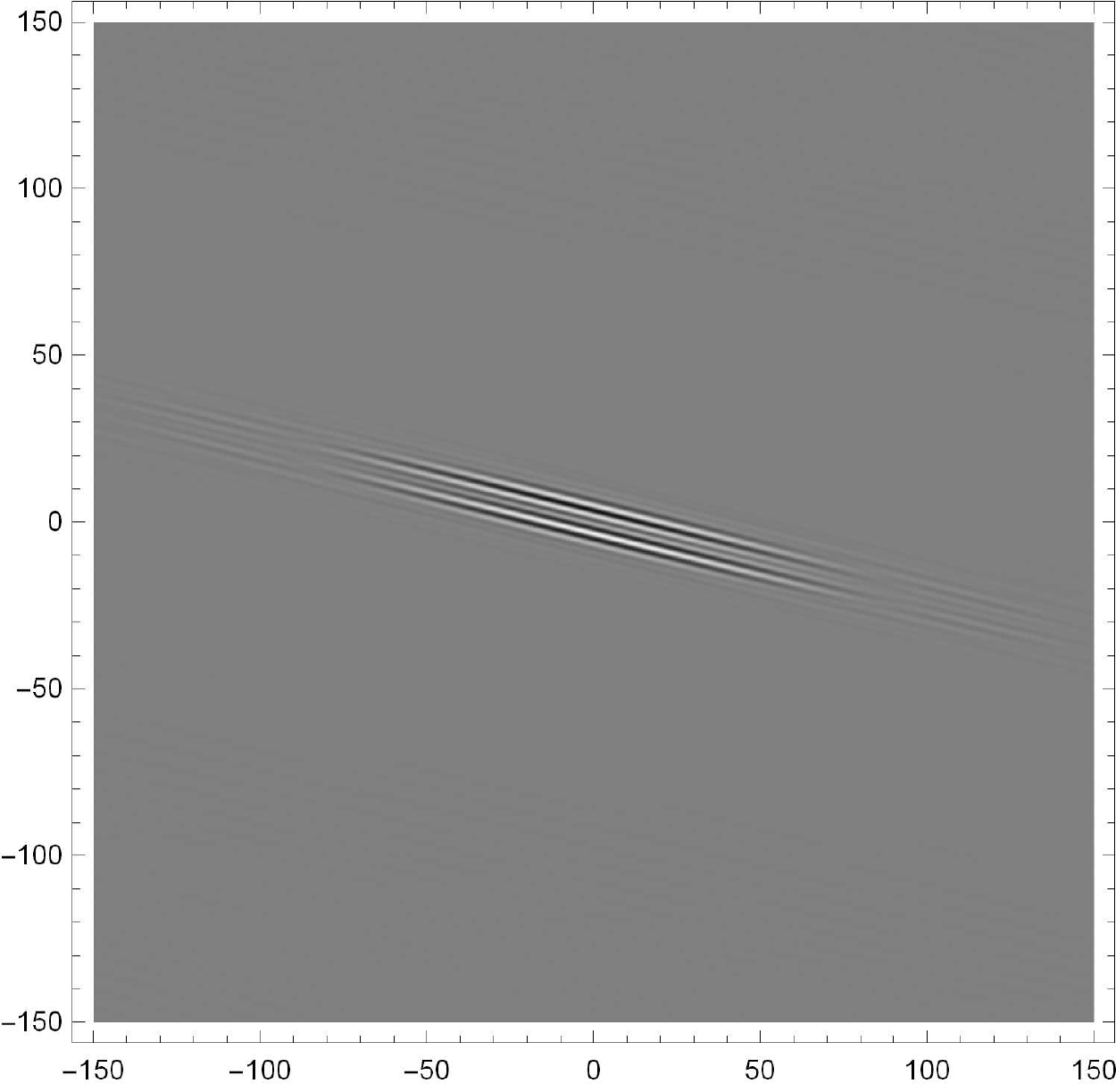}}
{${\scriptstyle\hspace{0.45cm}\tilde{x}}$}{-2mm}{\begin{rotate}{0}
\end{rotate}}{0mm}
\vspace*{-1.25em}\subcaption*{{\scriptsize (b) $\tilde{t}=1/2\tilde{t}^*$}}\vspace*{0.5em}
\end{subfigure}
\begin{subfigure}[t]{.49\textwidth}
\FigureXYLabel{\includegraphics[type=pdf,ext=.pdf,read=.pdf,width=0.85\textwidth]{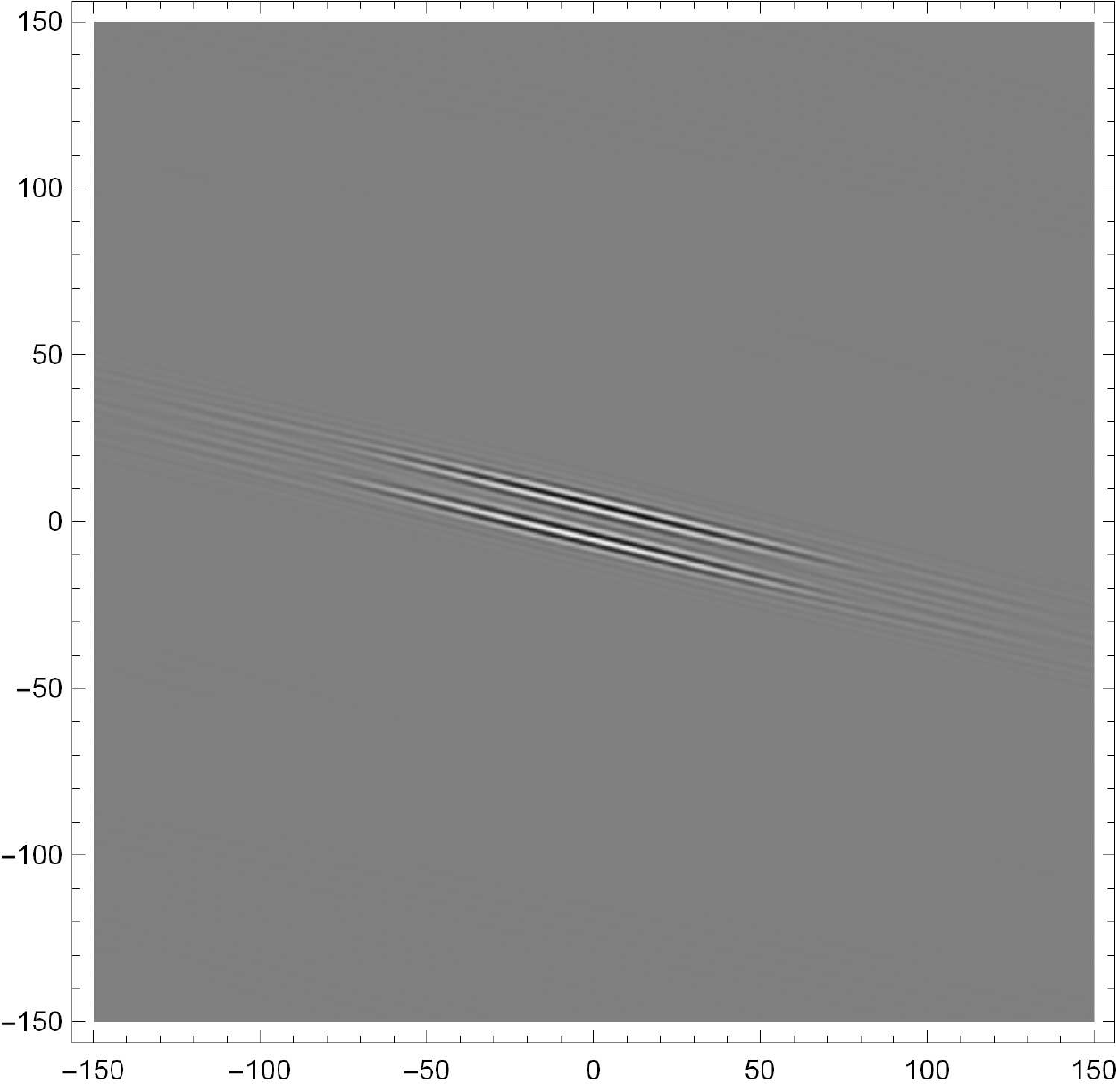}}
{${\scriptstyle\hspace{0.45cm}\tilde{x}}$}{-2mm}{\begin{rotate}{0}
\end{rotate}}{0mm}
\vspace*{-1.25em}\subcaption*{{\scriptsize (c) $\tilde{t}=\tilde{t}^*$}}\vspace*{0.5em}
\end{subfigure}
\hspace*{.2cm}
\begin{subfigure}[c]{0.0001\textwidth}
\vspace*{-6.5cm}${\scriptstyle \!\! \tilde{y}}$
\end{subfigure}
\begin{subfigure}[t]{.49\textwidth}
\FigureXYLabel{\includegraphics[type=pdf,ext=.pdf,read=.pdf,width=0.85\textwidth]{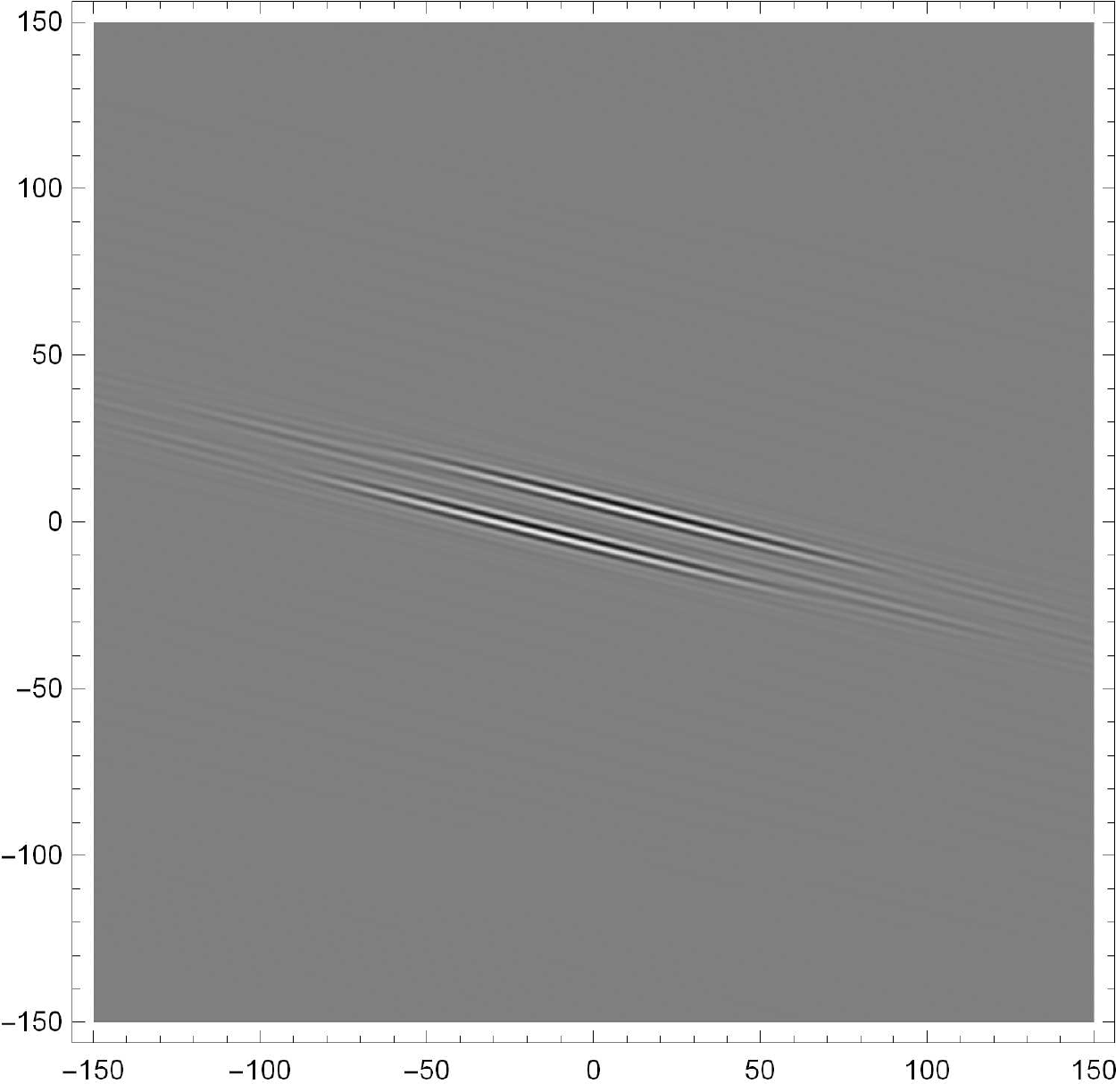}}
{${\scriptstyle\hspace{0.45cm}\tilde{x}}$}{-2mm}{\begin{rotate}{0}
\end{rotate}}{0mm}
\vspace*{-1.25em}\subcaption*{{\scriptsize (d) $\tilde{t}=3/2\tilde{t}^*$}}\vspace*{0.5em}
\end{subfigure}
\begin{subfigure}[t]{.49\textwidth}
\FigureXYLabel{\includegraphics[type=pdf,ext=.pdf,read=.pdf,width=0.85\textwidth]{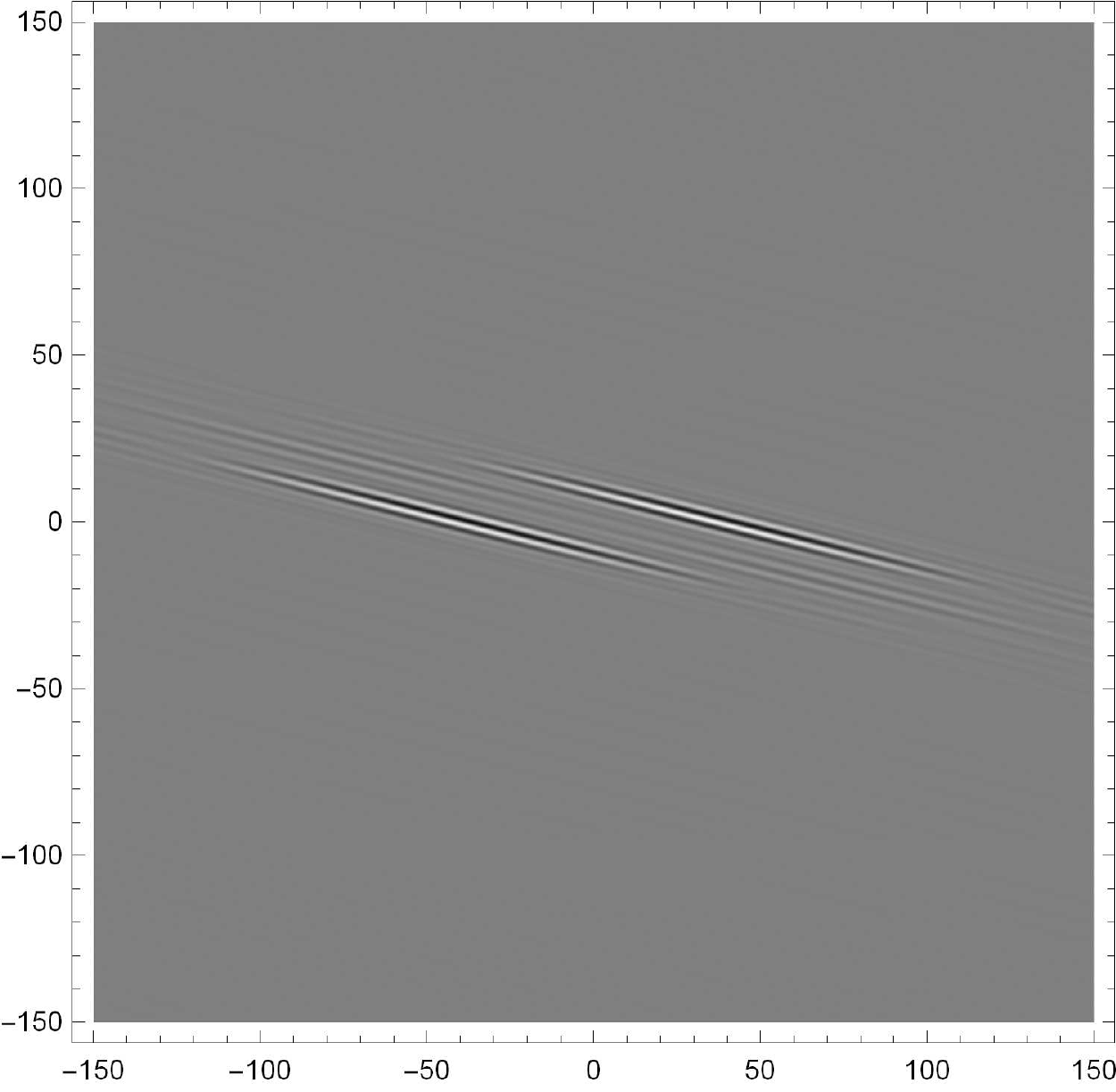}}
{${\scriptstyle\hspace{0.45cm}\tilde{x}}$}{-2mm}{\begin{rotate}{0}
\end{rotate}}{0mm}
\vspace*{-1.25em}\subcaption*{{\scriptsize (e) $\tilde{t}=2\tilde{t}^*$}}\vspace*{0.5em}
\end{subfigure}
\hspace*{.2cm}
\begin{subfigure}[c]{0.0001\textwidth}
\vspace*{-6.5cm}${\scriptstyle \!\! \tilde{y}}$
\end{subfigure}
\begin{subfigure}[t]{.49\textwidth}
\FigureXYLabel{\includegraphics[type=pdf,ext=.pdf,read=.pdf,width=0.85\textwidth]{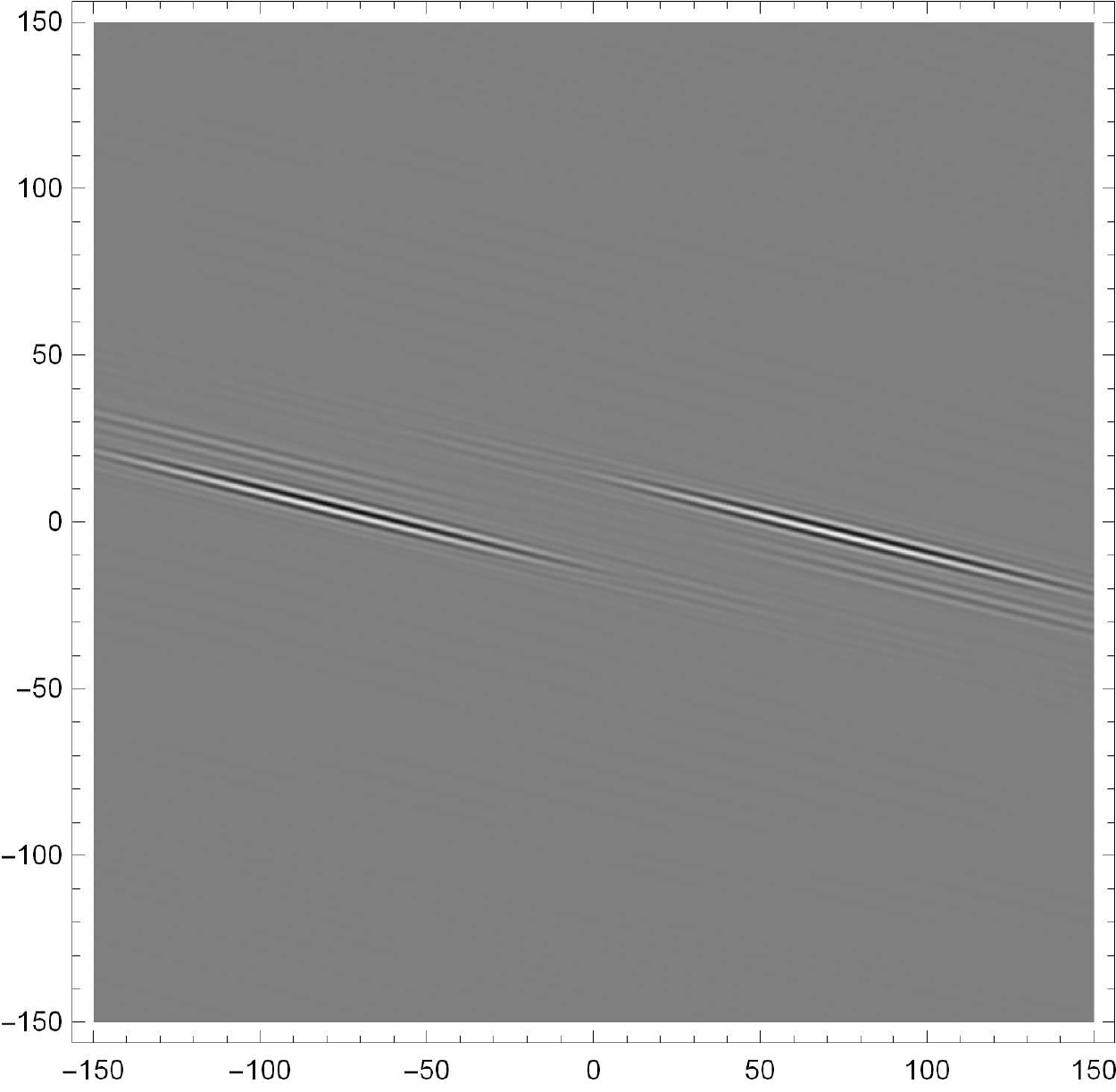}}
{${\scriptstyle\hspace{0.45cm}\tilde{x}}$}{-2mm}{\begin{rotate}{0}
\end{rotate}}{0mm}
\vspace*{-1.25em}\subcaption*{{\scriptsize (f) $\tilde{t}=5/2\tilde{t}^*$}}\vspace*{0.5em}
\end{subfigure}
\vspace*{-1em}\caption{{\footnotesize Evolution in physical space of the density field $\tilde{\rho}$ in the accelerated Kelvin frame,
synchronically corresponding to the evolution of its complementary density field in spectral space as shown in Fig.$\,$\ref{fig5}. Since the initial
condition \eqref{171129:1053}-\eqref{171129:1036} satisfies the reality constraint of the Fourier transform,
the density $\tilde{\rho}$ is a physical real-valued field, i.e. $\text{Im}(\tilde{\rho})= 0$, in contrast to the non-physical field of Fig.$\,$\ref{fig3}.
Although the geometrical structure of the solution in physical space nearly remains unchanged to Fig.$\,$\ref{fig3}, the density values, however, do differ significantly as shown in Table~\ref{tab1}.\label{fig7}}}
\end{figure}

\begin{figure}[H]
\begin{subfigure}[t]{.49\textwidth}
\FigureXYLabel{\includegraphics[type=pdf,ext=.pdf,read=.pdf,width=0.85\textwidth]{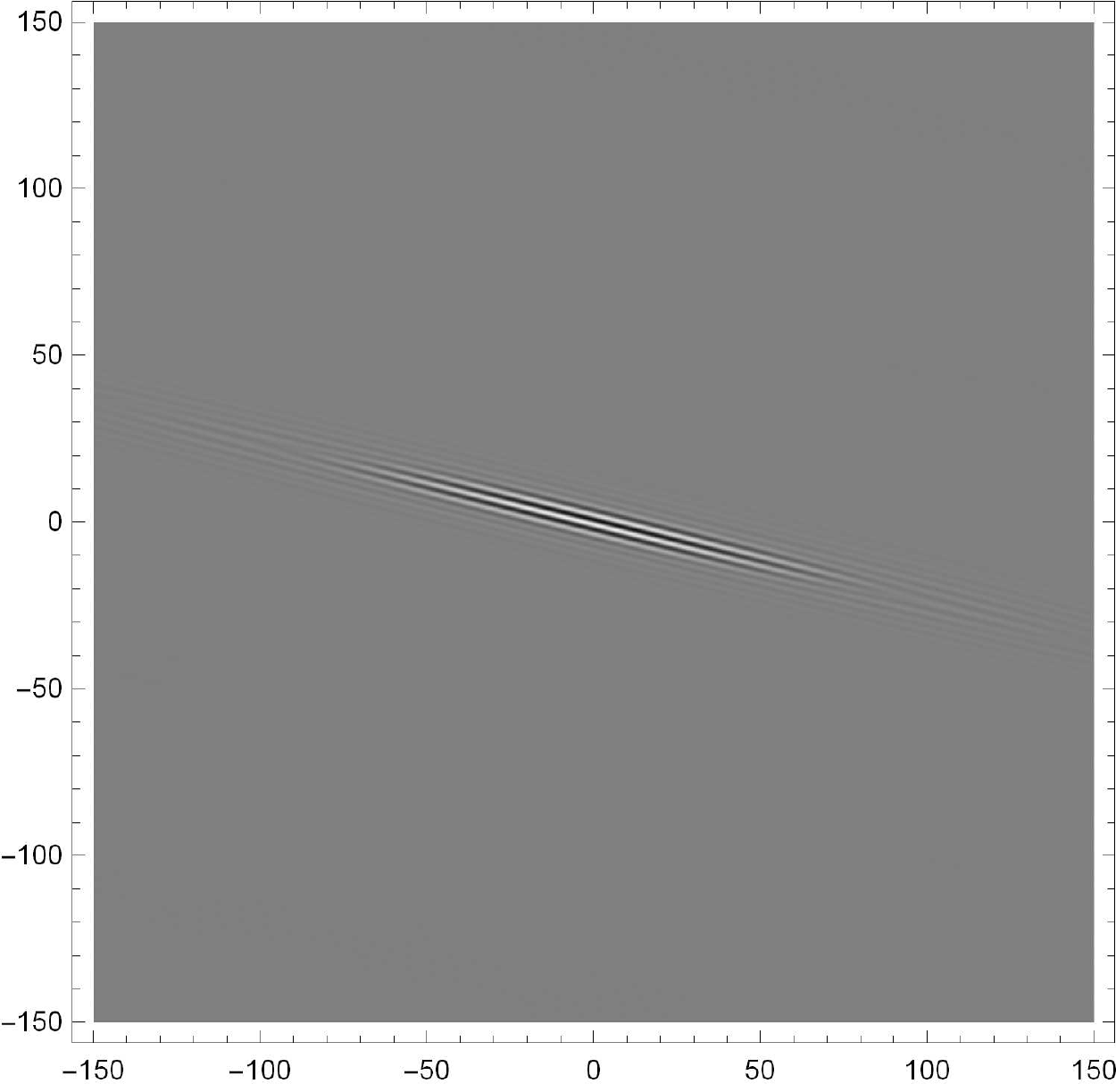}}
{${\scriptstyle\hspace{0.45cm}x}$}{-2mm}{\begin{rotate}{0}
\end{rotate}}{0mm}
\vspace*{-1.25em}\subcaption*{{\scriptsize (a) $t=0$}}\vspace*{0.5em}
\end{subfigure}
\hspace*{.2cm}
\begin{subfigure}[c]{0.0001\textwidth}
\vspace*{-6.5cm}${\scriptstyle \!\! y}$
\end{subfigure}
\begin{subfigure}[t]{.49\textwidth}
\FigureXYLabel{\includegraphics[type=pdf,ext=.pdf,read=.pdf,width=0.85\textwidth]{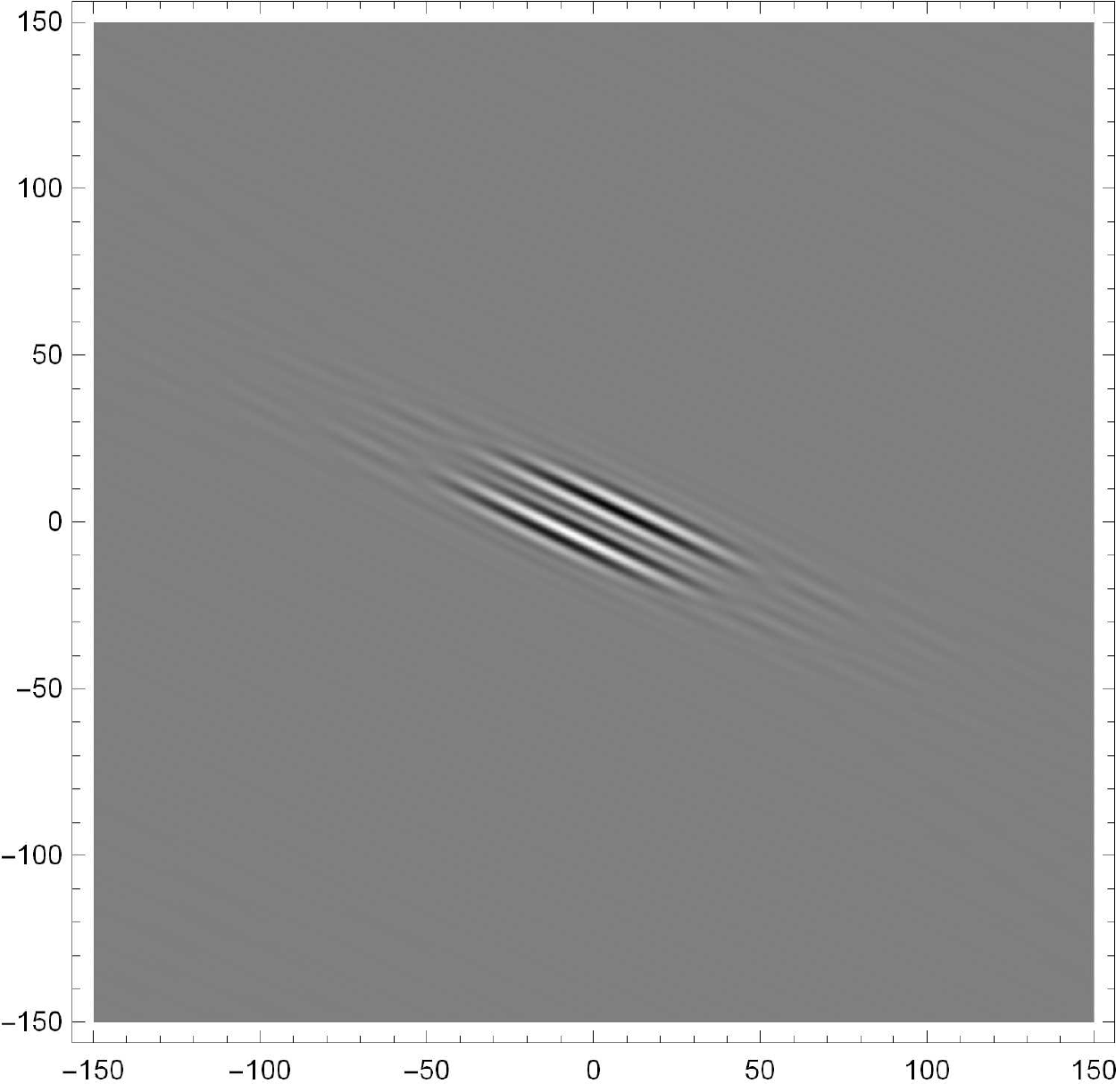}}
{${\scriptstyle\hspace{0.45cm} x}$}{-2mm}{\begin{rotate}{0}
\end{rotate}}{0mm}
\vspace*{-1.25em}\subcaption*{{\scriptsize (b) $t=1/2t^*$}}\vspace*{0.5em}
\end{subfigure}
\begin{subfigure}[t]{.49\textwidth}
\FigureXYLabel{\includegraphics[type=pdf,ext=.pdf,read=.pdf,width=0.85\textwidth]{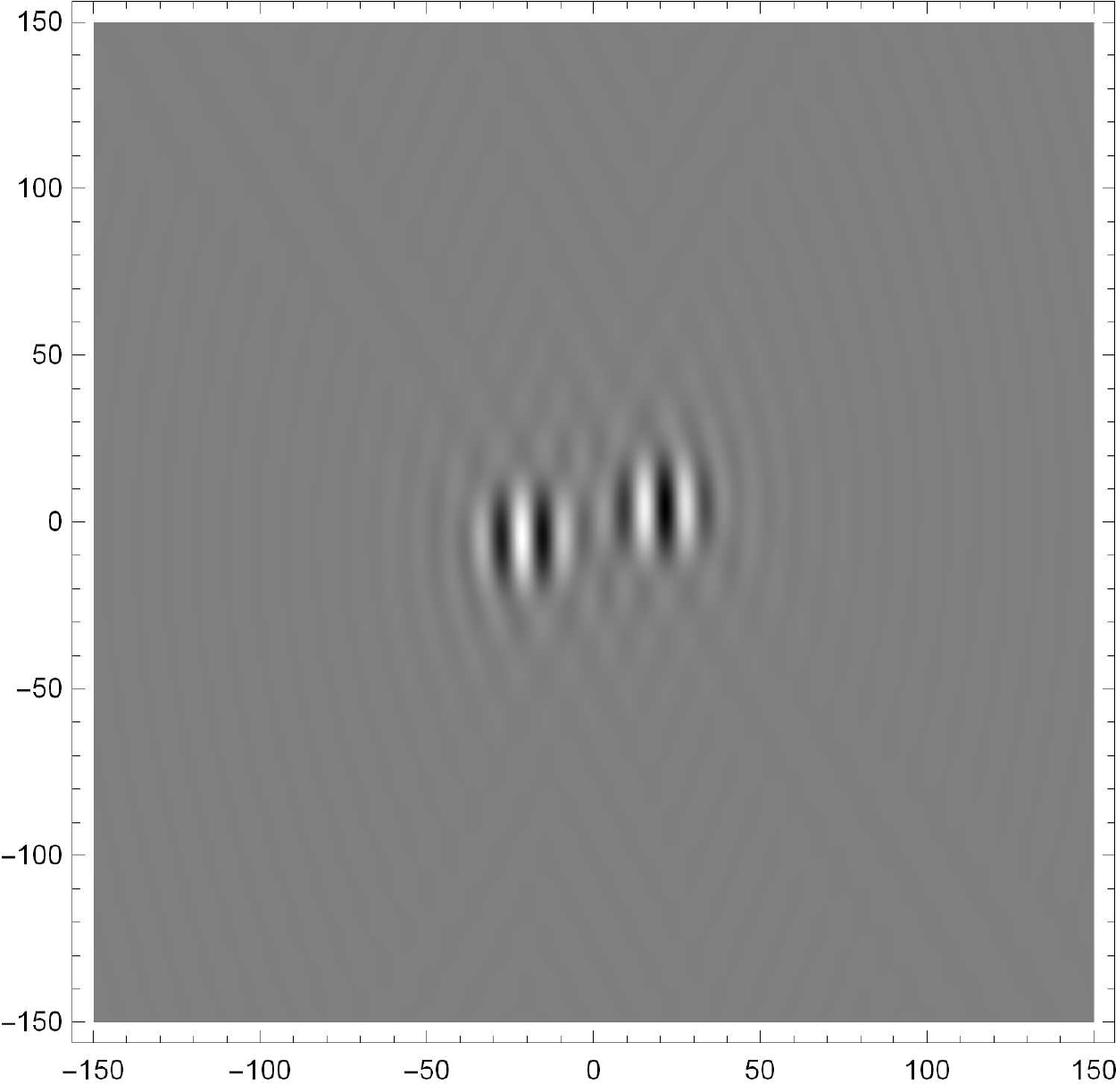}}
{${\scriptstyle\hspace{0.45cm}x}$}{-2mm}{\begin{rotate}{0}
\end{rotate}}{0mm}
\vspace*{-1.25em}\subcaption*{{\scriptsize (c) $t=t^*$}}\vspace*{0.5em}
\end{subfigure}
\hspace*{.2cm}
\begin{subfigure}[c]{0.0001\textwidth}
\vspace*{-6.5cm}${\scriptstyle \!\! y}$
\end{subfigure}
\begin{subfigure}[t]{.49\textwidth}
\FigureXYLabel{\includegraphics[type=pdf,ext=.pdf,read=.pdf,width=0.85\textwidth]{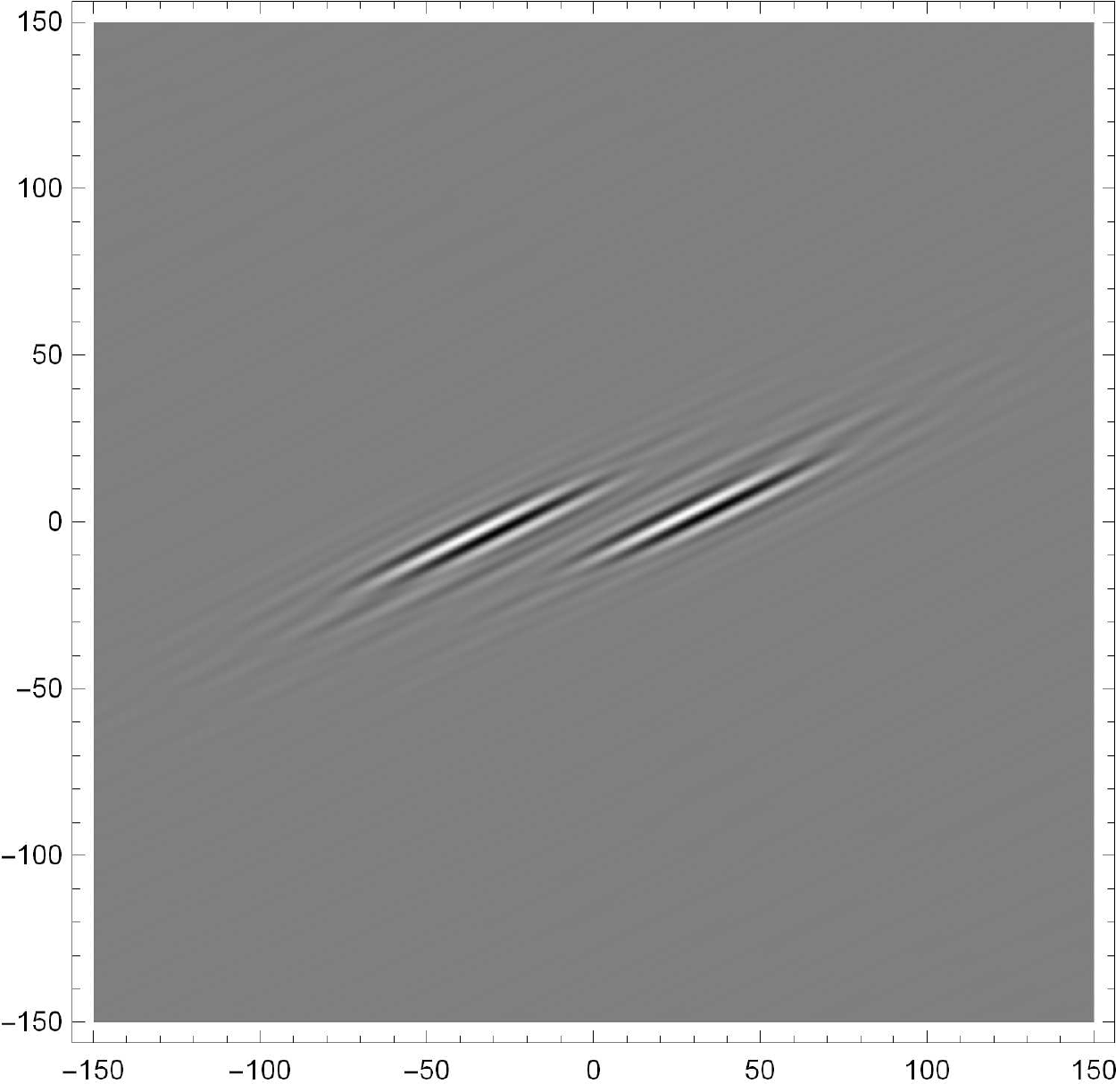}}
{${\scriptstyle\hspace{0.45cm} x}$}{-2mm}{\begin{rotate}{0}
\end{rotate}}{0mm}
\vspace*{-1.25em}\subcaption*{{\scriptsize (d) $t=3/2t^*$}}\vspace*{0.5em}
\end{subfigure}
\begin{subfigure}[t]{.49\textwidth}
\FigureXYLabel{\includegraphics[type=pdf,ext=.pdf,read=.pdf,width=0.85\textwidth]{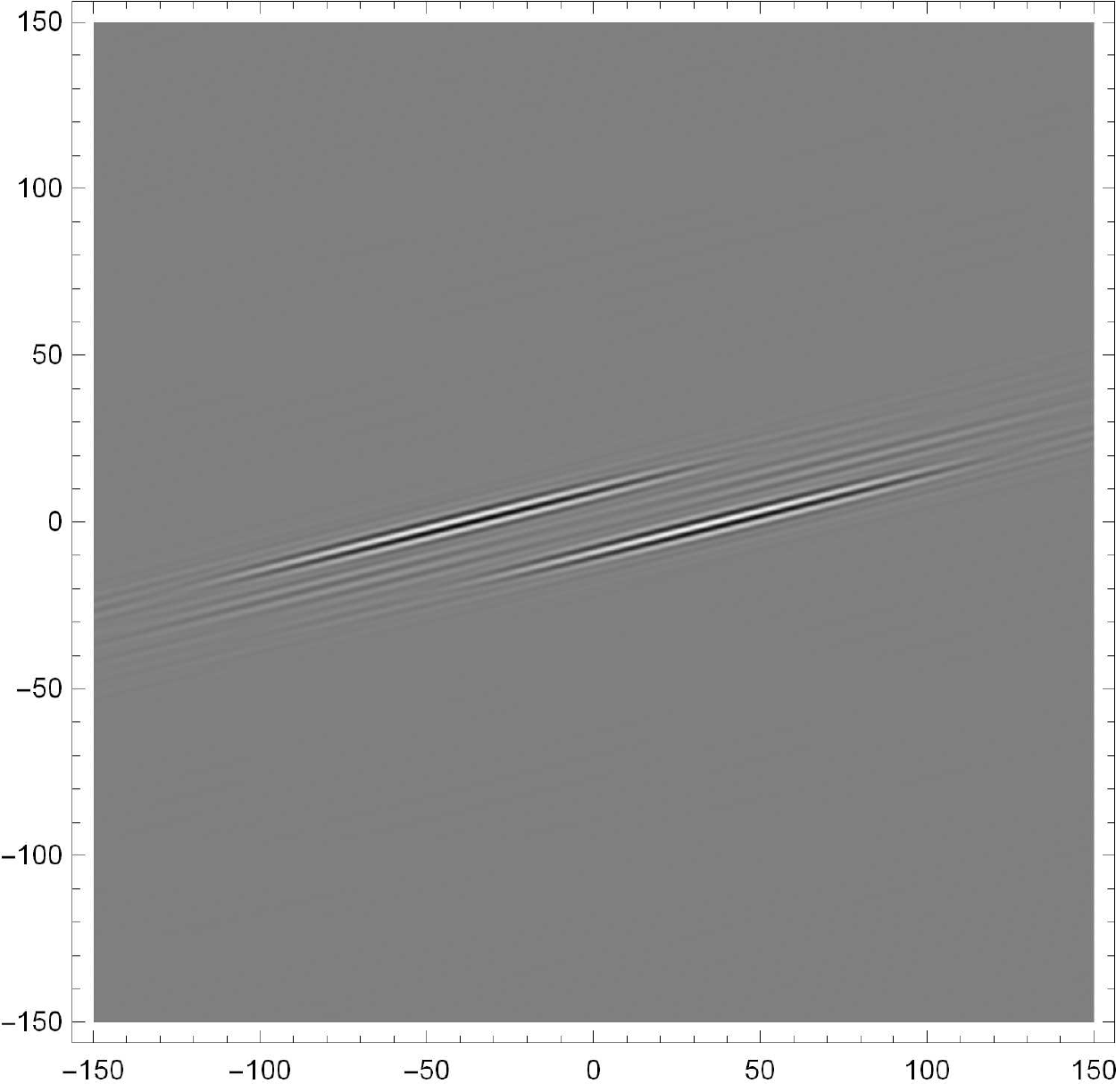}}
{${\scriptstyle\hspace{0.45cm} x}$}{-2mm}{\begin{rotate}{0}
\end{rotate}}{0mm}
\vspace*{-1.25em}\subcaption*{{\scriptsize (e) $t=2t^*$}}\vspace*{0.5em}
\end{subfigure}
\hspace*{.2cm}
\begin{subfigure}[c]{0.0001\textwidth}
\vspace*{-6.5cm}${\scriptstyle \!\! y}$
\end{subfigure}
\begin{subfigure}[t]{.49\textwidth}
\FigureXYLabel{\includegraphics[type=pdf,ext=.pdf,read=.pdf,width=0.85\textwidth]{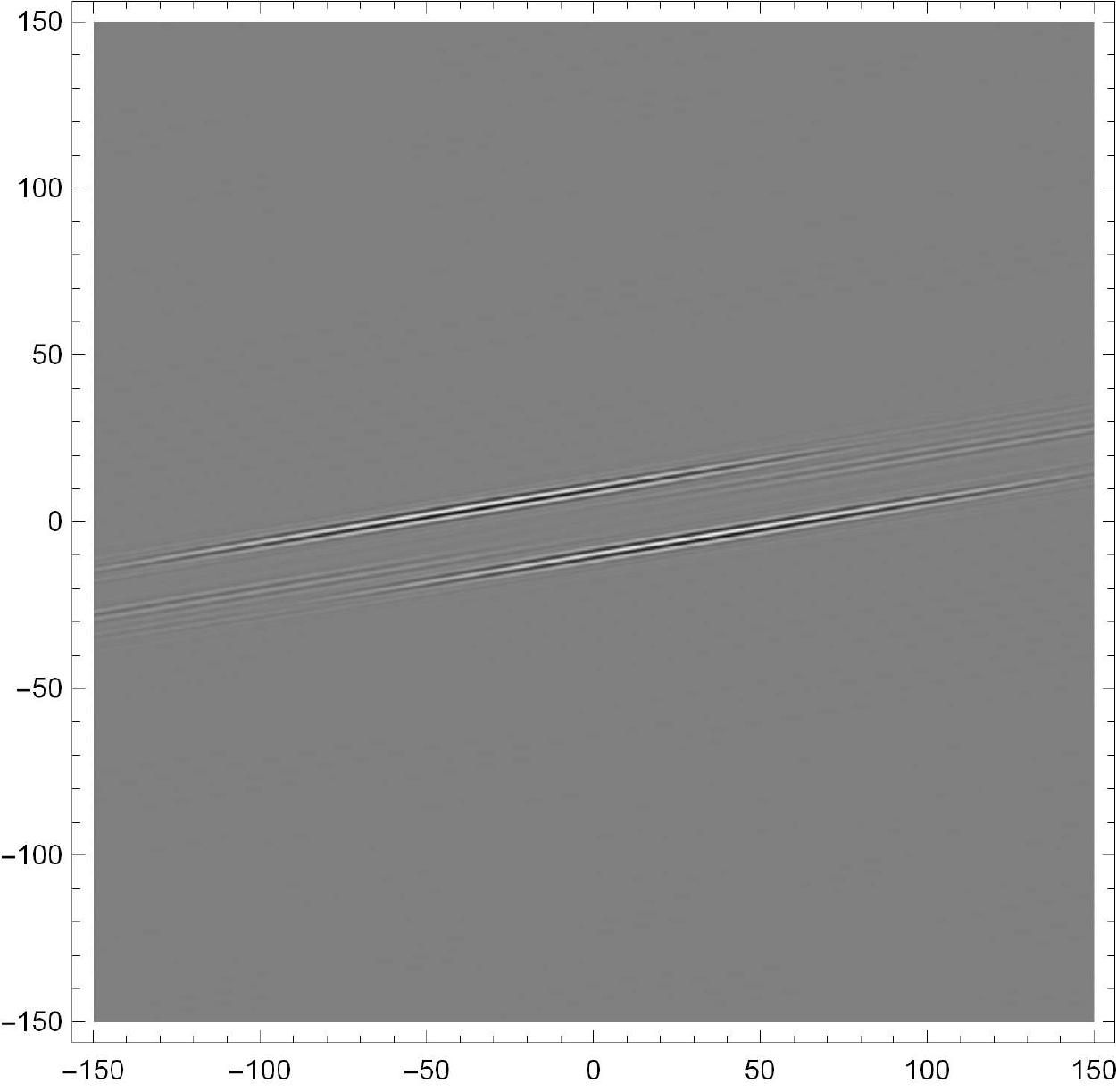}}
{${\scriptstyle\hspace{0.45cm} x}$}{-2mm}{\begin{rotate}{0}
\end{rotate}}{0mm}
\vspace*{-1.25em}\subcaption*{{\scriptsize (f) $t=5/2t^*$}}\vspace*{0.5em}
\end{subfigure}
\vspace*{-1em}\caption{{\footnotesize Evolution in physical space of the density field $\rho$ in the inertial frame,
synchronically corresponding to the evolution of its complementary density field in spectral space as shown in Fig.$\,$\ref{fig6}. Since the initial
condition \eqref{171129:0941}-\eqref{171129:1015} satisfies the reality constraint of the Fourier transform,
the density $\rho$ is a physical real-valued field, i.e. $\text{Im}(\rho)= 0$, in contrast to the non-physical field of Fig.$\,$\ref{fig4}.
Although the geometrical structure of the solution in physical space nearly remains unchanged to Fig.$\,$\ref{fig4}, the density values, however, do differ significantly as shown in Table~\ref{tab1}.
The figure shown above is the correction to Fig.$\,$[6.7] in \cite{Hau16}.\label{fig8}}}
\end{figure}

Instead of the unphysical initial condition \eqref{171123:1354} used before from the result of \cite{Hau16}, this section repeats the analysis for the initial condition
\begin{equation}
\left.
\begin{aligned}
\hat{u}(t,k_x,k_y)\big|_{t=0}&= c_s\Psi^{(+)}(k_x,k_y)\cos\varphi,\\[0.35em]
\partial_t\hat{u}(t,k_x,k_y)\big|_{t=0}&=Ak_x\partial_{k_y}\hat{u}(t,k_x,k_y)\big|_{t=0} -c_s^2\Psi^{(+)}(k_x,k_y)\sqrt{k_x^2+k_y^2}\,\sin\varphi,\label{171129:0941}
\end{aligned}
~~~\right\}
\end{equation}
that satisfies the reality constraint \eqref{171127:1622} of the Fourier transform at initial time $t=0$.\linebreak[4] This guarantees that the complementary fields in physical space are real-valued, and thus truly physical. The specific choice~\eqref{171129:0941} was guided by the objective to only adjust the unphysical initial condition \eqref{171123:1354}, as originally proposed by \cite{Hau16} in Appendix [D], such that, in the spirit of an earlier proposed initial condition Eq.$\,$[6.6], it turns into a physical initial condition. According to the inertial-frame solution \eqref{171124:2139}, the initial condition of the two remaining fields are then naturally fixed by \eqref{171129:0941} as
\begin{equation}
\hspace{1.5cm}\left.
\begin{aligned}
\hat{v}\big|_{t=0}=\; & \frac{1}{A^2+c_s^2k_x^2}\left[c_s^2k_xk_y\hat{u}\big|_{t=0}
-A\Big(\partial_t\hat{u}|_{t=0}-Ak_x\partial_{k_y}\hat{u}\big|_{t=0}\Big)\right],\\[0.5em]
\hat{\rho}\big|_{t=0}=\; & -\frac{i}{A}\Big(k_x\hat{v}\big|_{t=0}-k_y\hat{u}\big|_{t=0}\Big).
\label{171129:1015}
\end{aligned}
~~~~~~\right\}
\end{equation}
When transforming these initial conditions into the optimal Kelvin frame according to its defining spectral transformation $\hat{\mathsf{K}}$ \eqref{171123:2105}, they will take the more recognizable and simple form of the conditions Eq.$\,$[6.6]\footnote[2]{The initial conditions Eq.$\,$[6.6] in \cite{Hau16} refer to the solutions Eq.$\,$[5.45] \& Eq.$\,$[6.5], which exactly match with the herein presented Kelvin-frame solutions \eqref{171124:1828}. To note is that as Eq.$\,$[6.6] in \cite{Hau16} stands, it also violates the reality constraint \eqref{171127:1622} of the Fourier transform, and thus has to be adjusted too, in order to be~physical.} in \cite{Hau16}:
\begin{equation}
\left.
\begin{aligned}
\hat{\tilde{u}}(\tilde{t},k_{\tilde{x}},k_{\tilde{y}})\big|_{\tilde{t}=0}&= c_s\Psi^{(+)}(k_{\tilde{x}},k_{\tilde{y}})\cos\varphi,\\[0.35em]
\partial_{\tilde{t}}\hat{\tilde{u}}(\tilde{t},k_{\tilde{x}},k_{\tilde{y}})\big|_{\tilde{t}=0}&=-c_s^2\Psi^{(+)}(k_{\tilde{x}},k_{\tilde{y}})\sqrt{k_{\tilde{x}}^2+k_{\tilde{y}}^2}\,\sin\varphi,\label{171129:1053}
\end{aligned}
~~~\right\}
\end{equation}
and, respectively,
\begin{equation}
\left.
\begin{aligned}
\hat{\tilde{v}}\big|_{\tilde{t}=0}=\; & \frac{1}{A^2+c_s^2k_{\tilde{x}}^2}\Big(c_s^2k_{\tilde{x}}k_{\tilde{y}}\hat{\tilde{u}}\big|_{\tilde{t}=0}
-A\partial_{\tilde{t}}\hat{\tilde{u}}\big|_{\tilde{t}=0}\Big),\;\;\\[0.5em]
\hat{\tilde{\rho}}\big|_{\tilde{t}=0}=\; & -\frac{i}{A}\left(k_{\tilde{x}}\hat{\tilde{v}}\big|_{\tilde{t}=0}-k_{\tilde{y}}\hat{\tilde{u}}\big|_{\tilde{t}=0}\right).
\end{aligned}
~~~\right\}\hspace{-0.825cm}\label{171129:1036}
\end{equation}
Fig.$\,$\ref{fig5} shows the spectral solution of $|\hat{\tilde{\rho}}|$ and Fig.$\,$\ref{fig7} its associated physical-space solution of $\tilde{\rho}$ in the optimal Kelvin frame, while Fig.$\,$\ref{fig6} the corresponding spectral solution of $|\hat{\rho}|$ and Fig.$\,$\ref{fig8} its associated physical-space solution of $\rho$ in the inertial frame. To note here is that in both frames the associated physical-space solutions are now real-valued, i.e. $\text{Im}(\tilde{\rho})=\text{Im}(\rho)=0$, as it should be in order to constitute physical solutions. When comparing in physical space the non-physical solution (Fig.$\,$\ref{fig3} or Fig.$\,$\ref{fig4}) with the physical one (Fig.$\,$\ref{fig7} or Fig.$\,$\ref{fig8}), it is in both frames remarkable to see how robust the structure of this solution is against changes in the initial condition  --- but only in physical space, in spectral space such a structural invariance cannot be observed. Although the geometrical structure of the solution in physical space nearly remains unchanged under this variation in the initial condition, the density values, however, do change: For example, at the initial point $\tilde{t}=t=0$, the maximum value for the unphysical case is of size $|\text{Re}(\tilde{\rho})|_{\text{max}}=|\text{Re}(\rho)|_{\text{max}}\sim 5.6\cdot 10^{-4}$, while for the physical case it is reduced to $|\tilde{\rho}|_{\text{max}}=|\rho|_{\text{max}}\sim 5.0\cdot 10^{-4}$. In addition, before the critical time is reached ($t<t^*$), the physical solution\hfill is\hfill also\hfill not\hfill just\hfill monotonically\hfill decaying\hfill as\hfill it\hfill is\hfill the\hfill case\hfill for\hfill the\hfill non-physical\hfill solution,

\newgeometry{left=2.49cm,right=2.49cm,top=2.50cm,bottom=2.25cm,headsep=1em}

\noindent but instead shows a more complex and difficile variation as can be read off from Table~\ref{tab1} above. Also relevant for stability analysis is the nearly three times higher value in each time step after the critical time has been passed ($t>t^*$).

\setcounter{table}{2}
\begin{table}
\begin{center}
\begin{tabular}{p{0.0mm} c | c | c l}\cline{2-4}\\
& & {\it Non-physical solution} & {\it Physical solution}\\[0.5em]
& $\tilde{t}/\tilde{t}^*=t/t^*$ & $|\text{Re}(\tilde{\rho})|_{\text{max}}=|\text{Re}(\rho)|_{\text{max}}$ & $|\tilde{\rho}|_{\text{max}}=|\rho|_{\text{max}}$ &\\[0.5em]
& 0.0    & $5.6\cdot 10^{-4}$   & $\phantom{0}5.0\cdot 10^{-4}$ &\\[0.1em]
& 0.5    & $3.5\cdot 10^{-4}$   & $10.6\cdot 10^{-4}$ &\\[0.1em]
& 1.0    & $1.7\cdot 10^{-4}$   & $\phantom{0}4.8\cdot 10^{-4}$ &\\[0.1em]
& 1.5    & $3.4\cdot 10^{-4}$   & $\phantom{0}9.5\cdot 10^{-4}$ &\\[0.1em]
& 2.0    & $4.8\cdot 10^{-4}$   & $14.0\cdot 10^{-4}$ &\\[0.1em]
& 2.5    & $5.8\cdot 10^{-4}$   & $17.0\cdot 10^{-4}$ &
\end{tabular}
\caption{Maximum values of the physical-space solution for different times relative to the critical time $\tilde{t}=t\sim 8.5$, once for the non-physical solution shown in Fig.$\,$\ref{fig3} \& Fig.$\,$\ref{fig4}, and
once for the corresponding physical solution as shown in Fig.$\,$\ref{fig7} \& Fig.$\,$\ref{fig8}. The {\it non}-physical solution is the result of \cite{Hau16}, while the physical solution is its correction. To note is that all values in this table are invariant values, that is, they are the same both for the (non-optimal) inertial as well as for the (optimal) co-moving Kelvin frame, simply because within the defining transformation $\mathsf{K}$ \eqref{171123:1839} the time and the density field are transforming invariantly: $\tilde{t}=t$ and $\tilde{\rho}=\rho.$ Relevant for stability analysis is the significant difference in the variation of the solutions: (i) The physical solution is not monotonically decaying before $t/t^*=1$, and (ii) after $t/t^*=1$, it is nearly three times higher in each time step than the non-physical solution.} \label{tab1}
\end{center}
\vspace{-0.4em}\hrule
\end{table}
Nevertheless, even when adjusting the unphysical initial condition in \cite{Hau16} into a physical one, the solutions in physical space as shown in Fig.$\,$[6.5] \& Fig.$\,$[6.7] cannot be reproduced when following the construction process as described in \cite{Hau16}. To obtain solutions as Fig.$\,$[6.5] \& Fig.$\,$[6.7], a fundamentally different and highly specialized initial condition has to be employed. But such an initial condition is not mentioned in the text, since all initial conditions used in \cite{Hau16} are always referred to the {\it spectral} construction process outlined in Appendix [D]. But again, these initial conditions constructed from Appendix [D] cannot lead to such specialized results as shown in Fig.$\,$[6.5] \& Fig.$\,$[6.7], in particular, when recognizing the above mentioned fact of how robust the structure of the physical-space solution is against changes of the initial condition in spectral space.

\label{171201:0725}
The standard situation is that one always obtains two wave packets moving in opposite directions, one in positive and the other in negative shearwise direction without any upper or lower spatial bounds, both in the inertial as well as in the Kelvin frame. The existence of a single upper bound, as claimed in \cite{Hau16} and identified therein as a ``critical layer", can be misleading for two reasons: (i) In contrast to the critical layer obtained in the $\mathfrak{g}_1$-approach \citep{Campos99}, this newly defined critical layer within the $\mathfrak{g}_2$-approach is {\it not} associated to any flow singularity.\linebreak[4] (ii)~The~bounded motion as shown in Fig.$\,$[6.5] or Fig.$\,$[6.7] is just and only the result of a\linebreak[4] special, particularly arranged (yet for the reader still unknown) initial condition: Because when changing, for example, the internal wave-packet parameter $k_{y_0}$ to higher values, while leaving the parameters $A$ and $c_s$ (defining the packet's environment) and the Mach number $\mathcal{M}$ (defining the packet's relative motion to its environment) unchanged, will automatically lead to a greater range of motion in the positive $y$-direction\footnote[2]{This increased range of motion in the shearwise $y$-direction grows logarithmically with increasing value of $k_{y_0}$.\linebreak[4] This can be easily verified when following the path of the wave packet along the trajectory it sweeps out in physical space. The trajectory is determined by the wave's group velocity and is derived in \cite{Hau16} as Eq.$\,$[3.29a-b], or in \cite{Hau15} as Eq.$\,$[36-37]. It is clear that for a given particular set of initial parameters, the maximum extent $y^*$ of a {\it single} wave-packet in the $y$-direction is related to the critical time $t^*=k_{y_0}/(Ak_{x_0})$, i.e., $y^*=y^*(t^*)$. But to define this upper bound as a ``critical layer" can be misleading with regard to the genuinely existing critical layer of the $\mathfrak{g}_1$-approach, as correctly examined and discussed by \cite{Campos99}.} than shown in Fig.$\,$[6.5] or Fig.$\,$[6.7], with the result

\restoregeometry

\noindent that this new packet will inevitably cross and penetrate through the ``critical layer" set earlier by the smaller value $k_{y_0}$. --- As an analogon, this situation can be compared e.g. when throwing a ball up into the air: The height of the ball is determined by the ball's initial condition in position and velocity; changing the initial condition will change the maximum height. Now, from the reached maximum height of a single particular initial condition, it would be flawed to conclude that this particular height reached is due to an existing critical layer in the atmosphere that forces the ball back to the ground. Obviously, this reverse motion is caused by the presence of a gravitational force and not by some atmospheric phenomenon.

It is clear from this analogon that \cite{Hau16} does not identify his $\mathfrak{g}_2$-induced critical layer as a physically absorbing and non-penetrable layer in the fluid, but rather as the result of a bounded motion from a particularly arranged initial condition under a shear force. The $\mathfrak{g}_1$-induced critical layer, however, as correctly examined and discussed by \cite{Campos99}, is unfortunately misunderstood by \cite{Hau16} as such an absorbing and impenetrable fluid layer in physical space. According to \cite{Hau16}, the $\mathfrak{g}_1$-approach is a ``non-optimal approach" that bears the risk of generating incorrect and misleading results,
as now in the case of the critical layer which might be even compared to the well-known misconception of the ``Prandtl-Glauert singularity" [p.$\,$123] in aerodynamics. In contrast to the $\mathfrak{g}_2$-approach, which according to \cite{Hau16}, is the ``optimal approach" that allows to ``comprehend the phenomena of the wave over-reflection and the `critical layer'" in a correct way. But this conclusion of \cite{Hau16} is crucially misleading and thus incorrect for two reasons: (i) The genuine critical layer induced by the~$\mathfrak{g}_1$-approach simply cannot be compared to the ``critical layer" as defined by \cite{Hau16} via his preferred $\mathfrak{g}_2$-approach. They live in different, complementary spaces and thus cannot be compared. The~$\mathfrak{g}_1$-induced critical layer lives in a spectral Laplace-Fourier (complex) space representing a true singularity \citep{Campos99}, while the $\mathfrak{g}_2$-induced critical layer defined by \cite{Hau16} lives in a physical (real) space representing no singularity. (ii) As thoroughly explained and discussed in Section \ref{S3}, the $\mathfrak{g}_1$-approach is a non-redundant and thus a truly complementary approach to~$\mathfrak{g}_2$. It is misleading to call the former a ``non-optimal" and the latter a ``optimal" approach, because both give different information about the system's dynamics: The information obtained by the $\mathfrak{g}_1$-approach\footnote[2]{It is clear that the $\mathfrak{g}_1$-approach must also include the analysis of pseudo-spectra \citep{Trefethen05} in order to offer a complete modal description. But also this complete description does not change the fact that the $\mathfrak{g}_1$-induced singular critical layer exists.} is complementary to that of the $\mathfrak{g}_2$-approach, and should therefore not be played off against each other. A complete picture of the system's dynamics is given when viewing both approaches together.

To substantiate the two statements just made, a simple demonstrating example will be helpful: Let's consider as a {\it general} solution\footnote[3]{The function \eqref{171201:1233} is not a particular solution of the governing equations \eqref{171123:0928} considered in this section. But this circumstance is also irrelevant for the purpose to be demonstrated here. The function \eqref{171201:1233} should rather be regarded as a particular result of some devised equations that admit the symmetry algebras
$\mathfrak{g}_1$ and $\mathfrak{g}_2$.} in physical space the following normalized density field decaying sufficiently fast in time and space infinity
\begin{equation}
\rho(t,x,y)=\kappa e^{-at-b((x-Ayt)^2+y^2)}(y-ct)^2,\;\;\: \text{$t\geq 0,\,$ $a>0,\,$ $b>0$},\label{171201:1233}
\end{equation}
which represents a propagating density-field distribution localized about its center-of-mass, sweeping out in the $(x,y)$-plane the following smooth trajectory
\begin{equation}
\left.
\begin{aligned}
x(t)=\frac{\int_{-\infty}^\infty\! dxdy\,x\, \rho(t,x,y)}{\int_{-\infty}^\infty\! dxdy\, \rho(t,x,y)}&=-\frac{2Act^2}{1+2bc^2t^2},\\[0.75em]
y(t)=\frac{\int_{-\infty}^\infty\! dxdy\,y\, \rho(t,x,y)}{\int_{-\infty}^\infty\! dxdy\, \rho(t,x,y)}&=-\frac{2ct}{1+2bc^2t^2}.\label{171202:1000}
\end{aligned}
~~~\right\}
\end{equation}
The four parameters involved carry the physical dimensions $[a]=1/T$, $[b]=1/L^2$ and $[c]=L/T$, and $[\kappa]=1/L^2$. Now, in the $\mathfrak{g}_1$-approach (see Sec.$\,$\ref{S31}), the above general solution \eqref{171201:1233} would have the (non-local) Laplace-Fourier representation \eqref{170714:2338}
\begin{align}
\hat{\rho}^{(\text{m})}(\omega,k_x,y)&=\int_0^\infty \!dt\int_{-\infty}^\infty \frac{dx}{2\pi}\, \rho(t,x,y)\, e^{-(\omega t+ik_x x)},\;\;\: \text{$\omega\in\mathbb{C},\,$ $k_x\in\mathbb{R}$}\nonumber\\
&= \frac{\kappa e^{-\frac{k_x^2}{4b}-by^2}}{(\omega+a+iAk_x y)^3}\cdot\frac{y^2\big(\omega+a+iAk_x y\big)^2-2cy\big(\omega+a+iAk_x y\big)+2c^2}{2\sqrt{\pi b}},
\end{align}
showing a singularity and thus a genuine critical layer at $y=i(\omega+a)/(Ak_x)\in\mathbb{R}$, which, up to an offset in the complex-valued frequency $\omega$, is of the same structure as the critical layer discussed in \cite{Campos99}. In the $\mathfrak{g}_2$-approach (see Sec.$\,$\ref{S32}), however, the above general solution \eqref{171201:1233} would have the following (non-local) 2-D-Fourier representation \eqref{171122:1116}
\begin{align}
\hat{\rho}^{\text{(k)}}(t,k_x,k_y)\overset{\hat{\mathsf{K}}^{-1}}{\underset{\eqref{171124:0915}}=}\,\hat{\tilde{\rho}}^{(\text{k})}(\tilde{t},k_{\tilde{x}},k_{\tilde{y}})
&= \int_{-\infty}^\infty \frac{d\tilde{x}}{2\pi}\frac{d\tilde{y}}{2\pi}\, \tilde{\rho}(\tilde{t},\tilde{x},\tilde{y})
\, e^{-i(k_{\tilde{x}} \tilde{x}+k_{\tilde{y}} \tilde{y})},\;\;\: \text{$k_{\tilde{x}}\in\mathbb{R},\,$ $k_{\tilde{y}}\in\mathbb{R}$}\label{171201:1237}\\[0.5em]
&=\int_{-\infty}^\infty \frac{d\tilde{x}}{2\pi}\frac{d\tilde{y}}{2\pi}\, \kappa e^{-a\tilde{t}-b(\tilde{x}^2+\tilde{y}^2)}(\tilde{y}-c\tilde{t})^2
\, e^{-i(k_{\tilde{x}} \tilde{x}+k_{\tilde{y}} \tilde{y})}\nonumber\\[0.5em]
&=\: \kappa e^{-a\tilde{t}-\frac{k_{\tilde{x}}^2+k_{\tilde{y}}^2}{4b}}\cdot\frac{4b^2c^2\tilde{t}^{\,2}+2b\big(1+2ic\tilde{t}k_{\tilde{y}}\big)-k_{\tilde{y}}^2}{16\pi b^3}\nonumber\\[0.5em]
&\hspace{-3.1cm}\overset{\hat{\mathsf{K}}}{\underset{\eqref{171124:0914}}=}\kappa e^{-at-\frac{k_x^2+(k_y+Atk_x)^2}{4b}}\cdot\frac{4b^2c^2t^2+2b\big(1+2ict\big(k_y+Atk_x\big)\big)-\big(k_y+Atk_x\big)^2}{16\pi b^3},\nonumber
\end{align}
showing {\it no} singularity and thus {\it no} genuine critical layer. Hence, the {\it general} solution \eqref{171201:1233} may face a singularity or no singularity, depending on how it is represented and in which space it is living. According to \cite{Hau16}, the ``critical layer" in \eqref{171201:1233} would be defined from a particular initial value $(k_{x_0},k_{y_0})$ of the 2-D spectral variable $(k_x,k_y)$ in the  $\mathfrak{g}_2$-representation~\eqref{171201:1237}, as that particular $y$-value that the propagating distribution \eqref{171201:1233} reaches when passing the critical time $t=t^*$, which is defined when $k_{\tilde{y}_0}=k_{y_0}+At^*k_{x_0}=0$, i.e., when $t^*=-k_{y_0}/(Ak_{x_0})>0$ is reached. Now, when specifically choosing the initial values as $k_{x_0}=\sqrt{2b}$ and $k_{y_0}=A/c$, then the critical time $t^*$ defines a maximum value $y=y^*$ of the trajectory \eqref{171202:1000} in the shearwise direction. This maximum value $y^*(t^*)=1/\sqrt{2b}=1/k_{x_0}$ is then defined in \cite{Hau16} as the ``critical layer". It is obvious now, that this specifically prepared (non-singular) ``critical layer" $y^*=1/k_{x_0}$ obtained in the $\mathfrak{g}_2$-approach can and may not be compared to the genuine (singular) critical layer $y_c=i(\omega+a)/(Ak_x)$ obtained in the $\mathfrak{g}_1$-approach, as has been misleadingly done in \cite{Hau16}.
\label{171205:2313}

%
% Physical Solution %%%%%%%%%%%%%%%%%%%%%%%%%%%%%%%%%%%%%%%%%%%%%%%%%%%%%%%%%%%%%%%%%%%%%%%%%%%%%%%%%%%%%%%%%%%%%%%%%%%%%%%
%
\newpage

\newpage

\phantom{x}\newpage

\newpage
%
%%%%%%%%%%%%%%%%%%%%%%%%%%%%%%%%%%%%%%%%%%%%%%%%%%%%%%%%%%%%%%%%%%%%%%%%%%%%%%%%%%%%%%%%%%%%%%%%%%%%%%%%%
%

\section{Derivation of the Kelvin transformation in spectral space\label{SD}}

Given is the Kelvin transformation $\mathsf{K}$ \eqref{171123:1839} in real physical space
\begin{equation}
\phantom{\mathsf{K}^{-1}}\mathsf{K}\!:\quad\; \tilde{t}=t,\quad\; \tilde{x}=x-Ayt,\quad\; \tilde{y}=y,\quad\; \tilde{\psi}^{\{u,v,\rho\}}=\psi^{\{u,v,\rho\}},\label{171124:0820}
\end{equation}
with its inverse
\begin{equation}
\phantom{\mathsf{K}}\mathsf{K}^{-1}\!:\quad\; t=\tilde{t},\quad\; x=\tilde{x}+A\tilde{y}\tilde{t},\quad\; y=\tilde{y},\quad\; \psi^{\{u,v,\rho\}}=\tilde{\psi}^{\{u,v,\rho\}},\label{171124:0821}
\end{equation}
transforming the system of an unbounded linear shear flow from the inertial into the co-moving (accelerating) Kelvin frame. Given is also the 2-D Fourier transform \eqref{171123:1640} in the inertial frame
\begin{equation}
\psi^{\{u,v,\rho\}}(t,x,y)
= \int_{-\infty}^\infty dk_{x}dk_{y}\, \hat{\psi}^{\{u,v,\rho\}}(t,k_{x},k_{y})
\, e^{i(k_x x+k_y y)},\label{171124:0830}
\end{equation}
where $\hat{\psi}^{\{u,v,\rho\}}$ are the complementary (Fourier) fields to $\psi^{\{u,v,\rho\}}$ for the three dynamical fields $u$, $v$ and $\rho$, respectively. The inverse of \eqref{171124:0830} is given by
\begin{equation}
\hat{\psi}^{\{u,v,\rho\}}(t,k_x,k_y)
= \int_{-\infty}^\infty \frac{dx}{2\pi}\frac{dy}{2\pi}\, \psi^{\{u,v,\rho\}}(t,x,y)
\, e^{-i(k_x x+k_y y)}.\label{171124:0839}
\end{equation}
Now, let's assume that in the new Kelvin frame a 2-D Fourier transform also exists, i.e., let's assume the existence of the following invertible integral transformation for each of the three fields in this new frame
\begin{equation}
\hat{\tilde{\psi}}^{\{u,v,\rho\}}(\tilde{t},k_{\tilde{x}},k_{\tilde{y}})
= \int_{-\infty}^\infty \frac{d\tilde{x}}{2\pi}\frac{d\tilde{y}}{2\pi}\, \tilde{\psi}^{\{u,v,\rho\}}(\tilde{t},\tilde{x},\tilde{y})
\, e^{-i(k_{\tilde{x}}\tilde{x}+k_{\tilde{y}} \tilde{y})},\label{171124:0847}
\end{equation}
then transforming this relation according to $\mathsf{K}$ \eqref{171124:0820} will result to
\begin{align}
\hat{\tilde{\psi}}^{\{u,v,\rho\}}(\tilde{t},k_{\tilde{x}},k_{\tilde{y}})&=\int_{-\infty}^\infty \frac{d\tilde{x}}{2\pi}\frac{d\tilde{y}}{2\pi}
\tilde{\psi}^{\{u,v,\rho\}}(\tilde{t},\tilde{x},\tilde{y})e^{-i(k_{\tilde{x}} \tilde{x} +k_{\tilde{y}} \tilde{y})}
\nonumber\\[0.5em]
& = \int_{-\infty}^\infty |J|\, \frac{dx}{2\pi}\frac{dy}{2\pi}
\psi^{\{u,v,\rho\}}(t,x,y)e^{-i\big(k_{\tilde{x}}(x-Ayt) +k_{\tilde{y}}y\big)},\label{171124:0858}
\end{align}
where $J$ is the 2-D Jacobian, which for transformation \eqref{171124:0820} takes the value $|J|=1$. The only possible way now to restore the already existing Fourier-transform \eqref{171124:0839} in the inertial frame again,
is to define the spectral coordinate transformations as
$k_{\tilde{x}}=k_x$ and $k_{\tilde{y}}=k_y+Atk_x$, which then defines, according to \eqref{171124:0858}, the transformation rule for the new spectral fields as
\begin{align}
\hat{\tilde{\psi}}^{\{u,v,\rho\}}(\tilde{t},k_{\tilde{x}},k_{\tilde{y}})&=\int_{-\infty}^\infty |J|\, \frac{dx}{2\pi}\frac{dy}{2\pi}
\psi^{\{u,v,\rho\}}(t,x,y)e^{-i\big(k_x(x-Ayt) +(k_y+Atk_x)y\big)}
\nonumber\\[0.5em]
& =\int_{-\infty}^\infty \frac{dx}{2\pi}\frac{dy}{2\pi}
\psi^{\{u,v,\rho\}}(t,x,y)e^{-i(k_x x +k_y y)}\: \equiv\: \hat{\psi}^{\{u,v,\rho\}}(t,k_x,k_y).\label{170624:0906}
\end{align}
Hence, the to \eqref{171124:0820} complementary Kelvin transformation in spectral space is thus given as
\begin{equation}
\phantom{\mathsf{K}^{-1}}\hat{\mathsf{K}}\!:\quad\; \tilde{t}=t,\quad\; k_{\tilde{x}}=k_x,\quad\; k_{\tilde{y}}=k_y+Atk_x,\quad\;
\hat{\tilde{\psi}}^{\{u,v,\rho\}}=\hat{\psi}^{\{u,v,\rho\}},\label{171124:0914}
\end{equation}
with its inverse
\begin{equation}
\phantom{\mathsf{K}}\hat{\mathsf{K}}^{-1}\!:\quad\; t=\tilde{t},\quad\; k_x=k_{\tilde{x}},\quad\; k_y=k_{\tilde{y}}-A\tilde{t}k_{\tilde{x}},\quad\;
\hat{\psi}^{\{u,v,\rho\}}=\hat{\tilde{\psi}}^{\{u,v,\rho\}},\label{171124:0915}
\end{equation}
complementary to \eqref{171124:0821}. Worthwhile to note here is that the untransformed spectral coordinates $(k_x,k_y)$ are independent of time $t$, and the transformed ones $(k_{\tilde{x}},k_{\tilde{y}})$ independent of~$\tilde{t}$, that is, $\partial_tk_x=\partial_t k_y=0$ and $\partial_{\tilde{t}}k_{\tilde{x}}=\partial_{\tilde{t}}k_{\tilde{y}}= 0$, which is obvious, of course, since the wavenumber variables $(k_x,k_y)$ and $(k_{\tilde{x}},k_{\tilde{y}})$ are just the complementary variables to the spatial variables $(x,y)$ and $(\tilde{x}, \tilde{y})$, respectively. However,
$k_{\tilde{y}}$ is {\it not} independent of time $t$, and $k_y$ {\it not} independent of~$\tilde{t}$, although $\tilde{t}=t$ in \eqref{171124:0914}-\eqref{171124:0915}, that is, $\partial_t k_{\tilde{y}}\neq 0$ and
$\partial_{\tilde{t}}k_y\neq 0$, except for the specific case when~$k_{\tilde{x}}=k_x=0$.

\newpage

\bibliographystyle{jfm}
\bibliography{BibRef}

\end{document}